\documentclass{tlp}

\usepackage{aopmath,amsfonts}
\usepackage{latexsym}
\usepackage{epsfig}
\usepackage{picinpar}
\usepackage{url}

\newcommand{\osc}{$\cal OSC\:\:\:$}

\submitted{19 April 2005}
\revised{21 October 2005}
\accepted{3 July 2006}

\title[PALS: Efficient Or-Parallelism on Beowulf Clusters]{PALS: Efficient Or-Parallel Execution of Prolog on 
	Beowulf Clusters}

\author[E. Pontelli, K. Villaverde, H. Guo, G. Gupta]{
	ENRICO PONTELLI,  KAREN VILLAVERDE\\
	Department of Computer Science\\
	New Mexico State University\\
	\email{epontell | kvillave@cs.nmsu.edu}\\
	\and
	HAI-FENG GUO\\
	Department of Computer Science\\
	University of Nebraska at Omaha\\
	\email{haifengguo@mail.unomaha.edu}\\
	\and
	GOPAL GUPTA\\
	Department of Computer Science\\
	University of Texas at Dallas\\
	\email{gupta@utdallas.edu}}

\begin{document}
\renewcommand{\textfraction}{0.0001}

\maketitle

\begin{abstract}
This paper describes the development of the
\emph{PALS} system, an implementation of Prolog capable of efficiently
exploiting or-parallelism on \emph{distributed-memory}
platforms---specifically Beowulf clusters. PALS makes use of a novel
technique, called \emph{incremental stack-splitting}. The technique
proposed builds on the stack-splitting approach, previously described
by the authors and experimentally validated on shared-memory systems,
which in turn is an evolution of the
stack-copying method used in a variety of parallel logic and
constraint systems---e.g., MUSE, YAP, and Penny.
The PALS system is the first
distributed or-parallel implementation of Prolog based on the
stack-splitting method ever realized. The results presented confirm
the superiority of this method as a simple yet effective technique to
transition from shared-memory to distributed-memory systems. PALS
extends stack-splitting by combining it with incremental copying; the
paper provides a description of the implementation of PALS, including
details of how distributed scheduling is handled.  We also investigate
methodologies to effectively support order-sensitive predicates (e.g.,
side-effects) in the context of the stack-splitting scheme.
Experimental results obtained from running PALS on both Shared Memory
and Beowulf systems are presented and analyzed.
\end{abstract}


\section{Introduction}

The literature on parallel logic programming
(see  \cite{codognet-survey,parallellp-survey} for a general
discussion of parallel logic programming) underscores
the potential for achieving excellent speedups and
performance improvements from execution of logic
programs on parallel architectures, with little or
no programmer intervention. Particular attention has
been devoted over the years to the design of technology
for supporting \emph{or-parallel} execution of Prolog
programs on shared-memory architectures.

Or-parallelism (OP) arises from the non-determinism
implicit in the process of reducing a given subgoal
using different clauses of the program.
The non-determinism arising during the execution of
a logic program is commonly depicted in the form of a
\emph{search tree} (a.k.a.\ \emph{or-tree}). Each internal node represents
a \emph{choice-point}, i.e.,
an execution point where multiple clauses are available to reduce the
selected subgoal. Leaves of the tree represent either failure points
(i.e., resolvents where the selected subgoal does not have a
matching clause) or success
points (i.e., solutions to the initial goal).
A sequential computation boils down to
traversal of this search tree according to some predefined search
strategy---e.g., Prolog  adopts a fixed strategy based on a left-to-right,
depth-first traversal of the search tree.

While in a sequential 
execution the multiple clauses that match a subgoal are explored one at a time
via backtracking, in or-parallel execution we allow different instances
of Prolog engines (\emph{computing agents})---executing as separate processes---to 
concurrently explore these alternative clauses.
Different agents
concurrently operate on different branches of the or-tree, each attempting to
derive a solution to the original goal using a different sequence of derivation
steps. 
In this work we will focus on or-parallel systems derived from the 
multi-sequential model originally proposed by D.H.D. Warren \cite{warren-sri}. In
this model, the multiple agents traverse the or-tree looking
for unexplored branches. If an unexplored branch (i.e.,
an untried clause to resolve a selected subgoal) is found, the agent picks
it up and begins execution. This agent will stop either
if it fails (reaches a failing leaf), or if it finds a solution. In case
of failure, or if the solution found is not acceptable to the user,
the agent will {\it backtrack}, i.e., move back up in the
tree, looking for other choice-points with untried alternatives
to explore. The agents need  to synchronize if they
access the same node in the tree---to avoid repetition of
computations. In the rest of this work we will call \emph{parallel
choice-points} those choice-points from which we allow exploitation
of parallelism.

Intuitively, or-parallelism
allows concurrent search for solution(s)
to the original goal. The importance of the research
on efficient techniques for handling or-parallelism
arises from the generality of the problem---technology
originally developed for parallel execution of Prolog
programs has found application in contexts such
as  constraint programming (e.g.,
\cite{schu00,per99}) and non-monotonic reasoning (e.g.,
\cite{stab-kentucky,stable-padl}).
Efficient
implementation of or-parallelism has also 
been extensively investigated
in the context of AI systems \cite{kumar,sahni}.

In sequential implementations of search-based AI systems or Prolog,
typically one branch of the tree resides on the inference
engine's stacks at any given
time. This simplifies implementation quite significantly. However,
in case of parallel systems, multiple branches of the tree co-exist
at the same time, making parallel implementation quite complex.
Efficient management of these co-existing branches is quite a difficult
problem, and it is referred to as the {\it environment management
problem} \cite{gj:crit}.

Most research in or-parallel execution of Prolog
so far has focused on techniques aimed
at \emph{shared-memory multiprocessors (SMPs)}.
Relatively fewer efforts \cite{foong,araujo,daos,opera,benjumea-or,Dorpp}
have been devoted to implementing Prolog
systems on \emph{distributed-memory platforms (DMPs)}. 
Out of these efforts only a  small number
have been implemented as working prototypes, and even fewer
have produced acceptable speedups. 
Existing techniques developed for SMPs are inadequate for
the needs of DMP platforms. In fact, most 
implementation methods require sharing of data and control
stacks in a SMP context to allow for synchronization between
agents with minimal communication.
Even in those models, such as
\emph{stack copying} \cite{muse-journal}, where the
different agents maintain independent copies of the
various stacks (i.e., they do not physically share them), 
the requirement of sharing part of the control structure
is still present. For example,  in the MUSE implementation
of stack copying,  parts of each choice-point are 
maintained in a shared data structure,  to ensure that the agents reproduce
the same observable behavior as in a sequential execution (e.g.,
they do not duplicate computations already performed by another
agent). In the case of recomputation-based methods
\cite{delphi,delphi1}, the sharing appears in the form of
the use of a centralized controller (as in the Delphi model)
to handle the communication of the different branches of the
tree to the computation agents. 
The presence of these forms of sharing are believed to
lead to degradation of performance of these schemes on a 
distributed
memory platform, as the lack of shared memory imposes the
need for explicit communication between agents.

Experimental \cite{muse-journal} and theoretical studies \cite{op-ngc} have
demonstrated that \emph{stack-copying}, and in particular
\emph{incremental} stack-copying, is one of the most
effective implementation techniques devised for exploiting
or-parallelism.
Stack-copying allows sharing of work between parallel  agents by
copying the
state of one agent (which owns unexploited tasks) 
to another agent (which is currently idle).
The idea of \emph{incremental} stack-copying is to only copy the
\emph{difference}
between the state of two  agents, instead of 
copying the entire state each time.
Incremental stack-copying has been used to implement
or-parallel Prolog efficiently
in a variety of systems (e.g., MUSE \cite{muse-journal}, YAP \cite{YapOr},
Penny \cite{par-penny}),
as well as to exploit parallelism from 
non-monotonic reasoning systems
 \cite{stable-padl,stab-kentucky}.

In order to improve the performance of stack-copying and allow its
efficient implementation on DMPs,
we propose a new technique,
called \emph{stack-splitting} \cite{iclp99,iclp01}. Stack-splitting
is a variation of stack-copying, aimed at solving the environment
management problem and  improving stack-copying by 
reducing the need for communication between agents during the execution
of work. This is accomplished by making use of strategies that 
distribute the work available in a branch of the search tree
between two processors during each scheduling operation.
In this paper,
we describe stack-splitting in detail, and provide results
from the first ever concrete implementation of stack-splitting
on both shared-memory multiprocessors (SMPs) and  distributed-memory
multiprocessors (DMPs)---specifically, a Pentium-based Beowulf---along 
with a novel scheme to combine incremental copying with stack-splitting on DMPs.
The \emph{incremental stack-splitting} scheme is based on a procedure which
labels parallel choice-points and then compares the labels to determine the
fragments of data and control areas that need to be exchanged between
agents.
We also describe  scheduling schemes suitable for
our incremental stack-splitting scheme and variations of 
stack-splitting providing efficient handling of order-sensitive
predicates (e.g., side-effects).  Both the incremental
stack-splitting and the
scheduling schemes described have been 
 implemented in the \emph{PALS} system,
a message-passing or-parallel implementation of Prolog. In this
paper we present  performance results obtained from this
implementation. To our knowledge, PALS is the first ever or-parallel
implementation of Prolog realized on a Beowulf architecture
(built from off-the-shelf components). The techniques have already
been embraced by other developers of parallel Prolog systems
\cite{fernando}.
The techniques we propose
are also immediately applicable to other systems based on similar  underlying
models, e.g.,  non-monotonic
reasoning \cite{stable-padl} systems. Indeed, a distributed implementation 
of \emph{answer set programming} based on incremental stack splitting
has been reported in \cite{padl-stable}---note that the execution model of
answer set programming relies on a search-tree exploration (built using
Davis-Putnam's procedure) and is \emph{not} a straightforward Prolog
implementation.

The contributions of this paper can be summarized as follows:
\begin{itemize}
\item design of a novel methodology---stack splitting---to efficiently
	support or-parallelism on distributed memory systems;
\item enhancement of the methodology to support incremental copying
	behavior;
\item investigation of different splitting modalities, in particular,
	to facilitate the handling of side-effects;
\item implementation of these methodologies in an industrial-strength
	Prolog system (ALS Prolog) and evaluation of its performance.
\end{itemize}

In the rest of this work we will focus on the execution of Prolog
programs (unless explicitly stated otherwise); this means that we will assume
that programs are executed according to the computation and selection rules
of Prolog. We will also frequently use the term \emph{observable semantics} to
indicate the overall observable behavior of an execution---i.e., the order
in which all visible activities of a program execution take place (order of
input/output, order in which solutions are obtained, etc.). If a 
parallel computation respects the observable Prolog semantics, then this 
means that the user does not see any difference between such computation 
and a sequential Prolog execution of the same program---except
for improved performance. Our goal in this work is to develop parallel 
execution models that properly reproduce Prolog's observable semantics and
are still able to guarantee improved performance.

\subsection{Related Work}
A rich body of research has been developed to investigate 
methodologies for the exploitation of or-parallelism from
Prolog executions on SMPs. Comprehensive surveys describing
and comparing these methodologies have appeared, e.g.,
\cite{gupta-phd,parallellp-survey,codognet-survey}. 

A theoretical analysis of the properties of different methodologies
has been presented in~\cite{op-ngc,op-complang}. These works 
provide an abstraction of the environment representation problem
as a data structure problem on dynamic trees. These studies identify
the presence of unavoidable overheads in the dynamic management
of environments in a parallel setting, and recognize methods 
with constant-time environment creation and access as optimal 
methods for environment representation. Methods such as
stack-copying \cite{muse-journal}, binding arrays
\cite{aurora}, and recomputation \cite{delphi} meet such
requirements.

Distributed implementations
of Prolog have been proposed by several researchers
\cite{foong,araujo,daos}.
However, none of these systems are very effective in producing speedups
over a wide range of benchmarks. 
Foong's system \cite{foong} and Castro et al's 
system \cite{daos} are based directly on 
stack-copying and generate 
communication overhead due to the shared choice-points (no
real implementation exist for the two of them).  Araujo's
system uses recomputation \cite{delphi} rather than stack-copying. Using 
recomputation for maintaining multiple environments 
is inherently inferior to stack-copying. The stack frames that
are copied in the stack-copying technique 
capture the effect of a computation. In the recomputation technique
these stack-frames are reproduced by re-running the computation.
A computation may run for hours and 
yet produce only a single stack frame (e.g.,
a tail-recursive computation). 
Distributed implementations of Prolog
have been developed on Transputer systems (The Opera System \cite{opera}
and the system of Benjumea and Troya
\cite{benjumea-or}). Of these, Benjumea's system
has produced quite good results. However, both the Opera system and
the Benjumea's system have
been developed on now-obsolete Transputer hardware,
and, additionally, both rely on a stack-copying mechanism which will
produce poor performance in programs where the task-granularity is
small. 
A different
approach has been suggested by Silva and Watson with their
DORPP model \cite{Dorpp}, which extends the binding array
scheme \cite{aurora} to a distributed setting, relying on the
\emph{European Declarative System (EDS)} platform to 
support distributed computation (EDS provides a limited
form of distributed shared memory); good results have been
presented running DORPP on an EDS simulator. 

%Gopal: 
%A share-nothing platform implementation of Prolog based
%on stack-splitting is currently in progress.
Finally, 
the idea of stack-splitting bears some similarities
with some of the loop transformation techniques which are commonly
adopted for parallelization of imperative programming languages,
such as loop fission, loop tiling, and index set splitting \cite{wolfe-book}.

\subsection{Paper Organization}

The rest of the paper is organized as follows. Section \ref{lab-op} provides
an overview of the main issues related to or-parallel execution of Prolog.
Section \ref{extension} describes the stack-splitting scheme, while Section 
\ref{ssimpl} describes its implementation. Section \ref{sched} analyzes the
problem of guaranteeing efficient distribution of work between idle
agents. 
Section \ref{sideff}
describes how stack-splitting can be adapted to provide efficient
handling of order-sensitive predicates of Prolog (e.g., control constructs,
side-effects).
Section \ref{results} analyzes the result obtained from the
prototype implementation in the PALS system.
Section \ref{costs} offers a general discussion about possible
optimizations of the implementation of stack-splitting.  
%Section \ref{andpp} discusses possible uses of stack-splitting to
%handle and-parallelism.
Finally, Section \ref{concl}
provides conclusions and directions for future research.

The reader is assumed to be familiar with the basic notions of 
logic programming, Prolog, and its execution model (e.g., a basic
understanding of the Warren Abstract Machine) \cite{lloyd,apt-book,kaci}.

\section{Or-Parallelism}
\label{lab-op}

In this section, we survey the main issues related to the exploitation
of or-parallelism from Prolog programs, and we discuss the main ideas behind
the stack-copying method.

\subsection{Foundations of Or-Parallelism}

Parallelization of logic programs can be seen as a direct consequence
of Kowalski's principle \cite{Kowalski79}\\
\centerline{\em Algorithm = Logic + Control}
Program development separates the control component from the logical
specification of the problem, thus making the two
orthogonal.  The lack (or, at least, the
 limited presence)
of knowledge about control 
in the program allows the run-time systems to adopt
different execution strategies without affecting the declarative meaning
of the program (i.e., the set of logical consequences of the 
program). The same is true  of
search-based systems, where the order of exploration of the branches
of the search-tree is flexible (within the limits imposed by the
semantics of the search strategy---e.g., search heuristics).

\begin{figure}[tp]
\hrule
\centerline{\psfig{figure=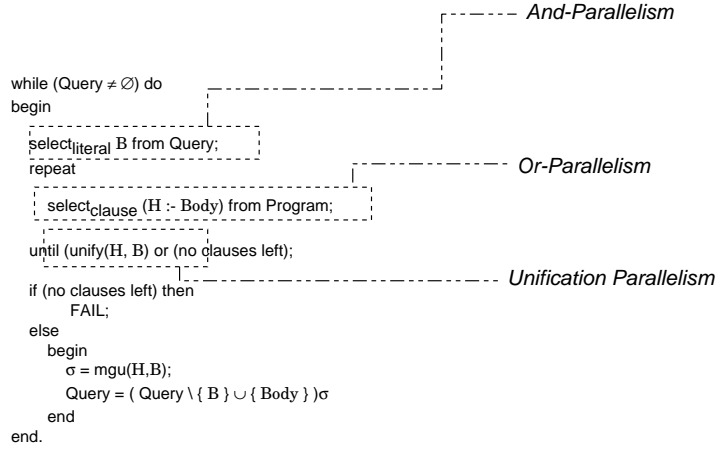,width=.75\textwidth}}
\hrule
\caption{Operational Semantics and Non-determinism}
\label{interpreter}
\end{figure}

Apart from the separation between logic and control,
from a programming languages perspective, logic programming offers two key
features which make exploitation of parallelism  more practical than in 
traditional imperative languages:
\begin{enumerate}
\item From an operational perspective,
	logic programming languages are 
	\emph{single-assignment} languages; variables
	are mathematical entities which can be assigned a value
	at most once during
	each derivation (i.e., along each branch of the or-tree)---this 
	relieves a parallel system from having to keep 
	track of complex flow dependencies such as those needed in parallelization of
	traditional programming languages \cite{zima}.
\item The operational semantics of logic programming, unlike imperative
	languages, makes substantial
	 use of \emph{non-determinism}---i.e., the operational semantics relies
	on the automatic exploration of a search tree. The alternative possible
	choices performed during such exploration (\emph{points of non-determinism})
	can be easily converted into parallelism without radically modifying
	the overall operational semantics. Furthermore, control in most
	logic programming languages is largely implicit, thus limiting
	programmers' influence on the development of the flow of execution.
\end{enumerate}
The second point is of particular importance: the ability to convert existing
non-determinism (and other ``choices'' performed during execution, such as
the choice of the subgoal to resolve)
into parallelism leads to
 the possibility of extracting parallelism
directly from the execution model, without requiring the programmer to 
perform any modifications of the original program and without requiring the
introduction of ad-hoc parallelization constructs in the source language
(\emph{implicit parallelization}).
The typical strategy adopted in the development of parallel logic
programming systems has been based on the translation of one
(or more) of the choices present in
the operational semantics (see Figure~\ref{interpreter})
 into parallel computations. This leads to 
the three ``classical'' forms of parallelism \cite{Coki81}:
\begin{itemize}
\item \emph{And-Parallelism}, which originates from parallelizing
        the selection of the next literal to be solved---thus allowing
        multiple literals to be solved concurrently. This can be
	visualized by imagining the operation {\tt select$_{\texttt{literal}}$}
	to return multiple literals that are concurrently processed
	by the rest of the algorithm.
\item \emph{Or-Parallelism}, which originates from parallelizing the
        selection of the clause to be used in the computation of the
        resolvent---thus allowing multiple clauses to be tried in parallel.
	This can be visualized by having the 
	{\tt select$_{\texttt{clause}}$} operation to select multiple
	clauses that are concurrently processed by the rest of the algorithm
\item \emph{Unification Parallelism}, which arises from the parallelization of
        the unification process.\footnote{By $mgu(a,b)$, in the Figure, we denote the
	 most general unifier of $a$ and $b$.}
\end{itemize}

Or-Parallelism originates from the parallelization of the
{\tt select$_{\texttt{clause}}$} phase in Figure~\ref{interpreter}.
Thus, {\it or-parallelism} arises when more than one rule defines 
a relation and a subgoal
unifies with more than one rule head---the corresponding bodies
can then be executed in parallel with each other, 
giving rise to or-parallelism.  
Or-parallelism is thus a way of efficiently searching
for solutions to the query, by exploring in parallel the
search space generated by the presence of multiple clauses
applicable at each resolution step. Observe that each parallel computation
is attempting to compute  a distinct solution to the original goal.

For example, consider the following simple logic program:

{\tt 
\begin{verbatim}

    f :- t(X, three), p(Y), q(Y).
    p(L) :- s(L, M), t(M, L).
    p(K) :- r(K).
    q(one).
    q(two).
    r(one).
    r(three).
    s(two, three).
    s(four, five).
    t(three, three).
    t(three, two).
\end{verbatim}
}

\noindent 
and the query {\tt ?- f.}
The calls to {\tt p}, {\tt s}, and
{\tt r} are non-deterministic and lead
to the creation of choice-points---while
the calls to
 {\tt t}, {\tt f}, and {\tt q} are deterministic. 
The multiple alternatives in these choice-points can be executed
in parallel. 

\begin{figure}[tp]
\hrule
\smallskip
\centerline{
\psfig{figure=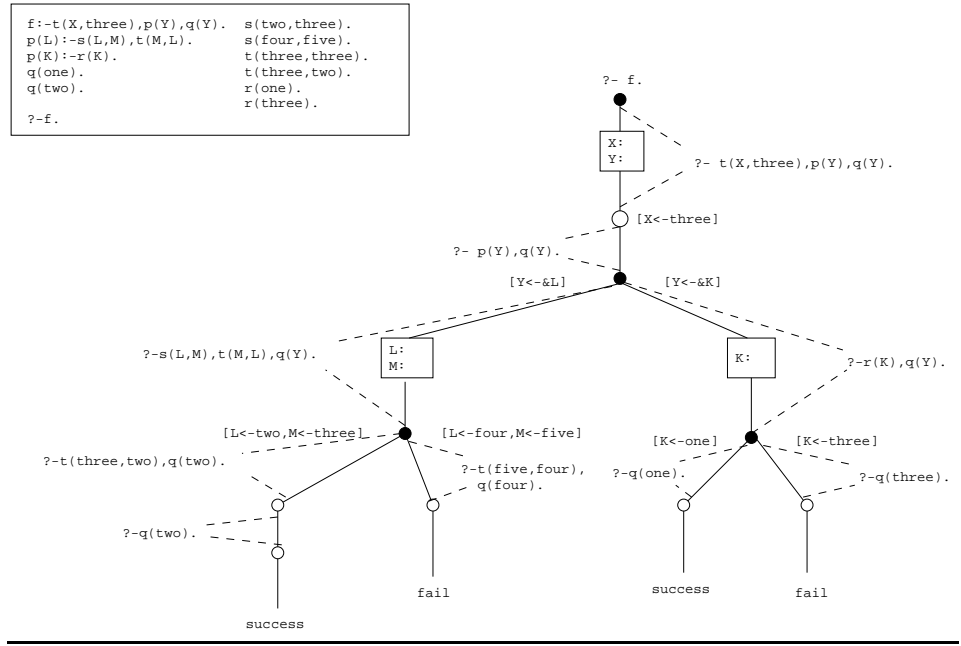,width=\textwidth}
}
\smallskip
\hrule
\caption{An Or-Tree}
\label{ortree}
\end{figure}

\noindent
A convenient way to visualize or-parallelism is through
the {\it or-tree}. 
Informally, an or-tree (sometimes referred to also
as \emph{search tree})
for a query $Q$ and logic program $LP$ is a tree of nodes, each with
an associated {\it goal-list}, such that:

\begin{enumerate}
\item The root node of the tree has $Q$ as its associated
        goal-list;

\item Each internal  node $n$ is created as a result of
        successful unification of the first goal in (the goal-list of)
        $n$'s parent node with the head of a 
        clause in $LP$, $$H \texttt{:-} B_1, B_2, \ldots, B_n$$ 
        The goal-list of node $n$ is
        $(B_1, B_2, \ldots, B_n, L_2, \ldots, L_m)\theta$,
        if the goal-list of the parent of $n$ is
        $L_1, L_2, \ldots, L_m$ and
        $\theta = mgu(H, L_1)$.
\end{enumerate}

\noindent Figure~\ref{ortree} shows the or-tree for the simple
program presented above. For the sake of readability, we have also
annotated the tree with the variables  created and the description
of the bindings performed.\footnote{This figure is for illustration purposes
only; e.g., in a real implementation, the order of variable bindings could
be different.} We have also introduced different notations
(empty nodes and filled nodes) to distinguish deterministic reductions
versus non-deterministic reductions. The boxes represent environments created
for a clause; the dotted lines are used to associate the segment of each
branch to the corresponding resolvent existing during that part of the computation;
variable bindings are indicated next to the node where the binding is computed.

Note that, since we are considering execution
of Prolog programs, the construction of the or-tree will
follow the operational semantics of Prolog---at each node we will
consider clauses applicable to the first subgoal, and the children of
a node will be considered ordered from left to right according to the
order of the corresponding clauses in the program. I.e., during
sequential execution the or-tree of Figure~\ref{ortree} is
built and explored in a left-to-right depth-first manner. 
However, if multiple agents are
available, then multiple branches of the tree can be constructed and
explored
simultaneously---although, as mentioned later, we will aim at still
constructing the same tree, i.e., reproduce the same observable 
semantics as sequential Prolog. Observe also that, if a fragment of a branch of
the or-tree contains multiple choice-points, and this is explored by
a single agent, then the agent will employ traditional backtracking
to search the various alternatives.

Or-parallelism frequently arises in applications that
explore a large search
space via backtracking. This is the typical case in application
areas such as expert systems, scheduling and optimization problems, 
and natural language processing. Or-parallelism
also arises during parallel execution
of deductive database systems \cite{ganguly,dbase}.

\subsection{The Environment Representation Problem}

Despite the theoretical simplicity and results, in practice 
\emph{implementation} of or-parallelism is difficult because keeping
the run-time and parallelism-related overheads small
is non-trivial due to the
practical complications which emerge from the sharing of nodes in the
or-tree. That is, given two nodes in two different
branches of the or-tree, all nodes above (and including)
the least common ancestor node of these two nodes are
shared between the two branches. A variable created in one
of these ancestor nodes might be bound differently in the
two branches. Thus, the environments of the two branches have to
be organized in such a fashion that, in spite of the ancestor
nodes being shared, the correct bindings applicable to each
of the two branches are easily discernible. 

Let us start by introducing some terminology. Whenever a new clause is 
applied to resolve a selected subgoal, an \emph{environment} is 
created. The environment plays a role analogous to that of the
activation record in the implementation of imperative languages---it
stores information to handle the execution of the clause (e.g., return
address) and it provides storage for the local variables introduced
by the clause. The boxes containing variables shown in Figure~\ref{ortree}
can be thought as representing a part of the environment of the clause.

During Prolog execution, variables might receive bindings. If a variable is
created before a choice-point but bound after the choice-point (e.g., variable
{\tt L} in Figure~\ref{ortree})---such a variable is refereed to as a 
\emph{conditional variable} in the literature---then the variable might be bound
differently in each branch of the choice-point. In a sequential execution,
conditional variables are handled using \emph{trailing}: whenever the
conditional variable is bound, the address of the variable is pushed on
a special stack (the \emph{trail stack}). During backtracking, the
content of the trail stack is used to determine which bindings should
be removed, thus clearing up (\emph{untrailing}) 
conditional variables and preparing them
for the new bindings they might receive in the alternative branches
explored. This mechanism allows the use of a single memory location to 
store the value of the variable (since the location can be reused across
different branches of the or-tree, by repeatedly clearing it via untrailing).

More generally, consider a variable {\tt V} in node {\tt n}$_1$, whose
binding {\tt b} has been created in node {\tt n}$_2$. If there are no
choice-points between {\tt n}$_1$ and {\tt n}$_2$, then the variable
{\tt V} will have the binding {\tt b} in every branch that is created
below {\tt n}$_2$. Such a binding can be stored in-place in {\tt
  V}---i.e., it can be directly stored in the memory location
allocated to {\tt V} in {\tt n}$_1$.  However, if there are choice-points
between {\tt n}$_1$ and {\tt n}$_2$, then the binding {\tt b}
cannot be stored in-place, since other branches created between nodes
{\tt n}$_1$ and {\tt n}$_2$ may impart different bindings to {\tt V}.
The binding {\tt b} is applicable to only those nodes that are below
{\tt n}$_2$.  Such a binding to a conditional variable
 is known as a \emph{conditional binding}.
  For example,
variable {\tt Y} in Figure~\ref{ortree} is a conditional variable. A
binding that is not conditional, i.e., one that has no intervening
choice-points between the node where this binding
was generated and the node containing the corresponding variable, is
termed \emph{unconditional}. The corresponding variable is called an
\emph{unconditional variable} (for example, variable {\tt X} in 
Figure~\ref{ortree}).

If the different branches are searched in or-parallel, then 
the conditional variables (e.g.,  variable {\tt  L}) 
receive different bindings in different branches of the tree, all
of which will be active at the same time. Storing and later accessing
these bindings efficiently is a problem.  In sequential execution the
binding of a variable is stored in the memory location allotted to
that variable. Since branches are explored one at a time, and bindings
are untrailed during backtracking, no problems arise.  In parallel
execution, multiple bindings exist at the same time, hence they cannot
be stored in a single memory location allotted to the variable.  This
problem, known as the {\it multiple environment representation
  problem} in the literature, is a major problem in implementing or-parallelism
\cite{op-ngc,parallellp-survey}.

The main problem in implementing
or-parallelism is the efficient representation of the multiple
environments that co-exist simultaneously in the or-tree corresponding to a program's execution---i.e., 
the development 
of an efficient way of associating the 
correct set of bindings to each branch of the 
or-tree.
Note that the main problem
in management of multiple environments is that of efficiently 
representing and accessing the conditional bindings;
the unconditional bindings can be treated as in normal sequential
execution of logic programs (i.e., they can be stored in-place).
The naive approach
of keeping a complete separate copy of the answer substitution
for each separate branch is highly inefficient, since it requires the
creation of complete copies of the substitution (which can be arbitrarily
large) every time a choice-point is created \cite{parallellp-survey,op-ngc}.
A large number of different methodologies have been proposed to address
the environment representation problem in OP \cite{parallellp-survey}. 

Variations of the same problem arise in many classes of search problems and
paradigms relying on non-determinism. 
For example, in the context of non-monotonic
reasoning under stable models semantics \cite{gelfond88,stable-padl}, 
the computation
needs to determine the possible belief sets of a logical theory; 
these are determined by guessing
the truth values of selected logical atoms, and deriving the consequences of such
guesses. In this case, the dynamic environment is represented by the truth
values of the various atoms along each branch of the tree. 

A more abstract view of the problem has been presented in 
\cite{op-ngc,op-complang}, where its theoretical properties have been
investigated. The theoretical results show that methodologies like stack
copying and stack recomputation are theoretically superior than other
schemes---i.e., in the formal abstraction of the environment representation
problem, these methods have a computational complexity that is better than
that of other proposed schemes.

\subsection{Stack-copying for Maintaining Multiple Environments}
\label{museintro}

Stack-copying \cite{muse-journal} is a successful approach
for environment representation in OP. In this approach, the 
environment representation problem is simply resolved by allowing
each agent to have its own copy of all the environments present in
the branch of the or-tree currently explored---this provides each
agent with its own copy of each conditional variable.

\begin{figure}[htb]
\centerline{\psfig{figure=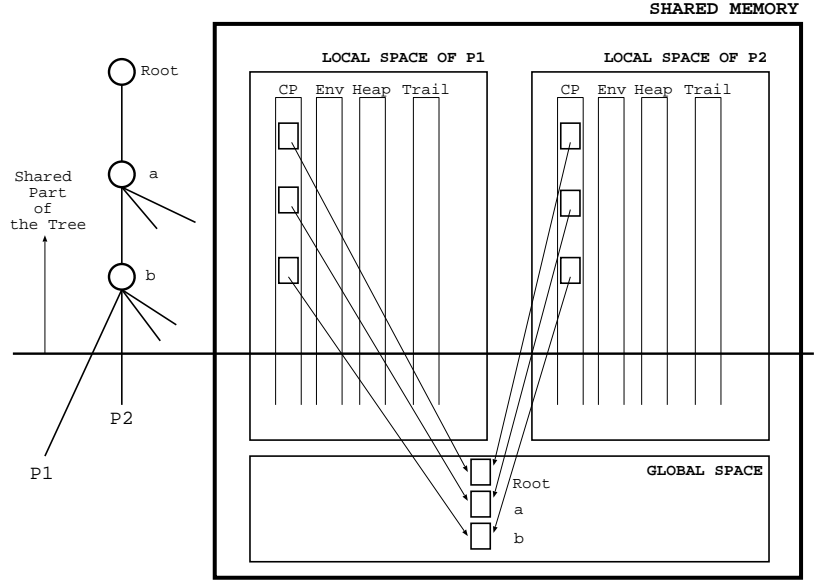,width=0.85\textwidth}}
\caption{Stack-copying based Or-parallelism}
\label{stack_copy}
\end{figure}

In this approach (originally developed in BC-machine
\cite{bcmachine} and successfully implemented in systems
like MUSE \cite{muse-journal,warren-93}
and YAP  \cite{YapOr}), agents
maintain a {\em separate} but {\em identical} address space---i.e.,
each agent is a process with its own address space, but separate
agents maintain exactly the same organization of the data structures
within their address space (i.e., they all locate data structures at the same logical
addresses).
Whenever
an agent $\cal A$
becomes idle (\emph{idle-agent}), 
it will start looking for unexplored alternatives 
generated by another
agent $\cal B$ (\emph{active-agent}). Once a choice-point $p$ 
is detected in the
tree ${\cal T}_{\cal B}$ generated by $\cal B$, 
$\cal A$ will create a local copy of
${\cal T}_{\cal B}$ and restart the  
computation by backtracking 
over $p$\/. 
 Since  all  or-agents 
maintain an identical logical
address space,\footnote{This design choice, adopted in MUSE,
simplifies the implementation in an existing Prolog system---though
it potentially limits the use of the model in thread-based
implementations.}
 the creation of a local copy of 
${\cal T}_{\cal B}$ is reduced to a simple
memory copying  (Figure~\ref{stack_copy})---without the need
for any explicit pointer relocation. Since each or-agent
owns a separate copy of the environments, the environment representation
problem is readily solved---each or-agent will store the locally
produced bindings in the local copy of the environments. Additionally,
each or-agent performs Prolog execution on a private copy of its
tree branch, thus relieving the need for sharing memory.
For this reason, stack-copying has been considered
highly suitable for execution on DMPs, where stack-copying
can be simply implemented using message passing between agents.

\smallskip

In practice, the stack-copying operation is  more involved than
simple memory copying,
as it is desirable to maintain a single copy of each 
choice-point, stored in a specialized area  accessible to
all agents. This is
important because the set of untried alternatives is now shared
between the two agents. If this set is not accessed in
mutual exclusion, then two agents may execute the
same alternative, leading to duplication of work. In
addition, the duplicate execution of the same alternative
will lead to an observable behavior which is different
from that of a sequential Prolog execution (e.g., if the
duplicated alternative contains a side-effect, this will
be seen repeated by the user).
Thus, after copying, parts of each choice-point in
${\cal T}_{\cal B}$ (specifically, the parts related to the
set of available alternatives) will
be transferred to a shared area---these will be called
\emph{shared frames}. 
Both  active and idle agents
will replace
their choice-points with  pointers to the corresponding shared
frames. Shared frames 
are accessed in  mutual exclusion.
This whole operation of obtaining work from another agent
 is usually termed
{\em sharing of or-parallel work}. This is illustrated
in Figure \ref{stack_copy}. Note that CP denotes the choice-point stack
and Env the environment stack in the figure. For illustration purposes we assume
that the choice-points and environments are 
allocated space in separate stacks even though this is not always the case;
choice-points and environments may be allocated space in a single common stack. 
In Figure \ref{stack_copy} the part of the tree labeled as
\emph{shared} has been copied from agent {\tt P1} to
agent {\tt P2}; the choice-points lying in this part of
the tree have been also moved to the shared space to avoid
repetition of work. 
In particular, agent {\tt P2} picks an
untried alternative from choice-point {\tt b}, created by {\tt P1}.
To begin execution along this alternative, {\tt P2} first transfers
the choice-points between the root node and {\tt b} (inclusive)
in a global area (accessible by all agents), and then copies
{\tt P1}'s local stacks from root node up to node {\tt b}. It
untrails the appropriate variables to restore the computation 
state that existed when {\tt b} was first created, and it
begins the execution of the alternative that was picked.

%In the Figure, we denote with CP the choice-point stack, with Env the
%environment stack, with Heap the WAM heap, and with Trail the trail stack (used to 
%support backtracking).\footnote{Note that certain WAM implementations do 
%not separate the CP and Env stacks.}

A major reason for the success of MUSE  and
YAP is that they effectively implement incremental
stack copying with 
{\it scheduling on bottom-most choice-point}.
Each idle agent
picks work from the bottom-most
choice-point of an or-branch. During the sharing operation all the choice-points
between the bottom-most  and the top-most
choice-points are shared between the two agents. This means that,
in each sharing operation, we try to
maximize the amount of work shared between the two agents.
The stack  segments upwards of
this choice-point are copied before the exploration
of this alternative is begun. The copied stack segments
may contain other choice-points with untried alternatives---which are
locally available without any further copying operation and with
very limited synchronization between processors, i.e.,
they become accessible via simple backtracking (modulo simple
use of locks for mutual exclusion).
Thus, a significant amount
of work becomes available to the copying agent
every time a sharing operation is performed. The cost of
having to copy potentially larger fragments of the  tree becomes relatively
insignificant considering that this technique drastically reduces
the \emph{number of sharing operations} performed. It is important
to observe that each sharing operation requires both the agents
involved to stop the regular computation and cooperate in the sharing.
Furthermore, to reduce the amount of information transferred 
during the sharing operation, copying is done {\em incrementally}, 
i.e., only the {\it difference} between ${\cal T}_{\cal A}$ and
${\cal T}_{\cal B}$ is actually copied.

\subsection{Incremental Stack-Copying}
\begin{figure}[htb]
\centerline{\fbox{\psfig{figure=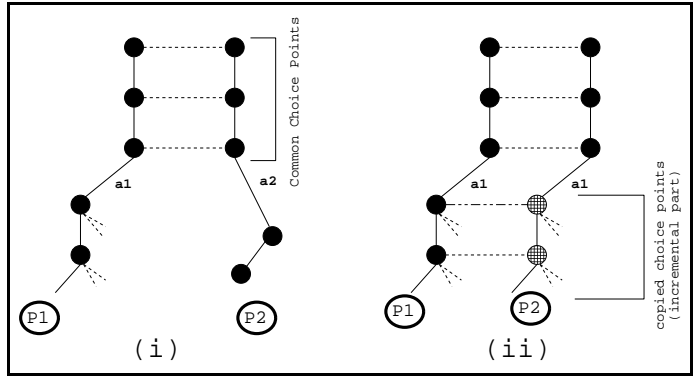,width=.7\textwidth}}}
\caption{Incremental Stack-Copying}
\label{increm}
\end{figure}

Traditional stack-copying requires agents which share work to transfer a complete
copy of the data structures representing the status of the computation. In the case of
a Prolog computation, this may include transferring most of the choice-points along with
copies of the other data areas (trail, heap, environments).
Since Prolog computations can make use of large quantities of memory (e.g., generate
large structures on the Heap), this copying operation can become quite expensive. 
MUSE introduced a variation of
stack-copying, adopted by many other stack-copying systems, called
\emph{Incremental Stack-Copying} \cite{muse-journal}, which allows to considerably
reduce the amount of data transferred during a sharing operation.  The idea is to compare
the content of the data areas in the two agents involved in a sharing operation, and transfer
only the difference between the state of the two agents. This is illustrated in Figure~\ref{increm}.
In Figure~\ref{increm}(i) we have two agents (P1 and P2) which have 3 choice-points in common (e.g.,
from a previous sharing operation). P1 owns two additional choice-points with unexplored
alternatives while P2 is out of work. If P2 obtains work from P1, then there is no need of copying
again the 3 top choice-points (Figure~\ref{increm}(ii)).

Incremental stack-copying, in a shared-memory context, is relatively simple to realize---the shared
frames can be used to identify which choice-points are common and which are not \cite{muse-journal}.
This is primarily because all the information needed
for performing incremental copying efficiently can be found in the shared frames---the
use of shared frames is essential to determine the bottom-most \emph{common}
choice-point between the two agents. The determination of such choice-point
is typically accomplished by analyzing the bitmaps stored in the various shared frames,
which are used to keep track of the agents which currently maintain a copy of the associated 
choice-point (each bit is associated to a different agent). An additional component required
by incremental stack-copying is the need for \emph{binding installation}. As illustrated
in Figure~\ref{increm}, the part of the environment stack corresponding to the three topmost
choice points is not copied. On the other hand, variables present in such environments
might have received bindings during the execution of the bottom part of the computation; these
bindings need to be explicitely installed after copying, in order to reflect the proper
computation state.

\section{Choice-point Splitting in the Stack-Copying Model}
\label{extension}
In this section,  we discuss the issues related to porting
the stack-copying model to a DMP platform, and we present the
basic idea behind the novel stack-splitting scheme.

\subsection{Copying on DMPs}
As mentioned earlier, to avoid duplication of work and to 
guarantee effective scheduling, during the copying operation
part of the content of each copied choice-point is transferred
to a shared memory area; the various agents access each shared
frame in mutual exclusion, thus synchronizing and guaranteeing
unique execution of each alternative.
This solution  works fine
on SMPs---where mutual exclusion is easily implemented
using {\it locks}.
However, on a DMP
this process is a source of significant overhead---access to
 the shared area becomes a bottleneck \cite{haren}. 
This is because sharing
of information in a DMP leads to frequent
exchange of messages and hence considerable overhead. Centralized
data structures, such as the shared frames, are
expensive to realize in a distributed setting. On the
other hand, stack copying appears to be more suitable
to support OP in a distributed-memory setting 
\cite{daos,foong,araujo,opera,benjumea-or}, since,
although the 
choice-points are shared, at least other data-structures representing
the computation---such as, in the case of Prolog, 
the environment, the trail, and the heap---are not. Other
environment representation schemes, e.g., the popular Binding
Arrays scheme \cite{aurora}, have been specifically designed for
SMPs and share most of the computation; the
communication overhead produced by these alternative schemes
on DMPs is likely to be prohibitive.\footnote{Researches
have also proposed to combine these methods with distributed
shared memory schemes \cite{Dorpp}.}
To avoid the problem of sharing choice-points in distributed
implementations, many 
implementors have reverted back to the 
{\it scheduling on top-most choice-point} strategy 
\cite{daos,foong}. The reason  is that
untried alternatives of a choice-point created higher up 
in the or-tree are more likely to generate large
subtrees, and sharing work from the highest choice-point 
leads to smaller-sized stacks 
being copied.
However, if the granularity does not turn out
to be large, then another  untried alternative has to
be picked and a new copying operation has to be
performed. In contrast,
in scheduling on bottom-most, more work could be found
via backtracking, since more choice-points
are copied during the same sharing operation. Additionally,
scheduling on bottom-most is closer to the depth-first
search strategy used by sequential  systems, and
facilitates support of Prolog semantics (e.g., support of
order sensitive predicates). Indeed, comparative studies
about scheduling strategies indicate that scheduling on bottom-most
is superior to scheduling on top-most \cite{warren-93}.
This is especially true for 
the stack-copying technique because: 
\begin{enumerate}
\item the number of copying
operations is minimized; and, 
\item  the alternatives
in the choice-points copied are ``cheap'' sources of
additional work, available via backtracking.
\end{enumerate}
However, the fact that these choice-points are shared 
is a major drawback for a distributed implementation
of copying. The question we consider is: can
we avoid sharing of choice-points while keeping
scheduling on bottom-most? The answer is affirmative, as 
is discussed next.

\subsection{Split Choice-point Stack Copying}
\label{split}
In Stack-Copying, the primary reason why
a choice-point has to be shared is because we want to
serialize the selection of untried alternatives,
so that no two agents can  pick the
same alternative. The shared frame is locked while
the alternative is selected to achieve this effect. 
However,
there are other simple ways of ensuring the same
property:
{\em the untried alternatives of a choice-point can be split
between the two copies of the choice-point stack}.
We call this operation {\it choice-point stack-splitting}
or simply {\it stack-splitting}. This will ensure that no 
two agents pick the same alternative. 

We can envision different schemes for splitting the
set of alternatives between shared choice-points---e.g.,
each choice-point receives half of the alternatives, or the
partitioning can be guided by information regarding
the unexplored computation, such as granularity and likelihood
of failure.
In addition, the need for a
shared frame, as a critical section to protect the alternatives
from multiple executions, has disappeared, as each stack copy has
a choice-point with a different set of
unexplored alternatives.
All the choice-points can be evenly split in this
way during the copying operation.  

\begin{figure}[htb]
\centerline{\psfig{figure=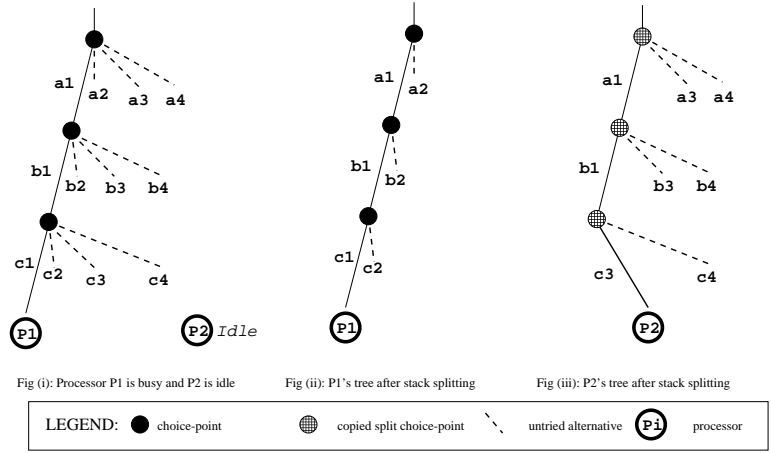,width=.8\textwidth}}
\caption{Horizontal stack-splitting based or-parallelism}
\label{stacksplit}
\end{figure}

The choice-point
stack-splitting
operation is illustrated in Figure~\ref{stacksplit}.
The strategy adopted in this example is what we call
\emph{horizontal splitting}: the remaining alternatives
in each of the shared choice-points are split between the
two agents.

A variation of choice-point stack-splitting relies on splitting
the content of the choice-point stack, instead of splitting the 
individual choice-points. This means that, during a sharing operation,
the list of available choice-points is partitioned between the two
agents. We will refer to this approach as \emph{vertical splitting}.
In this case, we can assume the availability of a 
partition function:
\[ part: CP^* \rightarrow CP^* \times CP^* \]
where $CP$ is the set of all possible choice-points and
$CP^*$ denotes a list of choice-points. The intuition is that,
given the sequence $B$ of choice-points in the branch to be shared,
$part(B)$ will return a partition of $B$ in two subsets
$\langle B_{keep}, B_{give} \rangle$, where $B_{keep}$ are the 
choice-points kept by the active agent and $B_{give}$ are the
choice-points given to the idle agent. 

\begin{figure}[htb]
\centerline{\psfig{figure=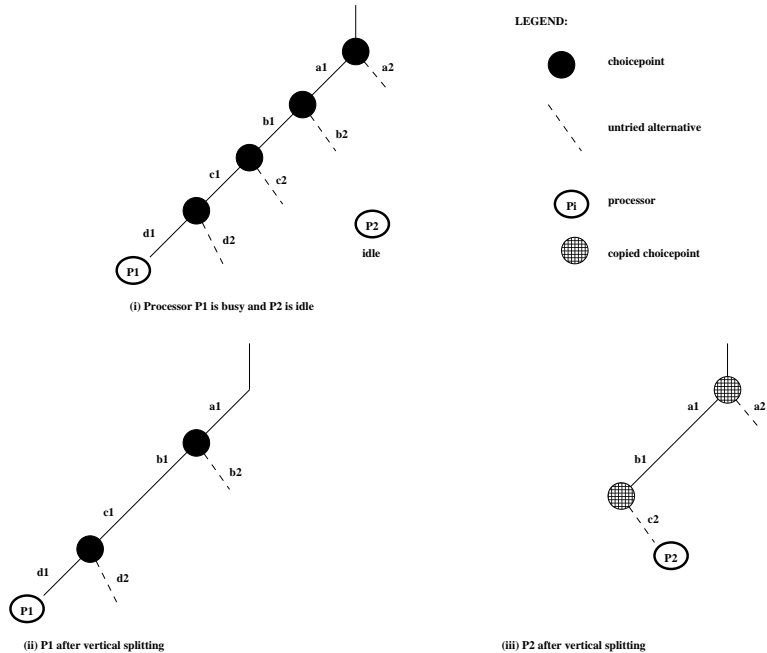,width=0.8\textwidth}}
\caption{Vertical Splitting of Choice-Points ($alternate$ strategy)}
\label{vert}
\end{figure}

\begin{figure}[htb]
\centerline{\psfig{figure=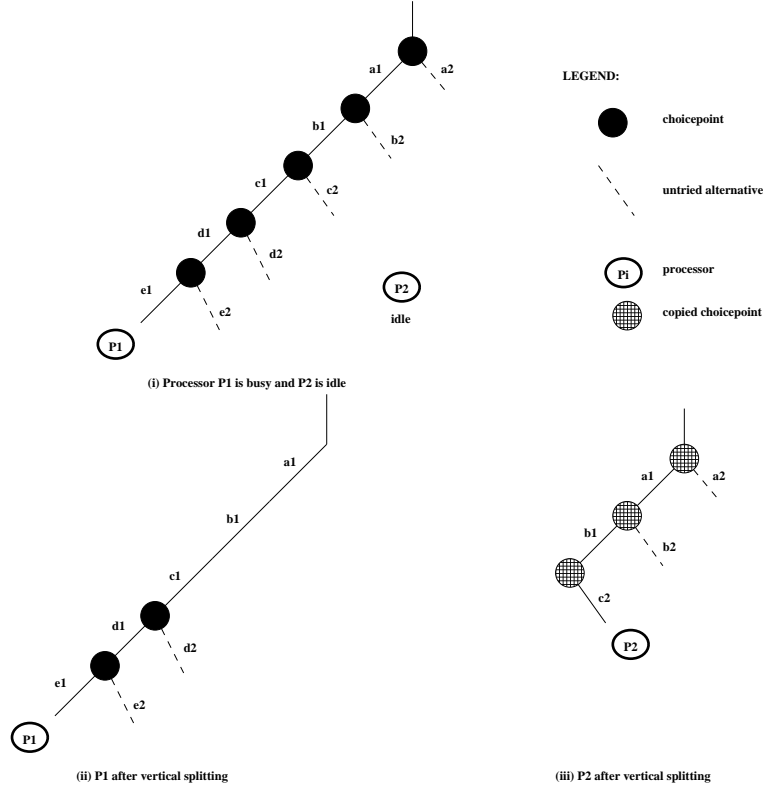,width=0.8\textwidth}}
\caption{Vertical Splitting of Choice-Points ($block$ strategy)}
\label{vert1}
\end{figure}

In the rest of this work we will consider two main strategies for
partitioning the choice-points:
\begin{itemize}
\item $alternate(a_1 a_2 a_3 a_4 \dots ) =
	\langle a_2 a_4 \dots \:,\: a_1 a_3 \dots \rangle$
	i.e., the choice-points in the even positions are kept
	while those in the odd positions are given away (see Figure
	\ref{vert}).

\item $block(a_1 a_2 \dots a_n) = \langle a_{i}\dots a_n \:,\: a_1 \dots a_{i-1}\rangle$
	i.e., the list of choice-points is cut in two segments, the
	first given to the idle agent, while the second is kept by
	the active agent (see Figure \ref{vert1}).
\end{itemize}
Observe that, in practice, all choice-points are copied---as it would be too
expensive to selectively copy only the required ones---and the ones that are
not needed are ``cleared'' of their alternatives; this is explained in detail
in the next section.

The idea of splitting the list of choice-points is
particularly useful  when the search tree is \emph{binary}---which is a frequent
situation in several  Prolog applications as well as in
other search problems (e.g.,
non-monotonic reasoning where the choice-points represent choices
of truth values). In these cases the use of horizontal 
splitting is rather ineffective.
 Splitting
of alternatives can be resorted to when very few choice-points
with many alternatives
are present in the stack. 

Different mixes of splitting of
the list of choice-points and choice-point splitting can be tried to 
achieve a good load balance---as discussed in
\cite{karen-phd,fernando,pals-se}. Eventually, the user could also
be given control regarding how the splitting is done---e.g., by allowing the
user to declare one of a set of splitting strategies for given predicates---although
our system does not currently support this option.

The major advantage of stack-splitting is that 
scheduling on bottom-most can still be used without
incurring huge communication overheads.
Essentially, after splitting
the different or-parallel agents become 
independent of each other, and hence communication
is minimized during execution. This makes the
stack-splitting technique highly suitable
for DMPs.
The possibility of parameterizing the splitting of the
alternatives based on additional semantic information
(granularity, non-failure, user annotations) can further
reduce the likelihood of additional communications due
to scheduling.

\section{Towards Practical Stack-Splitting and Incremental Stack Splitting}
\label{ssimpl}

%guo
%In the rest of the paper we describe a complete implementation of
%stack-splitting using a message passing platform, analyzing in
In the rest of the paper we describe the incremental stack-splitting scheme
and its implementation issues on a message passing platform, analyzing in
detail how the various problems mentioned earlier have been
tackled. In addition to the basic stack-splitting scheme, we
also
\begin{itemize}
\item analyze how stack-splitting can be extended to incorporate
\emph{incremental copying}, an optimization which has been deemed
essential to achieve speedups in various classes of applications, and
\item analyze how to handle order-sensitive predicates (e.g., side-effects)
	in the presence of stack-splitting.
\end{itemize}
 The solution we describe has been developed in a concrete
implementation, realized by modifying the engine of a commercial
Prolog system (ALS Prolog) and making use of the Message Passing
Interface (MPI) as a communication platform. The ALS Prolog system
is based on an implementation of the Warren Abstract Machine
(WAM).

\subsection{Data Structures for Stack-Splitting and Incremental Stack-Splitting}
The data structures employed by our distributed engine include all
the data areas of a standard Warren Abstract Machine (e.g.,
stack for the choice-points, stack for the environments,
a heap for the dynamic creation of terms, a trail to support undoing
of variable bindings during
backtracking). We assume that
the code-area is initially duplicated between all processors.

During stack-splitting, all WAM areas, except for the code area, are
copied from the agent giving work to the idle one.
Next, the parallel choice-points are split between the two agents.
Blindly copying all the stacks every time an agent shares work with
another idle agent can be wasteful, since frequently the two agents
already have parts of the stacks in common due to previous copying.
We can take advantage of this fact to reduce the amount of copying by performing
\emph{incremental copying}, as discussed earlier.
In our stack-splitting scheme, there are no shared frames, hence performing
incremental stack-copying will incur more overhead due to the communication
overhead involved.
In order to figure out the incremental part that only needs to be copied during
incremental stack-splitting,
parallel choice-points will be {\it labeled} in a certain way. The goal of  labeling
is to uniquely identify the original ``source'' of each choice-point (i.e.,
which agent created it), to allow unambiguous detection of copies of common
choice-points. Thus, the labels effectively replace the bitmaps used in the shared
memory implementations of stack-copying.
The labeling procedure
is described next.

To perform labeling, each agent maintains a counter.
Initially, the counter in each agent is set to \textit{1}.
The counter is incremented each time the labeling procedure is performed.
When a parallel choice-point is copied for the first time, a label for it is created.
The label is composed of three parts:
\begin{enumerate}
\item agent rank,
\item counter, and
\item choice-point address.
\end{enumerate}
The agent rank is the rank (i.e., id) of the agent which
created the choice-point. The counter is the current value of the
labeling counter for the agent generating the labels.
The choice-point address is the address of the choice-point which is being labeled.
The labels for the parallel choice-points are recorded in a separate
{\it label stack}, in the
order they are created---the choice-point address in the label maintains the
connection between the label (stored in the label stack) and the corresponding
choice-point (stored in the choice-point stack).
 Also, when a parallel choice-point is removed from the stack,
its corresponding label is also removed from the label stack.
Initially, the label stack in each agent is set to \textit{empty}.
The label stack keeps a record of the labels for the agent's 
shared choice-points. Observe that the choice of maintaining
labels in a stack---instead of associating them directly to the corresponding
choice-points---has been dictated by efficiency reasons.

\begin{figure}[htb]
\begin{center}
\begin{minipage}[c]{.35\textwidth}
\centering{\fbox{\psfig{figure=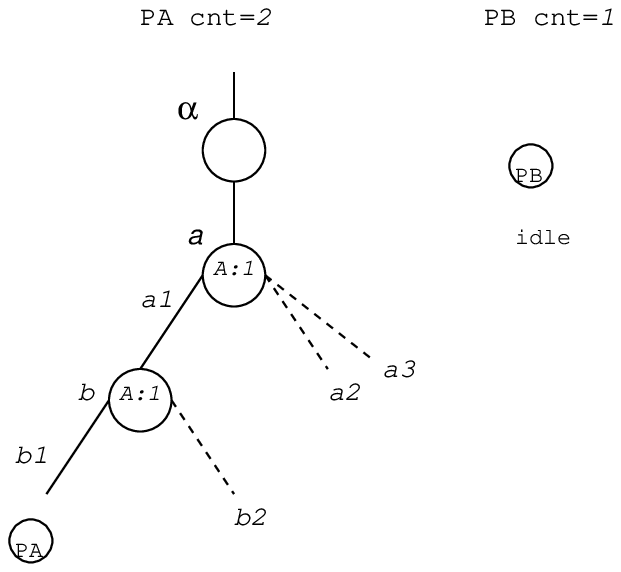,width=\textwidth}}}
\caption{Agent A Labels its two Parallel Choice-points}
\label{Al}
\end{minipage}
\hspace{.01\textwidth}
\begin{minipage}[c]{.6\textwidth}
\centering{\fbox{\psfig{figure=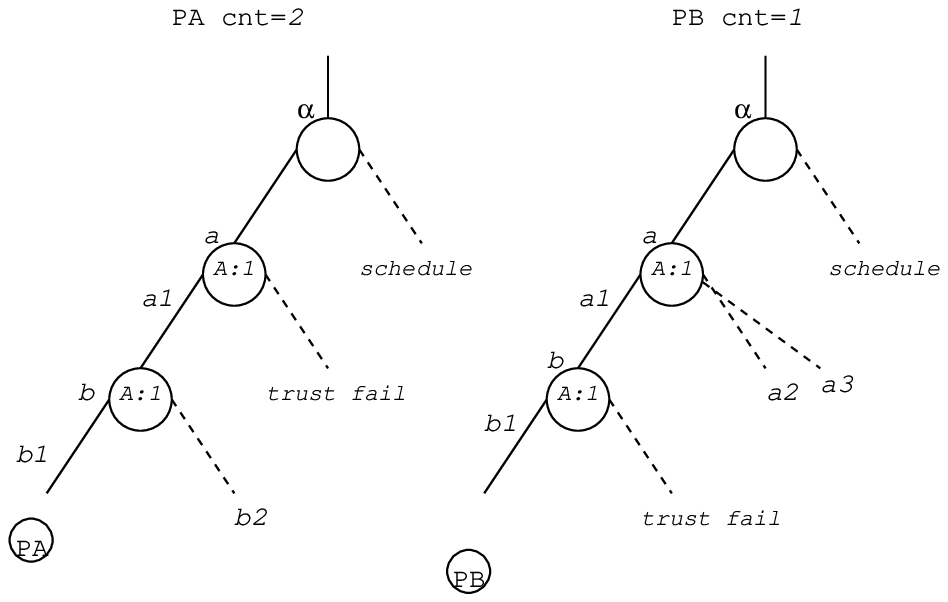,width=\textwidth}}}
\caption{Agent A Gave Work to Agent B}
\label{AgB}
\end{minipage}
\end{center}
\end{figure}

Let us illustrate stack-splitting accompanied by labeling with an example. In
the rest of the discussion we assume the use of vertical splitting strategy.
Suppose agent A has just created two parallel choice-points and agent B is idle.
Agents A and B have their counters set to \textit{1} and their label stacks set to \textit{empty}.
Then agent B requests work from agent A.
Agent A first creates labels for its two parallel choice-points.
These labels have their rank and counter parts as \textit{A:1}.
Agent A then pushes these labels into its label stack. This is 
illustrated in Figure~\ref{Al}; for simplicity, in our figures, we 
do not show the label stack 
explicitly but show each label
rank and counter parts inside the parallel choice-point being labeled.
Notice that agent A incremented its counter to 2 after the labeling procedure
was over. In the figure, $\alpha$ denotes the root of the tree.

The next step requires the actual execution of stack-copying.
Agent B receives a message that contains all the parallel 
choice-points of agent A, along with agent A's label stack. At
this point, it becomes possible to perform stack-splitting.
Agent A will keep the alternative \textit{b2} but not \textit{a2}
and \textit{a3},
and agent B will get the alternatives \textit{a2, a3} but not \textit{b2}.
We have designed a new WAM scheduling instruction (\textit{schedule}) 
which is placed in the next alternative field of the choice-point
above which there is no more parallel work. The execution of this
instruction forces the agent to enter scheduling, and it implements
the scheduling scheme described in Section \ref{sched}.
%We now discuss the details of the stack-splitting process.
%Process A keeps the alternative \textit{b2} of choice-point \textit{b} and changes the next alternative field of
%choice-point \textit{a} to new WAM instruction \textit{schedule} which will take process A into scheduling.
Agent A keeps the alternative \textit{b2} of choice-point \textit{b}, changes the next alternative field of
choice-point \textit{a} to WAM instruction \textit{trust\_fail} to avoid taking the original alternative of
this choice-point, and changes the next alternative field of the choice-point above \textit{a} to the new
WAM instruction \textit{schedule} which will take agent A into scheduling.\footnote{This is a common
technique used in other modifications of the WAM---e.g., 
the MUSE WAM \cite{muse-journal}.} The \textit{trust\_fail} instruction will simply
act as a filler to denote that the choice-point does not have any further alternatives. 
Observe that in practice it is possible to optimize
this scheme (e.g., in the example, we could have introduced the \textit{schedule}
instruction directly in the
choice-point \textit{a}).

In turn, agent B changes the next alternative field of choice-point \textit{b} to 
WAM instruction \textit{trust\_fail}, to
avoid taking the original alternative of this choice-point, keeps the alternatives \textit{a2, a3} of choice-point
\textit{a}, and changes the next alternative field of the choice-point above \textit{a} to the \textit{schedule}
instruction.  See  Figure \ref{AgB}.
Afterwards, agent B backtracks, removes choice-point \textit{b} along with its corresponding label in the label
stack, and then takes alternative \textit{a2} of choice-point \textit{a}.

\begin{figure}[htb]
\centering{\fbox{\psfig{figure=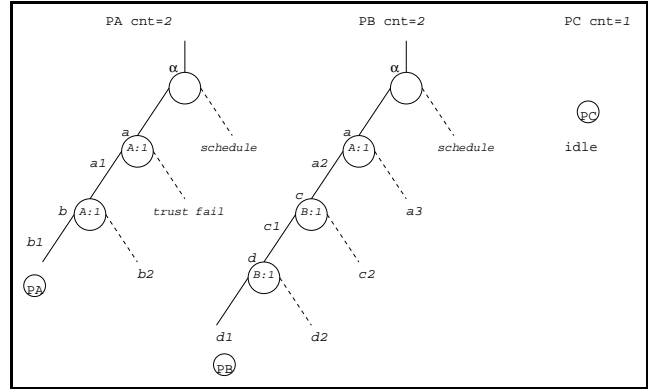,width=.65\textwidth}}}
\caption{Agent B Labels its Two New Parallel Choice-points}
\label{Bl}
\end{figure}

Suppose now that agent B creates two parallel choice-points and agent C is idle.
Agent C, with its counter set to \textit{1} and its label stack set to \textit{empty},
requests work from  B.
Agent B first creates labels for its two new parallel choice-points.
These labels have their rank and counter parts as \textit{B:1}.
Agent B then pushes these labels into its label stack.
See  Figure \ref{Bl}.
Notice that agent B incremented its counter to \textit{2}.

\begin{figure}[htb]
\centering{\fbox{\psfig{figure=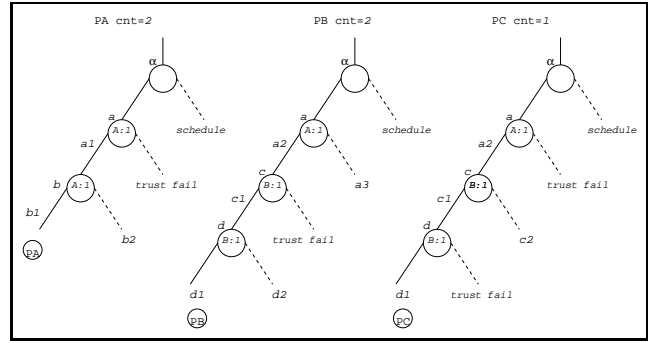,width=.65\textwidth}}}
\caption{Agent B Gives Work to Agent C}
\label{BgC}
\end{figure}

At this point in time,  stack-copying takes place.
Agent C gets all the parallel choice-points of agent B along with agent B label stack.
The stack-copying phase is followed by the actual
stack-splitting operation:
agent B will keep alternatives \textit{d2} and \textit{a3} but not \textit{c2},
and agent C will keep alternative \textit{c2} but not \textit{d2} nor \textit{a3}.
Notice that all three parallel choice-points of agent B have been split among B and C.
Agent B keeps the alternative \textit{d2} of choice-point \textit{d} and changes the next alternative field of
choice-point \textit{c} to WAM instruction \textit{trust\_fail}
to avoid taking the original alternative of this
choice-point, and keeps the alternative \textit{a3} of choice-point \textit{a}.
%Process C changes the next alternative field of choice-point \textit{d} to WAM instruction \textit{trust\_fail} to
%avoid taking the original alternative of this choice-point, keeps the alternative \textit{c2} of choice-point
%\textit{c}, and changes the next alternative field of choice-point \textit{a} to \textit{schedule}. This is illustrated in Figure~\ref{BgC}.
Agent C changes the next alternative field of choice-point \textit{d} to WAM instruction \textit{trust\_fail} to
avoid taking the original alternative of this choice-point, keeps the alternative \textit{c2} of choice-point
\textit{c}, changes the next alternative field of choice-point 
\textit{a} to WAM instruction \textit{trust\_fail}, 
and changes the next
alternative field of the choice-point above \textit{a} to \textit{schedule}. This is illustrated in Figure~\ref{BgC}.
Agent C backtracks, removes choice-point \textit{d} along with its corresponding label in the label stack,
and then takes alternative \textit{c2} of choice-point \textit{c}.

\subsection{Incremental Stack-splitting: The Procedure}

In this section we describe how the label stacks are used to compute the incremental part to be copied.
Let us assume that agent A is giving work to agent B.
Agent A will label all its parallel choice-points which have not been labeled before and will push them into its
label stack.
Agent A then increments its counter.

If the label stack of agent B is empty, 
then stack-copying will need to be performed followed by stack-splitting.
Agent A sends its complete choice-point stack and its complete label stack to agent B.
Then stack-splitting is performed on all the parallel choice-points of agent A.
Agent B then tries its new work via backtracking.

However, if the label stack of 
agent B is not empty, then agent B will send its label stack to agent A. The
objective is for agent A to locate the topmost label in common between A and B---and this is realized
by comparing the content of the two stacks until a match is found. Let us denote with
\textit{ch} the most recent choice-point with a common label between A and B.
In this way, agents A and B are guaranteed to have the same computation
\emph{above} the choice-point {\it ch}, while their computations will be
different below such choice-point.

If the choice-point \textit{ch}
does not exist, then (non-incremental)
stack-copying will need to be performed followed by
stack-splitting just as described before.
However, if choice-point \textit{ch} does exist, then agent
 B backtracks to choice-point \textit{ch}, and
performs incremental-copying.
Agent A sends its choice-point stack starting from
choice-point \textit{ch} to the top of its choice-point stack.
Agent A also sends its label stack starting from the label
corresponding to choice-point \textit{ch} to the top of its label stack.
Stack-splitting is then performed on all the parallel choice-points of agent A.
Afterwards, agent B tries its new work via backtracking.

\begin{figure}[htb]
\begin{center}
\begin{minipage}[t]{.48\textwidth}
\centering{\fbox{\psfig{figure=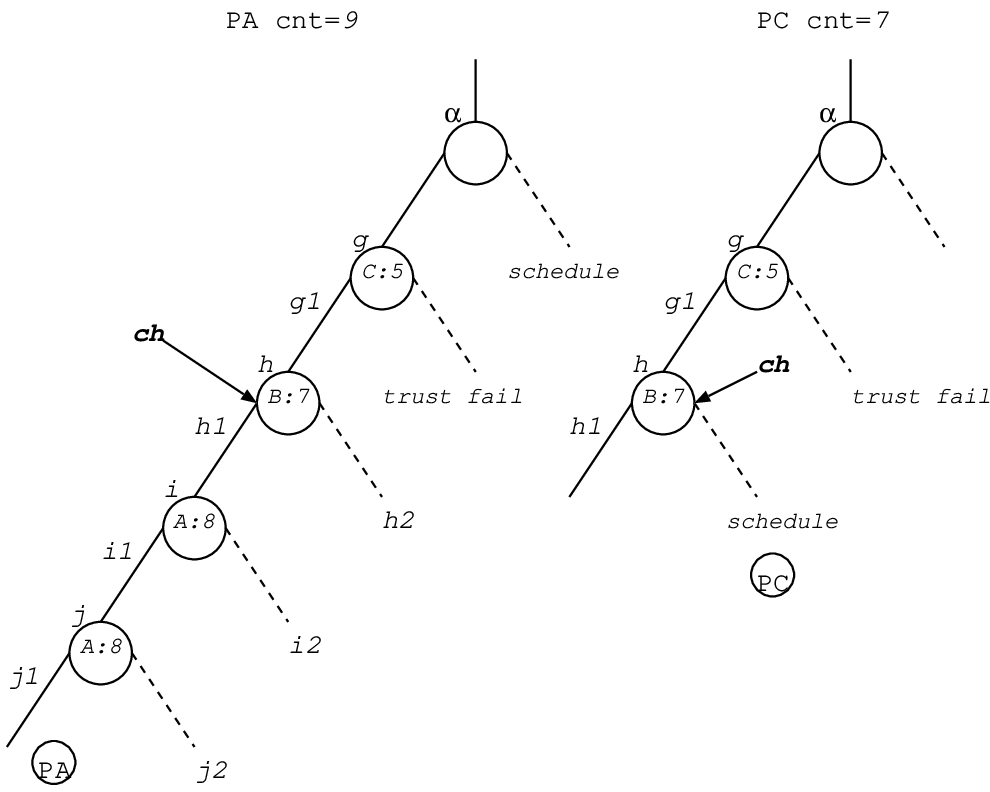,width=\textwidth}}}
\caption{Agent A Labels its Two New Parallel Choice-points and Compares Labels with Agent C}
\label{Al2}
\end{minipage}
\hspace{.01\textwidth}
\begin{minipage}[t]{.48\textwidth}
\centering{\fbox{\psfig{figure=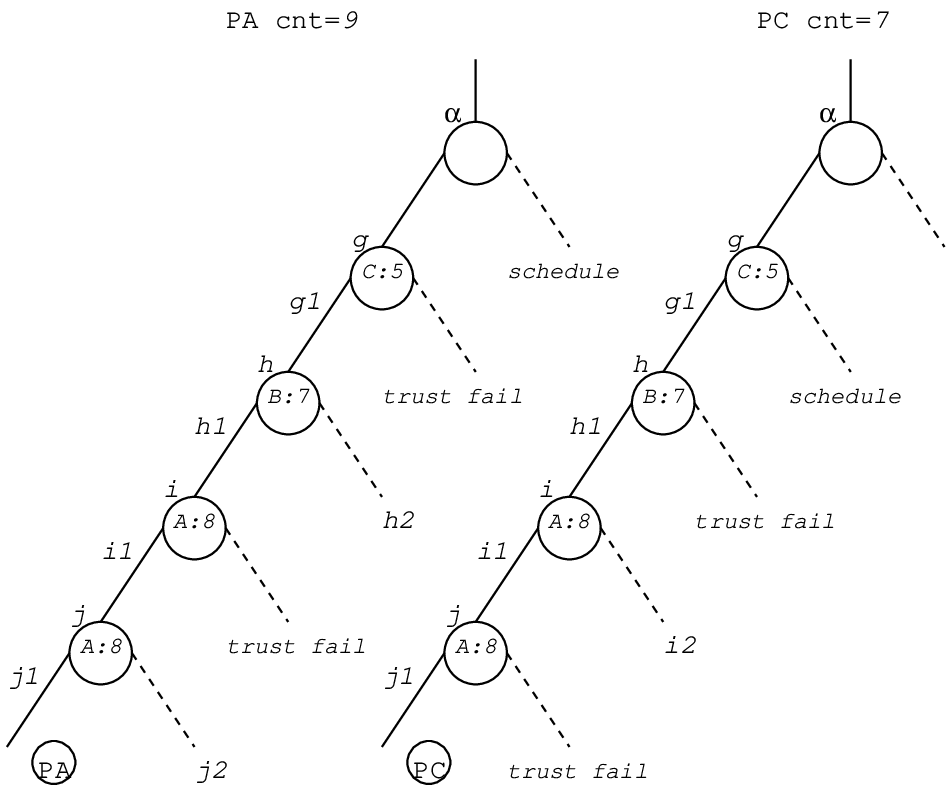,width=\textwidth}}}
\caption{Agent A Gave Work to Agent C}
\label{AgC}
\end{minipage}
\end{center}
\end{figure}

We illustrate the above procedure by the following example.
Suppose agent A has three parallel choice-points and agent C requests work from A.
Agent A first labels its last two parallel choice-points which have not been labeled before
and then increments its counter.
Afterwards, agent C sends its label stack to agent A.
Agent A compares its label stack against the label stack of agent C and finds the last choice-point \textit{ch}
with a common label.
Above choice-point \textit{ch}, the Prolog trees of agents A and C are equal.
Below choice-point \textit{ch}, the Prolog trees of agents A and C differ.
See Figure \ref{Al2}.

Now, agent C backtracks to choice-point \textit{ch}.
Incremental stack-copying can then take place. Agent A sends its choice-point
stack starting from choice-point \textit{ch} to the top of its choice-point stack.
Agent A also sends its label stack starting from the label corresponding to
choice-point \textit{ch} to the top of its label stack.
Then, stack-splitting takes place on the three parallel choice-points of agent A. See Figure \ref{AgC}.
Agent C backtracks to choice-point \textit{i} and takes alternative \textit{i2}.

\subsection{Incremental Stack-splitting: Challenges}

Four issues that were not discussed above and which are fundamental for the correct
implementation of the incremental stack-splitting scheme presented are discussed below.

\subsubsection{Sequential Choice-points}
The first issue is related to the management of \emph{sequential choice-points}.
Typically, only a subset of the choice-points present during the execution 
are suitable to provide work that can be effectively parallelized. 
These choice-points
are traditionally called \emph{parallel choice-points}, to distinguish them
from \emph{sequential choice-points}, whose alternatives are meant to be explored
by a single agent. Systems like PALS, MUSE, and Aurora allow the user to explicitly declare
predicates as parallel (while, by default, the others are treated as sequential).

The problem arises when sequential choice-points are located among the parallel choice-points
that will be split between two agents.
If the alternatives of these choice-points are kept in both agents, we may
have repeated, useless or wrong computations.
Hence, the alternatives of these choice-points should only be kept in one agent---e.g., the agent that
is giving work.  In our
current approach, we keep the alternatives of sequential choice-points in the agent 
giving work; as a consequence, 
the agent that is receiving work should change the next alternative field of all these choice-points to the WAM
instruction \textit{trust\_fail} to avoid taking the original alternatives of these choice-points.

\subsubsection{Installation Process}

The second issue has to do with the bindings of conditional variables (i.e.,
variables that may be bound differently in different or-parallel
branches)  which need to be copied too as part of the
incremental stack-splitting process.

\begin{figure}[htb]
\centering{\fbox{\psfig{figure=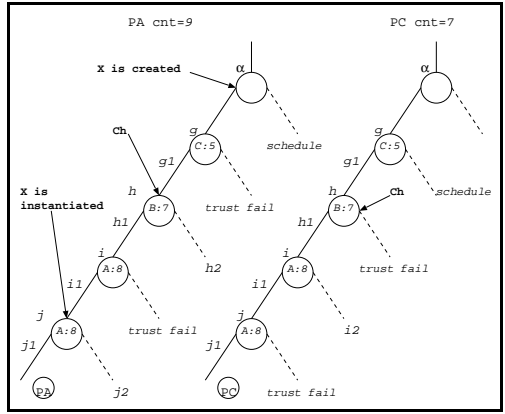,width=.50\textwidth}}}
\caption{The Binding of Conditional Variable \textit{X} Needs to be Copied}
\label{AiC}
\end{figure}

For example, suppose that in our last example, before agent A gives work to agent C, agent A created
the  variable \textit{X} before choice-point \textit{ch} was created, and the
variable \textit{X} was instantiated after the creation of \textit{ch}. This is shown in
Figure  \ref{AiC}.
We can see that the binding for \textit{X} was not copied during incremental stack-splitting. 
This is because \textit{X} is a conditional variable which was created before choice-point \textit{ch}, and the
incremental part of the heap or environment stack that was copied did not contain its binding. This means
that the receiving agent does not see \textit{X} becoming automatically instantiated thanks to the
copying of the heap or environment stack.

In order to solve the problem, we need to ensure that, during
the sharing operation, also the bindings of the conditional variables
created in the common part of the branch are transferred from the
agent giving work to the idle agent.
In the current implementation, we have tackled this problem by 
modifying the trail structure of the ALS WAM engine. The \emph{trail}
is a stack, maintained by the WAM, which records which conditional
variables have been bound along the current branch of execution. The trail
is used by the WAM to support removal of bindings during backtracking. In
our system, the trail has been modified to a \emph{value trail}
\cite{aurora}, thus maintaining
with each bound conditional variable also a reference to its value. The value
trail is employed by agent giving work to build a special message containing the
values of the bound conditional variables, sent to the idle agent during the sharing
operation. The idle agent will make use of this message and install the appropriate
bindings for the conditional variables existing in the common segment of the search
tree branch. 

Observe that a similar problem appears also in shared-memory implementations of
stack-copying \cite{muse-journal}---though they do not need to rely on 
value-trails, since each agent can directly retrieve the values of the 
bindings from the other agent's environments (which are in shared memory).

\subsubsection{Garbage Collection}

The third issue arises when garbage collection takes place. In the current 
implementation of the ALS system (the underlying WAM we modified for this
project), garbage collection occurs also on the choice-point stack, leading
to possible shifting of choice-points.
When this situation occurs, the labels in our label stack may no longer label
the correct parallel choice-points---since labels are connected to choice-points
by storing the address of the corresponding choice-points inside the label.
Therefore, we need to modify our labeling procedure so that when garbage collection on 
an agent takes place,
the label stack of this agent is invalidated.
This has been realized  by just setting its label stack to empty.
The next time this agent gives work, full stack-copying will have to take place.
This solution is analogous to the one adopted in the MUSE system \cite{muse-journal}
to address the similar problem in stack-copying.
Alternative solutions---e.g., use of indirect labels---would introduce costs in each
step of sharing, instead of an occasional additional cost during garbage collection, and
have not been used in our system.

\subsubsection{Next Clause Fields}

The fourth issue arises when the next clause fields of the parallel choice-points
between the first parallel choice-point \textit{first cp} and the last choice-point \textit{ch} with a common label
in the agent giving work are not the same compared to the ones in the agent
receiving work. This situation occurs after several copying and splitting operations---that caused the 
next clause field of some choice-points to be changed to \textit{trust\_fail}, while 
other agents still have active alternatives in such choice-points.
In this case, it is not correct to
 just copy the part of the choice-point stack between choice-point \textit{ch}
and the top of the stack and then perform the splitting. This is because the splitting will not
be performed correctly.

\begin{figure}[htb]
\begin{center}
\begin{minipage}[t]{.48\textwidth}
\centering{\fbox{\psfig{figure=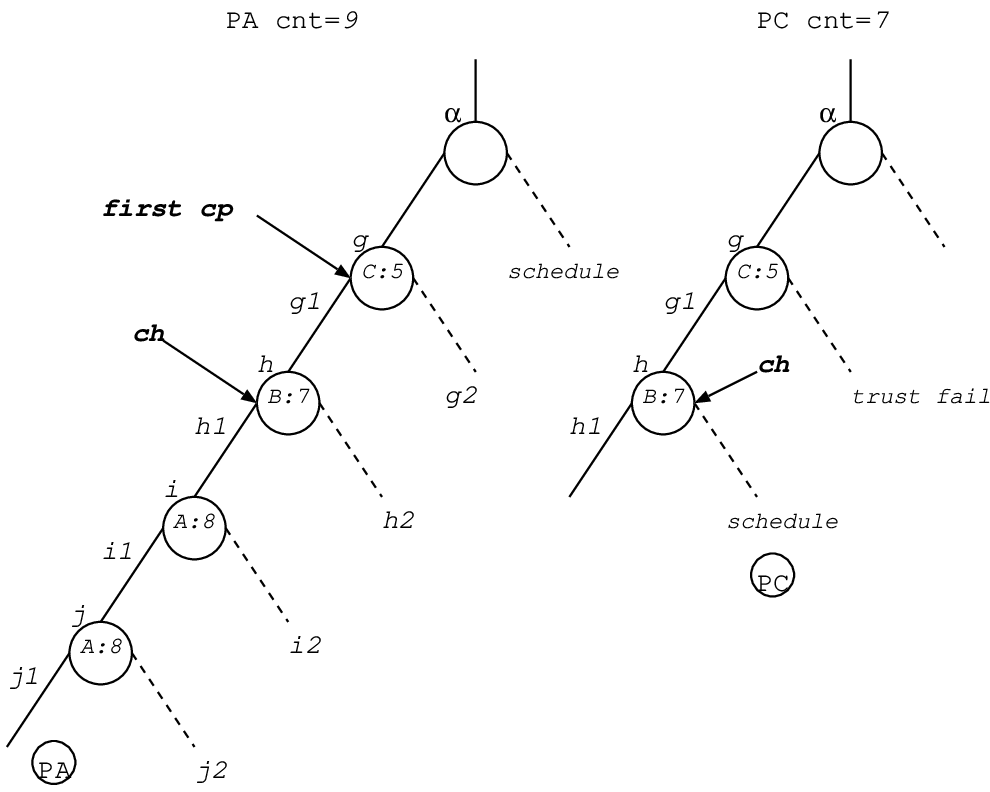,width=\textwidth}}}
\caption{Copy the Next-clause Fields between \textit{first cp} and \textit{ch}}
\label{AnC}
\end{minipage}
\hspace{.01\textwidth}
\begin{minipage}[t]{.48\textwidth}
\centering{\fbox{\psfig{figure=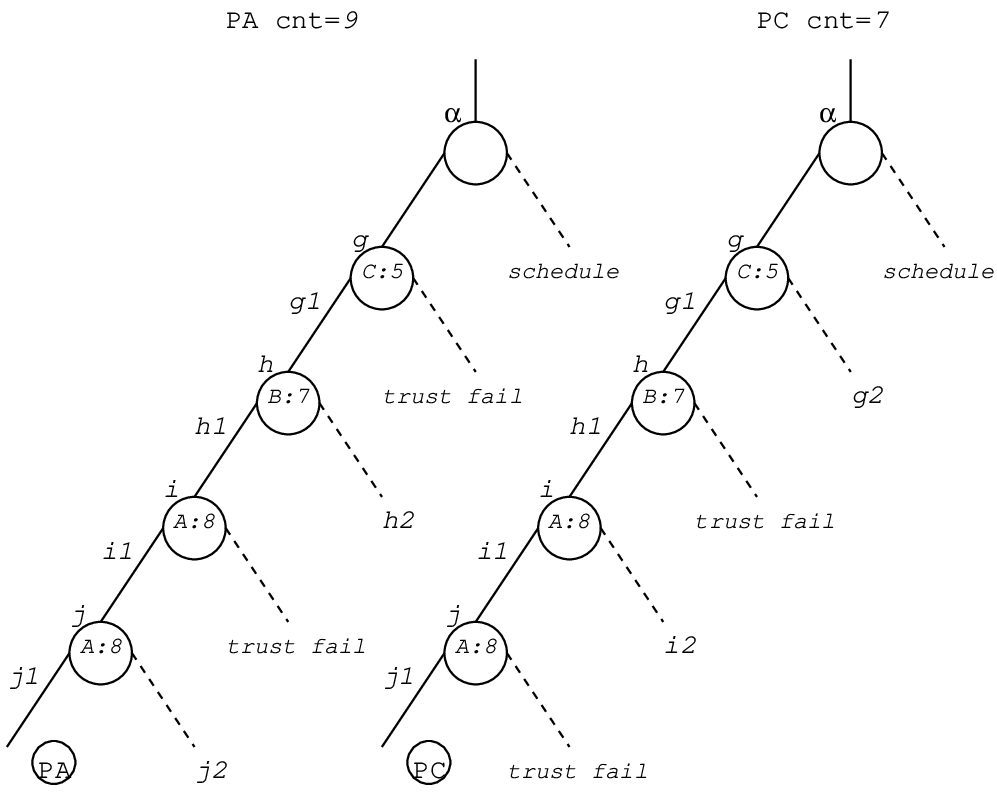,width=\textwidth}}}
\caption{Agent C Received the Next Clause Fields }
\label{AgnC}
\end{minipage}
\end{center}
\end{figure}

For example, suppose that in our previous example (see Fig.~\ref{AiC}), when agent C requests work
from agent A, we have this situation, as illustrated in Figure \ref{AnC}.
Let us assume that the scheduler decides to transfer
the choice-point \textit{g} to agent C. But agent C does
not have the right next clause field for this choice-point. Hence, we need to modify our
procedure once again.
This can be done by having the agent giving work send all the next clause fields between its
first parallel choice-point \textit{first cp} and choice-point \textit{ch} to the
agent receiving work. Then the splitting of all parallel choice-points can take place correctly.
See Figure \ref{AgnC}.

\section{Scheduling}
\label{sched}

Scheduling is an important aspect of any parallel system.
The scheduling strategy adopted largely determines the level
of speedup obtained for a particular parallel execution.
The main objective of a scheduling strategy is to
balance the amount of parallel work done by different agents.
Additionally, work distribution among agents
should be done with a minimum of
communication overhead.
These two goals are somewhat at odds with each other,
since achieving perfect balance may result in a very complex scheduling
strategy with considerable communication overhead,
while a simple scheduling strategy which re-distributes work
less often may  incur a  lower communication overhead but 
lead to a poorer balancing of work.
Therefore, it is obvious that there is an intrinsic
contradiction between distributing parallel work as evenly as possible
and minimizing the distribution overhead. Thus our main goal is to
find a trade-off point that results in a reasonable scheduling strategy.

We adopt a simple and fair distributed algorithm to implement a scheduling
strategy in the PALS system.
A new data
structure---the {\it load vector}---is introduced to provide
an \emph{approximated} view of the
work-loads of different agents. The work-load of an agent
is approximated by the number of parallel choice-points with
unexplored alternatives present in its
local computation tree. This is analogous to the approach
originally used by MUSE, and it can be efficiently implemented
within ALS; furthermore, the majority of examples we encountered
offer parallel choice-points with a small number of alternatives
(often just two), thus making our approximated notion of work-load
essentially equivalent to more refined versions. 
 Each agent keeps a work-load vector {\it V}
in its local memory, and the value of {\it V[i]} represents the 
estimated work-load of the agent with rank {\it i}. Based on the work-load vector,
an idle agent can request parallel work from other agent with
the greatest work-load, so that parallel work can be fairly distributed.
The load vector is updated at runtime. When stack-splitting is
performed, a {\tt Send\_LoadInfo} message with updated load information will be 
broadcasted
to all the agents so that each agent has the latest information of work-load
distribution.
Additionally, load information is attached with each incoming message. For
example: when a {\tt Request\_Work} message is received from agent $P_{1}$,
the value of $P_{1}$'s work-load, 0, can be inferred.

Based on its work-load each agent can be in
one of two states: {\it scheduling} state or {\it running} state.
When an agent has some work to do, it is in a running state, otherwise,
it is in a scheduling state. An agent that is running, occasionally checks whether there
are incoming messages. Two possible types of messages are checked by the running agent:
one is {\tt Request\_Work} message sent by an idle agent, and the other
is {\tt Send\_LoadInfo} message, which is sent when stack-splitting occurs. 
The idle agent in scheduling state is also called a scheduling agent.
An idle agent wants to get work as soon as possible from another agent,
preferably the one that has the largest amount of work.
The scheduling agent searches through its local load vector for the agent with
the greatest work-load, and then sends a {\tt Request\_Work} message to that agent
asking for work. If all the other agents are idle (in scheduling
state), then the
execution of the current query is finished and the agent halts.
When a running agent receives a {\tt Request\_Work} message,
stack-splitting will be performed if the running agent's
work-load is greater than a predefined threshold (the
\emph{splitting threshold}), otherwise, a {\tt Reply\_Without\_Work} message
with a positive work-load value will be sent as a reply. If a scheduling agent
receives a {\tt Request\_Work} message, a {\tt Reply\_Without\_Work} message with 
work-load 0 will be sent as a reply.

The distributed scheduling algorithm mainly consists of two parts: one is for
the scheduling agent, and the other is for the running agent.
The running agent's algorithm can be briefly described as follows:
{\small
\begin{verbatim}
1:        while (any incoming message) {
2:           get an incoming message;   
3:           switch (message type) {
4:           case Send_LoadInfo:
5:                update the corresponding agents' work-load;
6:                break;
7:           case Request_Work:
8:                if (local work-load > Splitting Threshold) {
9:                   reply a message of type Reply_With_Work and perform 
                                                         stack-splitting;
10:                   broadcast the updated work-load to all the agents;
11:                }
12:                else {
13:                   reply a message of type Reply_Without_Work
14:                         and the value of its own work-load;
15:                      set work-load of the message source to 0;
16:                   }
17:                break;
18:           }
19:        }
\end{verbatim}}
At fixed time intervals (which can be selected at initialization of the
system) the agent examines the content of its message queue for
eventual pending messages. {\tt Send\_LoadInfo} messages are quickly
processed (lines {\tt 4-6})
to update the local view of the overall load in the system. Messages
of the type {\tt Request\_Work} are handled as described above 
(lines {\tt 7-17}). If stack-splitting is realized (line
{\tt 9}), then the agent will also notify the whole system of
the new work-loads (line {\tt 10}).
 
We should remark that
the implementation concretely checks for the presence of the two types
of messages with different frequency---i.e., request for work messages are
considered less frequently than requests for load update. All messages are
handled asynchronously; {\tt Send\_LoadInfo} messages are given 
higher priority by the receiving agents (i.e., they are processed before
any other types of messages), 
to ensure that the work-load vector remains as much up-to-date as possible.
The reason of keeping work-load vector up-to-date as much as possible 
for each agent is that 
when a scheduling agent is looking for work, it is able to obtain work 
from the agent with the highest work-load immediately. We have 
observed worse performance by giving higher priority to other types
of messages. This is because if work-loads are not up-to-date, an agent thought
to have the highest work-load may turn out to have work-load lower than others,
reducing the granularity of work obtained and increasing the 
number of splitting operations performed.

The scheduling agent's algorithm can be briefly described as follows:
{\small
\begin{verbatim}
1:        while (1) {
2:           D = the rank of the agent with the greatest work-load;
3:           if (D's work-load == 0) and termination detection returns true 
                                    then halt;   /* The whole work is done */
4:           send a Request_Work message to D;
5:           matched = false;
6:           while (!matched) {
7:              get an incoming message;
8:              switch (message type) {
9:                 case Reply_With_Work:
10:                    stack-splitting with the agent which sent the message;
11:                    update the corresponding work-load;
12:                    simulate failure and go to execute the split work;
13:                    return;
14:                 case Reply_Without_Work:
15:                    if (source of message is D) matched = true;
16:                    V[message sender Id] = work-load of agent which sent 
                                                                 the message;
17:                    break;
18:                 case Request_Work:
19:                    reply a message of type Reply_Without_Work and
20:                         its work-load 0 to the source of incoming message;
21:                    V[message sender Id] = 0;
22:                    break;
23:                 case Send_LoadInfo:
24:                    update the corresponding agents' work-load;
25:                    break;
26:              }
27:          }
\end{verbatim}}
Observe:
\begin{itemize}
\item a Request\_Work message is sent to the agent with the greatest work-load according to
the local load vector (lines 2 and 4); an optimization to avoid some communication overhead 
is that if the greatest work-load is below the splitting threshold value, the Request\_Work message
can be delayed until there exists some agent that has work-load higher than the threshold; in other words,
if all the other agents have low work-load, no stack-splitting takes place in our
strategy; 
\item the  loop {\tt 6-27} is repeated until a reply is received from the
	agent contacted in line {\tt 4};
\item if a reply is positive, then the scheduling phase is left and execution
	restarted; if the reply is negative, then another iteration of the
	outermost loop is performed;
\item during scheduling, requests for work from other agents are denied (and
	this is used to update to zero the work-load of the requesting agent), 
	as shown in lines {\tt 18-22};
\item messages containing new work-load information are used to update the
	work-load vector (lines {\tt 23-25});
\item if the work-load vector contains only zeros (line {\tt 3}), then the
	scheduler initiates a procedure to verify global termination. The
	global termination process is based on a fairly standard 
	black-white token ring scheme~\cite{termination}.
\end{itemize}
Let us point out that the scheduling procedure bears some similarities
with the Argonne scheduler used by Aurora \cite{Arg-Sched}. In
 our experiments on both shared-memory as well as distributed-memory
platforms we did not perceive the problems noticed in other
similar schedulers (e.g., see \cite{warren-93})
with this approach (e.g., the ``honey-pot'' problem, where every worker
tries to grab the same piece of work).

%%%%%%%%%%%%%%%%%%%%%%%%%%%%%%%%%%%%%%%%%%%%%%

\section{Supporting Prolog's Sequential Semantics}
\label{sideff}

In this section, we discuss how the stack-splitting
scheme can be adapted to support the correct semantics
during parallel execution of programs containing
side-effects and other order-sensitive predicates.

\subsection{Order Sensitive Predicates}
A parallel Prolog system that maintains Prolog semantics
reproduces the behavior of  a sequential system (same solutions, 
in the same order, and with the same termination
properties). 
Sequential Prolog systems include
 features that allow
the programmer to introduce a component of sequentiality in the
execution. These may be in the form of facilities to express
side-effects (e.g., I/O) or constructs
to control the order of construction
 of the computation  (e.g., pruning
operations, user-defined search strategies). In a parallel
system, such \emph{Order Sensitive
Components ($\cal OSC$)}---i.e., built-in predicates whose
semantics is tied to the sequential operational semantics
of Prolog---need to be performed in
the \emph{same order as in a sequential execution}; if this requirement
is not met, the parallel computation may  lead to an observable
semantics different
from the one indicated by the programmer \cite{parallellp-survey}.

In the context of Prolog, there are three different
 classes of $\cal OSC$:
\emph{side-effects} predicates (e.g., I/O), 
\emph{meta-logical} predicates (e.g., test
 the instantiation state of variables), 
and \emph{control} predicates (e.g., for pruning branches of the search tree).
In the context of or-parallelism 
only certain classes of \osc
require sequentialization across parallel computations---only side-effects
 and control predicates.
The presence of \osc does not require a 
sequentialization  of the whole execution involved,
only the \osc
themselves need to be sequentialized.
If the \osc are infrequent and spaced 
apart, good speedups
can be obtained, even in a DMP.
The correct order of execution of \osc corresponds
to an in-order  traversal of the computation tree. A specific
\osc $\alpha$
can be executed only if all the \osc
that precede $\alpha$ in the traversal 
have been completed (this assumes also that we do not
have infinite branches in the computation tree). 
Detecting when all the  \osc
 to the left have  been executed is an
undecidable problem,\footnote{It is a fairly simple exercise
to show that the ability to detect precedence of side effects
can be used to decide termination of computations---a known
undecidable problem.} thus
requiring the use of  approximations.
The most commonly used approximation is to execute an \osc only when 
the branch containing it becomes the left-most branch 
in the  tree \cite{aurora-cutseffs}. Thus, we approximate the termination
of the preceding \osc by verifying the termination of the
\emph{branches} that contain them.
Most of the
schemes proposed \cite{se-acta,parallellp-survey}
rely on traversals of the tree, where the
computation attempting an \osc walks up its branch
verifying the termination of all the branches
to its left. These approaches can be  realized
\cite{aurora-cutseffs,muse-journal,peter-dyn} in  presence of a shared 
representation of the computation tree---required to check 
the status of other
executions without  communication. 
These solutions do not scale to  the case of DMP, 
where a shared representation of the computation tree
is not available. 
Simulation of a shared representation is infeasible,
as it leads to unacceptable bottlenecks \cite{karen-phd}. 
Some attempts to generalize mechanisms to handle \osc
to DMPs have been made \cite{araujo-sidef}, but only
at the cost of  sub-optimal scheduling mechanisms.
It is unavoidable to introduce a communication component
to handle \osc in a distributed setting.
We demonstrate that 
 stack-splitting  can be modified to solve this
problem with minimal communication \cite{pals-se}. 
The modification is inspired by the optimal algorithms for
   \osc
studied in \cite{se-acta}. In particular, in the context of
this work we focus on side-effect predicates; we believe
these results can provide the foundations to handle also
cut and pruning operators,  but their 
effective management requires more significant 
changes, e.g.,  to the scheduling 
strategies, and they are not addressed in the scope of this
work.

\subsection{Optimal Algorithms for Order-sensitive Executions}
\label{optimal}
\begin{figure}[htbp]
\centerline{\psfig{figure=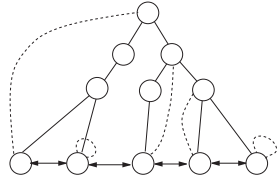,width=.3\textwidth}}
\caption{Data Structure for Order-Sensitive Computations}
\label{optim-fig}
\end{figure}

The problem of efficiently handling \osc
during parallel executions has been pragmatically tackled in a 
variety of proposals \cite{parallellp-survey}. 
Nevertheless, only recently the problem has
been formally studied, 
deriving  solid theoretical foundations regarding the inherent
complexity of testing for leftmostness  in a dynamically changing
tree \cite{se-acta}. 
Let ${\cal T} = \langle N, E \rangle$ be the computational tree
(where $N$ are its nodes and $E$ the current edges). The computation
tree is dynamic; the modifications to the tree can be described by 
two operations: {\sf expand} which adds a (bounded) number of children
to a leaf, and {\sf delete} which removes a leaf from the tree.
Whenever a branch encounters a side-effect, it must check
if it can execute it.
This check boils down to verifying that the branch containing the side-effect
is currently the leftmost active computation in the tree.
If $n$ is the current leaf of the branch where the side-effect is encountered,
its computation is allowed to continue only if
$\mu(n) = \:${\sf root}, where 
$\mu(n)$
indicates the highest node $m$ in the tree  (i.e., closest to the root)
such that $n$ is in the leftmost branch of the subtree rooted 
at $m$. $\mu(n)$ is also known in the parallel logic programming community as 
the {\em subroot node} of $n$ \cite{aurora-cutseffs}. 
%Observe that 
%for the time being, we consider \osc different from cut; the proposed
%approach can be easily modified to accommodate to handle the cut as well. 
Thus, checking if a side-effect can be executed requires the ability of
performing the  operation {\sf find\_subroot}$(n)$ which, given a leaf $n$, 
computes the	node $\mu(n)$.

The work presented in \cite{se-acta} studies the data structure problem 
leading to the following result:
any
sequence of {\sf expand}, {\sf delete}, and {\sf find\_subroot} 
operations can be performed in $O(1)$ time per operation on 
pure pointer machines---i.e., without the need of complex
arithmetic (i.e., the
 solution does not rely on the use of ``large'' labels).
The data structure used to support this optimal solution is based on
maintaining a dynamic list---i.e., a list which allows arbitrary
insertions and deletions to be performed at run-time---which represents the frontier of the tree
(the solid arrows in Figure~\ref{optim-fig}). 
The dynamic list can be updated in $O(1)$ time each time leaves are
added or removed (i.e., when expanding a branch and performing
backtracking). Subroot nodes can be efficiently maintained for each 
leaf (these are depicted by dotted lines in the Figure)---in particular,  each delete operation
affects the subroot node of at most one other leaf.
Identification of the computations 
 an \osc $\alpha$
 depends on can be simply accomplished by
traversing the list of leaves right-to-left from $\alpha$.
Executability (i.e., leftmostness) 
can be verified in constant time by simply checking whether
the subroot of the leaf points to the root of the tree \cite{se-acta}. 
Although the use of an explicit list to maintain the frontier of 
the computation tree has been suggested in other works
(e.g., in the Dharma scheduler \cite{Dharma}), the data
structure which allows its management in constant-time
was  proposed for the first time in \cite{se-acta}. The reader
is referred to \cite{se-acta} for more details.
 
This solution is  feasible in a shared memory
context but requires adjustment in a distributed-memory context. In
the rest of this section we show how stack-splitting can incorporate
a good solution to the problem, following the spirit of this optimal
scheme.

\subsection{Stack-Splitting and Order-sensitive Computations}
\label{sidefex}
Determining the executability of  an \osc $\alpha$ in
a distributed-memory setting
requires two coordinated activities: 
\emph{(a)} determining \emph{what are} the computations to the left of
$\alpha$ in the computation tree---i.e., which agents
have acquired work in branches to the left of $\alpha$\/;
\emph{(b)} determining what is the \emph{status} of the computations to the
left of $\alpha$\/.
On DMPs, both steps require exchange of messages between agents. The main
difficulty is represented by step \emph{(a)}---without the help of a shared
data structure, discovering the position of the different agents 
requires arbitrary  localization messages exchanged between the 
agent in charge of $\alpha$ and all the other agents. 
What we
propose is a shift in perspective, directed from the ideas presented
in Section \ref{optimal}: through a simple modification in the
strategy for stack-splitting, we can guarantee that agents are aware
of the position of their subroot nodes.\footnote{Note that it is practically
infeasible to have all processors know the location of all shared nodes.}
 Thus, instead of having to locate
the subroot nodes whenever an \osc occurs, these are
implicitly located (without added communication) whenever a sharing operation
is performed (a very infrequent operation, compared to the frequency of
$\cal OSC$ steps). Knowledge of the position of the subroot nodes
allows agents to maintain an approximation of the ordering of the
leaves of the tree, which in turn can be used to support the execution of 
step \emph{(b)} above.

In the original stack-splitting procedure---using vertical
splitting (Section \ref{split})---during a sharing operation
the parallel choice-points 
are alternatively  split between two agents.
The agent that is giving the work keeps the bottom-most choice-point, 
the third bottom-most choice-point, the fifth bottom-most choice-point, 
etc. The agent that
receives the work keeps the second bottom-most 
choice-point, the fourth bottom-most choice-point, etc.
In our previous works \cite{iclp99,icpp01} we have demonstrated that this
splitting strategy is effective and leads to good speedups for large classes
of representative benchmarks. The alternation in the distribution of
choice-points is aimed at reducing the danger of focusing
 a particular agent
on a set of fine-grained computations.

This strategy for splitting a computation branch between two agents
has a significant drawback w.r.t. execution of $\cal OSC$,
since the two agents, through backtracking, may arbitrarily move left
or right of each other. This makes it impossible to know a-priori whether one
agent affects the position of the subroot node
of other agents, preventing the detection of the
position of agents in the frontier of the tree.
From Section \ref{optimal} we learn that an agent operating
on a leaf of the computation tree can affect other agents'
subroot nodes only in a limited fashion. The idea can be
easily generalized: if an agent limits its activities to 
the bottom part of a branch, then the number of leaves affected by
the agent is limited and well-defined. This observation leads
to a modified splitting strategy, where the agent giving work
keeps the lower segment of its branch as private, while the
agent receiving work obtains the upper segment of the branch.
This modification guarantees that
the agent receiving work will be always to the 
right of the agent giving the work. 
Since the result of a sharing operation is 
always broadcasted to all the agents---to allow agents to 
maintain an approximate view of the distribution of work---this 
method  also allows each agent to have
an approximate view of the composition of the frontier of the
computation tree.

Observe that this modification to the splitting strategy leads
to a scheduling strategy different from the traditional 
bottom-most scheduling mentioned earlier. Nevertheless, as discussed
in the experimental evaluation section, this modification does
not harm parallel performance in applications with presence
of $\cal OSC$, and it does not relevantly degrade performance
in absence of $\cal OSC$.

The next sections show how this
new splitting strategy can be made effective to support
\osc without losing parallel performance.

\subsubsection{Implementation}

\paragraph{Data Structures:}
In order to support the new splitting strategy and use it to support 
$\cal OSC$
steps, 
each agent will require only two additional data structures:
\emph{(1)} the \emph{Linear Vector} and
\emph{(2)} the \emph{Waiting Queue}.
Each agent keeps and updates a {\it linear vector} which
 consists of an array of agent Ids
that represents the linear ordering of the agents in the 
search tree---i.e., the respective position of the agents within
the frontier of the computation tree (section \ref{optimal}).
The idea behind this {\it linear vector} is 
that whenever an agent wants to execute an $\cal OSC$,
it first waits until there are no agents Ids to its left on 
the {\it linear vector}. Such a status indicates that all the agents that
were operating to the left have completed their tasks and moved to the right
side of the computation tree, and the subroot node has been
pushed all the way to the root of the tree. Once this happens,
the agent can safely execute the $\cal OSC$, being
 left-most in the search tree. 
Initially, the linear vector of all agents contains only the 
Id of the first running agent. In the original bottom-most scheduler 
developed for stack-splitting (Section \ref{ssimpl}),
every time a sharing operation is performed,
a {\tt Send\_LoadInfo} message is broadcast to all agents;
this is used to inform all agents of the change in the
workload and of the agents involved in the sharing.
For every {\tt Send\_LoadInfo} message, each agent updates
 its linear vector by moving the Id of the agent that received
work immediately to the right of the Id of the agent giving
work.
Each agent also maintains a
{\it waiting queue} of  Ids, representing all the
agents that are waiting to execute an \osc
but are located to the right of this
agent.  
Whenever an agent enters
the {\it scheduling} state to ask for work, it  informs
 all agents in its waiting queue 
that they no longer need to wait on it to execute their
$\cal OSC$.

%guo
\paragraph{The Procedure:}
In stack-splitting  (Section \ref{ssimpl}), an agent  can only 
be in one of two states:
\emph{running state} or \emph{scheduling state}. 
In order to handle  $\cal OSC$, we need another state: 
the {\it order-sensitive} state. 
All agents wanting to execute an \osc
will enter
this state until it
is safe for them to execute their $\cal OSC$. 
%The algorithms to handle the $3$ states are illustrated in Fig. \ref{runn_agent},
%\ref{sched_agent}, \ref{side_agent}.
The transition between the  states requires the
introduction of three  types of messages:
{\sf (1)} {\tt Request\_OSC},
{\sf (2)} {\tt OSC\_Acknowledgment}, and 
{\sf (3)} {\tt Reply\_In\_OSC}.
Their detailed explanations are shown in the following scheduling algorithms.

%A {\tt Request\_OSC} will be used by an agent that wants 
%to execute an $\cal OSC$, and sent to all agents
% to its left in its linear vector. 
%When an agent receives a message of this type, and 
%it is not located to the 
%left of the sending agent,
%a reply message with tag {\tt OSC\_Acknowledgment} will be sent back. 
%This message informs the \osc agent that
%it no longer needs to wait on this agent.
%Agents in the order-sensitive
%state are not allowed to share work; requests to share work
%are denied with the {\tt Reply\_In\_OSC} message.

We update the distributed scheduling algorithms as follows to support
handling $\cal OSC$. Only those parts related to handling OSC are presented
in the algorithms. The ignored parts (denoted by {\tt ... ...}) can be found
from the previous algorithms presented in Section~\ref{sched}. 
The scheduling algorithm for an agent in an order-sensitive state 
is described as follows:
{\small
\begin{verbatim}
    send a Request_OSC message to all the agents whose Ids
      are on the left of its own Id in the linear vector;
    while (its own Id is not on the leftmost in the linear vector) {
       get an incoming message; 
       switch (message type) {
          case Request_OSC:
              update the requesting agent's work-load;
              consult the linear vector;
              if (the requesting agent Id is on the right of its own Id)
                  enqueue the requesting agent Id in the waiting queue;
              else
                  reply a message of type OSC_Acknowledgment;
              break;
          case OSC_Acknowledgment:
              update the sending agent's work-load;
              remove the message sender Id from the linear vector;
              break;
          case Send_LoadInfo:
              update the splitting agents' work-load;
              update the linear vector by placing the Id of the agent 
                who receives work to the right of the agent Id giving work;
              if (the agent Id who receives work is on the left of its own ID)
                 send a Request_OSC message to the agent;
              break;
          case Request_Work:
              remove the requester Id from the linear vector;
              reply a message of type Reply_In_OSC;
              V[the requester ID] = 0;
              break;
         }
      }   
   change to the running state to perform the OSC;
   send a Send_LoadInfo message to all other agents;
\end{verbatim}
}

Once an agent arrives to the order-sensitive state, it first 
sends a {\tt Request\_OSC} to all the agents to its left in 
its linear vector. It then waits for
 {\tt OSC\_Ac\-know\-led\-gment} messages from each of them.
An  {\tt OSC\_Acknowledgment} is sent by an agent
when it is no longer to the left of the agent wanting to execute the 
$\cal OSC$. When this message  is received, the Id of the agent
sending it will be removed from the linear vector. The position
of the sending agent will be re-acquired when such agent acquires
more work in the successive scheduling phase.
Notice that when the agent is waiting for these messages, it may
receive {\tt Send\_LoadInfo} messages. If this happens, the agent has to 
update its linear vector. In particular, if due to this sharing operation 
an agent moves to its left, a {\tt Request\_OSC} message needs 
to be sent to this agent as well. Once the agent receives 
{\tt OSC\_Acknowledgment} messages from all these agents, it can safely
perform the $\cal OSC$. And, finally, after the OSC is successfully performed,
a {\tt Send\_LoadInfo} message will be broadcasted to all other agents
with the precise work-load information.

In addition, an agent in an order-sensitive
state is not allowed to share work; requests to share work
are denied with the {\tt Reply\_In\_OSC} message.
Its linear vector can be easily updated by removing the Id 
of the agent requesting work.
Just as we attach load information to messages in the traditional
stack-splitting scheduling algorithm,
we also attach updated load information to these three new messages.

The updated scheduling algorithm for a running agent is described as follows:
{\small
\begin{verbatim}
        while (any incoming message) {
           get an incoming message;   
           switch (message type) {
           case Send_LoadInfo:
                update the linear vector by placing the Id of the agent 
                  who receives work to the right of the agent Id giving work;
                ... ...
           case Request_Work:
                if (local work-load > Splitting Threshold) {
                   update the linear vector by placing the requesting
                     agent Id to the right of its own Id;
                   ... ...      % stack-splitting
                 }
                 else { % no stack-splitting
                   remove the requester Id from the linear vector;
                   ... ...
                 }
                 break;
            case Request_OSC:
                 consult the linear vector;
                 if (the requesting agent Id is on the right of its own Id)
                    enqueue the requesting agent Id in the waiting queue;
                 else
                    reply a message of type OSC_Acknowledgment;
                 break;
            }
         }
\end{verbatim}}

%When an agent is in running state and 
%gives work to another agent, 
%a  {\tt Send\_LoadInfo} message is sent to all other agents informing
%them of  the new load information. 
%Each  agent receiving a {\tt Send\_LoadInfo} message 
%updates its own linear vector by placing the Id of agents
%that received the work to the right of the agent giving work. 

%When an agent is in  scheduling state and 
%receives work from another agent $p$, it will 
%also update its linear vector by placing 
%its Id to the immediate right of $p$.

When an agent is in running state and receives 
a {\tt Request\_OSC} message,
it consults its linear vector and reacts in the following way.
If the Id of the agent wanting to execute an \osc
is to its right in the linear vector,  
the Id of the requesting agent
 is inserted in the waiting queue.
When the running agent runs out
 of work and moves to the scheduling state, an 
 {\tt OSC\_Acknowledgment} message will be
sent back to the agent wanting to execute the $\cal OSC$.
If the Id of the agent wanting to execute 
the \osc is to its left,
an {\tt OSC\_Acknowledgment} message 
is immediately sent back to the agent wanting to 
execute the $\cal OSC$.
This means that the running agent is no longer
to the left of the agent
wanting to execute the $\cal OSC$.

The updated scheduling agent's algorithm can be briefly described as follows:
{\small
\begin{verbatim}
        dequeue all the agent Ids from the waiting queue and
          send an OSC_Acknowledgment to all of them;
        while (1) {
           ... ...
           while (!matched) {
              get an incoming message;
              switch (message type) {
                 case Reply_With_Work:
                      update the linear vector by placing the own Id
                        to the right of the message sender Id;
                      ... ...
                 case Reply_Without_Work:
                      if (the work-load of the message sender is 0)
                         remove the message sender Id from the linear vector;
                      ... ...  
                 case Request_Work:
                      remove the requester Id from the linear vector;
                      ... ...
                 case Send_LoadInfo:
                      update the linear vector by placing the Id of
                        the agent who receives work to the right of the
                        agent Id giving work;
                      ... ...
                 case Reply_In_OSC:
                      update the work-load of the message sender to 1;
                      break;       
                 case Request_OSC:
                      update the work-load of the message sender;
                      reply a message of type OSC_Acknowledgment;
                      break;
               }
           }
\end{verbatim}}

When an agent enters the scheduling state, it dequeues 
all the Ids from its waiting queue and
sends an {\tt OSC\_Acknowledgment}  
to all these agents, informing them that 
it is no longer to their left.
When a scheduling agent receives a {\tt Reply\_In\_OSC},
which means the current agent with the highest work-load is in
an order-sensitive state,
it then updates the work-load of that agent to {\tt 1} 
so that in the next round the agent will choose another agent 
with high work-load to request work from. 
The precise work-load will be updated later after the agent in the
order-sensitive state becomes a running-state agent.

\subsubsection{Implementation Details}

\paragraph{Partitioning Ratios:}
The stack-splitting modification  divides 
the stack of parallel choice-points
into two contiguous partitions, where the bottom partition
 is kept by the agent giving work and the upper 
partition is given away.
This stack-splitting modification guarantees that the
agent that receives work will be to the immediate right 
of the other agent.
The question is what is the partitioning ratio 
that will produce the best results? 
We first tried using a  partition 
where the agent that is giving work keeps the bottom 
half of the branch and
only gives away the top half.
After experimenting with lots of different partition ratios,
we found out that with a partition ratio of $3/4-1/4$ 
where the agent that is giving work keeps the bottom $3/4$
of the parallel choice-points and gives away the top $1/4$ 
of the parallel choice-points, 
our benchmarks without side-effects obtain excellent 
speedups---similar to our original alternating splitting \cite{icpp01}. 
When we run our benchmarks with side-effects, the partition ratio
of $3/4-1/4$ performed superior to the partition ratio of $1/2$. One
of the reasons is that it is common to have more side-effects towards
the bottom part of the computation tree; thus, using the proposed
partition we assign smaller chunks of work, but with a greater
probability of not encountering side-effects.
Additionally, keeping larger numbers of side-effects
locally reduces the number of interactions.

\begin{figure}[htb]
\centerline{\psfig{figure=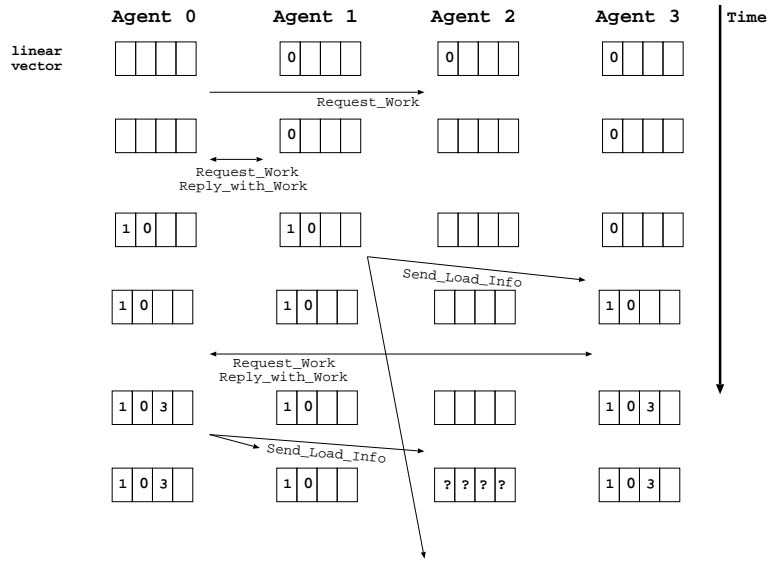,width=.8\textwidth}}
\caption{Example of Messages Out of Order}
\label{outof}
\end{figure}

\paragraph{Messages Out of Order:}
{\tt Send\_LoadInfo} messages may arrive out of order 
and then the linear vectors may be outdated.
E.g., agent 2 receives  from 
agent 0 a {\tt Request\_Work} message but decides not
to share work.
Since agent 0 is requesting work,
agent 2 removes 0 from its linear vector.  Later on, agent 0 
gets work from agent 1,
and agent 1 broadcasts a  {\tt Send\_LoadInfo} message.
Afterwards,
agent 0 gives work to agent 3 and also
 broadcasts a {\tt Send\_LoadInfo}
message. Now, suppose that agent 2 receives the 
second {\tt Send\_LoadInfo} 
message first and the first {\tt Send\_LoadInfo} next. 
When
agent 2 tries to insert 3 to the immediate right
 of 0 in the linear vector,
0 is not located
and therefore 3 cannot be inserted (see Figure~\ref{outof}). 
MPI (used in our system for agent communication) does not 
guarantee that two messages sent from different agents at different
times will arrive in the order that they were sent. 
The scenario presented above can be avoided if, in every sharing operation,
both involved agents
broadcast a {\tt Send\_LoadInfo} 
message to all the other agents. In this case every agent
will be informed that a sharing operation occurred either by the giver 
or by the receiver of
work. Agent 2 in the above scenario will first know 
that agent 0 obtained work
from agent 1, and then will know that agent 0 gave work to agent 3.
Duplication of {\tt Send\_LoadInfo} messages is handled
through the use of
two dimensional arrays 
$send1$ and $send2$ of size $N^2$\/, where $N$ is the total number of 
agents; $send1[i][j]$ ($send2[i][j]$) is incremented
when a sharing message  from $i$ to $j$ is received
from agent $i$ ($j$). Thus, $send1[i][j]$ and $send2[i][j]$ keep track
of how many times $i$ and $j$ have shared work; $send1$ records how many
times $i$ notified of a sharing with $j$ and $send2$ records how many
times $j$ notified of a sharing with $i$.
The linear vector will be updated only if $send1[i][j] > send2[i][j]$ 
($send2[i][j] > send1[i][j]$) and the message comes from agent $i$ ($j$).

\iffalse
\subsection{Discussion}

The modified stack-splitting strategy described in this
section provides a strategy to bias the exploration
of the branches in the search tree in the left-to-right order.
As far as exploring the search space
with a bias towards exploring the branches to the 
left is concerned, it will depend on the choice-point
splitting strategy used. Consider the choice-point with
alternatives {\tt a1} through {\tt a5} shown
in Figure~\ref{distr}(i). Two possible
splittings are shown in Figures~\ref{distr}(ii) and \ref{distr}(iii).  
In the first one (Figure~\ref{distr}(ii)), the 
list of alternatives is split
in the middle: agent P1 
will be  working on the left half
of the tree rooted at this choice-point {\tt a}, agent
P2 on the right half. 
In contrast,
in Figure~\ref{distr}(iii), the untried alternatives are distributed
alternately between the two choice-points. This splitting strategy is more
likely to produce a search that is biased to the left. This suggests
that modifications similar to the one presented in this work
can be extended to the case of horizontal splitting.

\begin{figure}[htb]
\centerline{\psfig{figure=distr.ps,width=0.75\textwidth}}
\caption{Distribution of Unexplored Alternatives}
\label{distr}
\end{figure}
 \fi

%%%
%%%
%%%

%%%%%%%%%%%%%%%%%%%%%%%%%%%%%%%%%%%%%%%%%%%%%%

\section{Performance Results}
\label{results}

In this section, we present experimental results and their evaluations
obtained from two implementations of the proposed methodologies---one
developed on a shared-memory platform and one on a Beowulf platform.
All the timings proposed have been obtained as an average over 
10 consecutive runs (excluding the lowest and highest times), executed
on lightly loaded machines.

\subsection{Shared Memory Implementation}

The stack-splitting procedure has been implemented on top of the
commercial ALS Prolog system using the MPI library for
message passing---specifically, the MPI-1 library 
natively provided by Solaris 5.9 (HPC 4.0). 
The whole system runs on a Sun Enterprise 4500
with fourteen processors (Sparc 400Mhz with 4GB of memory). 
While the Sun Enterprise is a SMP, it should
be noted that all communication---during scheduling, copying, splitting, 
etc.--- is done using messages. This has  enabled an easy migration
of the system to a Beowulf machine.
The timing results in seconds from our incremental stack-splitting system
on the 14 processor Sun enterprise are presented in Table~\ref{incre}. This system is based on the
scheduling strategy described in Section \ref{sched}.

The benchmarks that we have used to test our system are the following.
The \emph{9 Costas} and \emph{8 Costas} benchmarks compute the 
Costas sequences\footnote{Costas sequences are special numeric series 
used in signal processing.}
of length 9 and 8 respectively. 
The \emph{Knight} benchmark consists of finding a path of knight-moves on a
chess-board of size 5, starting at (1,1) and finishing at (1,5),
and visiting every square on the board just once.
The \emph{Stable} benchmark is a simple engine to compute the models of a logic
program with negation.
The \emph{Send More} benchmark consists of solving the classical crypto-arithmetic
puzzle.
The \emph{8 Puzzle} benchmark is a solution to the puzzle involving a
3-by-3 board with 8 numbered tiles.
The \emph{Bart} benchmark is a simulator used to test the safety of the controller for a train.
The \emph{Solitaire} benchmark is a solution to the standard game involving
a triangular board with pegs and one
empty hole.  
The \emph{10 Queens} and \emph{8 Queens} benchmarks consist of placing a number of queens on a chessboard
so that no two queens attack each other.
The \emph{Hamilton} benchmark consists of finding a closed path through
a graph such that all the nodes of the graph are visited once.
The \emph{Map Coloring} benchmark consists of coloring a planar map.

The \emph{9 Costas},\emph{8 Costas}, \emph{Knight}, \emph{Stable}, \emph{10 Queens}, \emph{8 Queens}, 
\emph{Hamilton}, and \emph{Map Coloring} benchmarks compute all the possible solutions.  
The \emph{Send More}, \emph{8 Puzzle}, \emph{Bart}, and \emph{Solitaire} benchmarks 
stop at the first 
solution (observe that \emph{Bart} actually has a unique solution).\footnote{An ad-hoc
pruning mechanism is used to cut at the first solution.}
The \emph{9 Costas}, \emph{8 Costas}, and \emph{Bart} benchmarks are fairly large programs, while
the rest are simpler.
However, all benchmarks provide sufficiently
different program structures to extensively test the behavior of the parallel engine.

\begin{table}[htb]
{\footnotesize
\begin{center}
\begin{tabular}{|c|c|c|c|c|c|}
\hline\hline
{\bf Benchmark} & \multicolumn{5}{c|}{\bf \# Agents}\\
 & {\bf 1} & {\bf 2} & {\bf 4} & {\bf 8} & {\bf 14}\\
\hline\hline
\emph{9-Costas}       & 715.369 & 368.298 (1.94) & 184.141 (3.88) & 92.165 (7.76)& 53.453 (13.38) \\
\hline
\emph{Stable}         & 653.705 & 368.943 (1.77) & 185.474 (3.52)& 92.811 (7.04)& 53.860 (12.13)\\
\hline
\emph{Knight}         & 275.737 & 141.213 (1.95)& 70.528 (3.9)& 35.539 (7.75)& 22.403 (12.3)\\
\hline
\emph{Send More}      & 115.183 & 65.271 (1.76)& 31.447 (3.66)& 16.496 (6.98)& 9.686 (11.89)\\
\hline
\emph{8-Costas}       & 66.392 &  34.281 (1.93)& 17.192 (3.86)& 8.680 (7.64)& 5.202 (12.76)\\
\hline
\emph{8-Puzzle}       & 52.945 & 29.601 (1.78)& 15.026 (3.52)& 7.845 (6.74)& 4.754 (11.13)\\
\hline
\emph{Bart}           & 25.562 & 15.411 (1.65)& 6.868 (3.72)& 3.577 (7.14)& 2.144 (11.93)\\ 
\hline
\emph{Solitaire}      & 12.912 & 7.598 (1.69)& 3.813 (3.38)& 2.029 (6.36)& 1.335 (9.67)\\
\hline
\emph{10-Queens}      & 7.575  & 3.922 (1.93)&  2.087 (3.62)&  1.378 (5.49) & 1.141 (6.63) \\
\hline
\emph{Hamilton}       & 6.895  & 3.879 (1.77)&  1.940 (3.55)&  1.151 (5.99) & 0.761 (9.06)\\
\hline
\emph{Map Coloring}   & 2.036  & 1.298 (1.56)&  0.696 (2.92)&  0.479  (4.25)& 0.430 (4.73)\\
\hline
\emph{8-Queens}       & 0.306  & 0.198 (1.54)&  0.143 (2.13)&  0.157 (1.94) & 0.149 (2.05)\\
\hline\hline
\end{tabular}
\end{center}}
\caption{Incremental Stack-splitting on Shared Memory (time in seconds and speedups)}
\label{incre}
\end{table}

\begin{figure*}
\begin{center}
\begin{minipage}[b]{.31\textwidth}
\psfig{figure=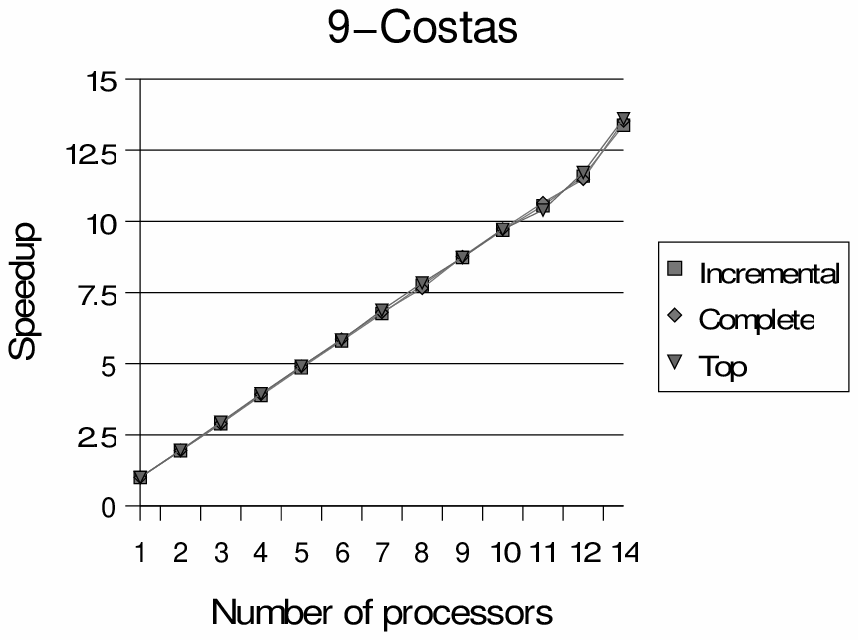,width=\textwidth}
\end{minipage}
\begin{minipage}[b]{.31\textwidth}
\psfig{figure=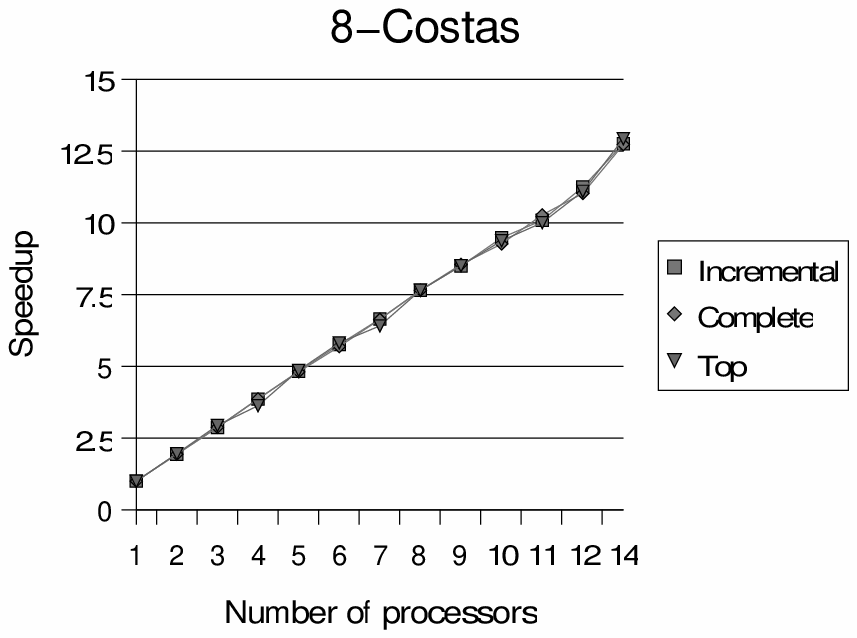,width=\textwidth}
\end{minipage}
\begin{minipage}[b]{.31\textwidth}
\psfig{figure=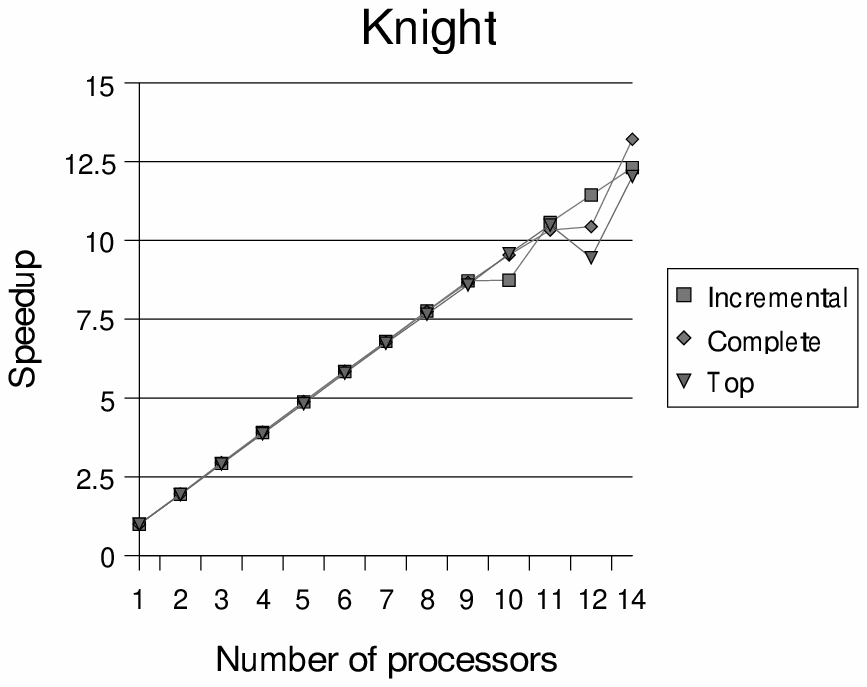,width=\textwidth}
\end{minipage}
\begin{minipage}[b]{.31\textwidth}
\psfig{figure=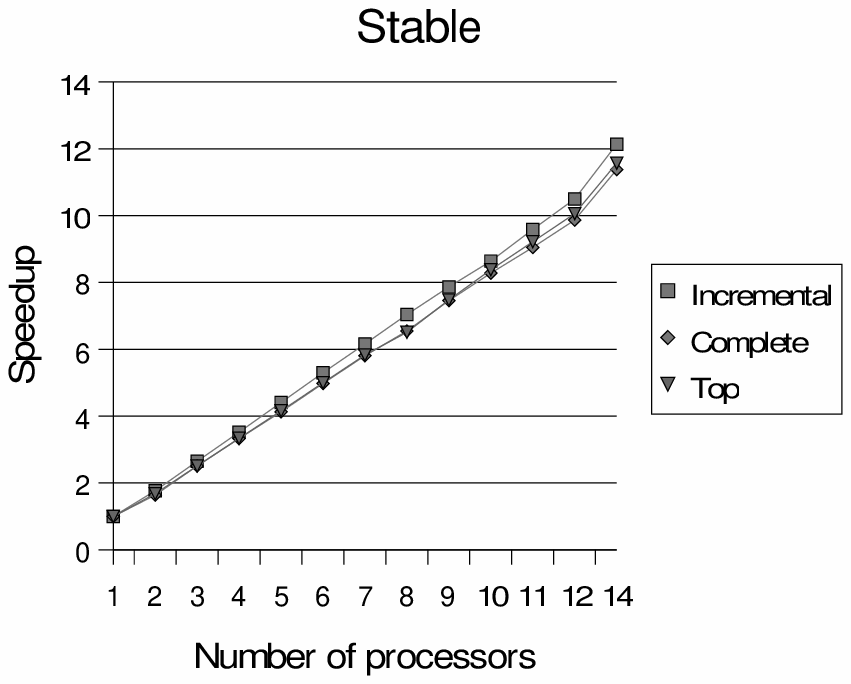,width=\textwidth}
\end{minipage}
\begin{minipage}[b]{.31\textwidth}
\psfig{figure=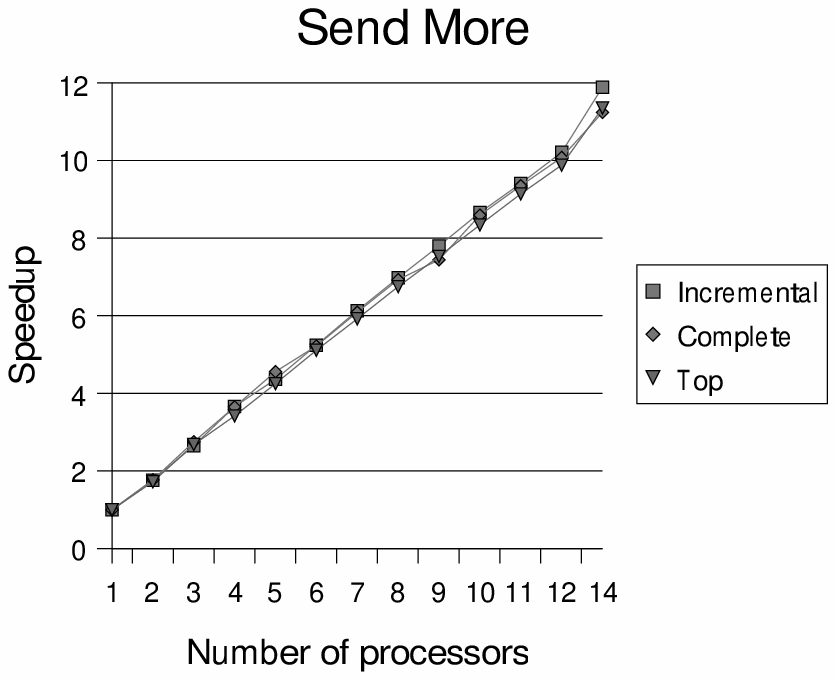,width=\textwidth}
\end{minipage}
\begin{minipage}[b]{.31\textwidth}
\psfig{figure=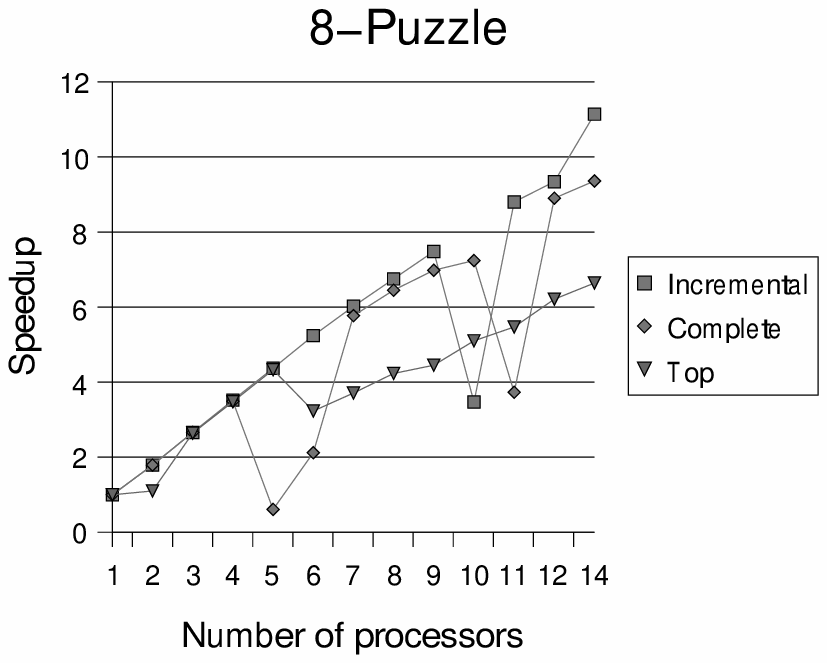,width=\textwidth}
\end{minipage}
\begin{minipage}[b]{.31\textwidth}
\psfig{figure=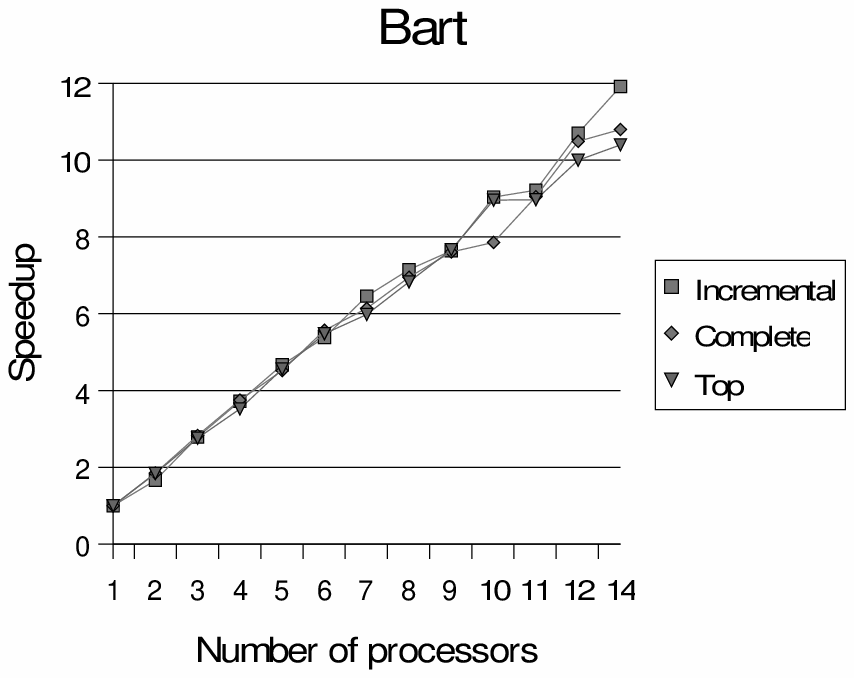,width=\textwidth}
\end{minipage}
\begin{minipage}[b]{.31\textwidth}
\psfig{figure=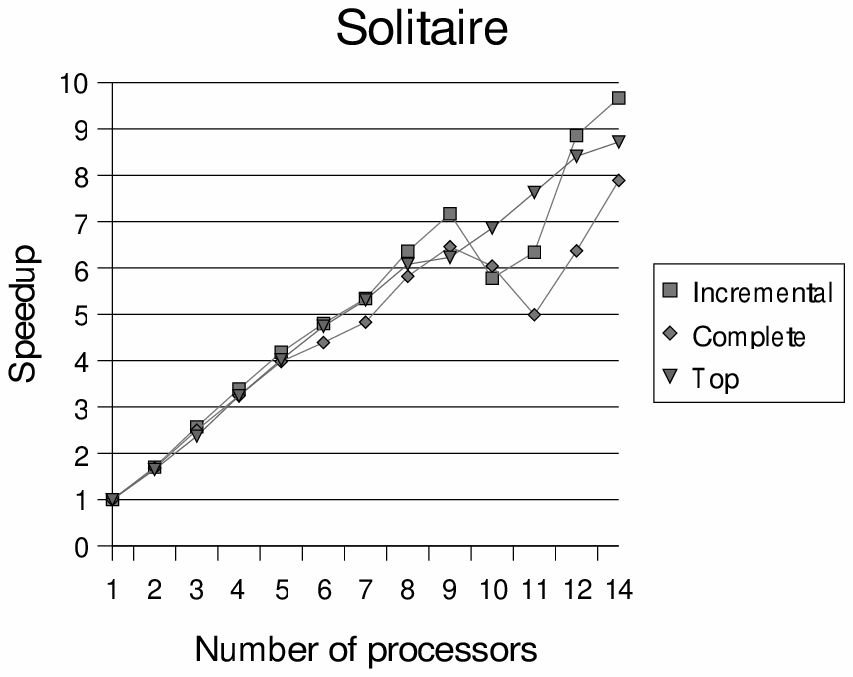,width=\textwidth}
\end{minipage}
\begin{minipage}[b]{.31\textwidth}
\psfig{figure=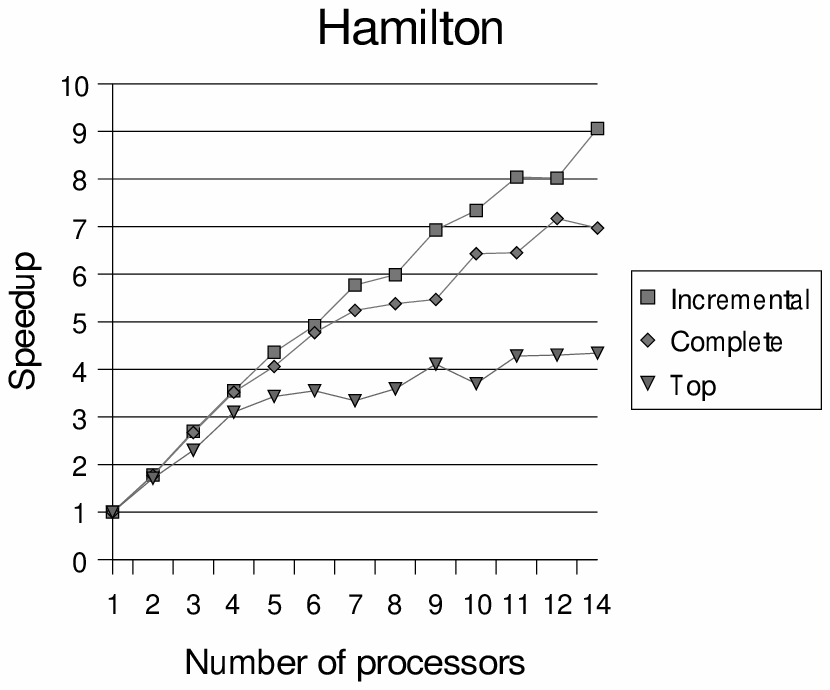,width=\textwidth}
\end{minipage}
\begin{minipage}[b]{.31\textwidth}
\psfig{figure=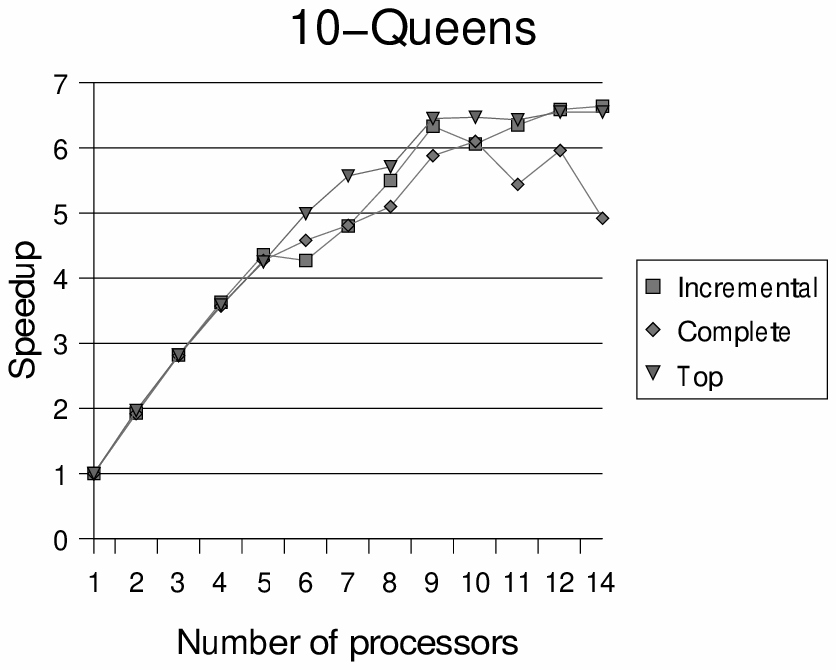,width=\textwidth}
\end{minipage}
\begin{minipage}[b]{.31\textwidth}
\psfig{figure=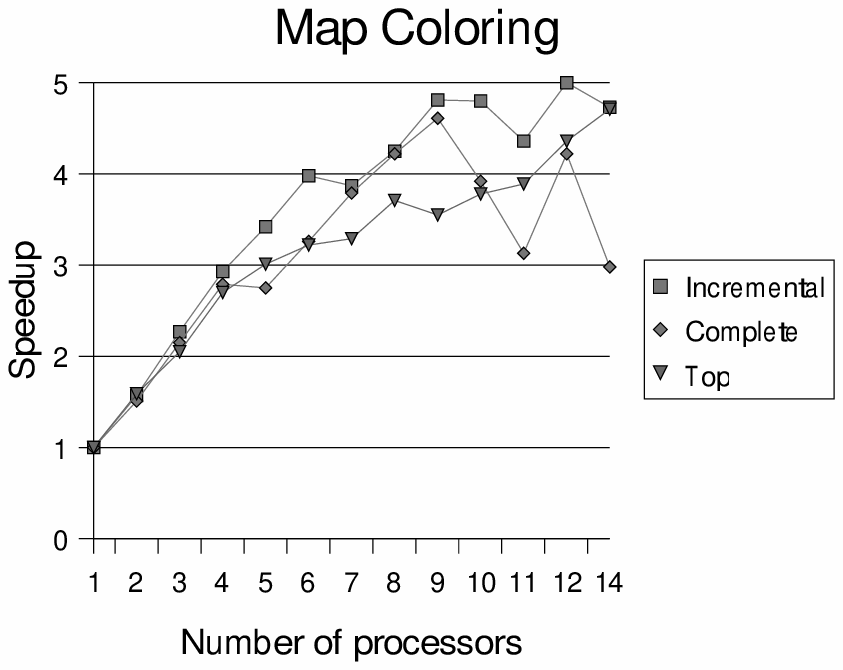,width=\textwidth}
\end{minipage}
\begin{minipage}[b]{.31\textwidth}
\psfig{figure=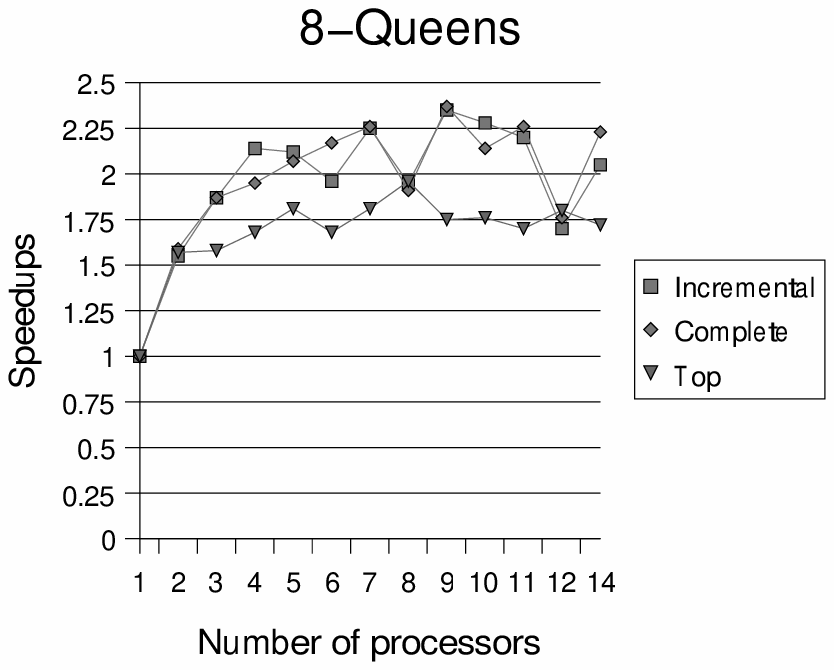,width=\textwidth}
\end{minipage}
\caption{Comparison of Speedups using Complete copying, Incremental Copying, and Top Most Scheduling (Shared Memory)}
\label{speedups2}
\end{center}
\end{figure*}

We observe that for benchmarks with substantial running time (\emph{large benchmarks}),
i.e., \emph{9-Costas}, \emph{8-Costas}, \emph{Knight}, and \emph{Stable},
the speedups are very good. 
We also observe that for benchmarks with not so substantial but also not very small
running time (\emph{medium benchmarks}), i.e., \emph{Send More}, \emph{8-Puzzle}, 
\emph{Bart}, \emph{Solitaire}, and \emph{Hamilton},
the speedups are still quite good.
See Figure~\ref{speedups2} under the label \emph{Incremental}.
Nevertheless, our system is reasonably efficient, given that even for
small benchmarks it can produce reasonable speedups.

In order to compare our incremental stack-splitting system we have also
implemented two other techniques using \emph{non-incremental stack-copying}:
we copy the entire WAM data areas when sharing work instead of
copying them incrementally as described above. One of these
techniques is based on stack-splitting, and the other is based on 
{\it scheduling on
top-most choice-point}: this methodology transfers
between agents only the highest (i.e., closer to the root)
choice-point  in the computation tree which 
contains unexplored alternatives. Observe that we employed non-incremental
copying with top-most scheduling since our previous experiments did not 
indicate a significant impact of incremental copying in presence of top-most
scheduling.
The timing results in seconds
from these other systems are presented in Tables~\ref{comple} and ~\ref{top}.
These two systems also used the scheduling strategy described above.
The speedups for these systems are shown in
Figure~\ref{speedups2} under the labels \emph{Complete} and \emph{Top}.

Most  benchmarks show that the incremental stack-splitting system obtains higher
speedups than the non-incremental systems. Between the non-incremental systems,
the stack-splitting system performs better in most of the benchmarks than the 
scheduling on top-most choice-point system. This is particularly evident in
the case of the \emph{Hamilton} benchmark (Figure~\ref{speedups2}).
Some of the benchmarks (\emph{9-Costas}, \emph{8-Costas}, and \emph{Knight}) show almost no difference in
performance among the three systems. 
One of the reasons why this is happening is that during the execution of these benchmarks
there are  only very few parallel choice-points which are given away or split per sharing; in
particular, by analyzing the source code for these benchmarks, we can see that in the 
three benchmarks just one parallel choice-point contains all the parallel work.

\iffalse
Observe that in Figure \ref{speedups2} some benchmarks show an irregular 
behavior for larger number of processors---we believe this is due to the inability
of the Solaris' scheduler to properly utilize all CPUs available.
\fi

Finally, the incremental stack-splitting system introduces a
reasonably small overhead with respect to the original sequential ALS Prolog system. Our
PALS system, on a single agent, is on average 5\% slower than the
sequential ALS system.

\begin{table}[htb]
{\footnotesize
\begin{center}
\begin{tabular}{|c|c|c|c|c|c|}
\hline\hline
{\bf Benchmark} & \multicolumn{5}{c|}{\bf \# Agents}\\
 & {\bf 1} & {\bf 2} & {\bf 4} & {\bf 8} & {\bf 14}\\
\hline\hline
\emph{9-Costas}       & 715.963 & 366.385 (1.95)& 182.654 (3.91)& 93.602 (7.64)& 52.901 (13.53)\\
\hline
\emph{Stable}         & 614.582 & 374.259 (1.64)& 184.404 (3.33)& 93.884 (6.54)& 54.022 (11.37)\\
\hline
\emph{Knight}         & 276.849 & 141.118 (1.96)& 70.568 (3.92)& 35.741 (7.74)& 20.958 (13.2)\\
\hline
\emph{Send More}      & 116.518 & 65.936 (1.76)& 31.892 (3.65)& 16.882 (6.9)& 10.364 (11.24)\\
\hline
\emph{8-Costas}       & 66.221 &  34.053 (1.94)& 17.126 (3.86)& 8.656 (7.65)& 5.202 (12.72)\\
\hline
\emph{8-Puzzle}       & 52.909 & 29.615 (1.78)& 15.148 (3.49)& 8.206 (6.44)& 5.654 (9.35)\\
\hline
\emph{Bart}           & 25.734 & 13.898 (1.85)& 6.863 (3.74)& 3.704 (6.94)& 2.382 (10.8)\\ 
\hline
\emph{Solitaire}      & 12.676 & 7.552 (1.67)& 3.910 (3.24)& 2.177 (5.82)& 1.606 (7.89)\\
\hline
\emph{10-Queens}      & 7.557  & 3.935 (1.92)&  2.116 (3.57)&  1.483  (5.09)& 1.535  (4.92)\\
\hline
\emph{Hamilton}       & 6.908  & 3.910 (1.76)&  1.963 (3.51)& 1.284 (5.38)  & 0.991 (6.97)\\
\hline
\emph{Map Coloring}   & 2.009  & 1.332 (1.5)&  0.721 (2.78)& 0.476 (4.22)  & 0.675 (2.97)\\
\hline
\emph{8-Queens}       & 0.308 & 0.194 (1.58)& 0.158 (1.94)& 0.161 (1.91)& 0.138 (2.23) \\
\hline\hline
\end{tabular}
\end{center}}
\caption{Complete Stack-splitting on Shared Memory (time in seconds and speedups)}
\label{comple}
\end{table}

\begin{table}[htb]
{\footnotesize
\begin{center}
\begin{tabular}{|c|c|c|c|c|c|}
\hline\hline
{\bf Benchmark} & \multicolumn{5}{c|}{\bf \# Agents}\\
 & {\bf 1} & {\bf 2} & {\bf 4} & {\bf 8} & {\bf 14}\\
\hline\hline
\emph{9-Costas}       & 756.785 & 385.251 (1.96)& 192.157 (3.93) & 96.560 (7.83)& 55.602 (13.61)\\
\hline
\emph{Stable}         & 644.989 & 384.961 (1.67)& 192.991 (3.34)& 99.071 (6.51)& 55.764 (11.56)\\
\hline
\emph{Knight}         & 270.672 & 139.307 (1.94)& 69.951 (3.86)& 35.338 (7.65)& 22.504 (12.02)\\
\hline
\emph{Send More}      & 111.345 & 64.650 (1.72)& 32.562 (3.41)& 16.504 (6.74)& 9.806 (11.35) \\
\hline
\emph{8-Costas}       & 70.362 &  35.899 (1.95)& 19.383 (3.63)&  9.197 (7.65)& 5.441 (12.93)\\
\hline
\emph{8-Puzzle}       & 53.843 & 48.754 (1.1)& 15.490 (3.47)& 12.731 (4.22)& 8.111 (6.63)\\
\hline
\emph{Bart}           & 26.419 & 14.378 (1.83)& 7.513 (3.51)& 3.870 (6.82)& 2.540 (10.4) \\  
\hline
\emph{Solitaire}      & 11.883 & 7.187 (1.65)& 3.664 (3.24)& 1.955 (6.07)& 1.363 (8.71)\\
\hline
\emph{10-Queens}      & 7.595 & 3.857 (1.96)& 2.117 (3.58)& 1.330 (5.71)& 1.160 (6.54) \\
\hline
\emph{Hamilton}       & 6.964 & 4.061 (1.71)& 2.246 (3.1)& 1.941 (3.58)& 1.606  (4.33)\\
\hline
\emph{Map Coloring}   & 2.207 & 1.389 (1.58)& 0.816 (2.7)& 0.595 (3.7)& 0.469  (4.7)\\
\hline
\emph{8-Queens}       & 0.304 & 0.194 (1.56)& 0.181 (1.67)& 0.155 (1.96)& 0.177 (1.71)\\
\hline\hline
\end{tabular}
\end{center}}
\caption{Top-most Scheduling on Shared Memory (time in seconds and speedups)}
\label{top}
\end{table}

\subsection{Beowulf Implementation}
\subsubsection{Stack-Splitting}

The stack-splitting procedure has been implemented by modifying the
commercial ALS Prolog system, using the MPI library  for
message passing---i.e., the mpich MPI-1 installation natively
supported by Myrinet (an instance of mpich 1.2.5). The whole system runs on a distributed-memory 
machine (a network of Xeon 1.7GHz nodes connected by Myrinet-2000 Switches).  
All communication---during scheduling, copying, splitting, etc.---is done 
using explicit message passing via MPI.

\begin{figure}
\begin{center}
\begin{minipage}[b]{\textwidth}
\psfig{figure=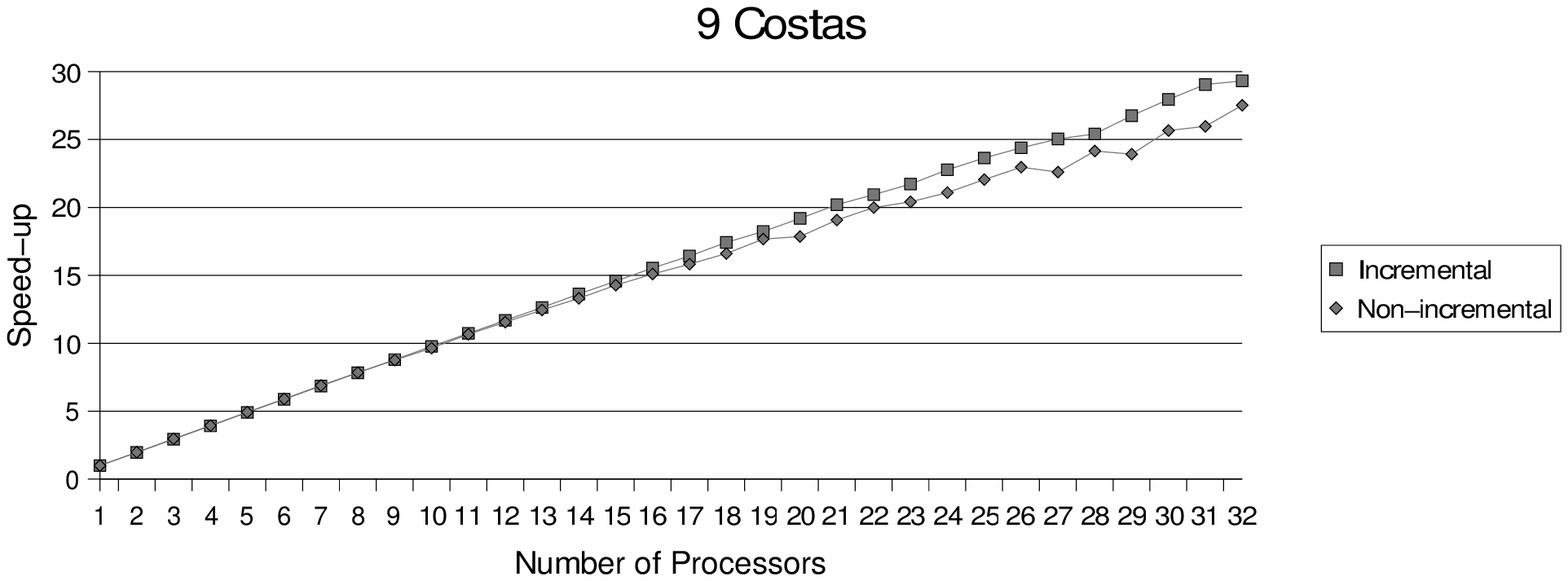,width=.53\textwidth}
\psfig{figure=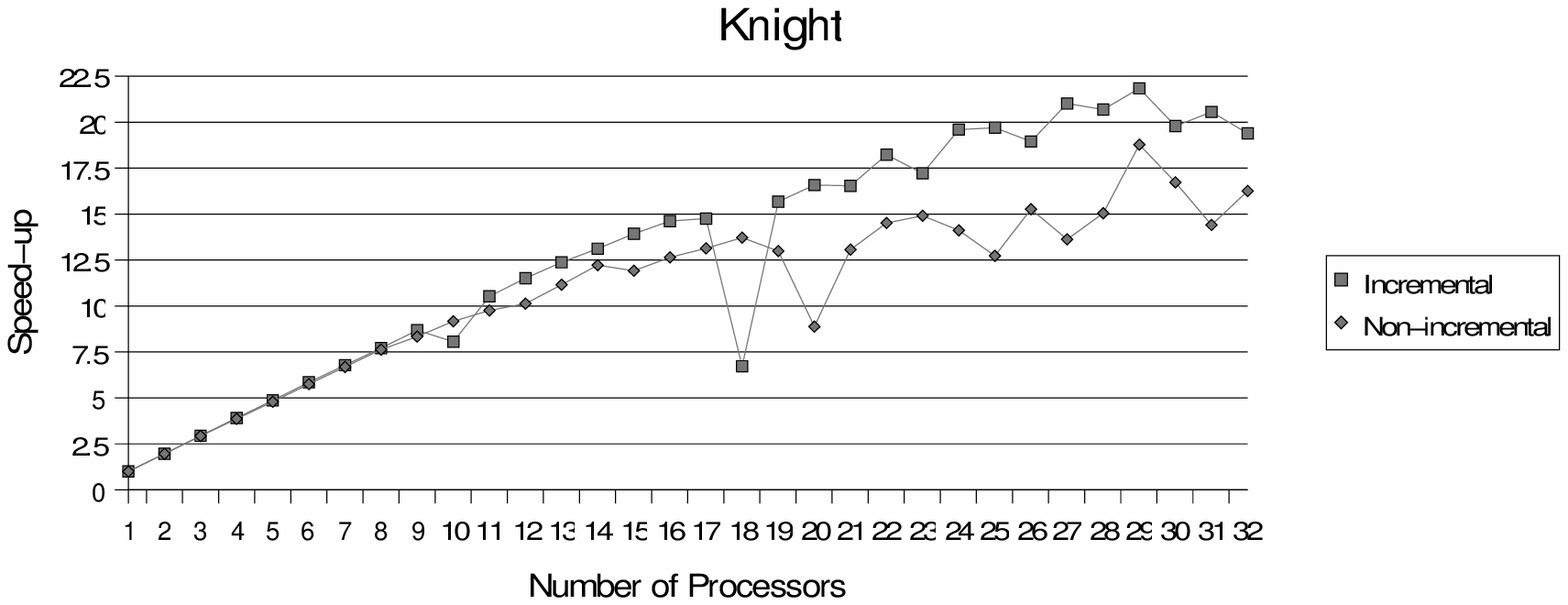,width=.53\textwidth}
\end{minipage}
\begin{minipage}[b]{\textwidth}
\psfig{figure=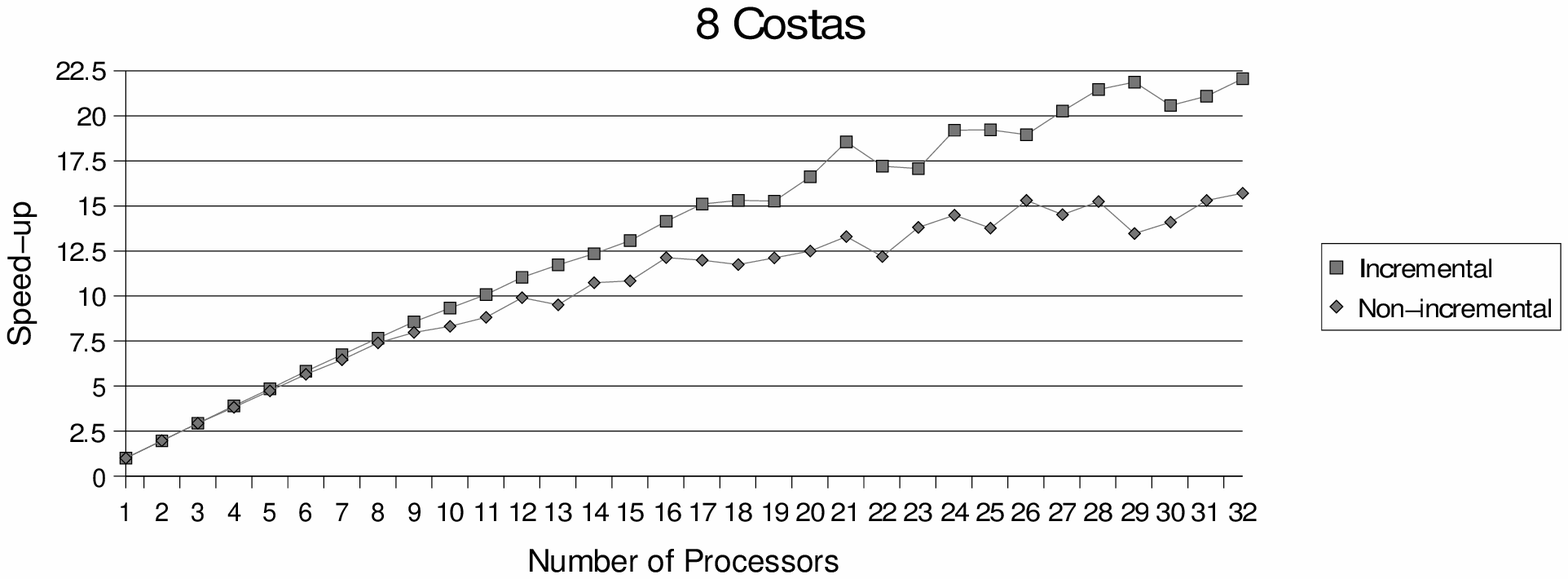,width=.53\textwidth}
\psfig{figure=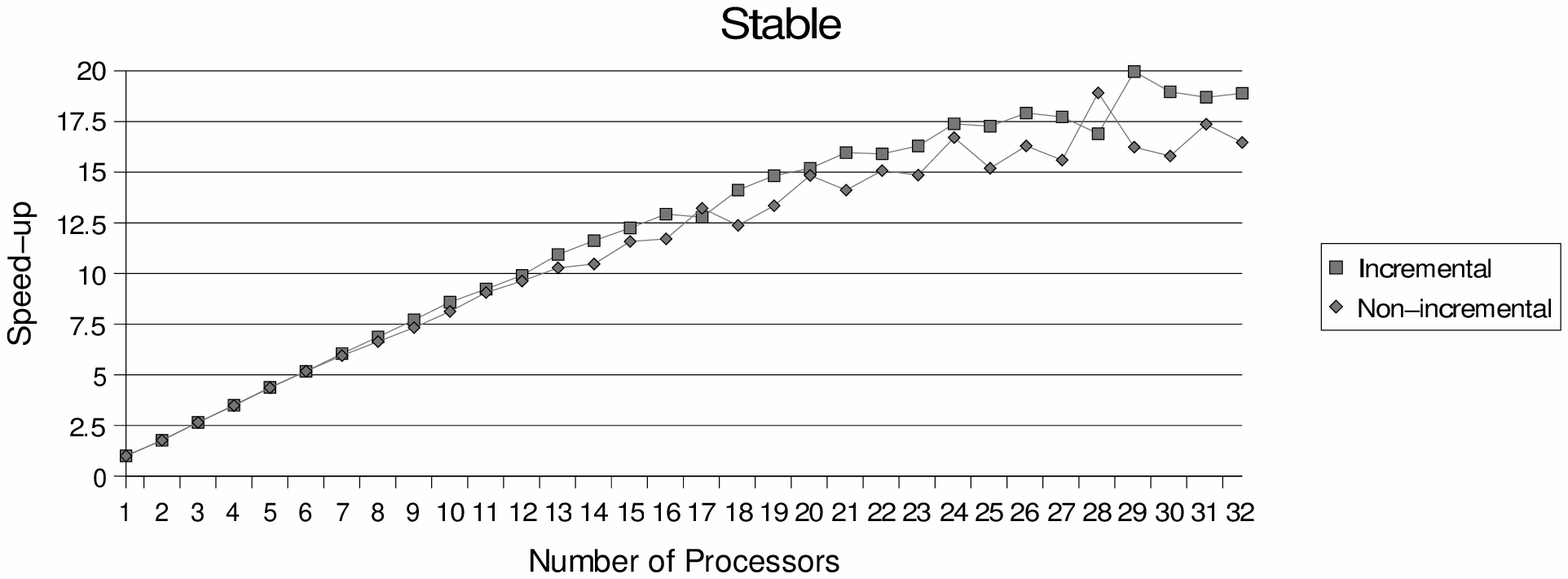,width=.53\textwidth}
\end{minipage}
\begin{minipage}[b]{\textwidth}
\psfig{figure=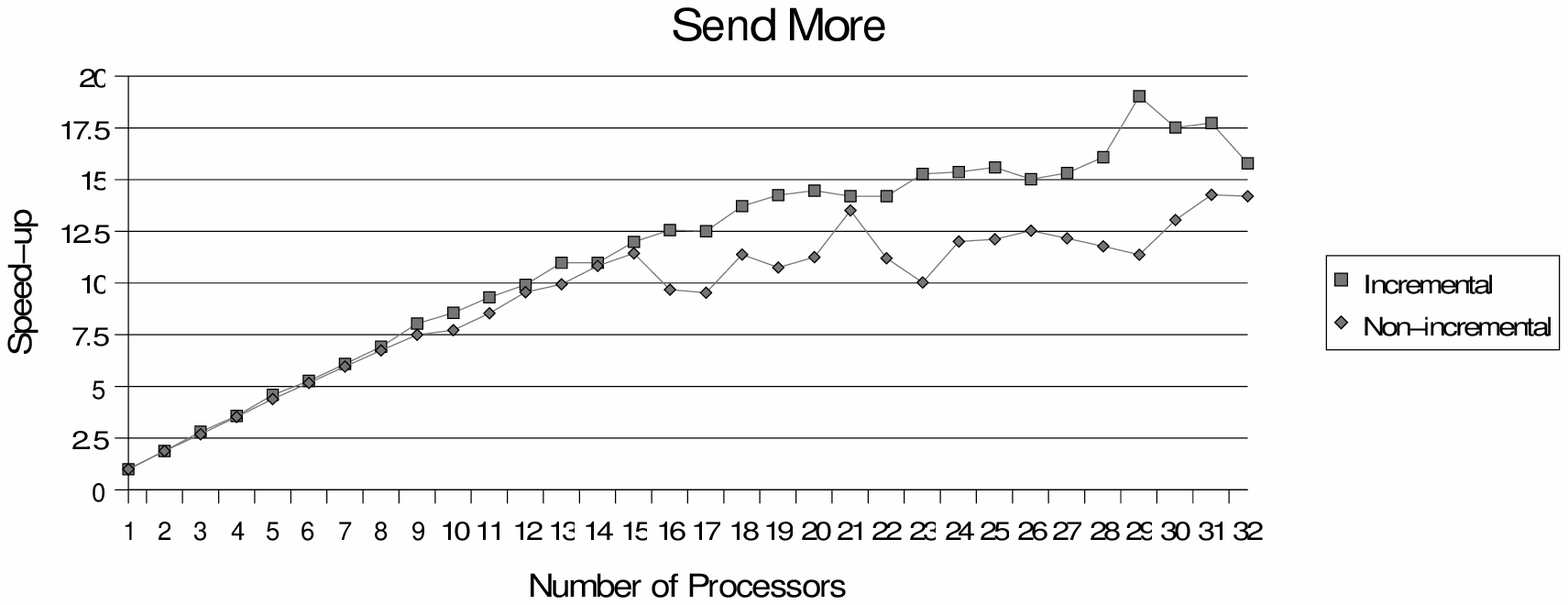,width=.53\textwidth}
\psfig{figure=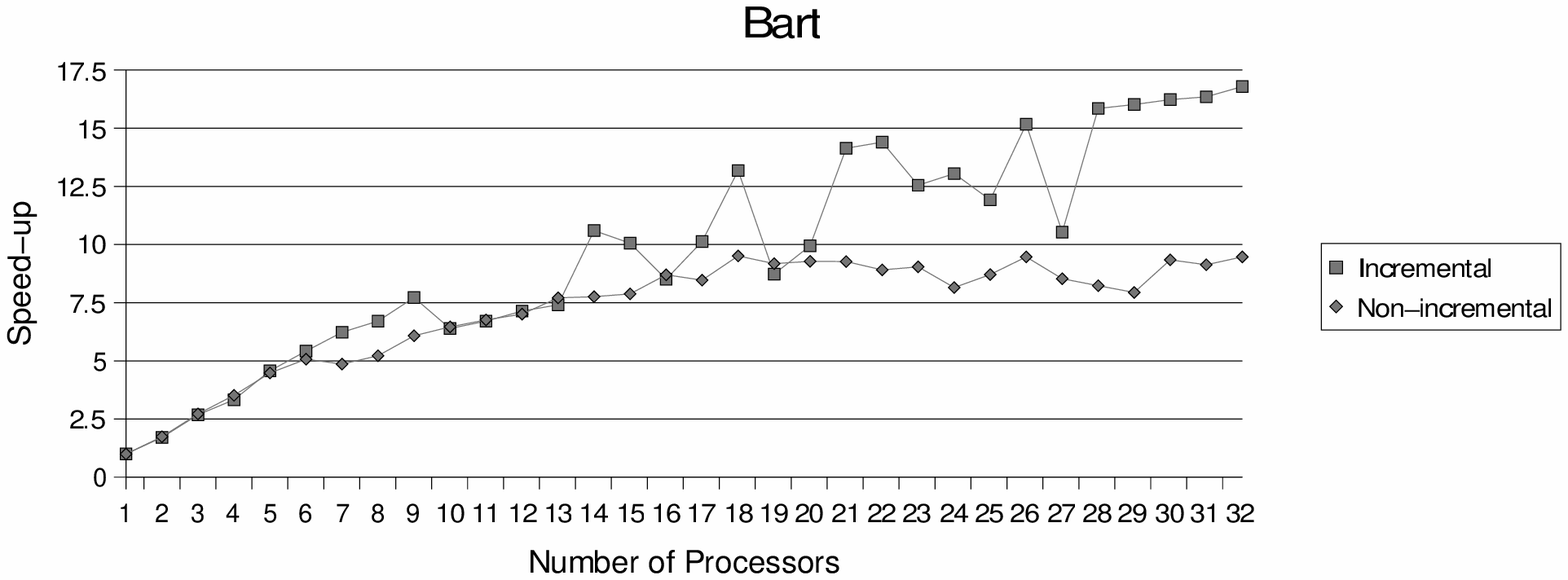,width=.53\textwidth}
\end{minipage}
\begin{minipage}[b]{\textwidth}
\psfig{figure=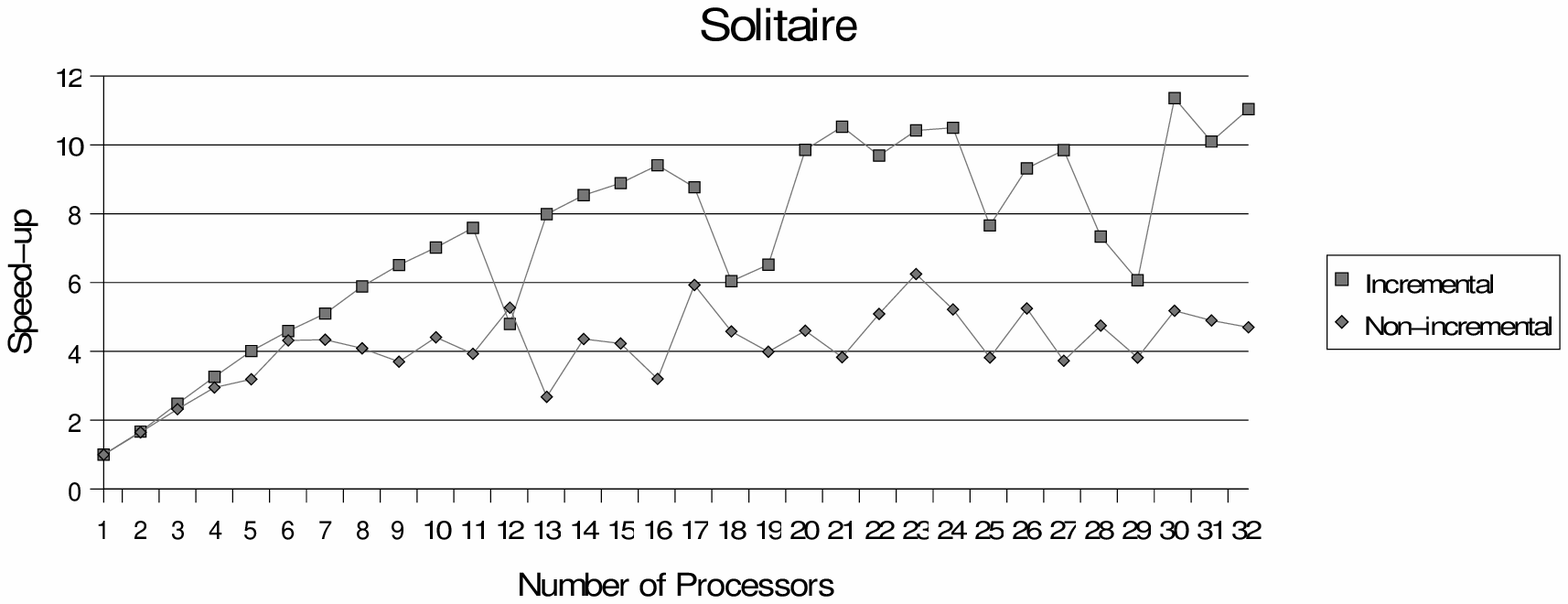,width=.53\textwidth}
\psfig{figure=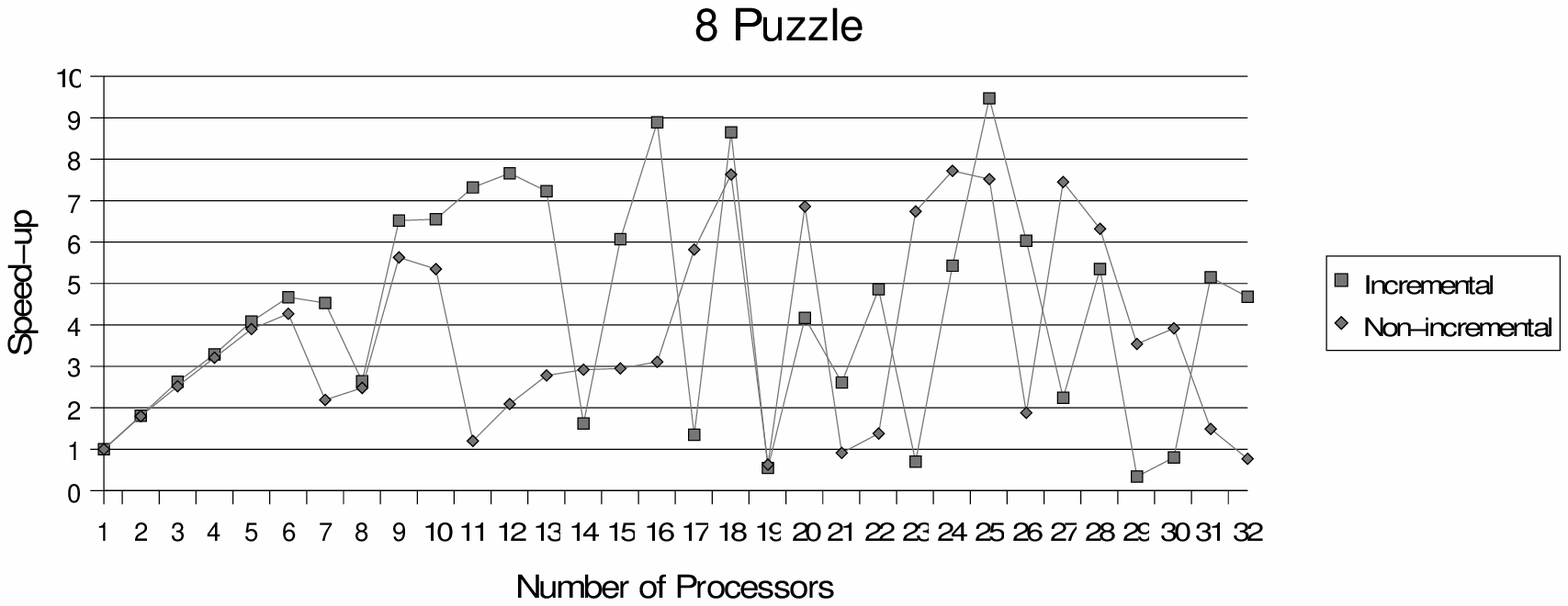,width=.53\textwidth}
\end{minipage}
\begin{minipage}[b]{\textwidth}
\psfig{figure=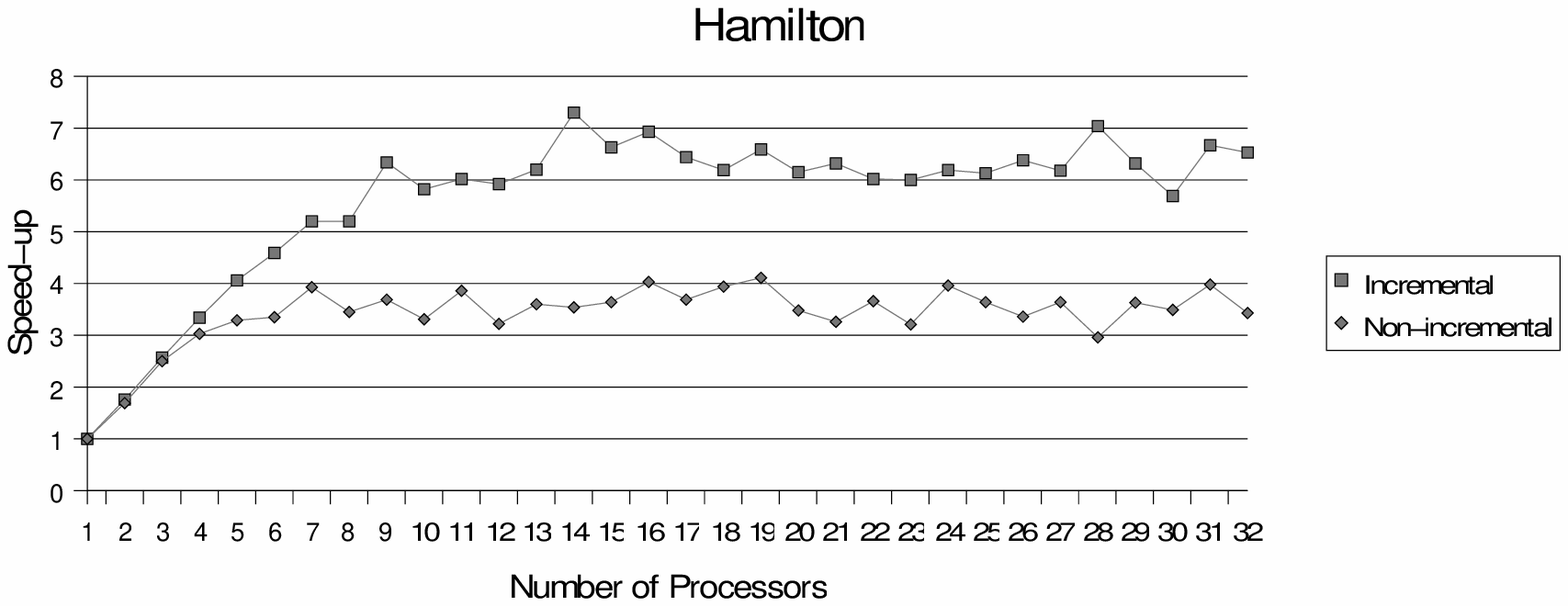,width=.53\textwidth}
\psfig{figure=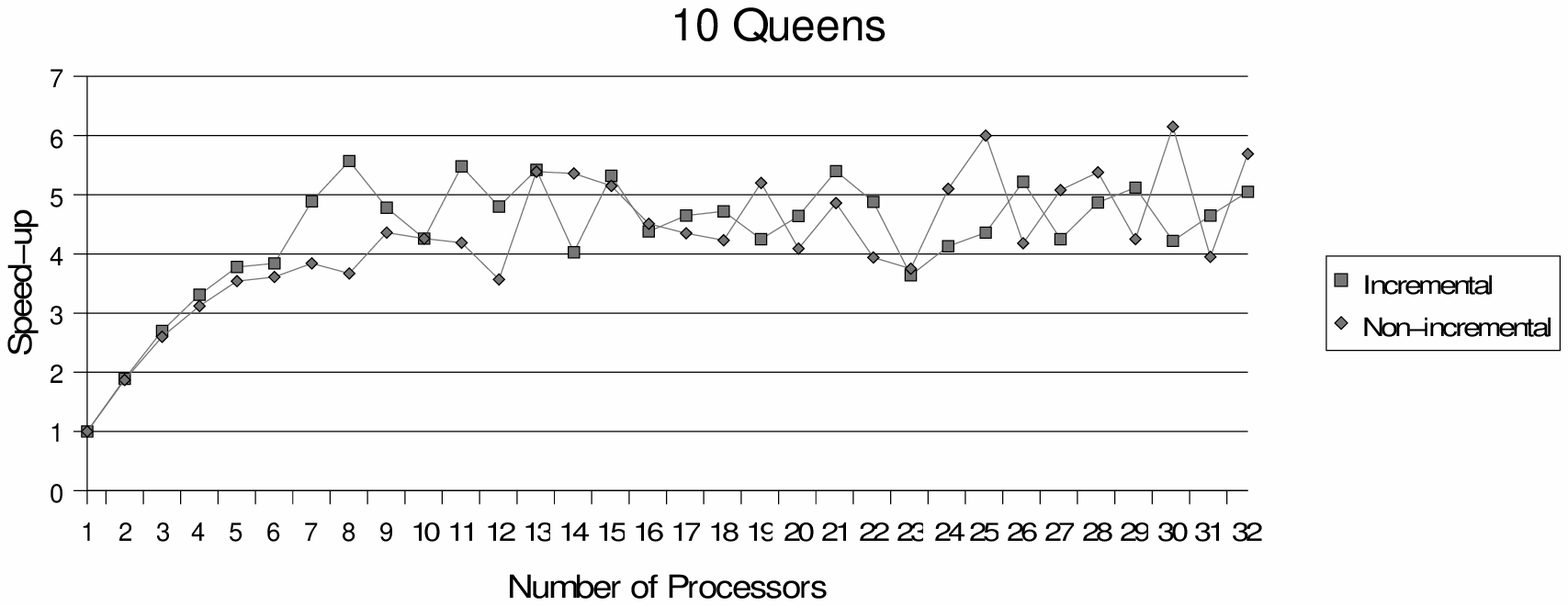,width=.53\textwidth}
\end{minipage}
\begin{minipage}[b]{\textwidth}
\psfig{figure=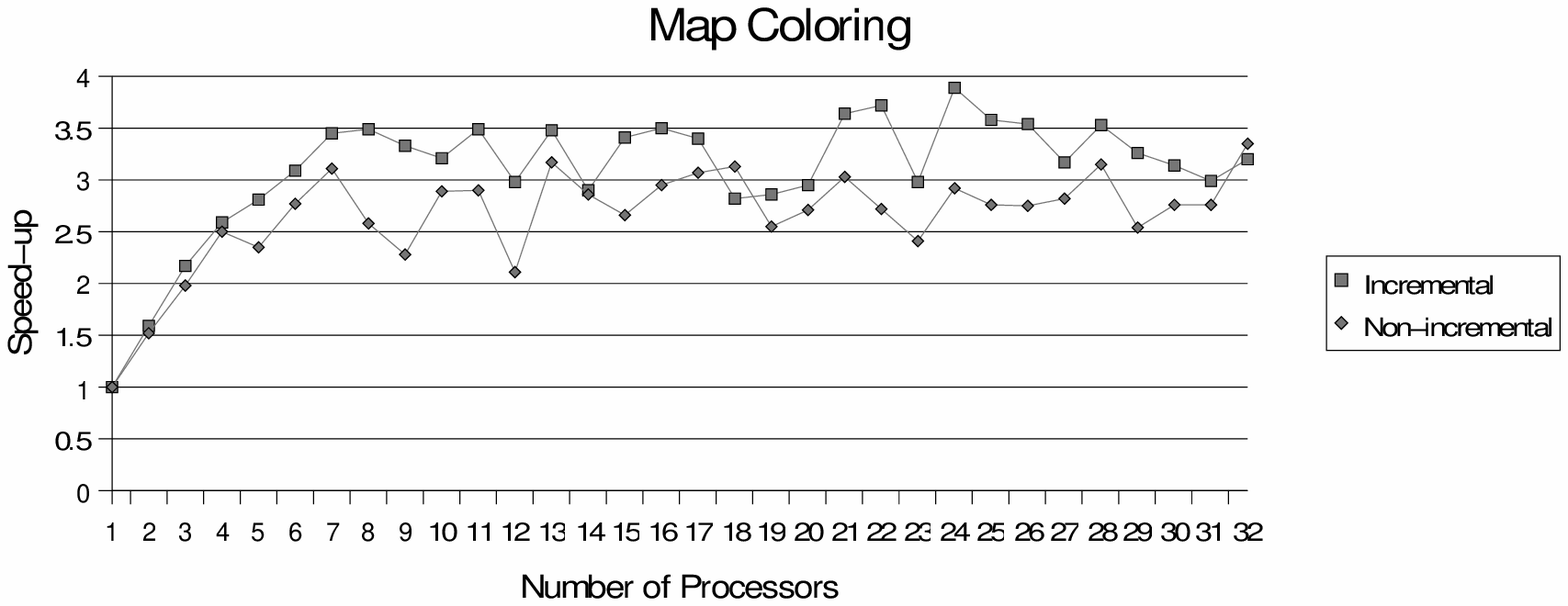,width=.53\textwidth}
\psfig{figure=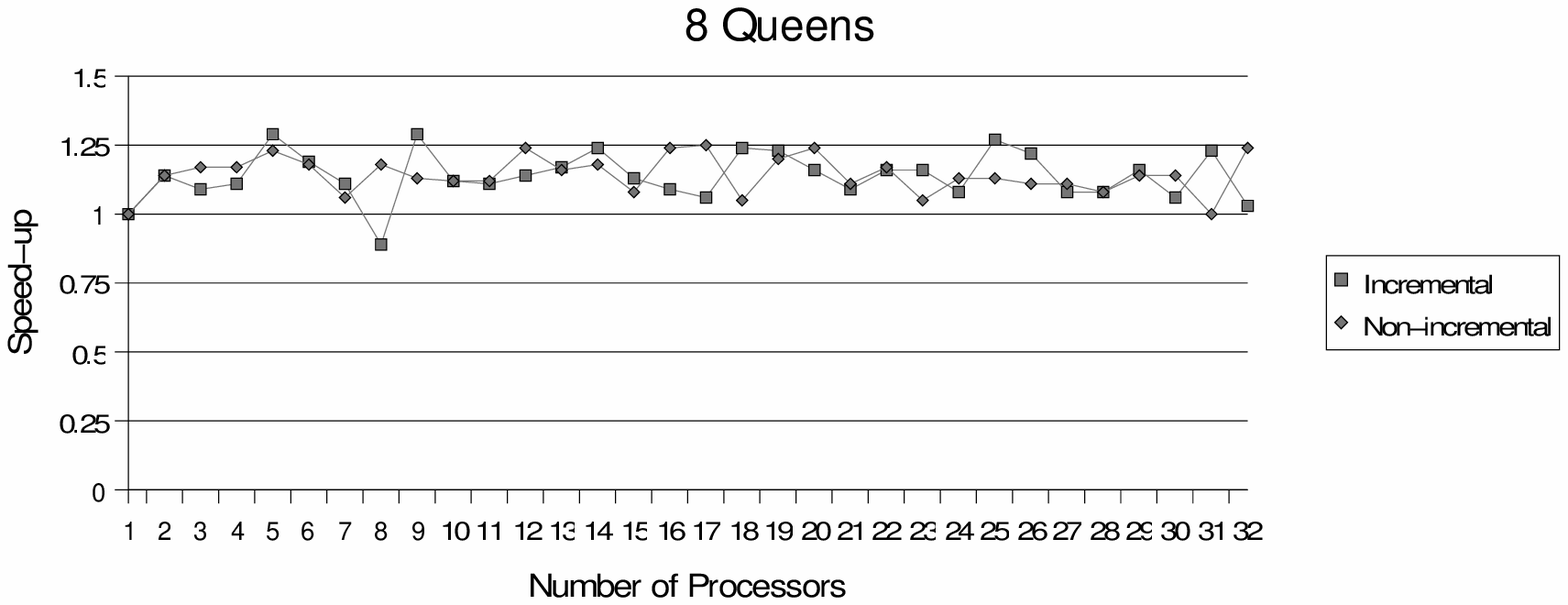,width=.53\textwidth}
\end{minipage}
\caption{Incremental Stack-Splitting vs. Non-Incremental Stack-Splitting }
\label{incre-non-incre}
\end{center}
\end{figure}

The timing results in seconds from our incremental stack-splitting system
are presented in Table \ref{incre-time}. The modifications made to the
ALS WAM are very localized and reduced to the minimum necessary. This has
allowed us to keep a very clean design---that, we hope, can be easily ported
to other WAM-based implementations---and to keep under control the parallel
overhead---our engine (in its incremental stack-splitting version)
 running on a single processor is on average only
$10\%$ slower than the ALS WAM.\footnote{The overhead in the non-incremental
stack-splitting engine are slightly lower.}
The corresponding speedups are presented in Figure~\ref{incre-non-incre}.
under the label \emph{incremental}.

\begin{table}[htb]
{\footnotesize
\begin{center}
\begin{tabular}{|c|c|c|c|c|c|}
\hline\hline
{\bf Benchmark} & \multicolumn{5}{c|}{\bf \# Agents}\\
                      &{\bf 1}&{\bf 2}&{\bf 8}&{\bf 16}&{\bf 32}\\
\hline\hline
\emph{9 Costas}        & 412.579 & 210.228 (1.96)&  52.686 (7.83)& 26.547 (15.54)& 14.075 (29.31) \\
\hline
\emph{Knight}          & 159.950 & 81.615 (1.95)& 20.754 (7.7)& 10.939 (14.62)& 8.248 (19.39)\\
\hline
\emph{Stable}          & 62.638  & 35.299 (1.77)&  9.117 (6.87)& 4.844 (12.93)& 3.315 (18.89)\\
\hline
\emph{Send More}       & 61.817 & 32.953 (1.87)&  8.931 (6.92)& 4.923 (12.55)& 3.916 (15.78)\\
\hline
\emph{8 Costas}        & 38.681 & 19.746 (1.95)& 5.052 (7.65)& 2.733 (14.15)& 1.753 (22.06)\\
\hline
\emph{8 Puzzle}        & 27.810 & 15.387 (1.8)&  10.522 (2.64)& 3.128 (8.89)& 5.940 (4.68)) \\
\hline
\emph{Bart}            & 13.619 & 7.958 (1.71)&  2.031 (6.7)& 1.600 (8.51)& 0.811 (16.79)\\ 
\hline
\emph{Solitaire}       & 5.909 & 3.538 (1.67)&  1.003 (5.89)& 0.628 (9.4)& 0.535 (11.04)\\
\hline
\emph{10 Queens}       & 4.572 & 2.418 (1.89)&  0.821 (5.56)& 1.043 (4.38)& 0.905 (5.05)\\
\hline
\emph{Hamilton}        & 3.175 & 1.807 (1.75)&  0.610 (5.2)& 0.458 (6.93)& 0.486 (6.53)\\
\hline
\emph{Map Coloring}    & 1.113 & 0.702 (1.58)&  0.319 (3.48)& 0.318 (3.5)& 0.348 (3.19)\\
\hline
\emph{8 Queens}        & 0.185 & 0.162 (1.14)&  0.208 (0.88)& 0.169 (1.09)& 0.180 (1.02)\\
\hline\hline
\end{tabular}
\end{center}}
\caption{Timings for Incremental Stack-Splitting (Time in sec.)}
\label{incre-time}
\end{table}

We observe that for \emph{large benchmarks}  (\emph{9-Costas}, \emph{Knight}, 
\emph{8-Costas}, \emph{Stable}, and \emph{Send More})
the speedups are very good.
We also observe that for \emph{medium benchmarks}
(\emph{Bart}, \emph{Solitaire}, \emph{8-Puzzle}, and \emph{Hamilton}) the speedups are still quite good.
Note that for the benchmarks with small running time (\emph{10-Queens}, \emph{Map Coloring} and \emph{8-Queens})
the speedups deteriorate.
This is consistent
with our belief that DMP implementations should be used for
parallelizing programs with coarse-grained parallelism. For programs with
small-running times, there is not enough work to offset the cost of
exploiting parallelism using a distributed communication model. Nevertheless,
our system is reasonably efficient, given that even for small benchmarks
it can produce some  speedups. It is also interesting to observe that in no
cases we have observed slow-downs due to parallel execution---thanks to the simple
granularity control mechanisms embedded in the scheduler (i.e., the use
of splitting thresholds, as mentioned in Section~\ref{sched}).

Note that the \emph{8-Puzzle} benchmark shows a very irregular behavior; we believe
this is	due to the small number of parallel choice-points created, and to the patterns
of communication that arise in presence of different number of processors (for certain
patterns, a successful distribution of work takes places, for others it does not).

One of the objectives of the experiments performed is to validate the effectiveness
of incremental stack-splitting as a methodology for efficient exploitation of 
parallelism on DMPs. In particular, there are two aspects that we were interested in 
exploring: \emph{(i)} verifying the effectiveness of stack-splitting versus a 
more ``direct'' implementation of the stack-copying method as implemented
in MUSE \cite{muse-journal} (i.e., keeping single 
copies of choice-points around the system); \emph{(ii)} verifying
the impact of \emph{incremental} splitting. 

Validity of stack-splitting vs. stack-copying can be inferred from the experiments
described in Section \ref{schedu}:  a direct implementation of stack-copying (where
we simulate shared frames by keeping ``ownership'' of choice-points to specific
processors) would produce an amount of communication that is at least as
high as in the case of centralized scheduling
described in Section \ref{schedu}.

In order to evaluate the impact of incrementality, we have measured the performance
of the system without the use of incremental splitting---i.e.,
each time a sharing operation takes place, a complete copy of the WAM data areas
is performed. The results obtained from this experiment are reported in 
Figure~\ref{incre-non-incre}: the figure compares the speedups observed with and
without incremental splitting.
We can observe that  
our incremental stack-splitting system obtains higher
speedups than the non-incremental stack-copying system. As expected, the difference
is more significant in those benchmarks where a large number of parallel choice-points
is generated, as there is an increased possibility of applying incremental splitting. It
is also important to observe that in the majority of the cases the incremental behavior 
has lead to an improvement in performance w.r.t. non-incremental splitting.

\subsubsection{Scheduling}
\label{schedu}

One of the major reasons to adopt stack-splitting, as described earlier, is the ability 
to perform scheduling on the bottom-most choice-point. Other DMP implementations
of or-parallelism have reversed to the use of scheduling on the top-most choice-point
(e.g., \cite{araujo,opera,Dorpp}, where
during a sharing operation only the oldest choice-point with unexplored alternatives is
exchanged between agents. Top-most scheduling will share only one choice-point at the
time, thus relieving the engine from the need of controlling access to shared choice-points.

\begin{figure}
\begin{center}
\begin{minipage}[b]{\textwidth}
\psfig{figure=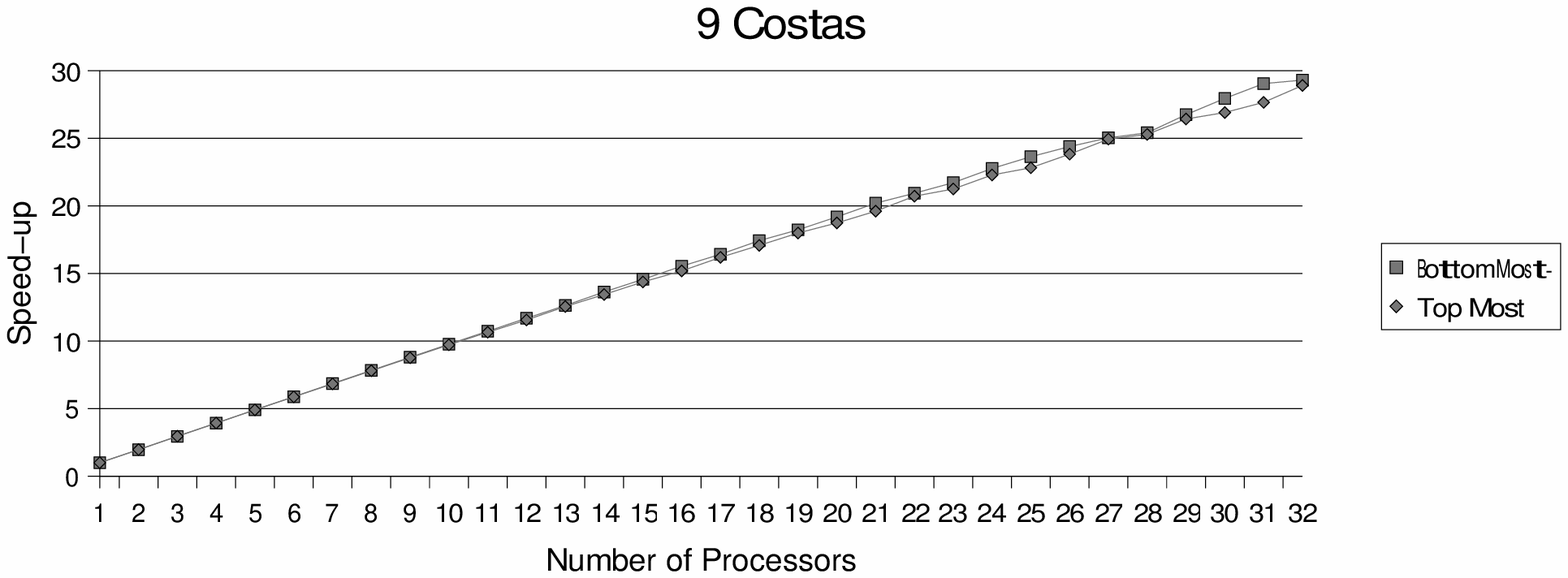,width=.52\textwidth}
\psfig{figure=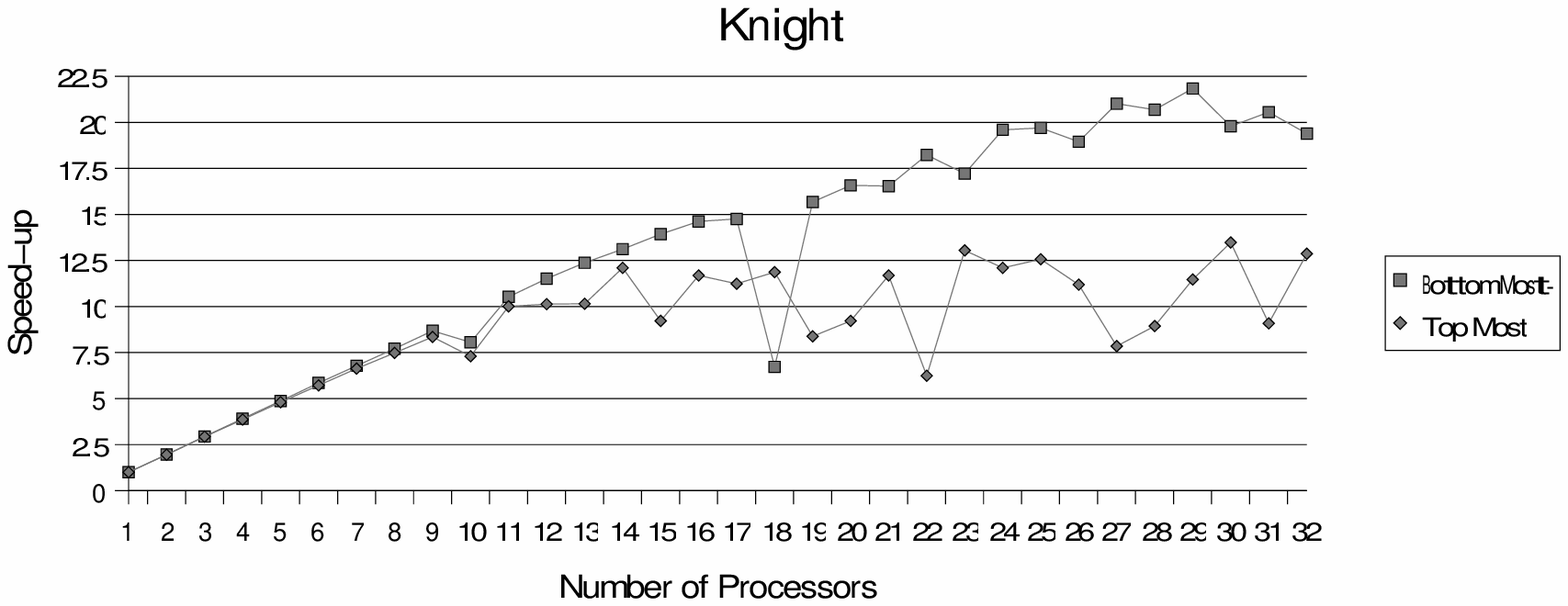,width=.52\textwidth}
\end{minipage}
\begin{minipage}[b]{\textwidth}
\psfig{figure=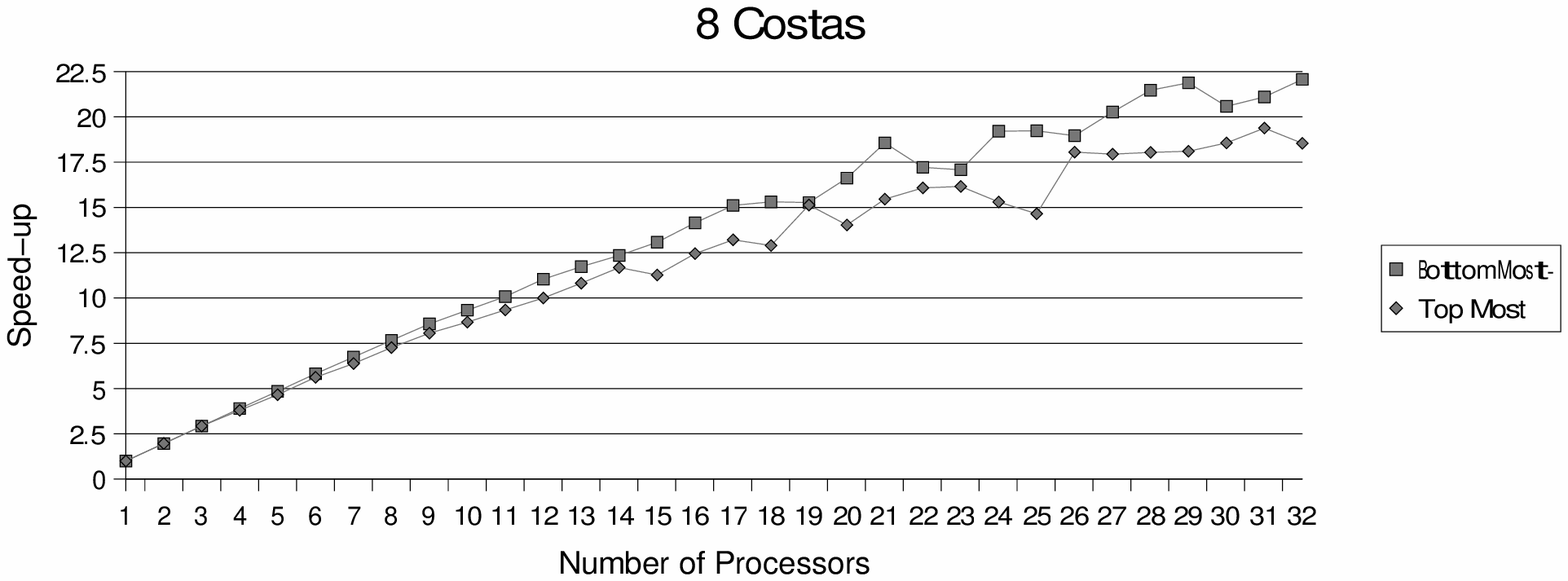,width=.52\textwidth}
\psfig{figure=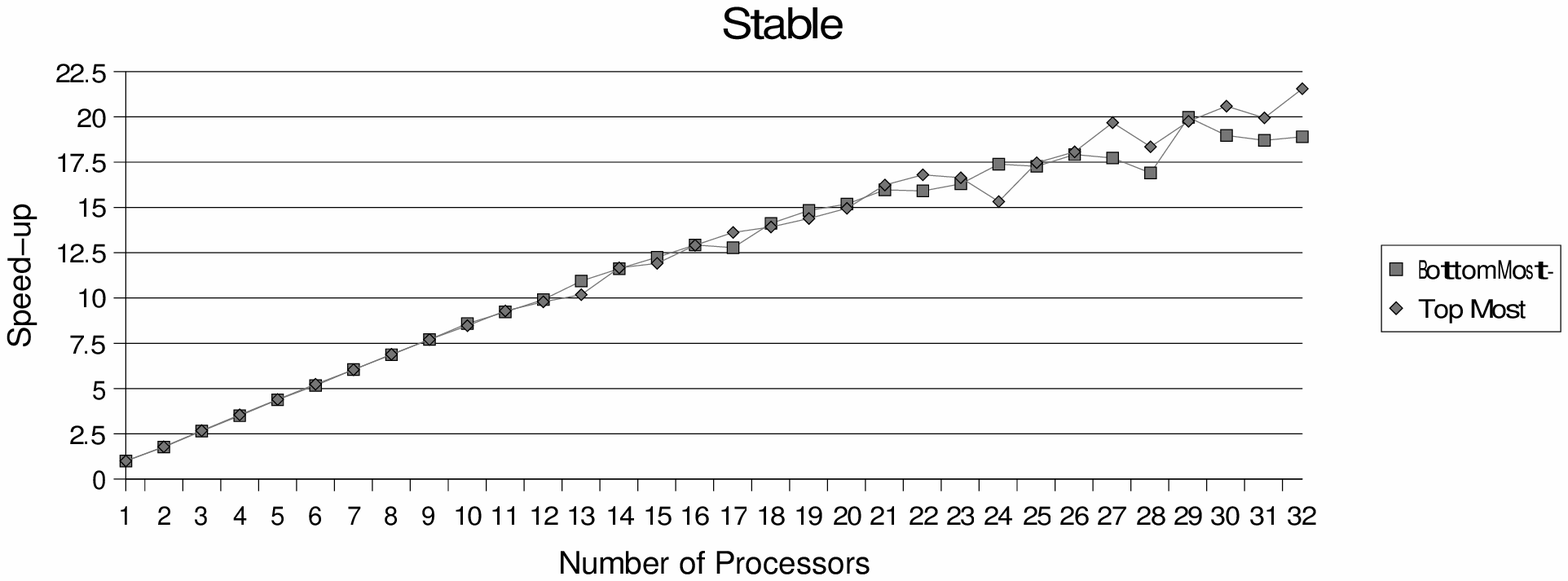,width=.52\textwidth}
\end{minipage}
\begin{minipage}[b]{\textwidth}
\psfig{figure=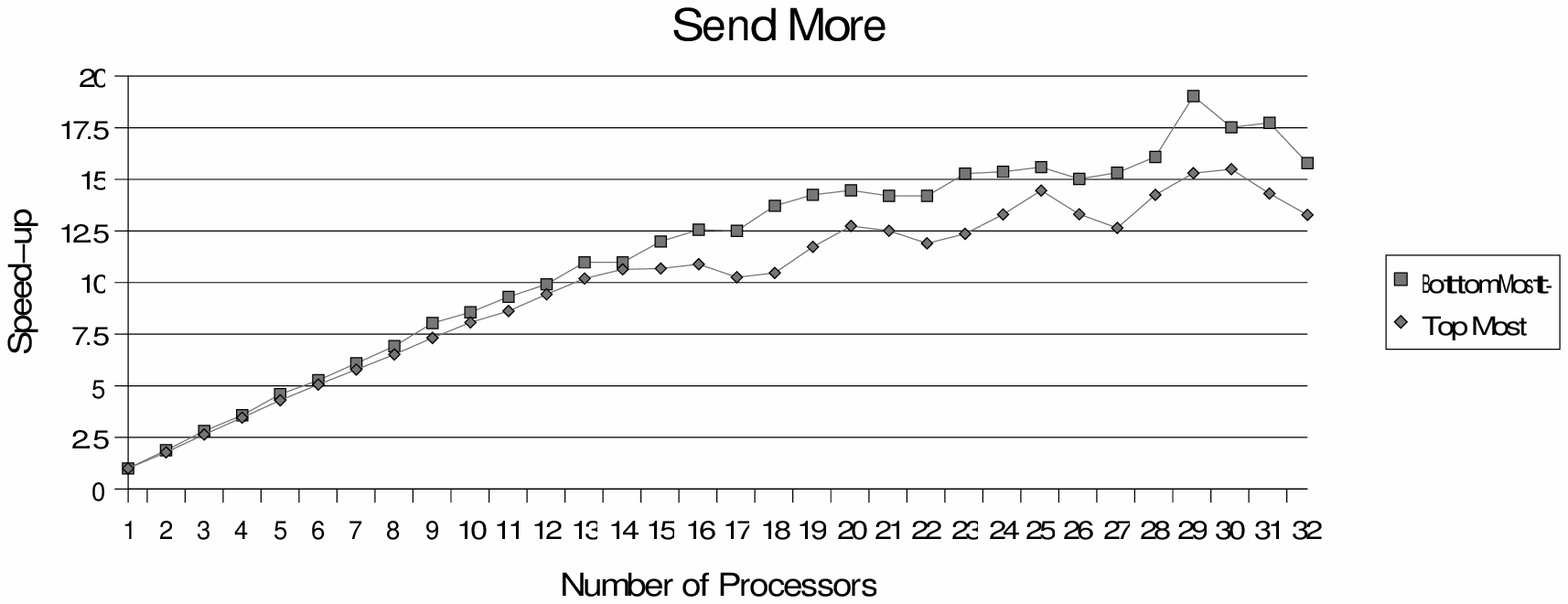,width=.52\textwidth}
\psfig{figure=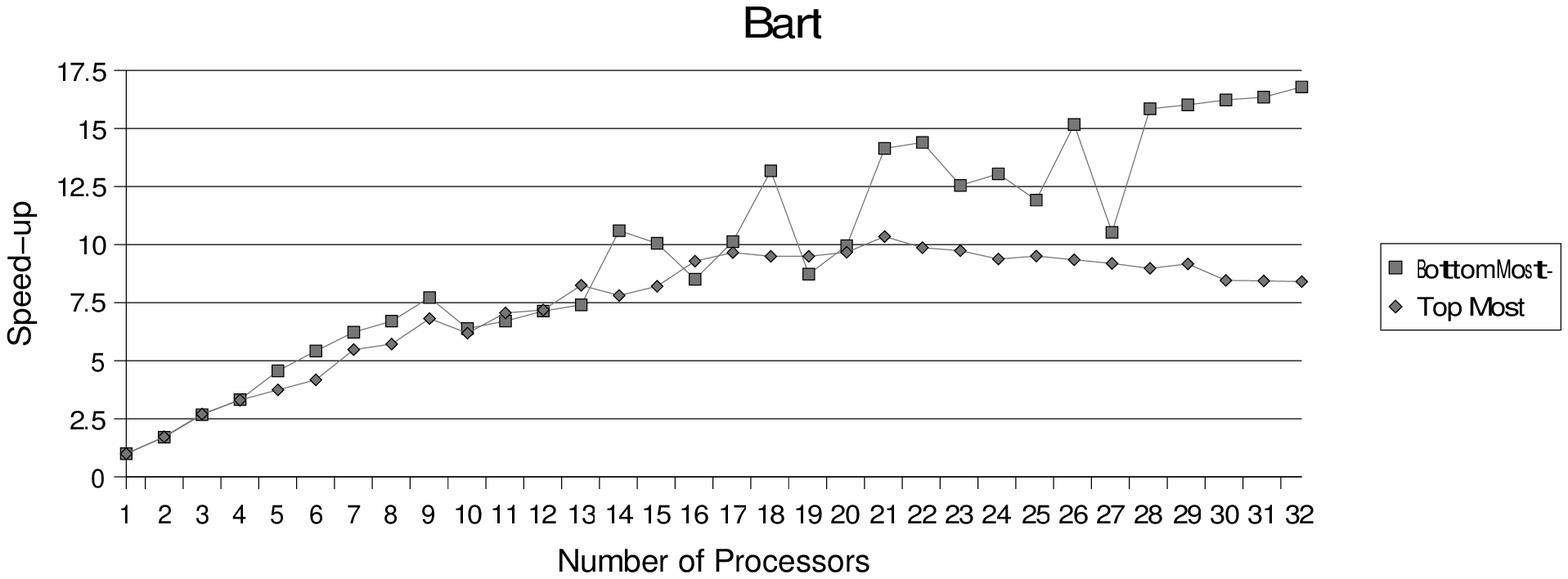,width=.52\textwidth}
\end{minipage}
\begin{minipage}[b]{\textwidth}
\psfig{figure=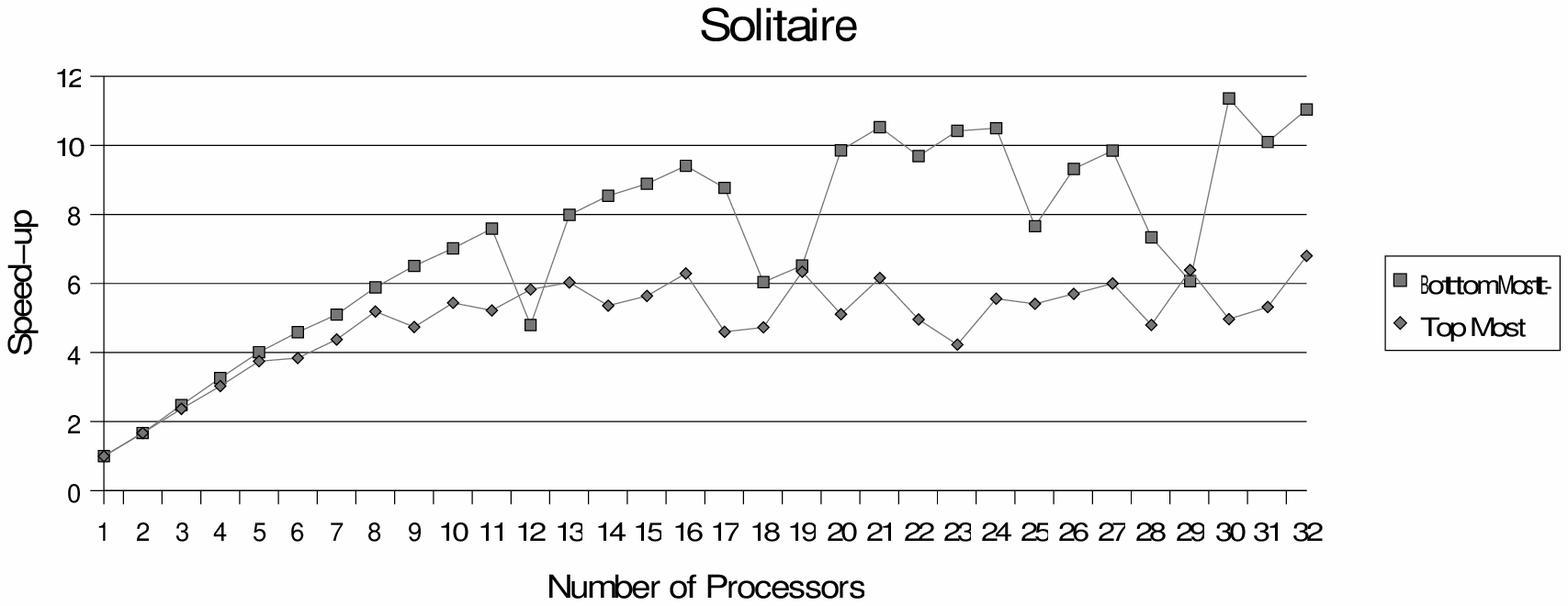,width=.52\textwidth}
\psfig{figure=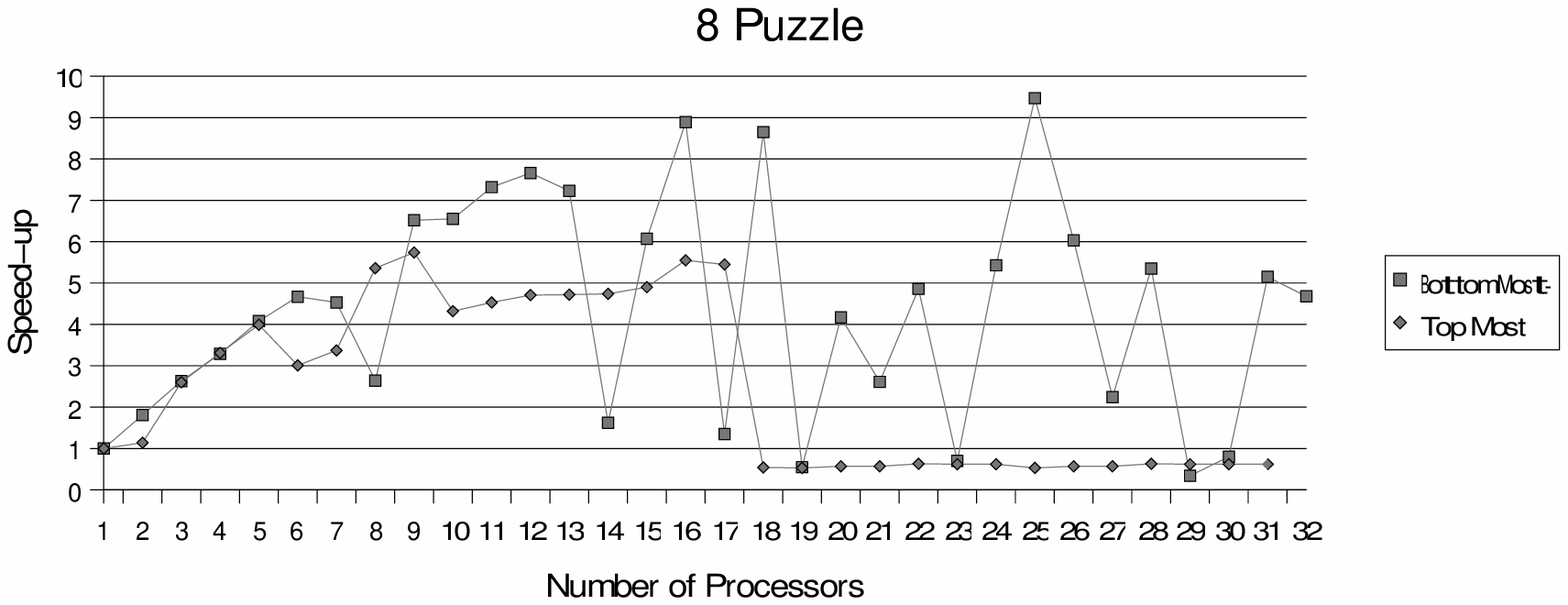,width=.52\textwidth}
\end{minipage}
\begin{minipage}[b]{\textwidth}
\psfig{figure=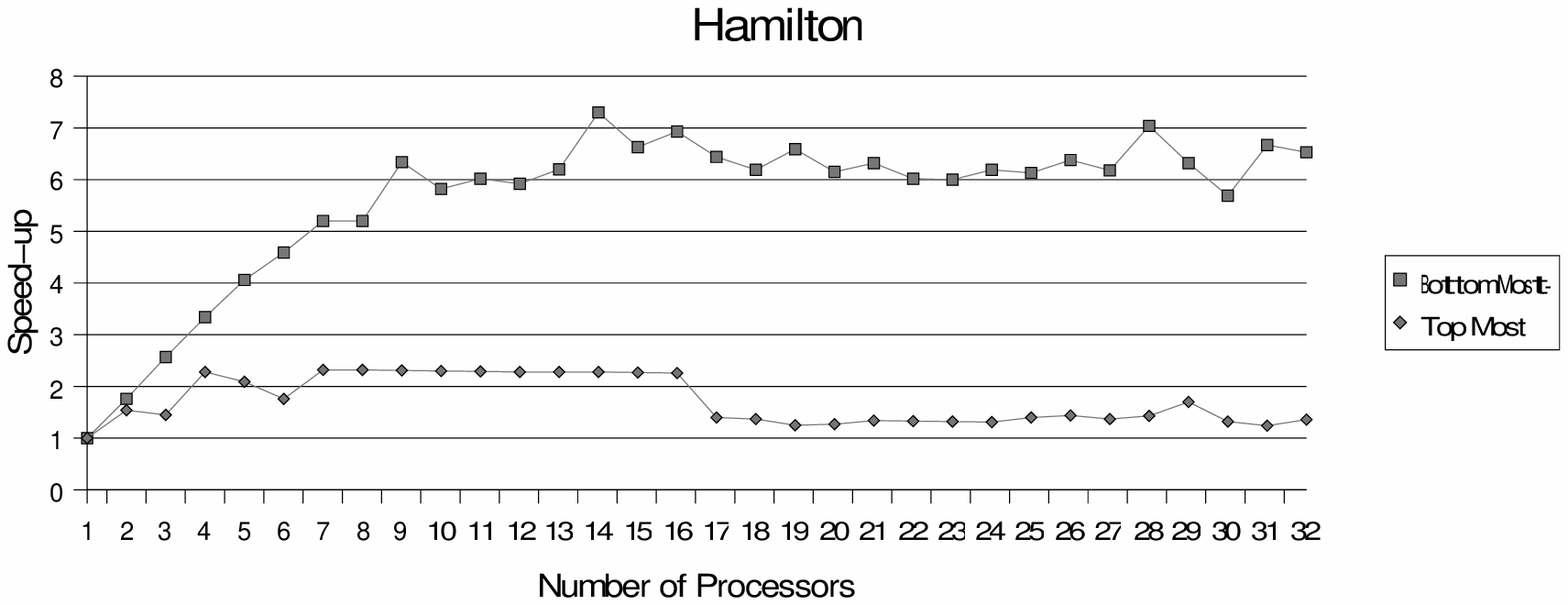,width=.52\textwidth}
\psfig{figure=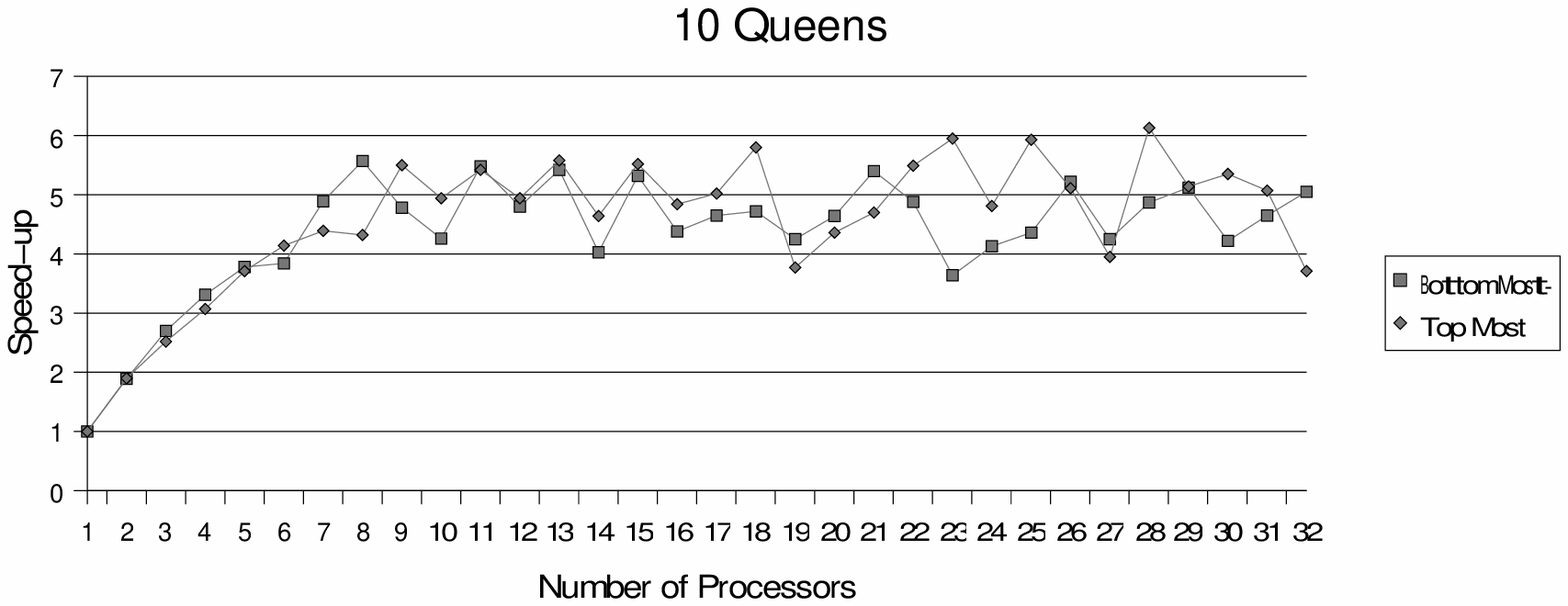,width=.52\textwidth}
\end{minipage}
\begin{minipage}[b]{\textwidth}
\psfig{figure=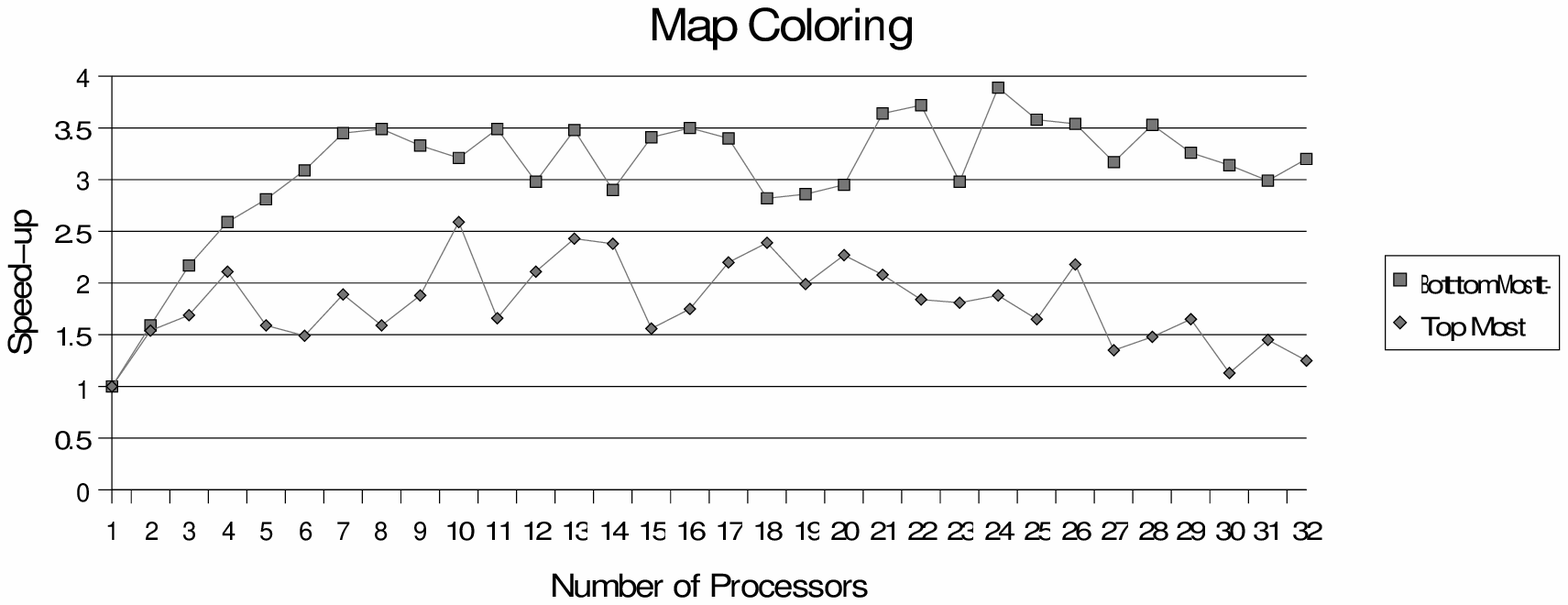,width=.52\textwidth}
\psfig{figure=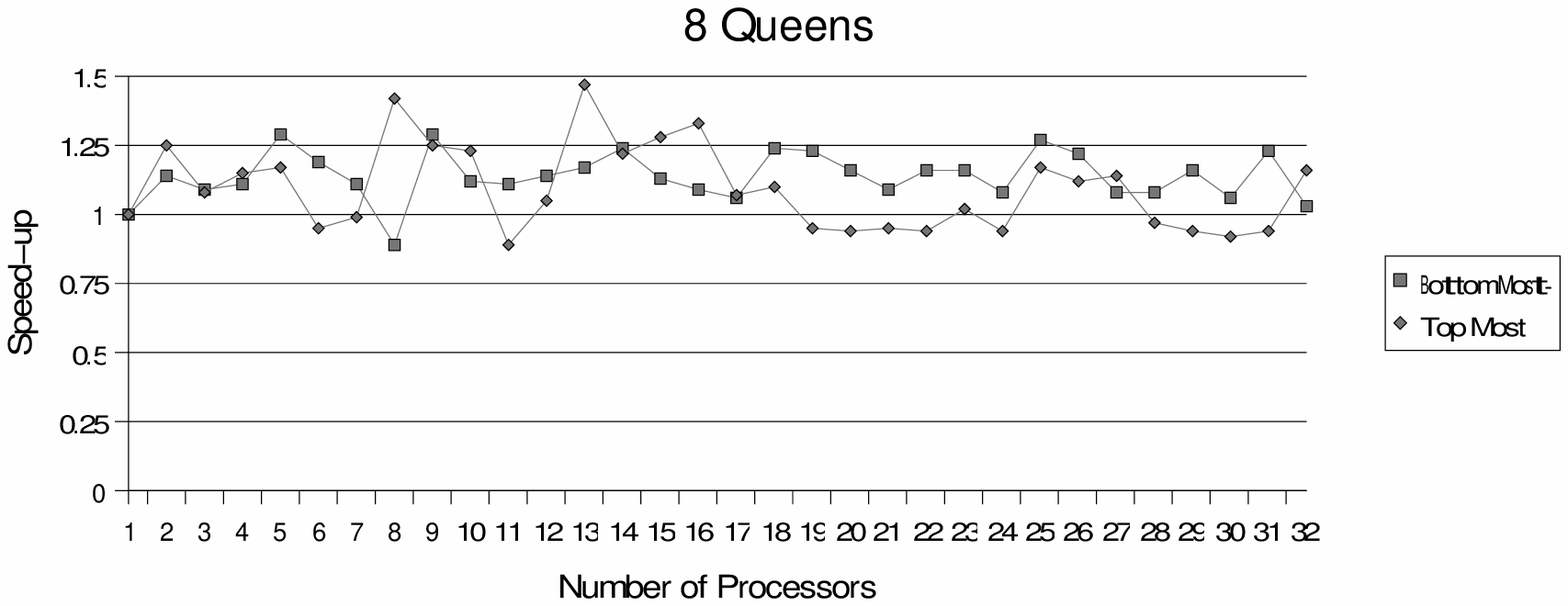,width=.52\textwidth}
\end{minipage}
\caption{Bottom Most vs. Top Most Scheduling}
\label{incre-top}
\end{center}
\end{figure}

To validate the effectiveness of our claim, we have developed a top-most scheduler for
our incremental stack-splitting system, and compared its 
performance with that of the incremental stack-splitting
with bottom-most scheduling.\footnote{The top-most scheduler used here is a different
implementation than the one described in the previous section, 
though based on the
same principles.}
Figure~\ref{incre-top} compares the speedups observed using the two different schedulers. 
As we can observe from Figure~\ref{incre-top}, in most benchmarks 
bottom-most scheduling provides a sustained speedup considerably higher than top-most scheduling.
For example, in \emph{Hamilton} we have a large number of choice-points (which can be
easily and quickly found), each with relatively 
small alternatives; the top-most scheduling forces an excessive number of interactions between
agents---since agents quickly run out of work and they require additional sharing operations.
This situation derives from 
 the reduced number of calls to the scheduler performed during the execution---agents
are busy for a longer period of time than using top-most scheduling.
In the remaining benchmarks, top-most and bottom-most scheduling provide similar results, as a
small number of choice-points are created and only one at a time is shared between agents.

\begin{figure}
\begin{center}
\begin{minipage}[c]{.495\textwidth}
\centerline{\psfig{figure=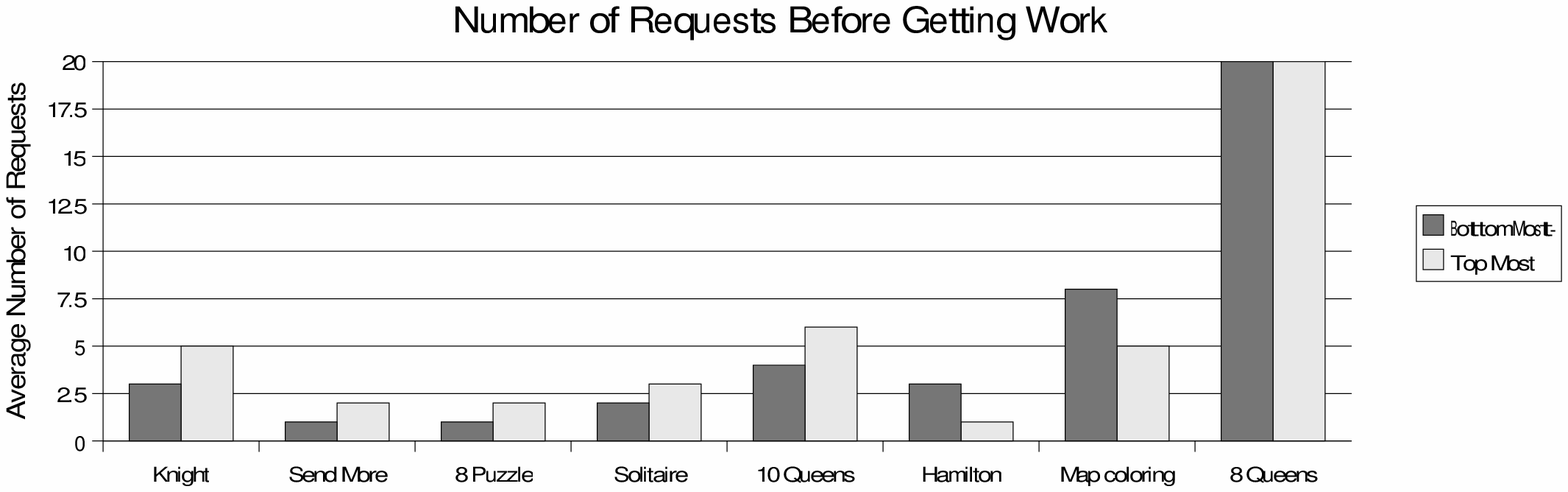,width=\textwidth}}
\end{minipage}
\begin{minipage}[c]{.495\textwidth}
\centerline{\psfig{figure=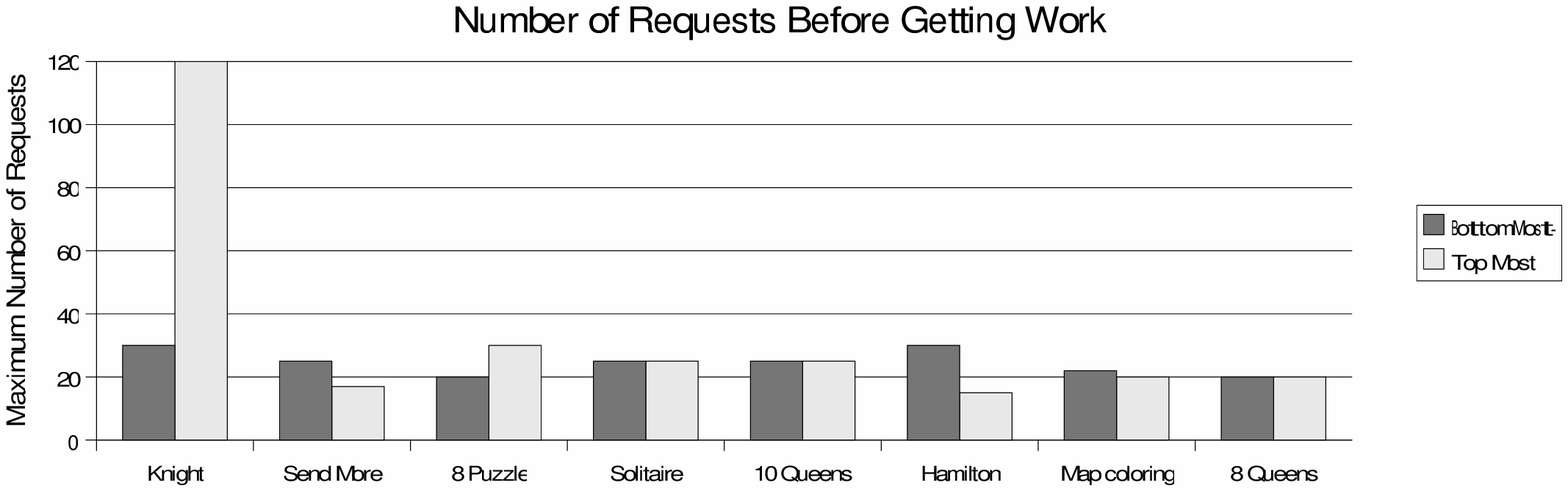,width=\textwidth}}
\end{minipage}
\end{center}
\caption{Average and Maximum Number of Tries to Acquire Work}
\label{tries}
\end{figure}

Another aspect of our implementation that we are interested in validating is the performance of the
distributed scheduler. As mentioned in Section \ref{sched}, our scheduler is based on keeping in
each agent an ``approximated'' view of the load in each other agent.
The risk that this method may encounter is that an agent may have out-of-date information
concerning the load in other agents, and as a consequence it may try to request work from
idle agents or ignore agents that may have unexplored alternatives. Figure~\ref{tries} provides
some information concerning the number of attempts that an agent needs to perform before
receiving work. The figure on the left measures the average number of requests that an agent
has to send (experiments performed using an 8-agent run); as we can see, the number is very small (in most
cases 1 to 3 requests are sufficient) and 
such number is generally better if we adopt bottom-most scheduling. The figure on the right shows
the maximum number of requests observed; these numbers tend to grow towards the end of the computation
(when less work is available)---nevertheless, typically only one or two agents achieve these
maximum values, while the majority of the agents remain close to the average number of attempts.

\begin{figure}
\begin{center}
\begin{minipage}[b]{\textwidth}
\psfig{figure=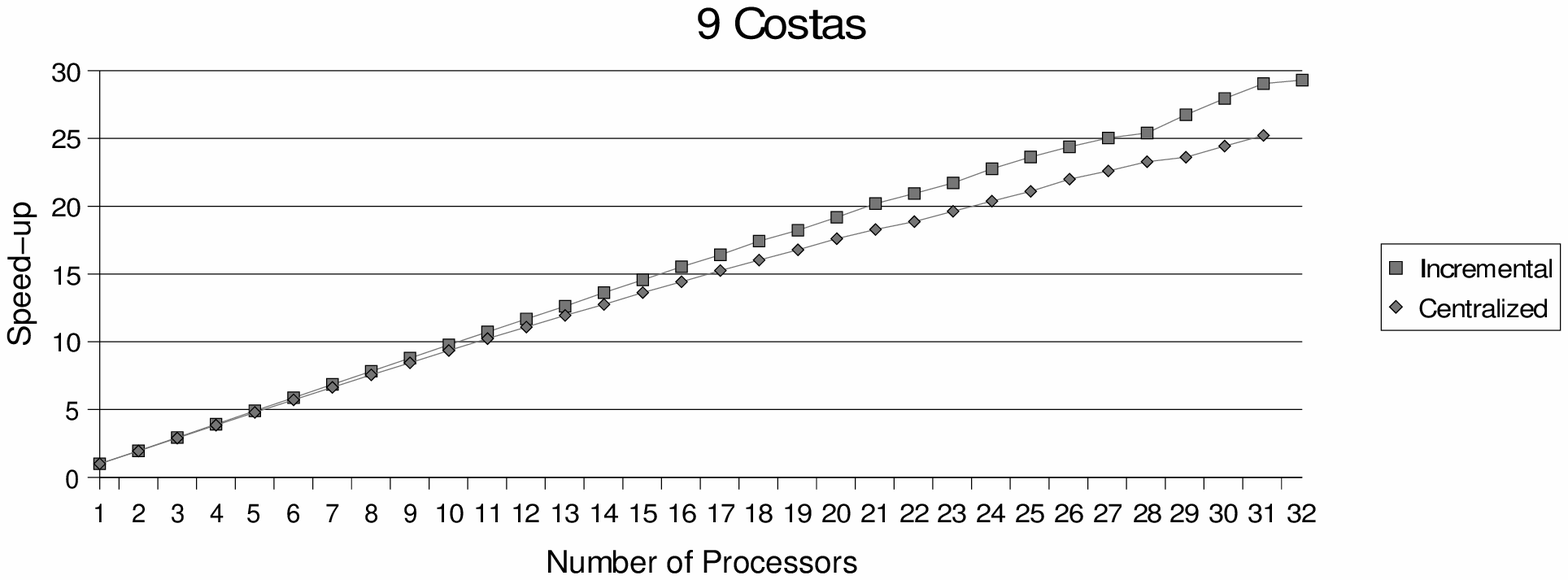,width=.52\textwidth}
\psfig{figure=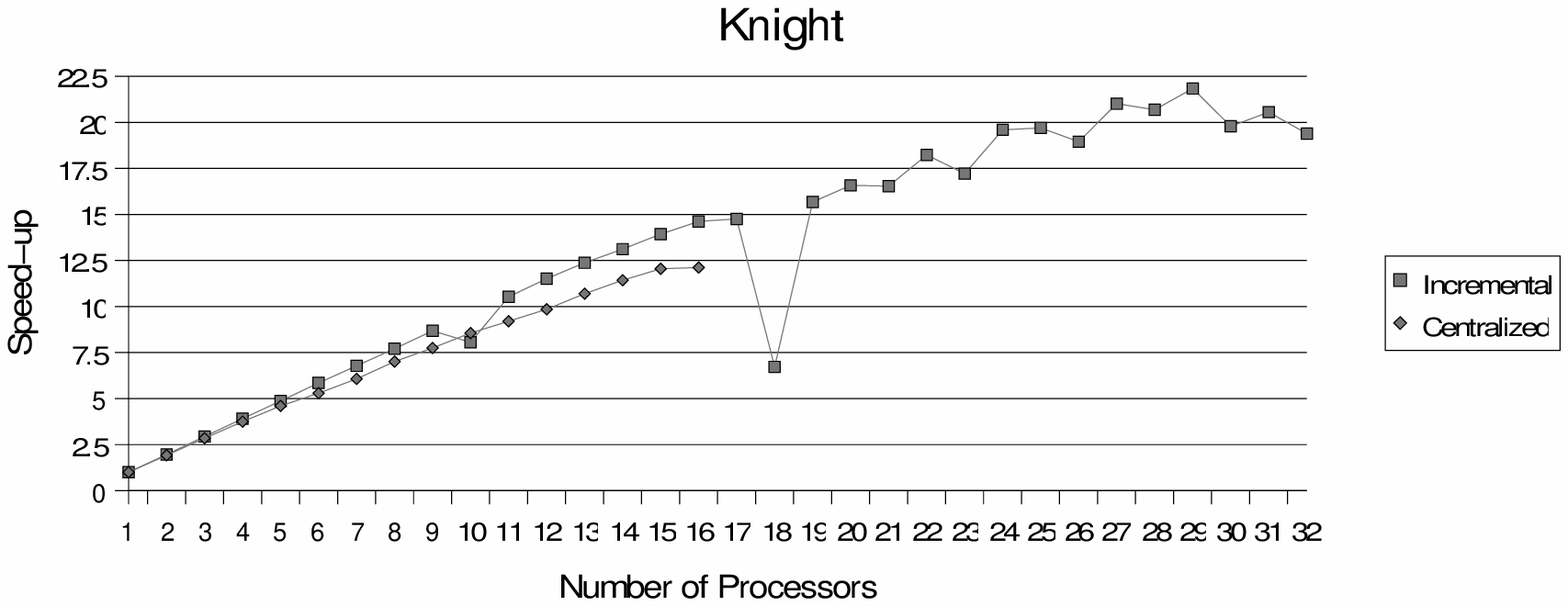,width=.52\textwidth}
\end{minipage}
\begin{minipage}[b]{\textwidth}
\psfig{figure=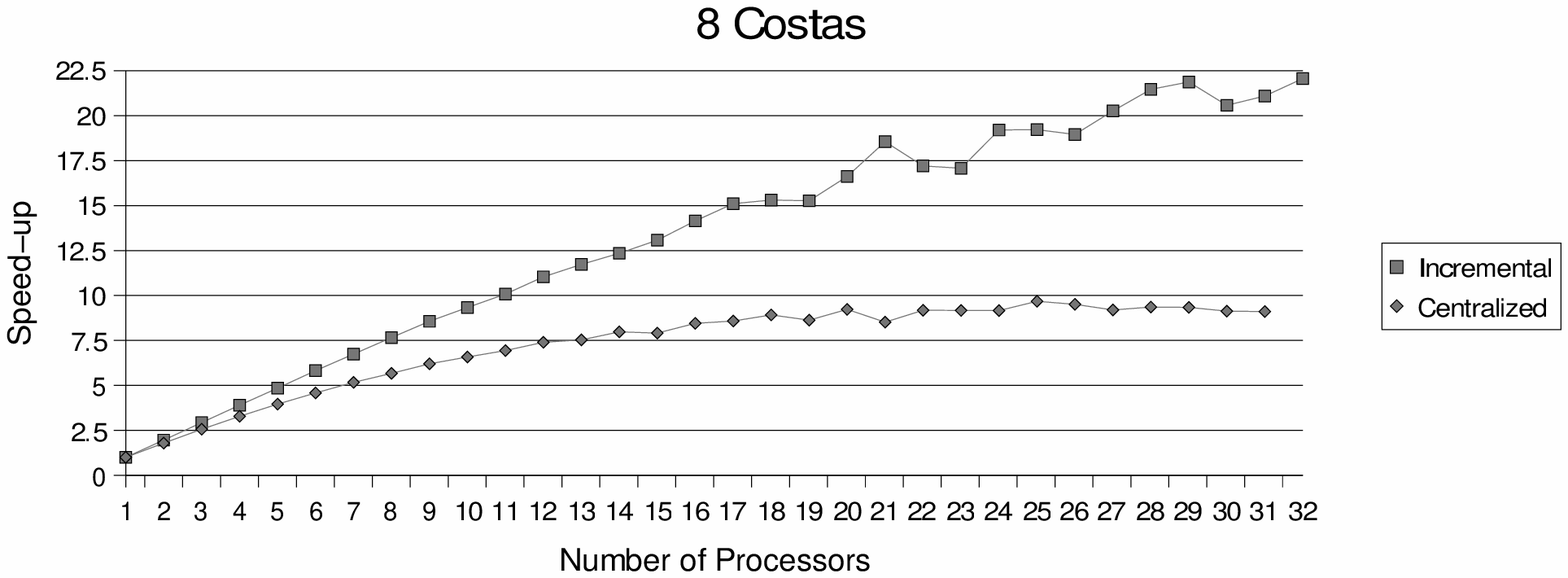,width=.52\textwidth}
\psfig{figure=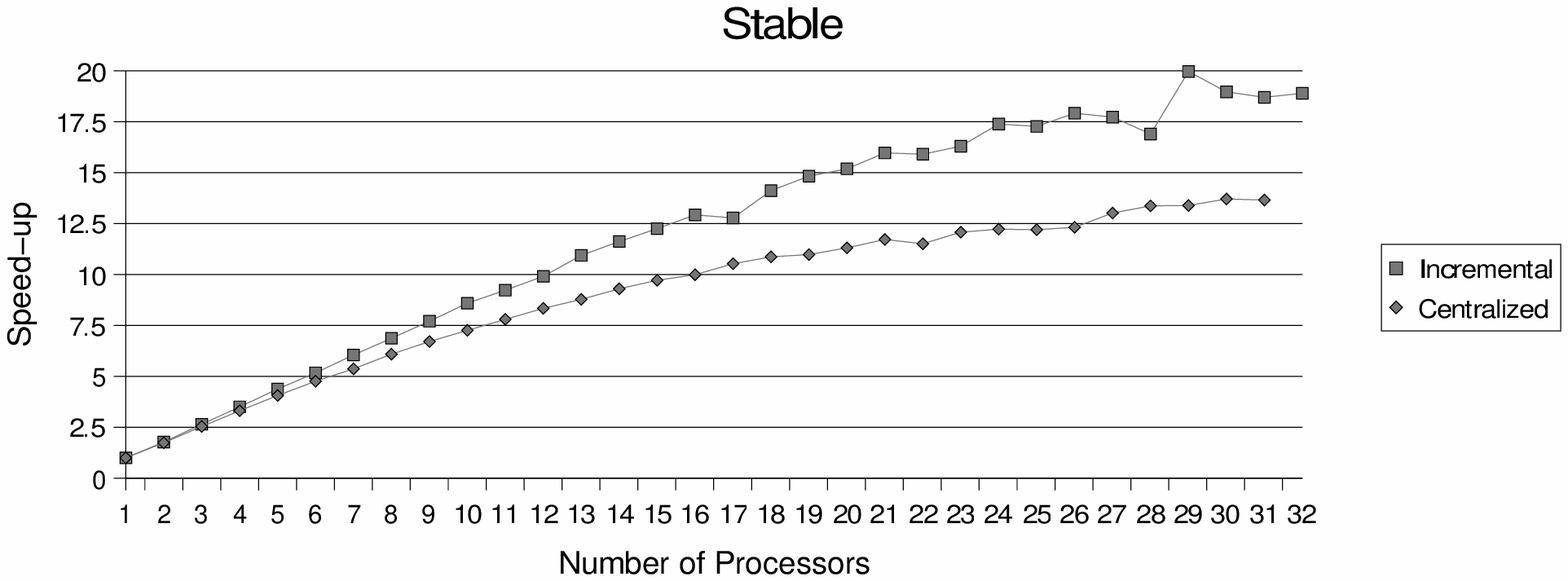,width=.52\textwidth}
\end{minipage}
\begin{minipage}[b]{\textwidth}
\psfig{figure=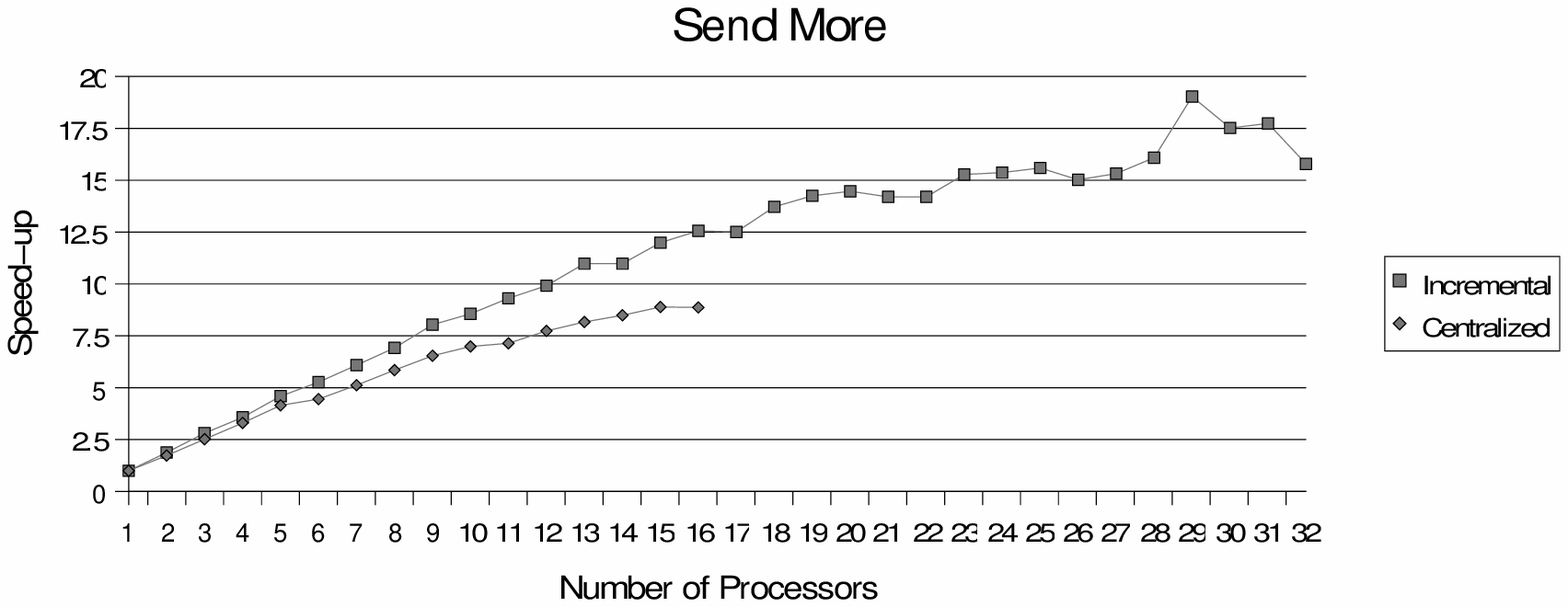,width=.52\textwidth}
\psfig{figure=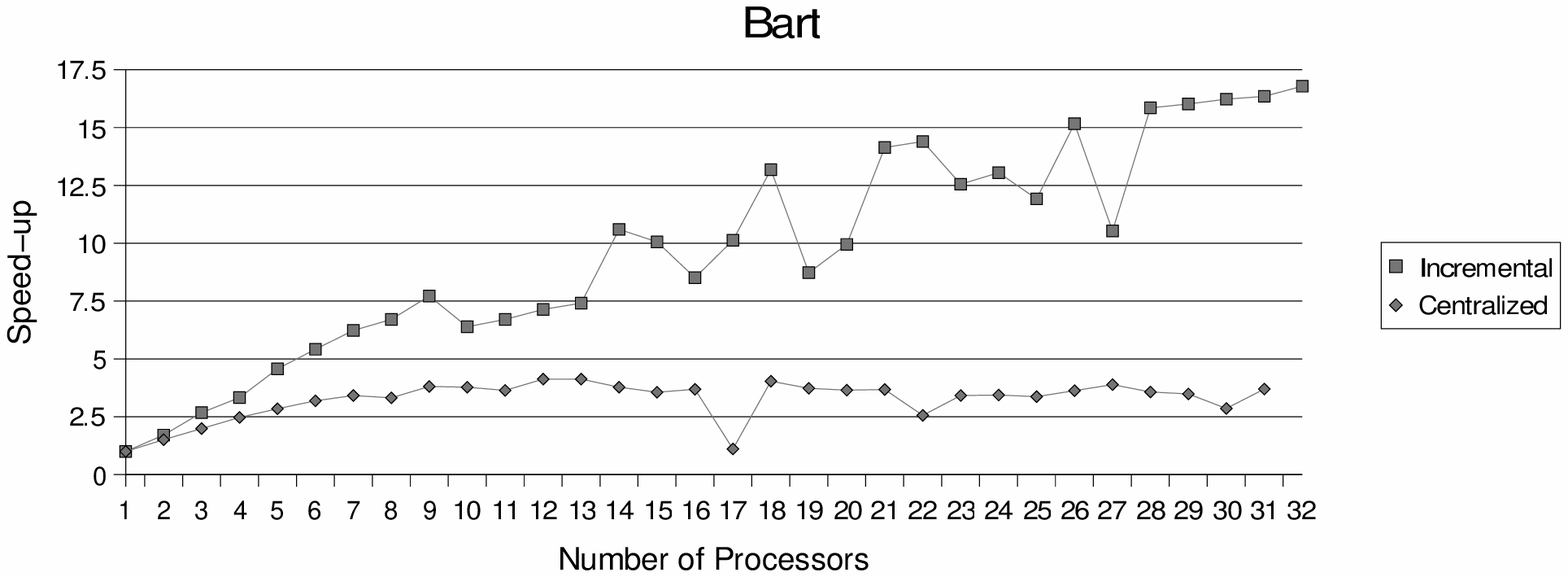,width=.52\textwidth}
\end{minipage}
\begin{minipage}[b]{\textwidth}
\psfig{figure=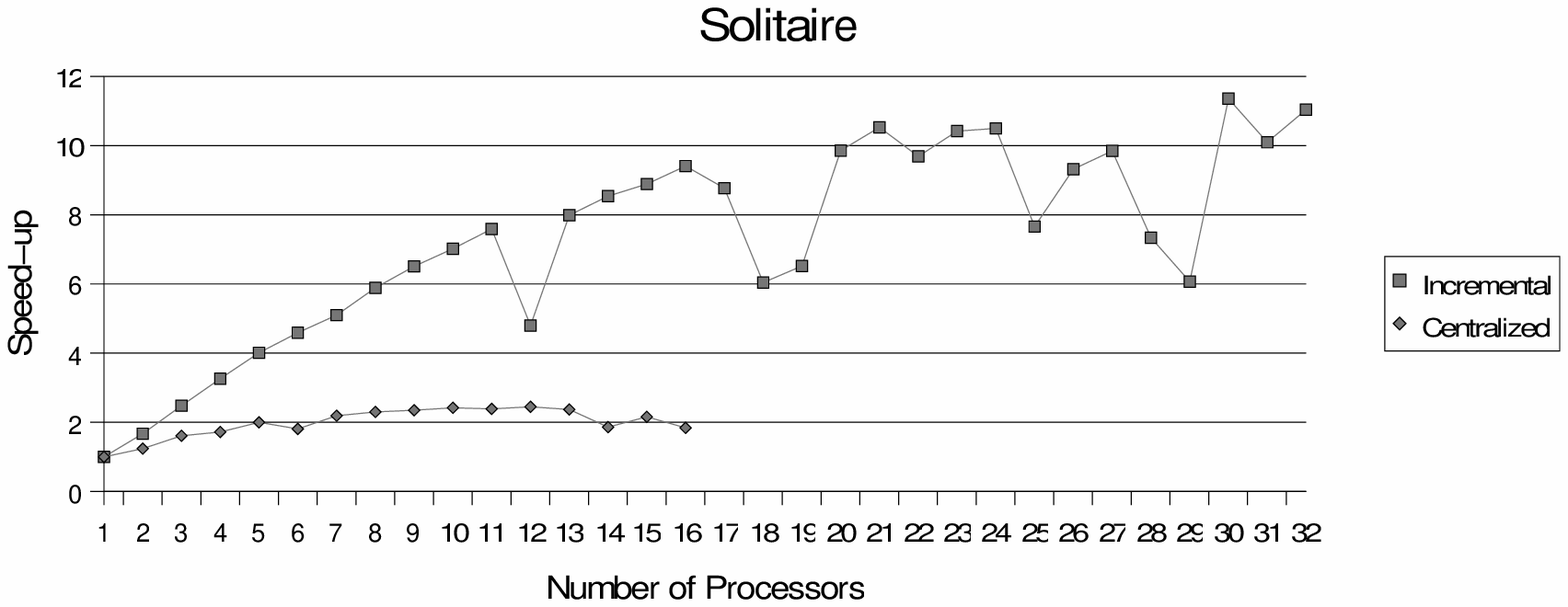,width=.52\textwidth}
\psfig{figure=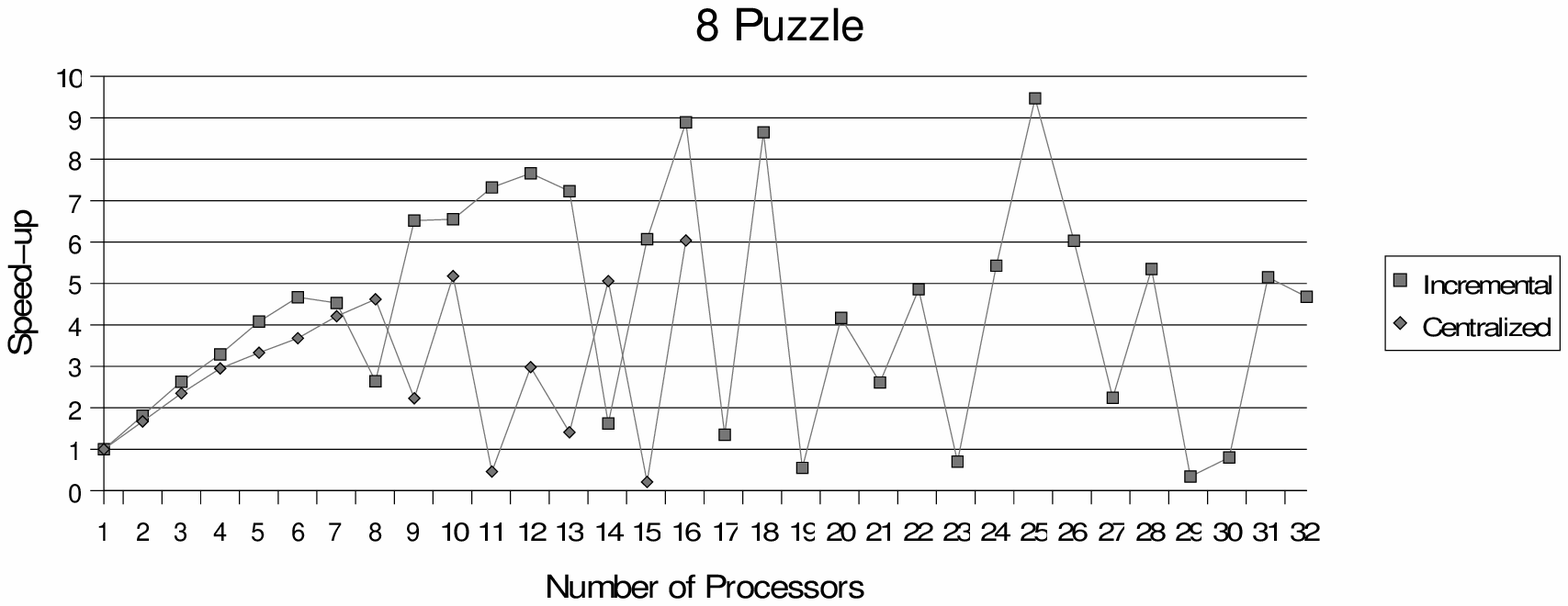,width=.52\textwidth}
\end{minipage}
\begin{minipage}[b]{\textwidth}
\psfig{figure=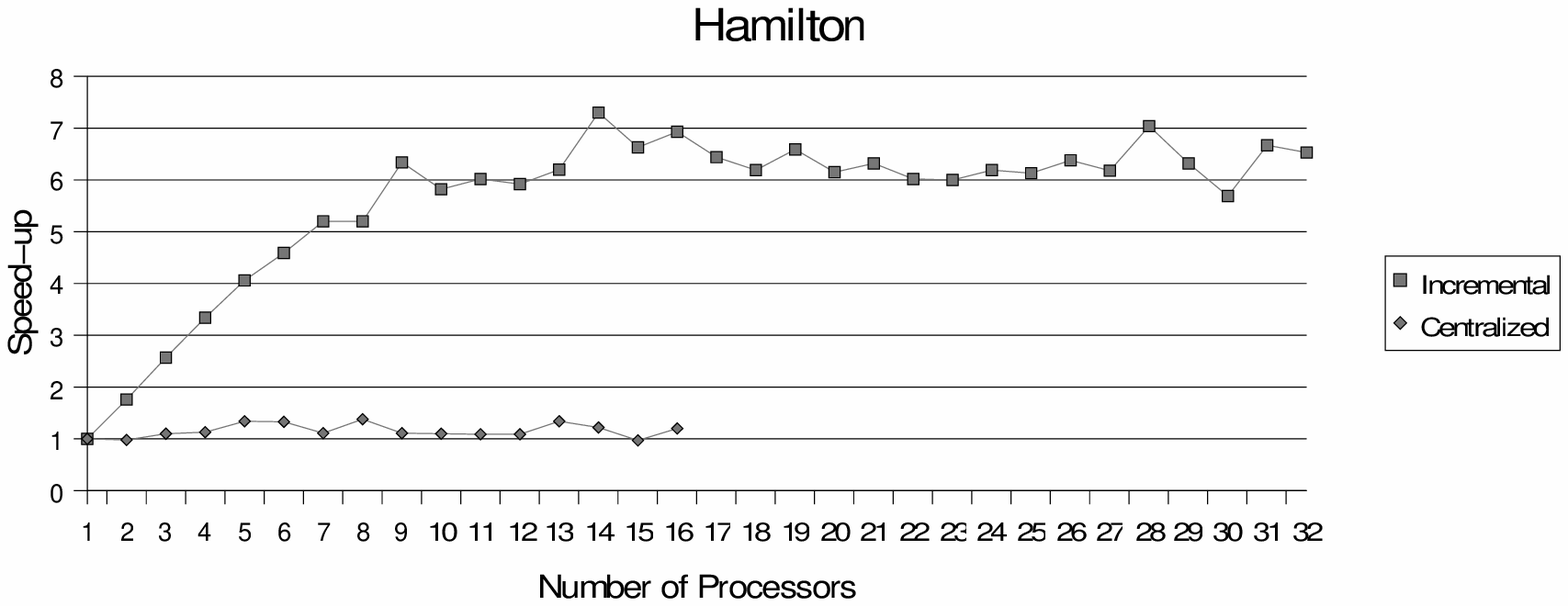,width=.52\textwidth}
\psfig{figure=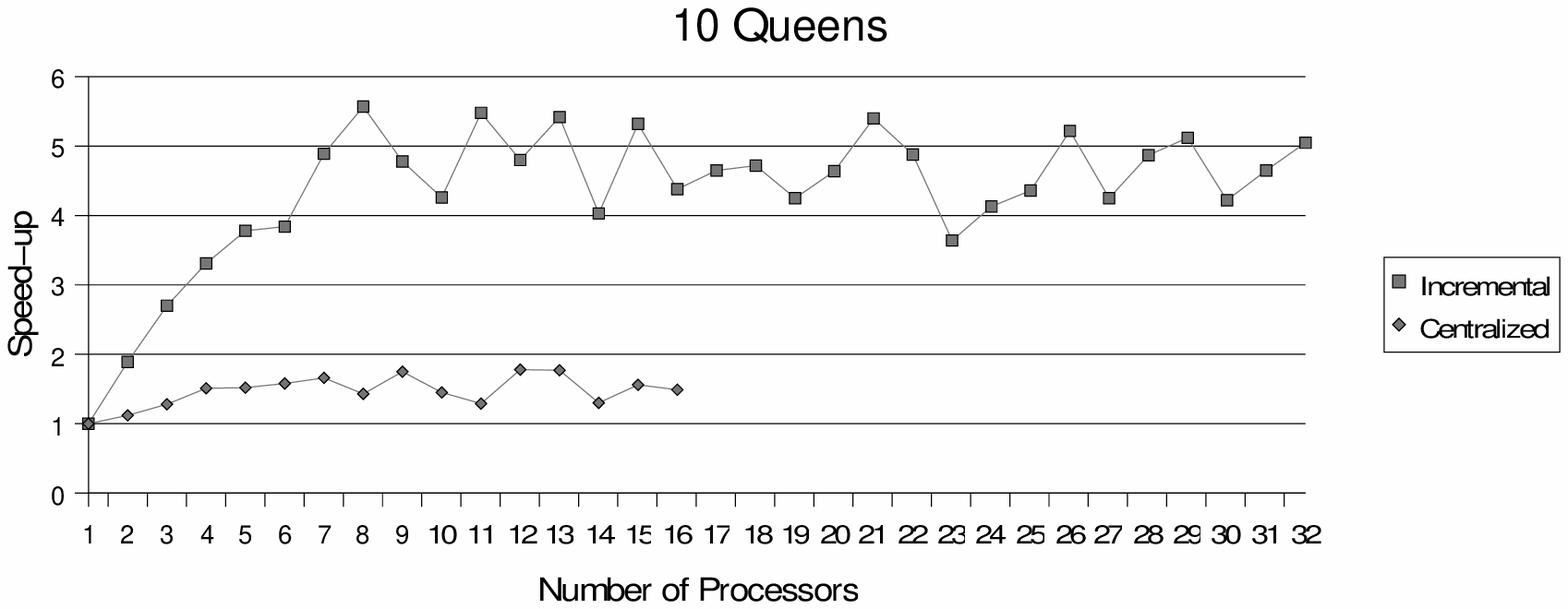,width=.52\textwidth}
\end{minipage}
\begin{minipage}[b]{\textwidth}
\psfig{figure=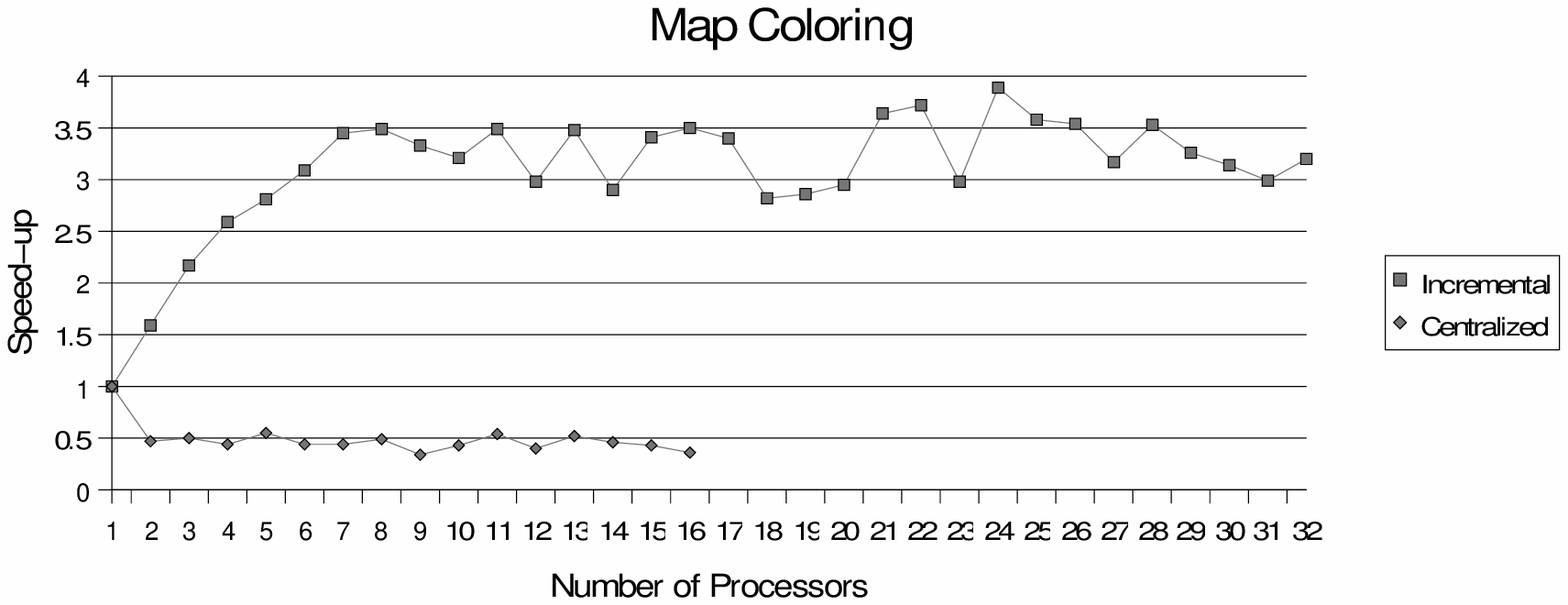,width=.52\textwidth}
\psfig{figure=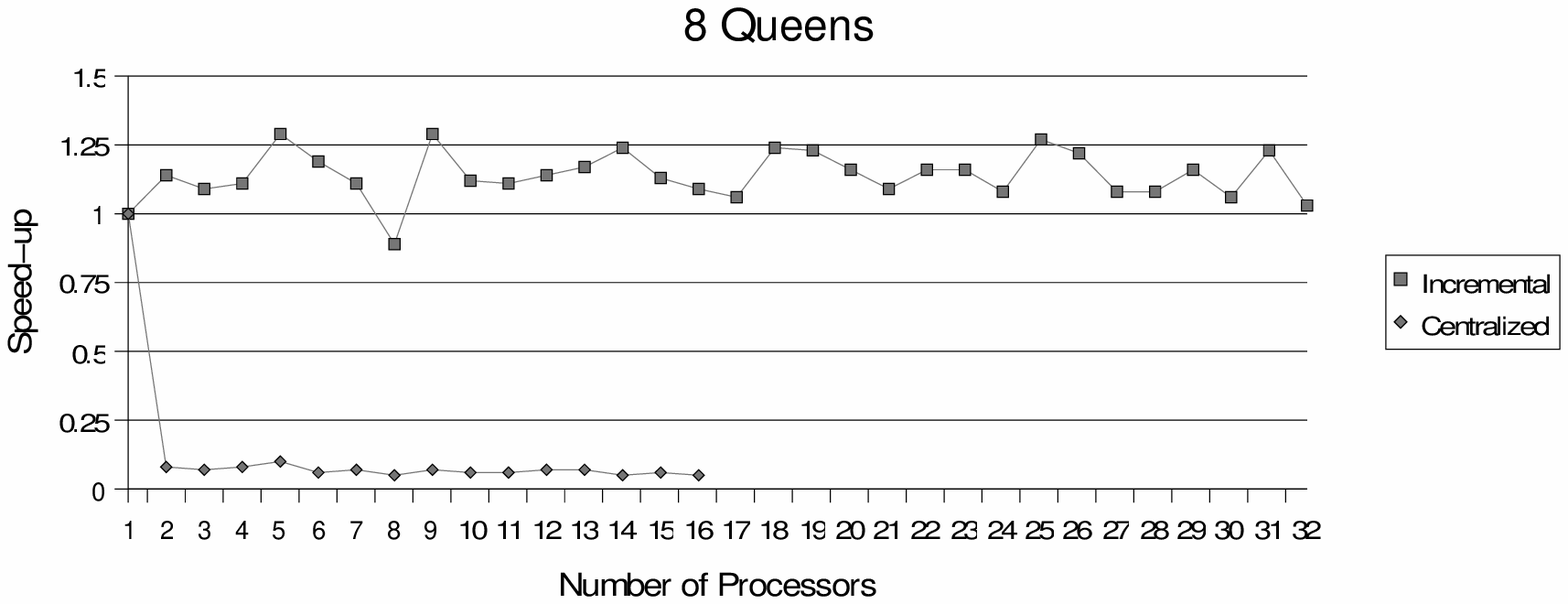,width=.52\textwidth}
\end{minipage}
\caption{Distributed vs. Centralized Scheduling}
\label{incre-central}
\end{center}
\end{figure}

To further validate our scheduling approach, we have compared it with an alternative 
scheduling scheme developed in PALS. This alternative scheme is an implementation of 
a centralized scheduling algorithm, designed following the guidelines of the scheduler
used in the Opera system \cite{opera}. 
In the centralized scheduler approach,
one agent, called \emph{central}, does not perform actual computation, but
it is only in charge of keeping track of the load information. 
Idle agents send their requests for work directly to the central agent. In turn, the
central agent is in charge of implementing a matchmaking algorithm between idle and busy 
agents. The central agent matches requests from idle agents with busy agents
with highest load. The central agent is also in charge of detecting termination. When stack-splitting
occurs, only the central agent is informed about the load information update.  
Figure~\ref{incre-central} compares the speedups achieved using centralized scheduling with the
speedups observed using the distributed scheduling approach.\footnote{We had to limit the experiments
to a smaller number of CPUs due to unavailability of half of the machine at that time.}
 As evident from the figure, in many benchmarks (mostly those with medium and small size
computations) the
speedups observed in centralized scheduling are almost negligible---this is due to the inability
of the scheduling method to promptly respond to the requests for new work. Also, the use of
a reasonably fast interconnection network (Myrinet) leads to the creation of a severe bottleneck
at the level of the centralized scheduler. From our experiments we can observe that the centralized
scheduler is a feasible solution only if very few coarse-grained tasks are generated. For benchmarks
such as \emph{Hamilton}, where a fairly large number of choice-points is generated, the centralized
scheduler leads to a considerable loss of performance.

\begin{figure}
\begin{center}
\begin{minipage}[b]{\textwidth}
\psfig{figure=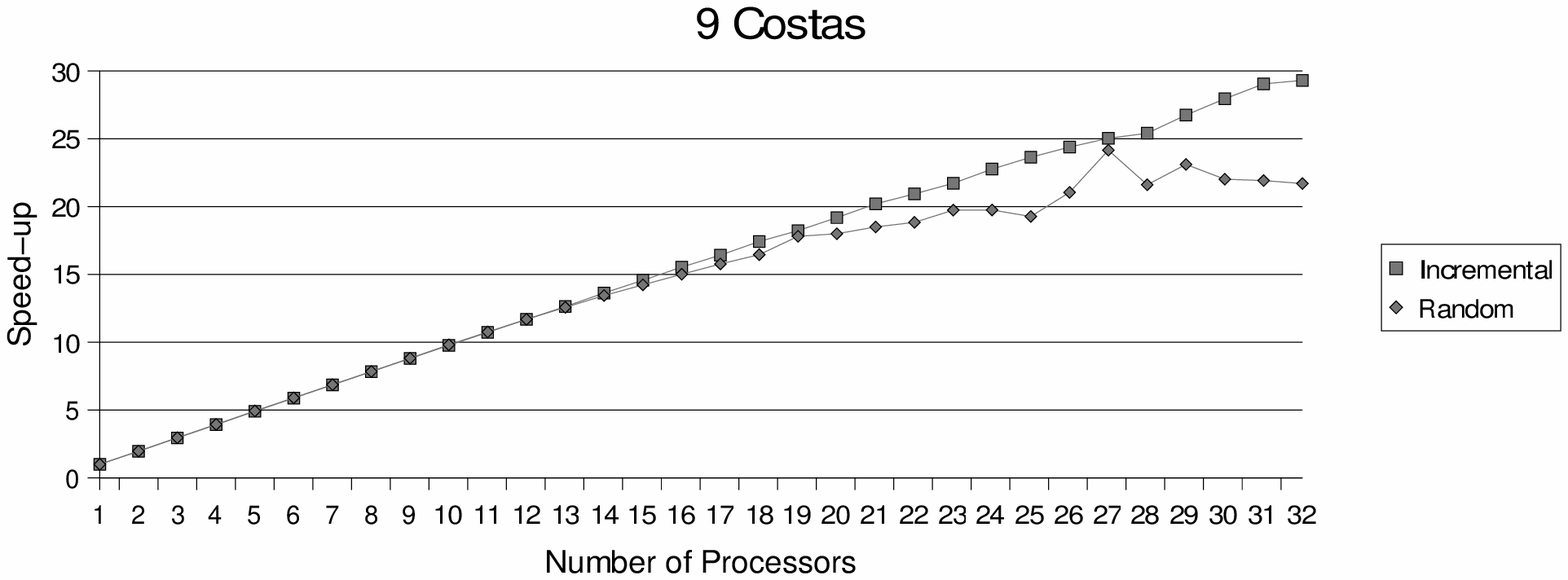,width=.52\textwidth}
\psfig{figure=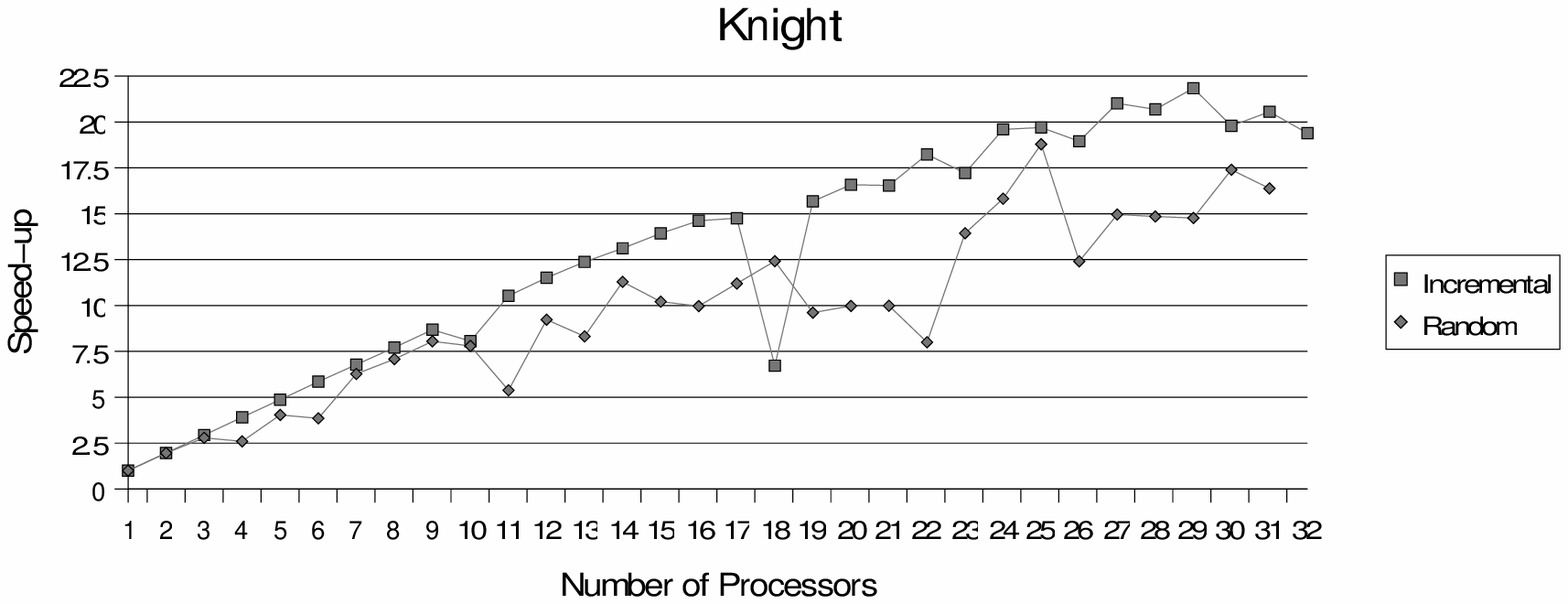,width=.52\textwidth}
\end{minipage}
\begin{minipage}[b]{\textwidth}
\psfig{figure=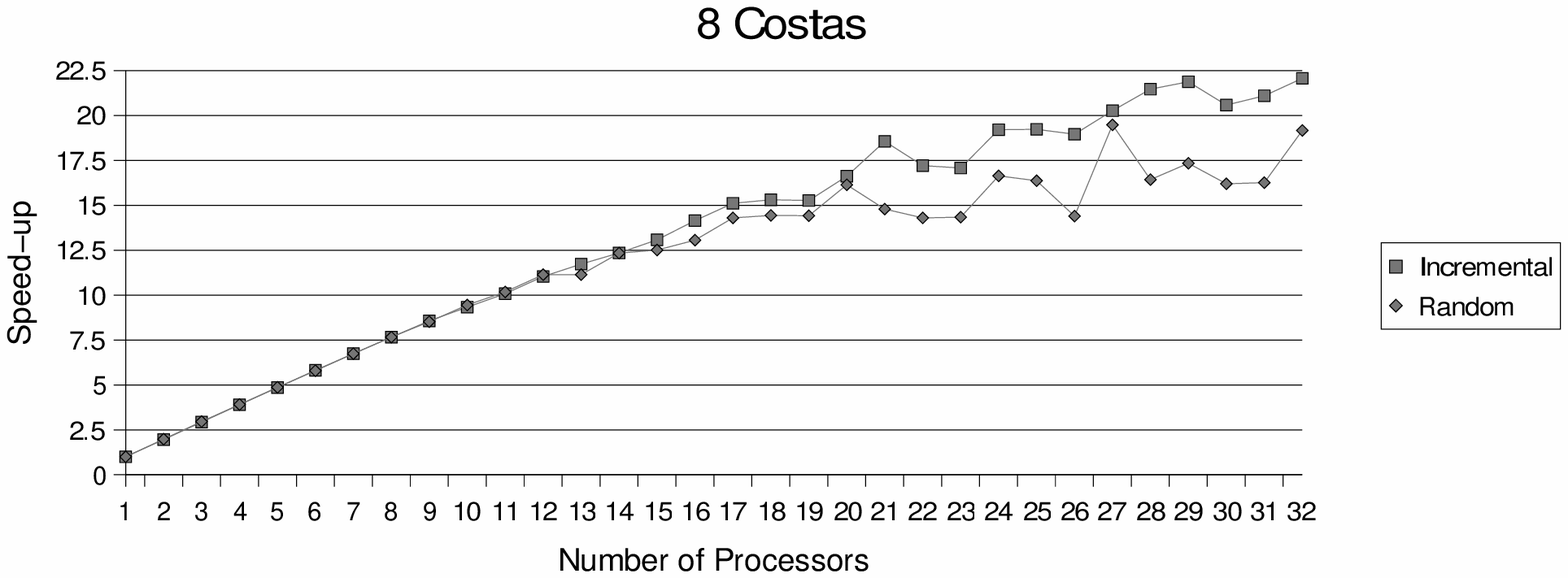,width=.52\textwidth}
\psfig{figure=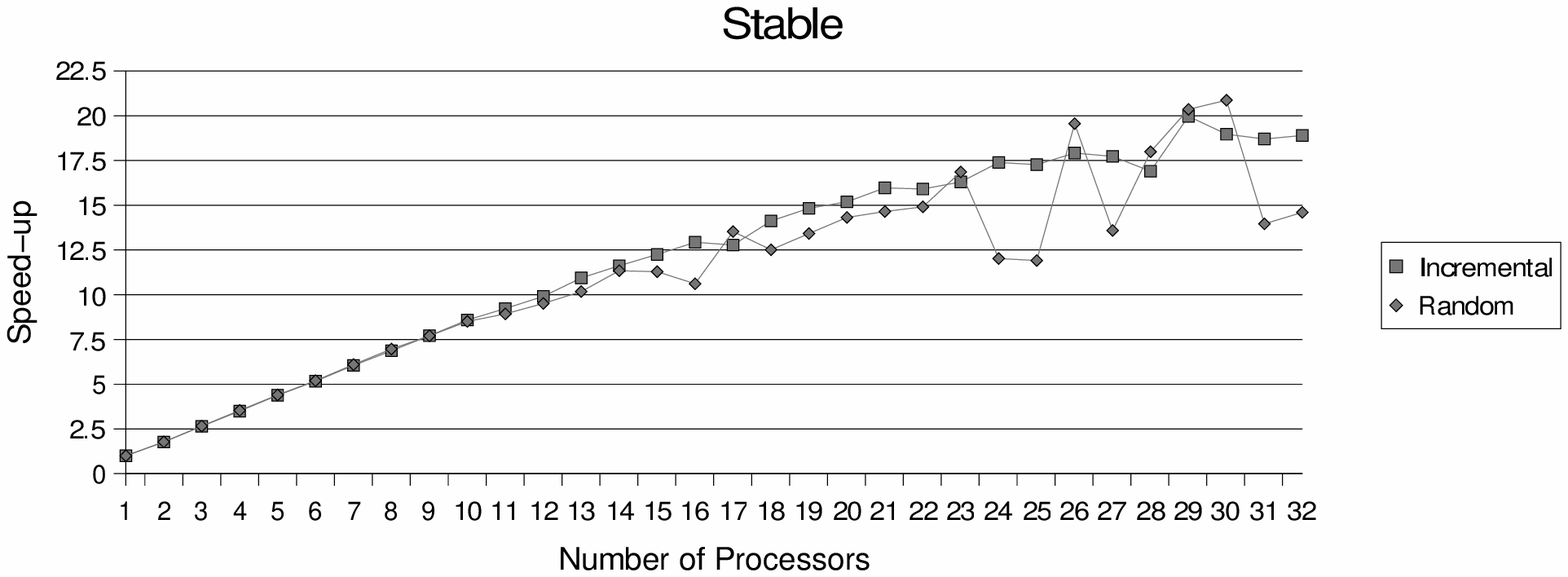,width=.52\textwidth}
\end{minipage}
\begin{minipage}[b]{\textwidth}
\psfig{figure=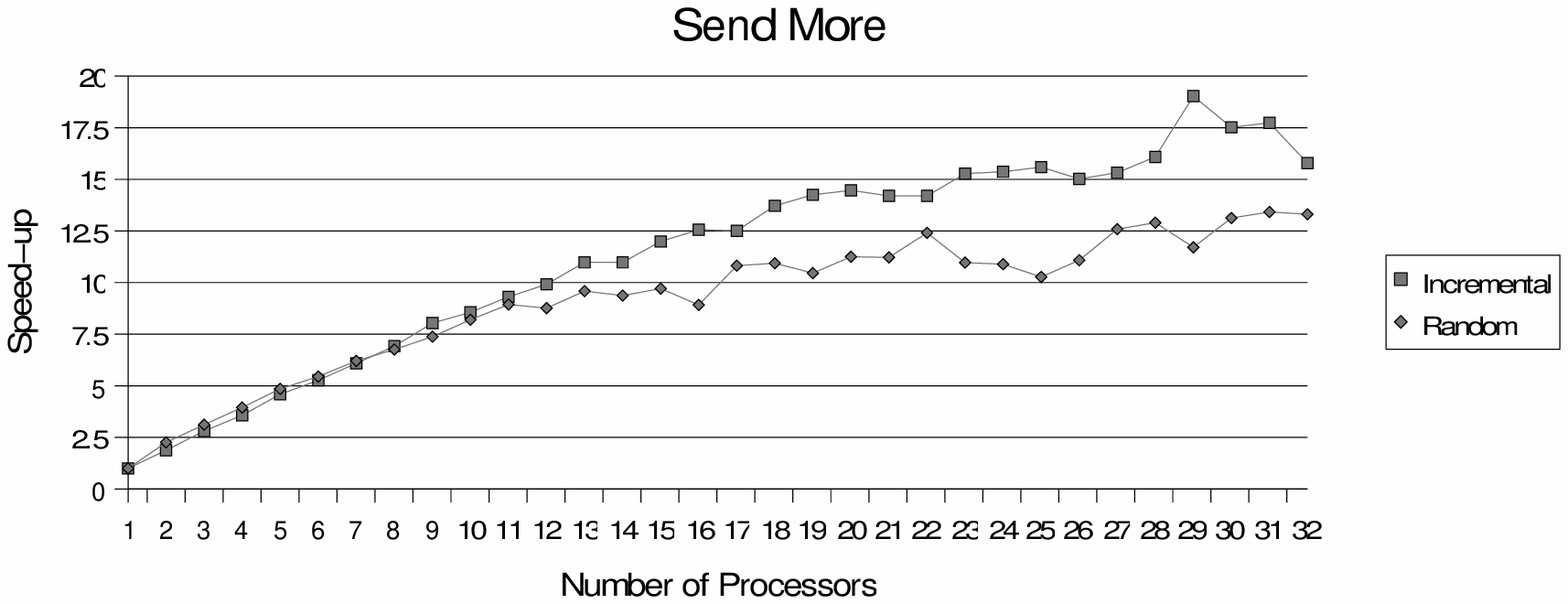,width=.52\textwidth}
\psfig{figure=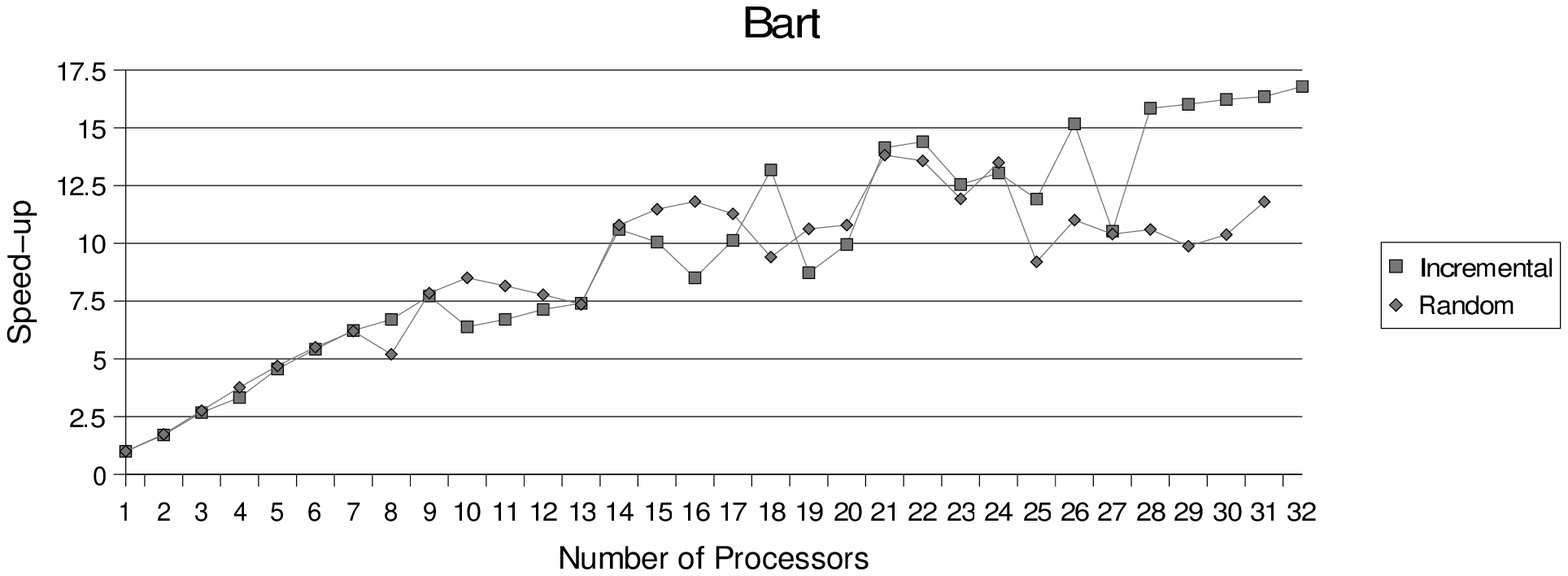,width=.52\textwidth}
\end{minipage}
\begin{minipage}[b]{\textwidth}
\psfig{figure=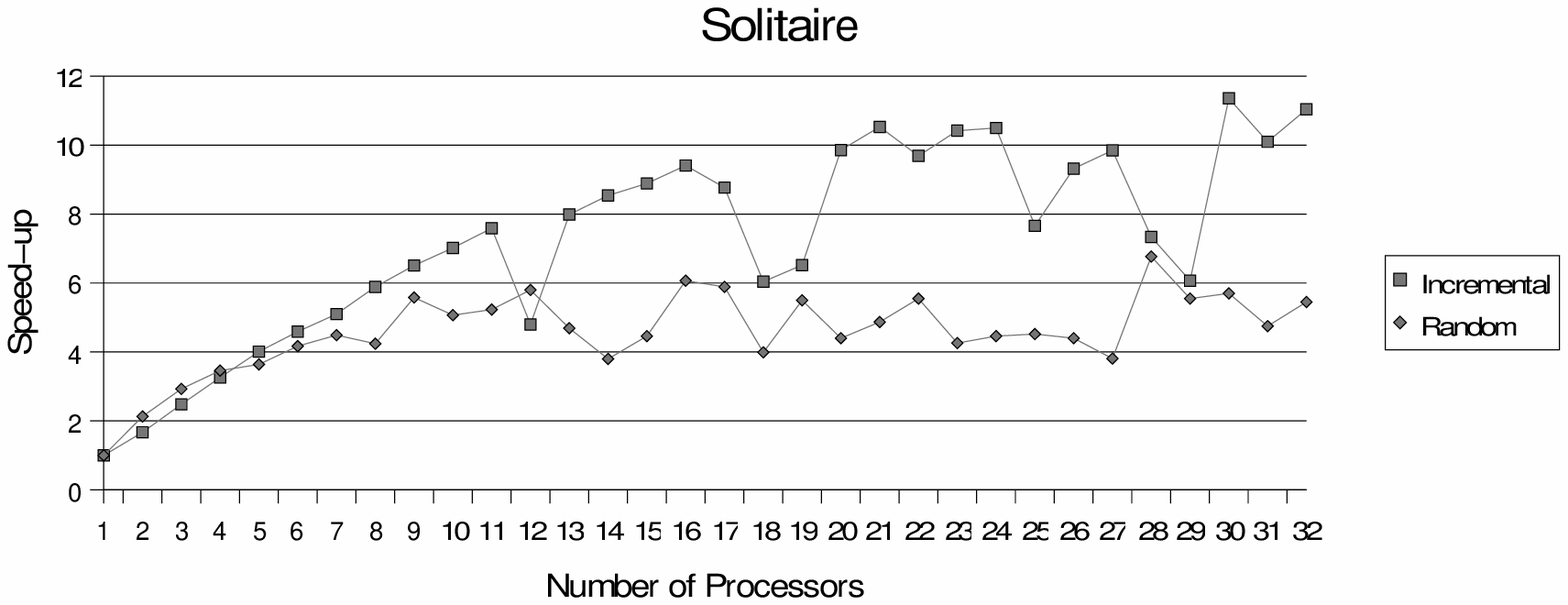,width=.52\textwidth}
\psfig{figure=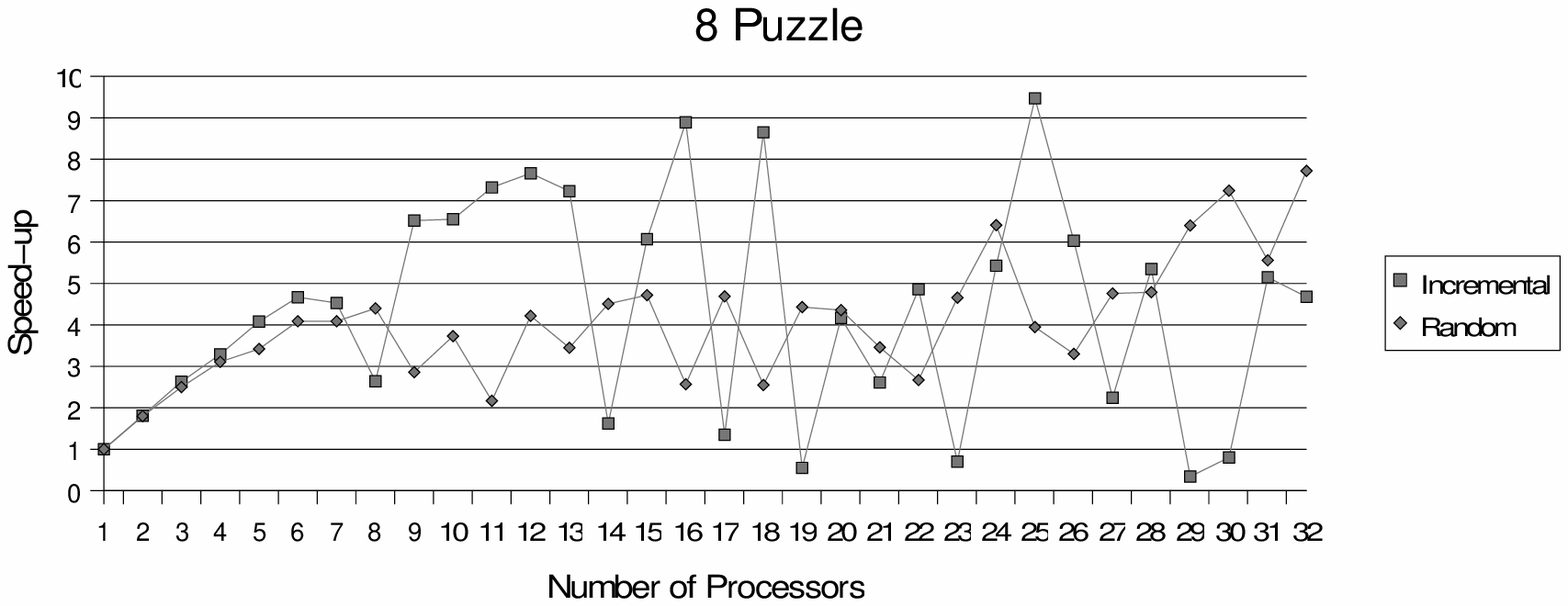,width=.52\textwidth}
\end{minipage}
\begin{minipage}[b]{\textwidth}
\psfig{figure=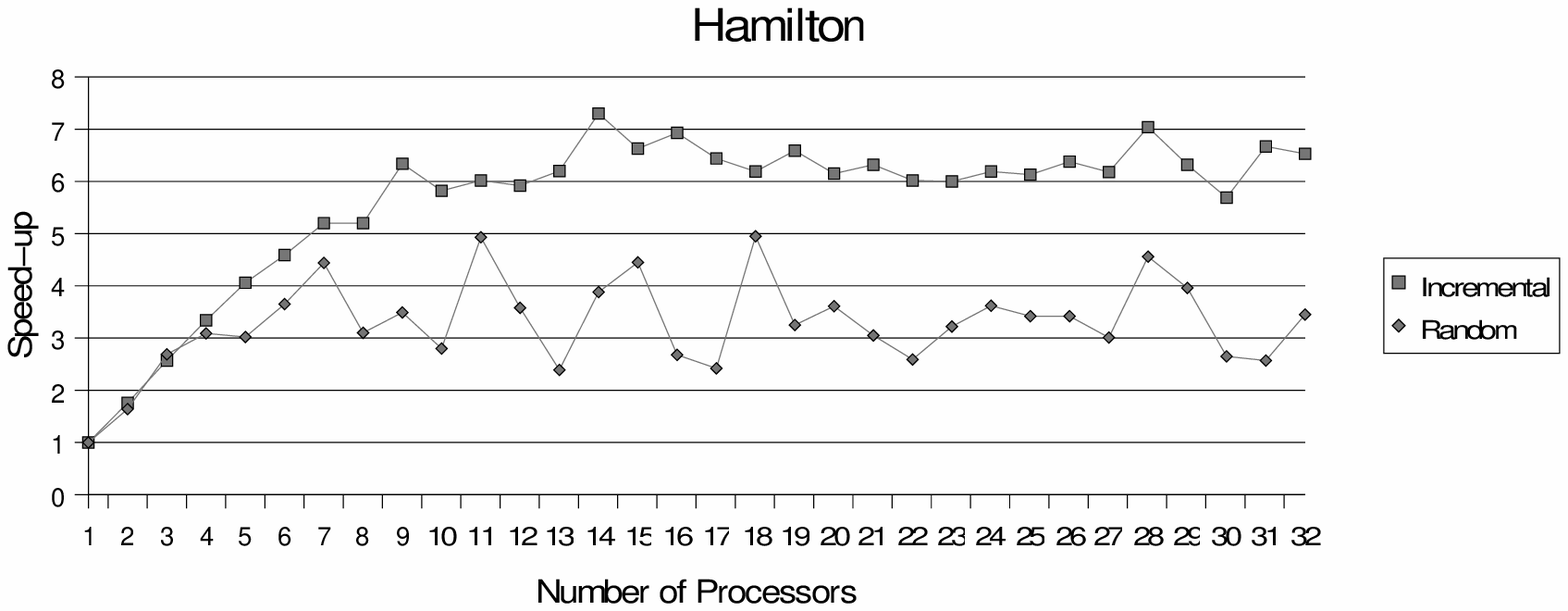,width=.52\textwidth}
\psfig{figure=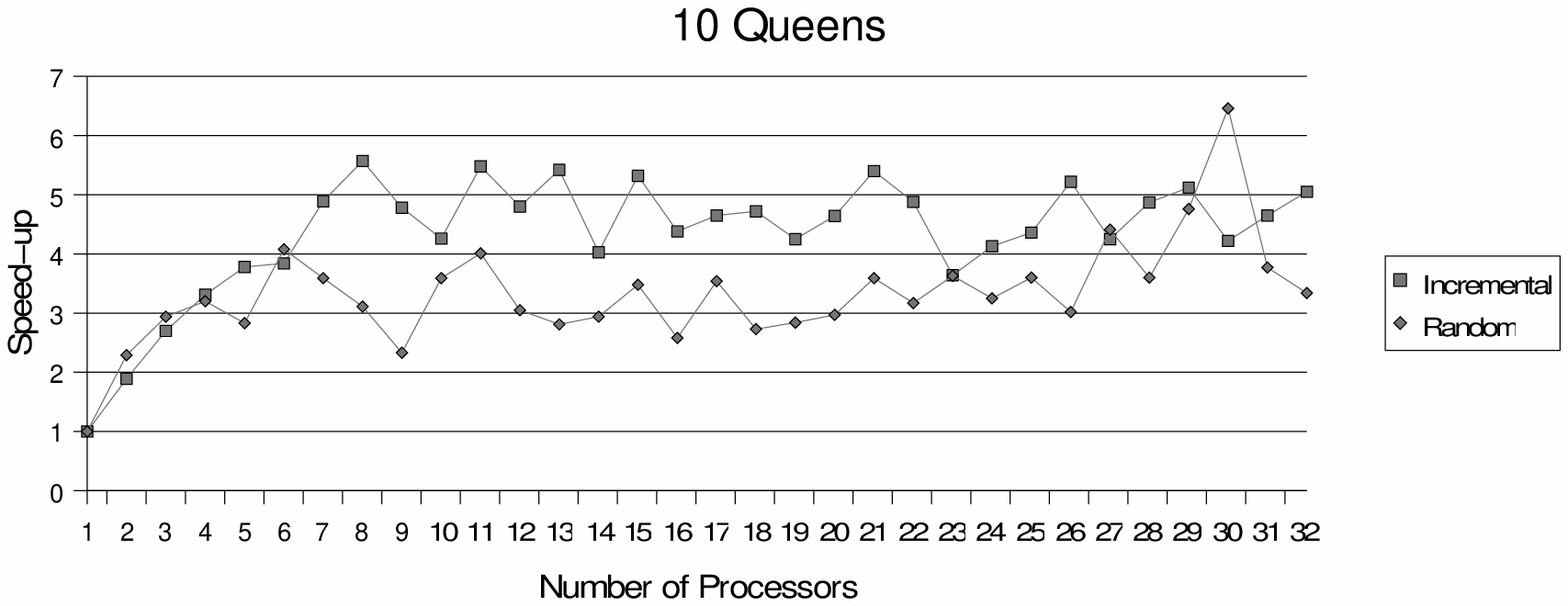,width=.52\textwidth}
\end{minipage}
\begin{minipage}[b]{\textwidth}
\psfig{figure=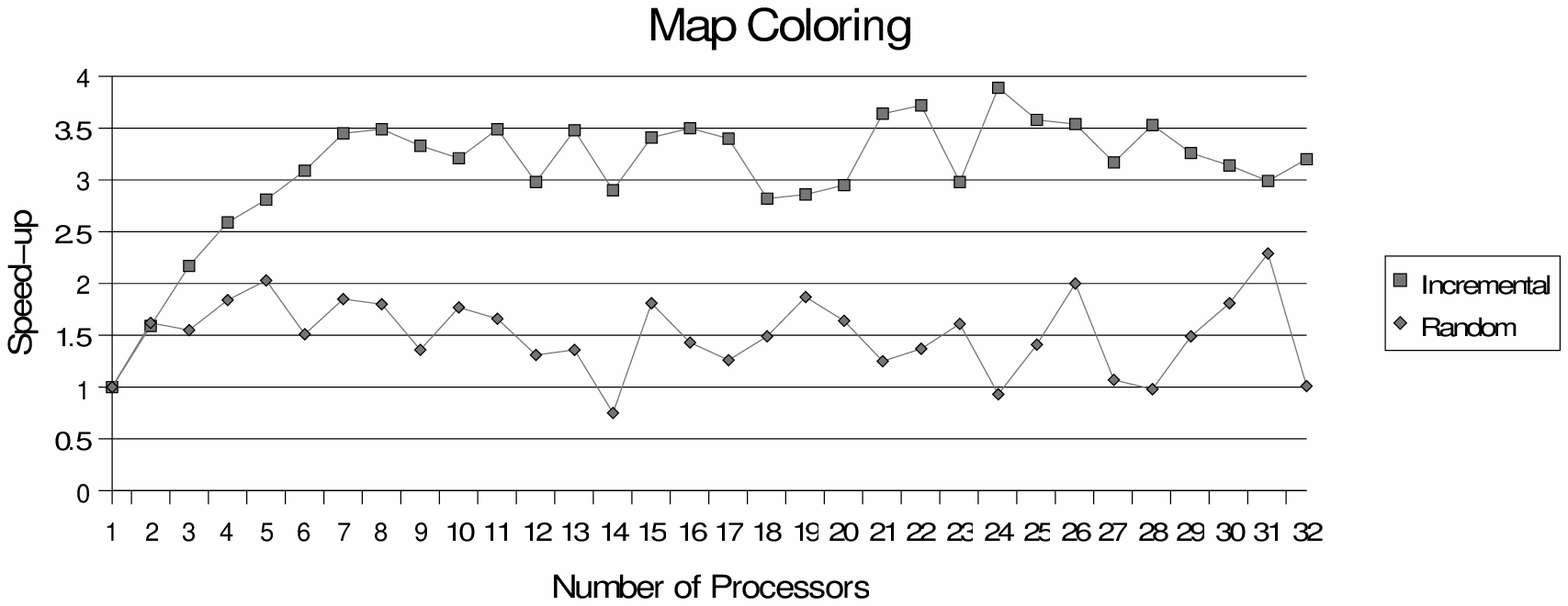,width=.52\textwidth}
\psfig{figure=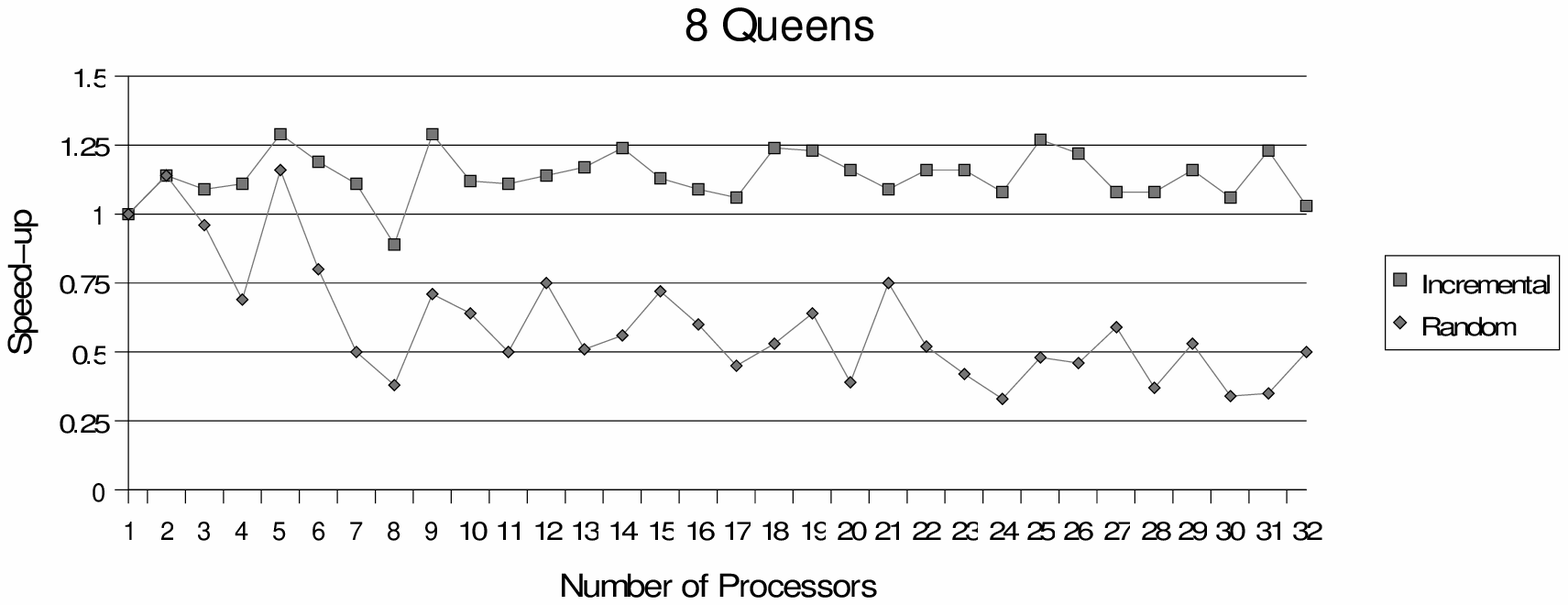,width=.52\textwidth}
\end{minipage}
\caption{Load-based  vs. Random Scheduling}
\label{incre-random}
\end{center}
\end{figure}

The results presented in \cite{warren-93} suggest that random selection of work may provide
a simple and effective alternative when searching for or-parallel work. We have experimented with
this idea, by modifying the scheduler to select any busy agent for scheduling instead of the
one with the highest load. The idea is to avoid bottleneck situations where multiple idle agents
are concentrating their requests for work towards the same busy agent. We have named
this new version of the scheduler \emph{Random Scheduler}.
In this version, an idle agent searches its load vector for the next agent with load
greater than a given small  threshold (effectively performing a round-robin management). Figure~\ref{incre-random} compares the speedups observed in
the Random scheduler with those from the standard bottom-most scheduling with selection of agent
with highest load. The results   indicate that the Random scheduler is less effective. 
This suggests that selecting work from the agent with highest load is not a severe bottleneck
and sending requests to possibly lightly loaded agents may increase the number of calls to
the scheduler.

\begin{figure}
\begin{center}
\psfig{figure=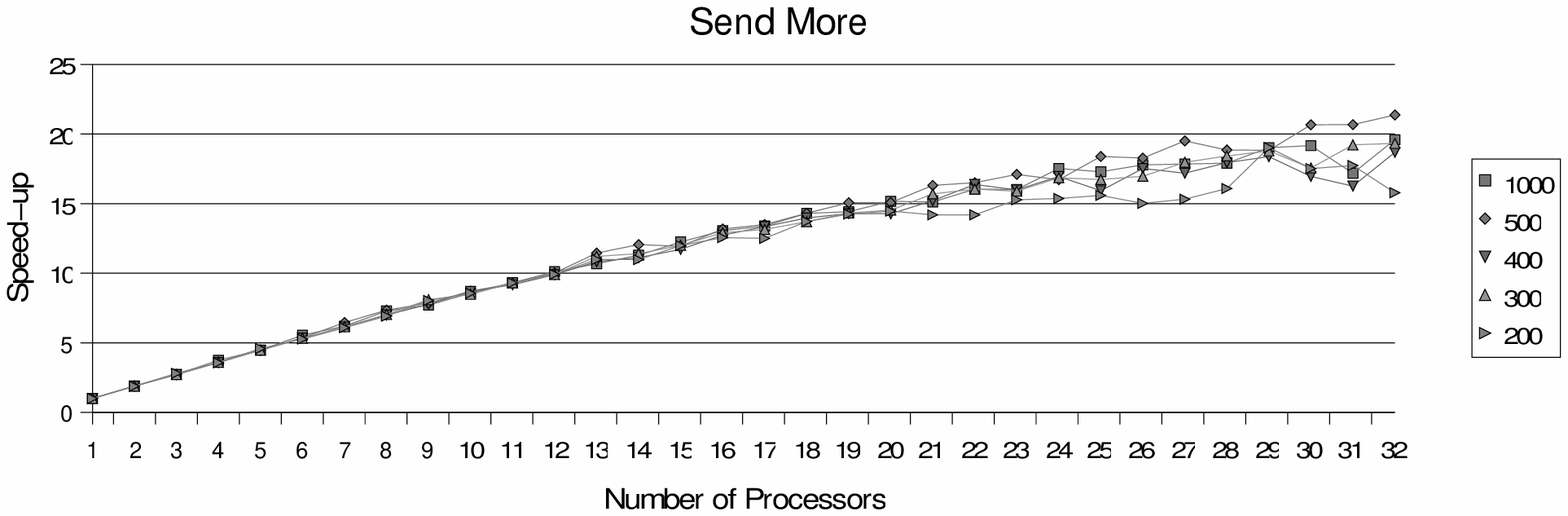,width=.62\textwidth}\\
\psfig{figure=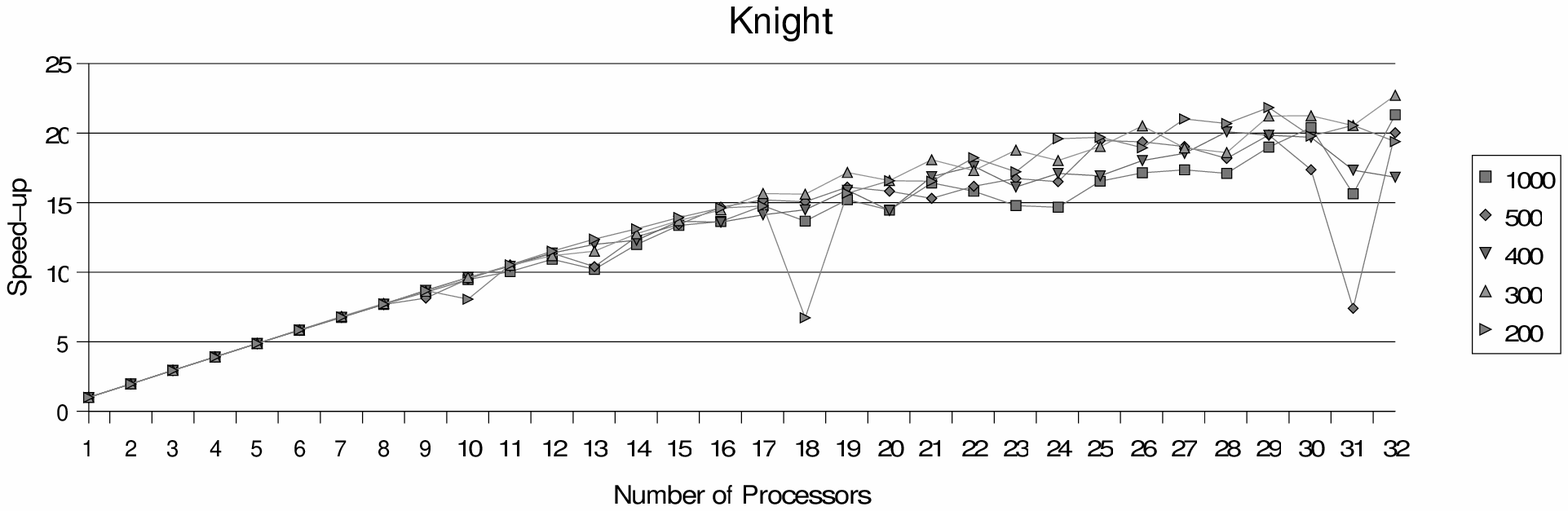,width=.62\textwidth}\\
\psfig{figure=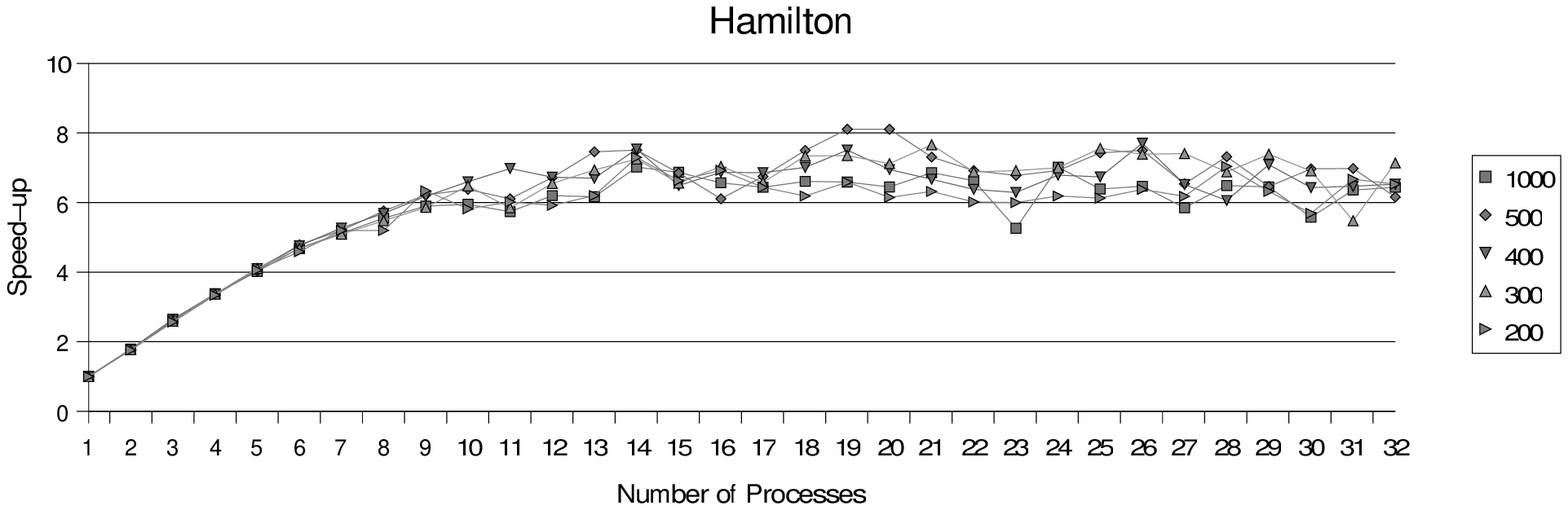,width=.62\textwidth}\\
\psfig{figure=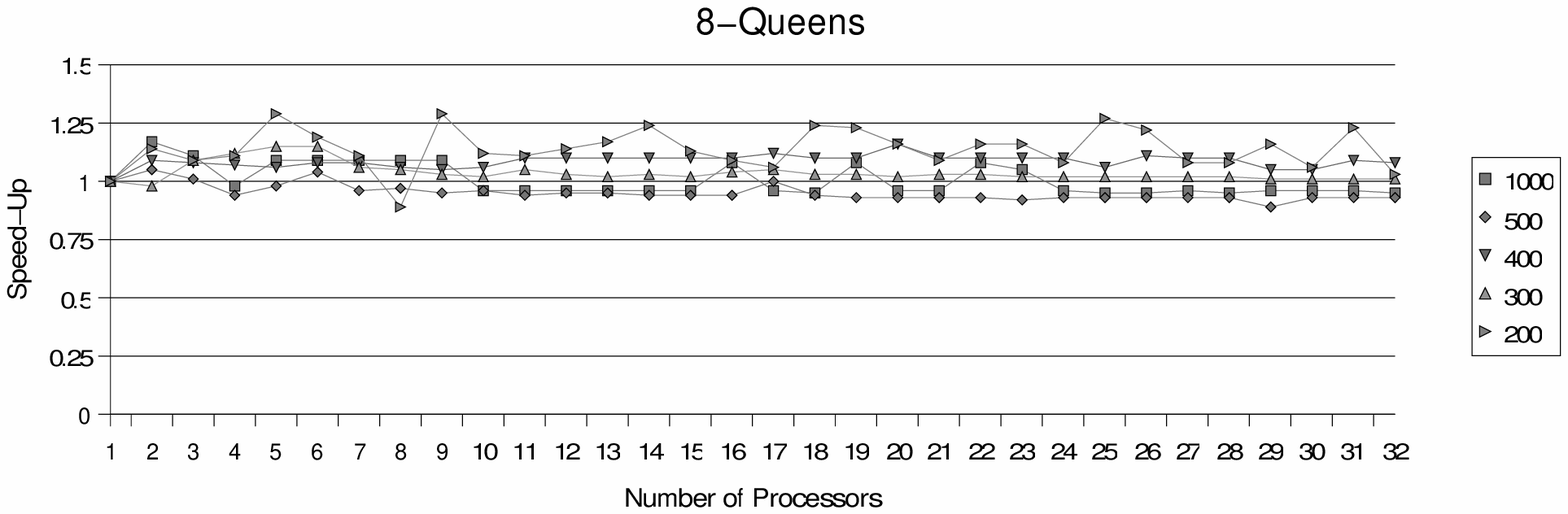,width=.62\textwidth}\\
\psfig{figure=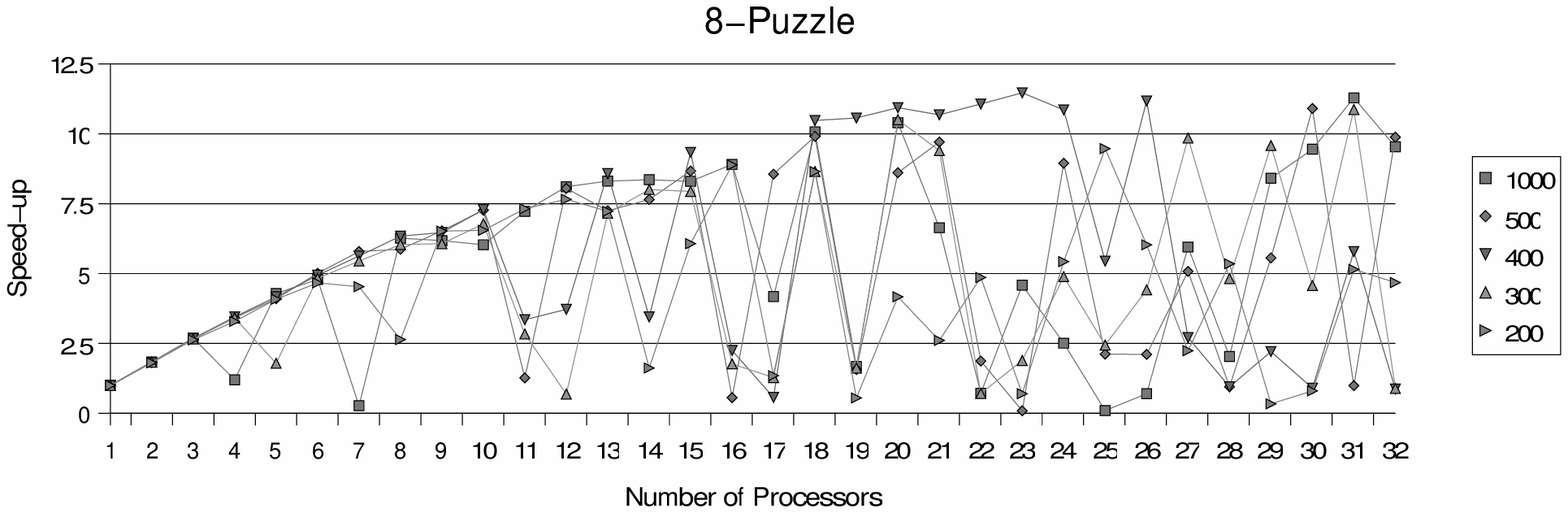,width=.62\textwidth}\\
\psfig{figure=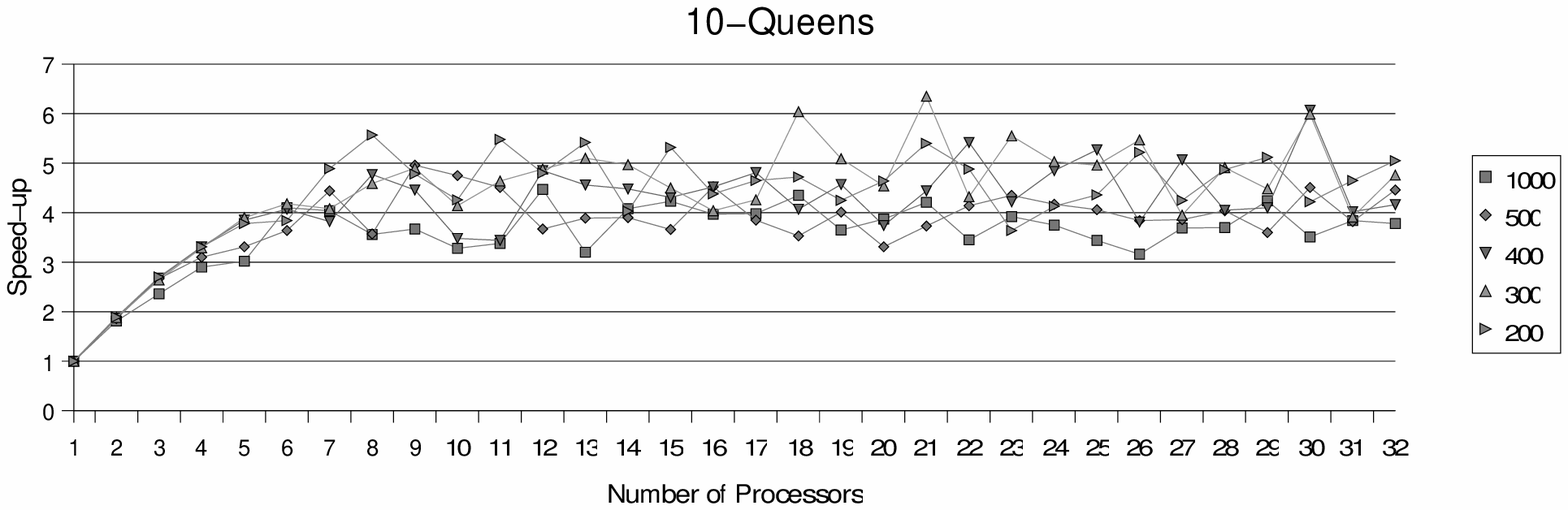,width=.62\textwidth}
\caption{Incremental Stack-Splitting Message Checking Frequencies (1)}
\label{incre-freq1}
\end{center}
\end{figure}

\begin{figure}
\begin{center}
\psfig{figure=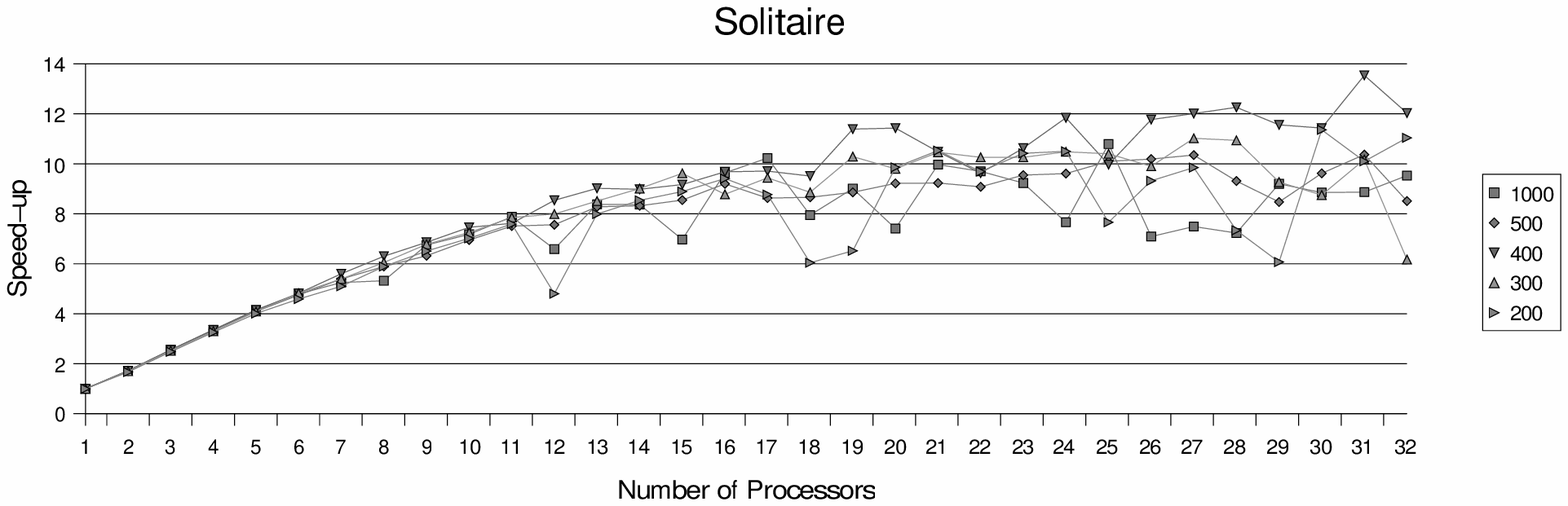,width=.62\textwidth}\\
\psfig{figure=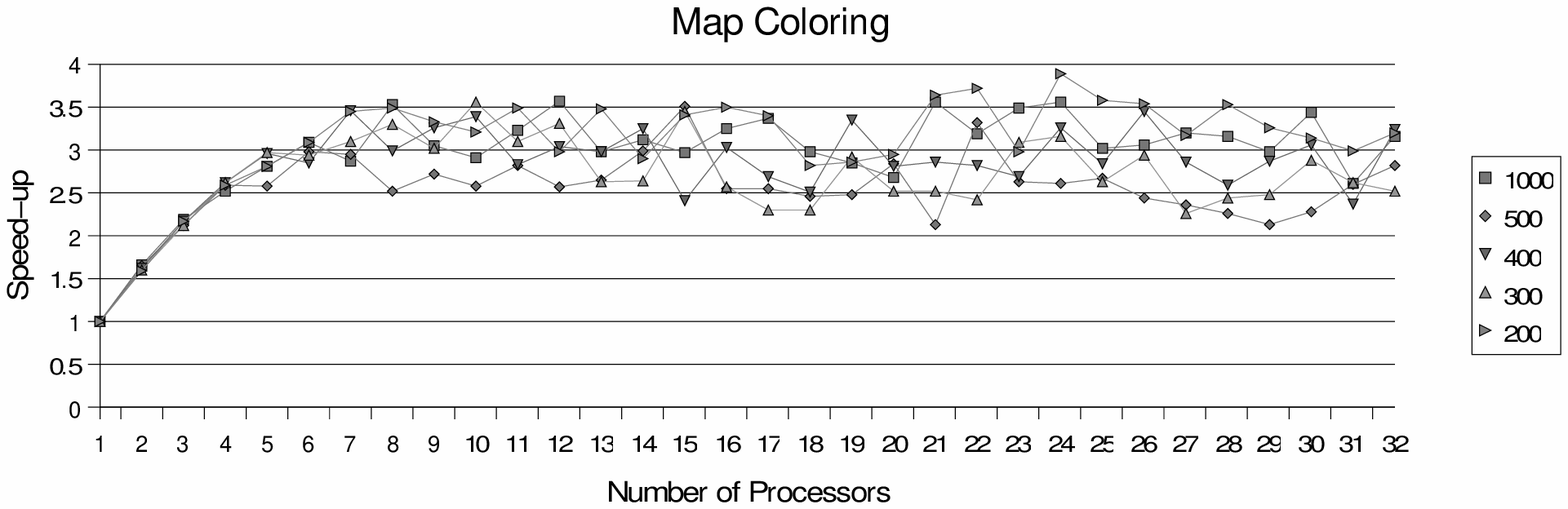,width=.62\textwidth}\\
\psfig{figure=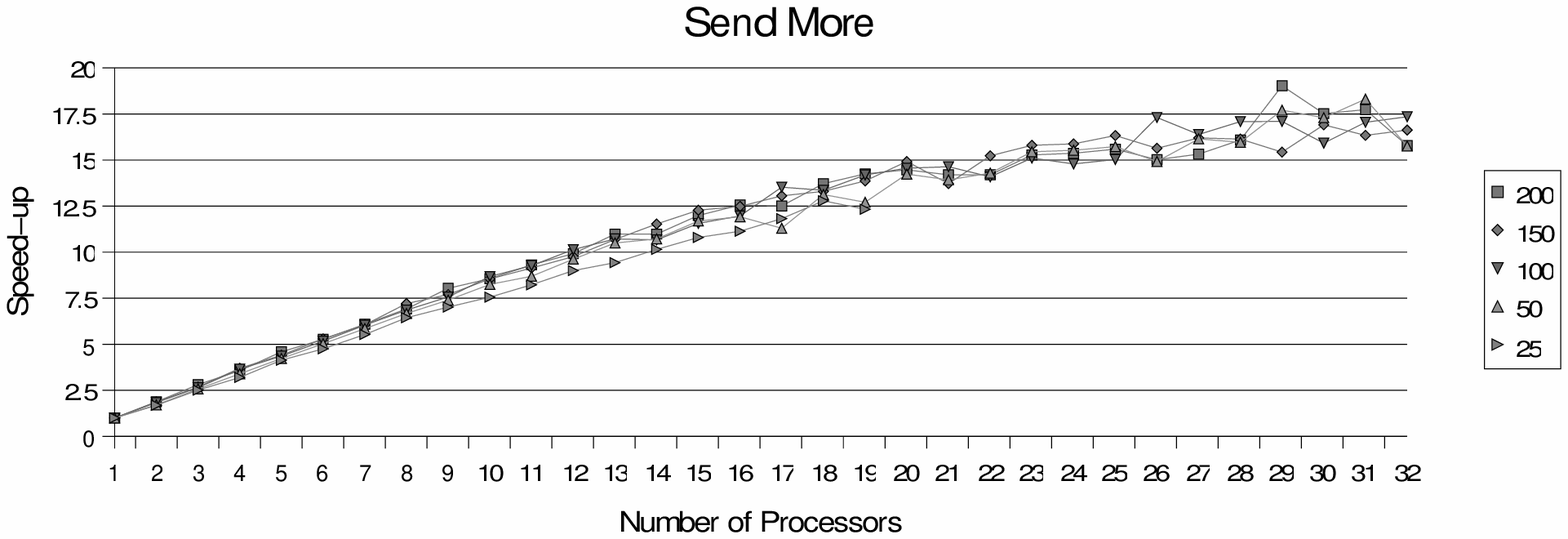,width=.62\textwidth}\\
\psfig{figure=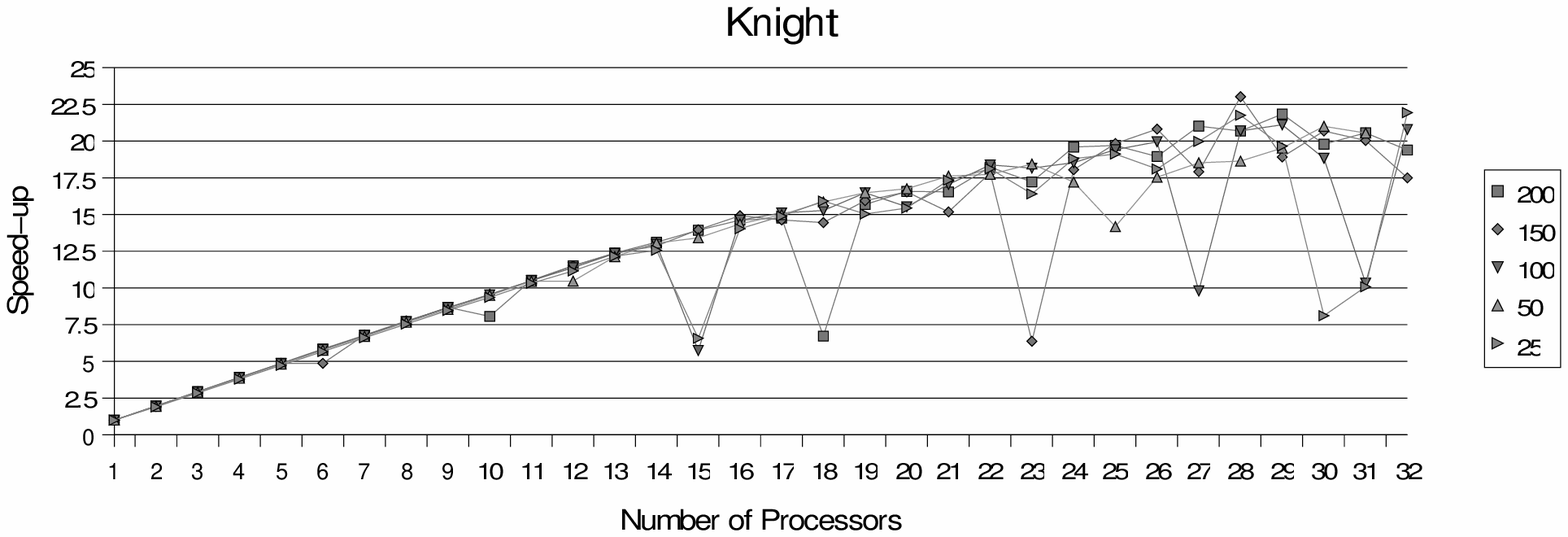,width=.62\textwidth}\\
\psfig{figure=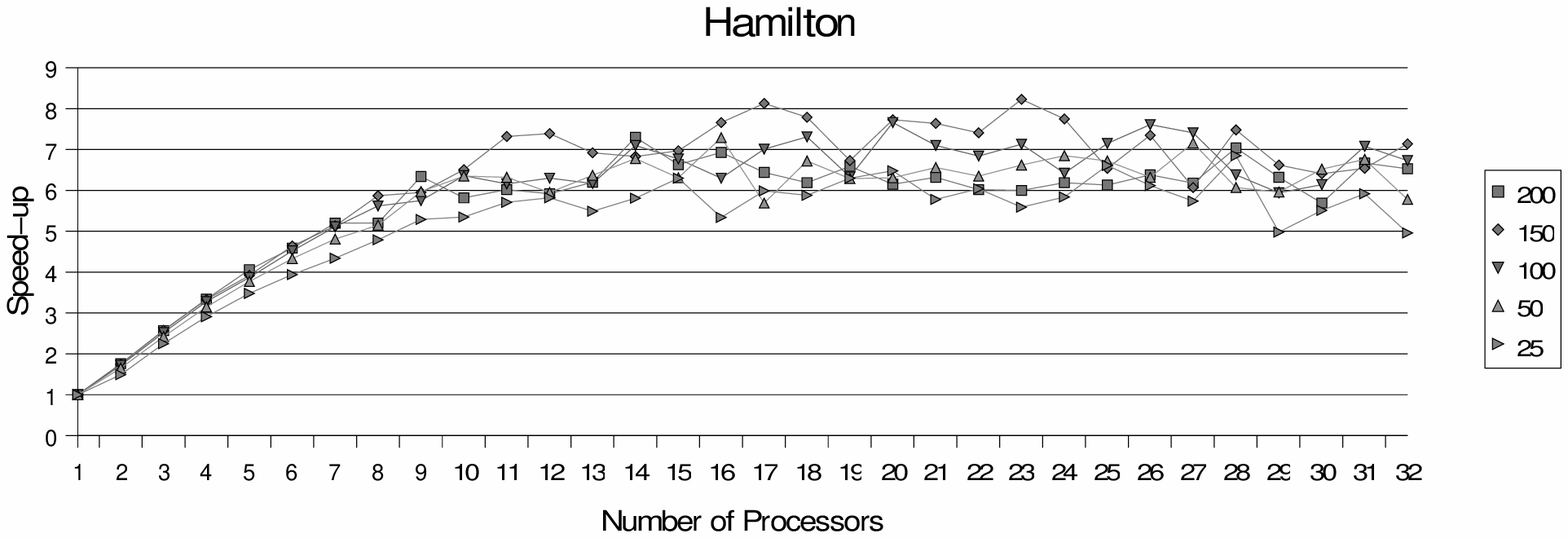,width=.62\textwidth}\\
\psfig{figure=figures/8queens_if1.eps,width=.62\textwidth}
\caption{Incremental Stack-Splitting Message Checking Frequencies (2)}
\label{incre-freq2}
\end{center}
\end{figure}

\subsubsection{Tuning the System}

The implementation of stack-splitting depends on a number of parameters, such
as \emph{(1)} the frequency at which each agent checks for incoming requests, and 
\emph{(2)} the
frequency of propagation of load information. We have performed a number of 
experiments to study the impact of these parameters on the overall performance.

\begin{figure}
\begin{center}
\psfig{figure=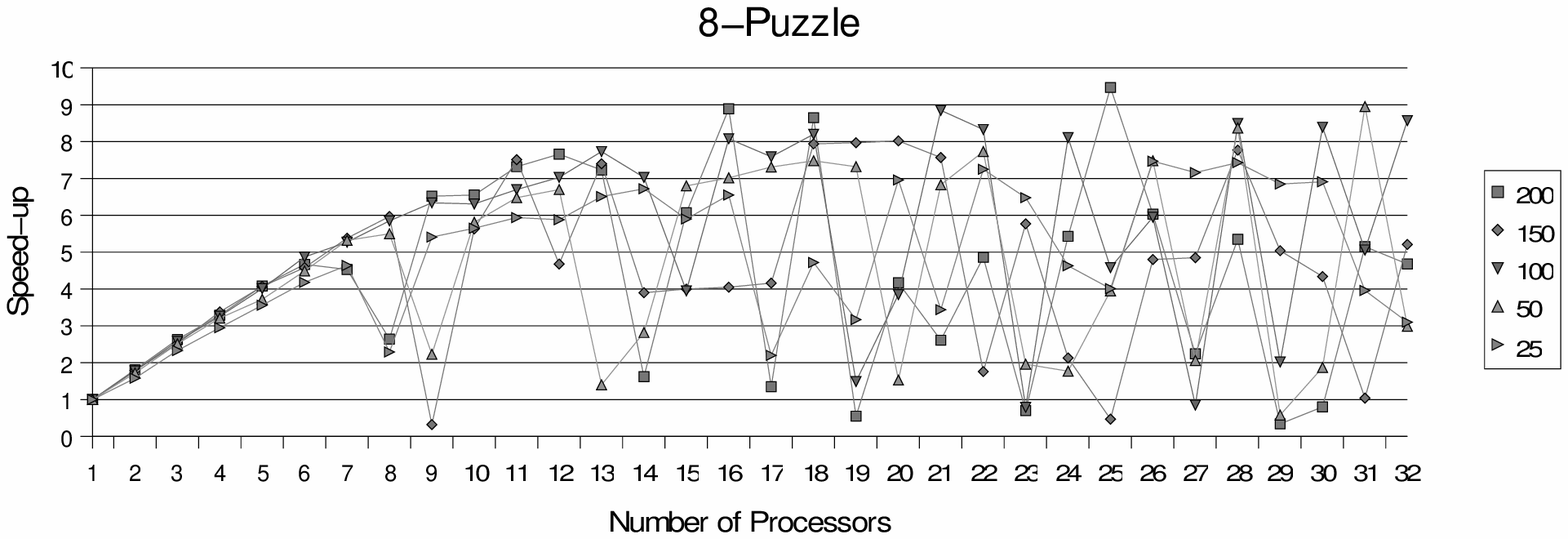,width=.65\textwidth}\\
\psfig{figure=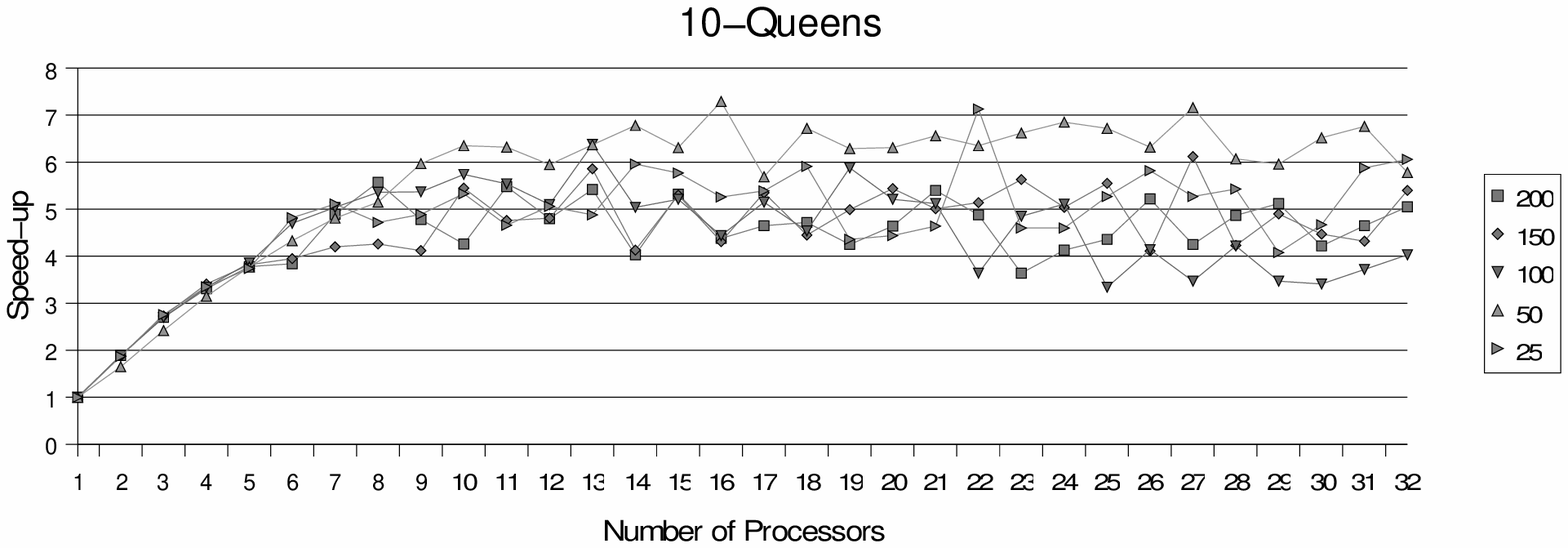,width=.65\textwidth}\\
\psfig{figure=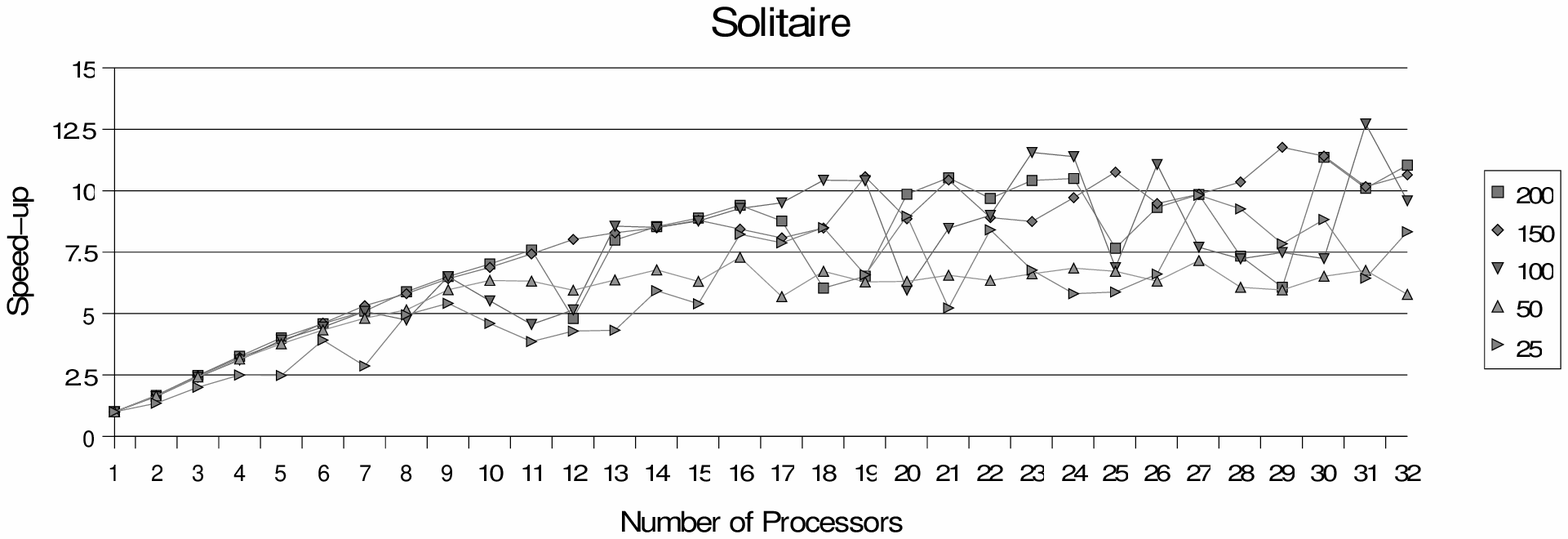,width=.65\textwidth}\\
\psfig{figure=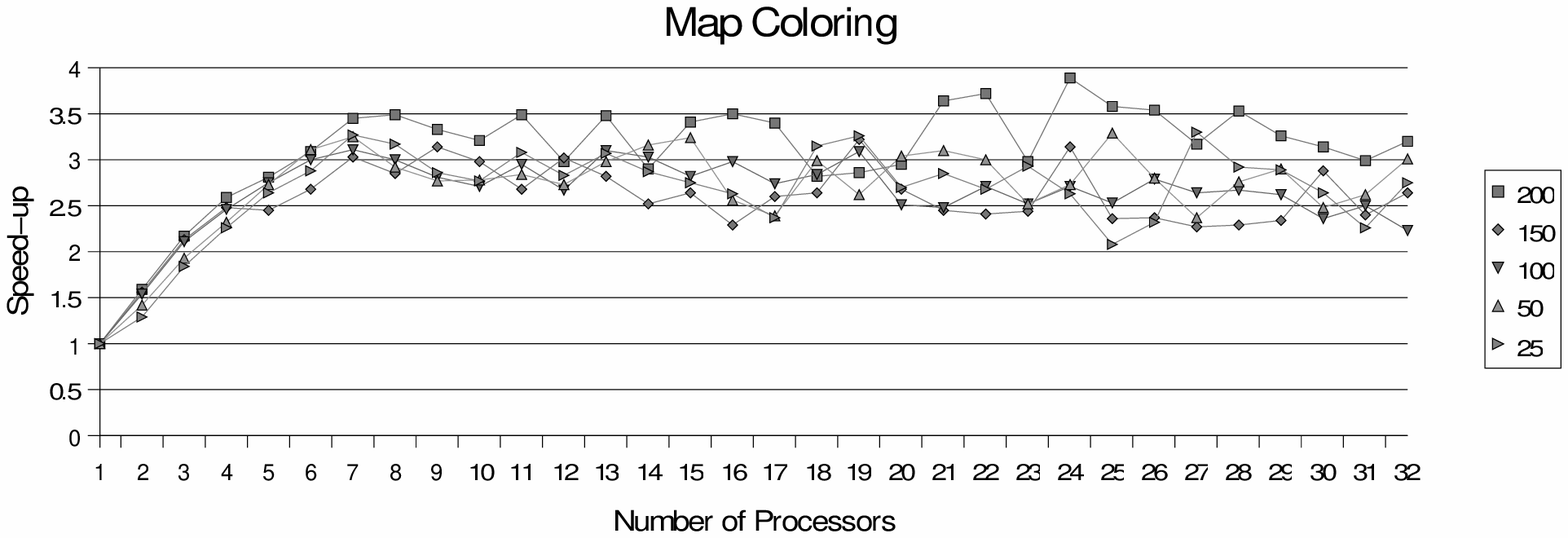,width=.65\textwidth}
\caption{Incremental Stack-Splitting Message Checking Frequencies (3)}
\label{incre-freq3}
\end{center}
\end{figure}

Regarding the first parameter, the previously presented results make use of a
frequency of one test every 200 procedure calls. 
Figs.~\ref{incre-freq1}-\ref{incre-freq3} show that
this choice was the best, although in some benchmarks only
minimal differences can be observed for different
frequency values.

\begin{figure}
\begin{center}
\begin{minipage}[b]{\textwidth}
\psfig{figure=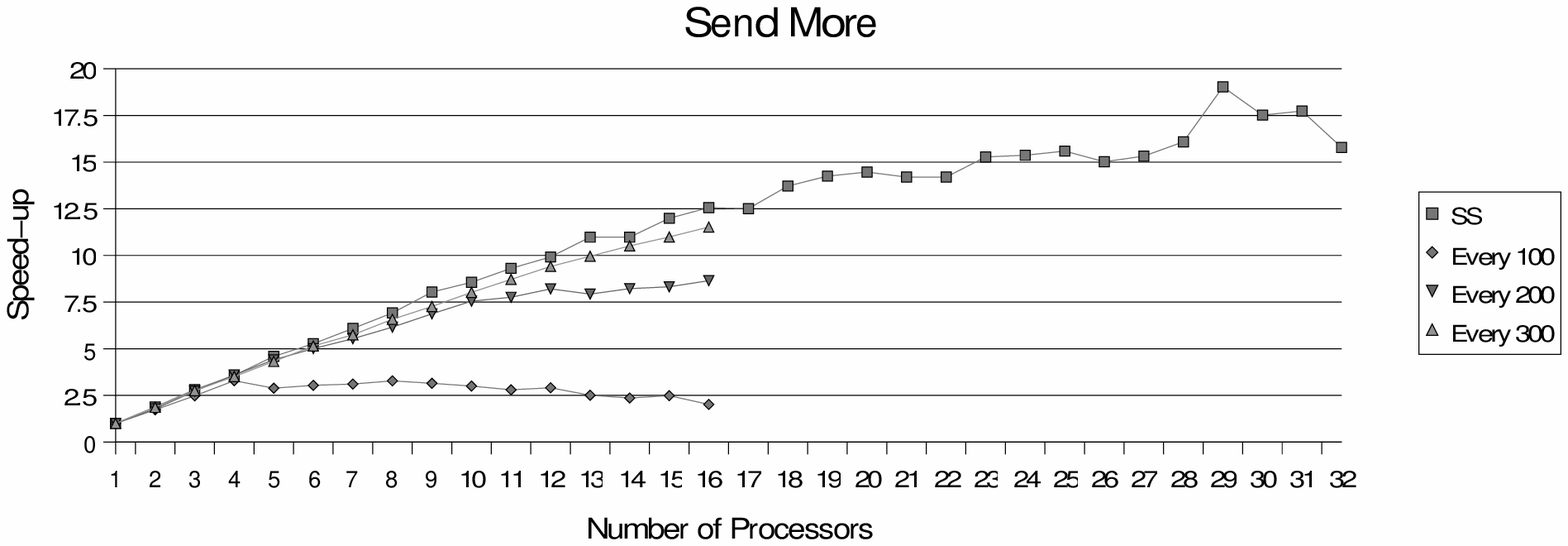,width=.53\textwidth}
\psfig{figure=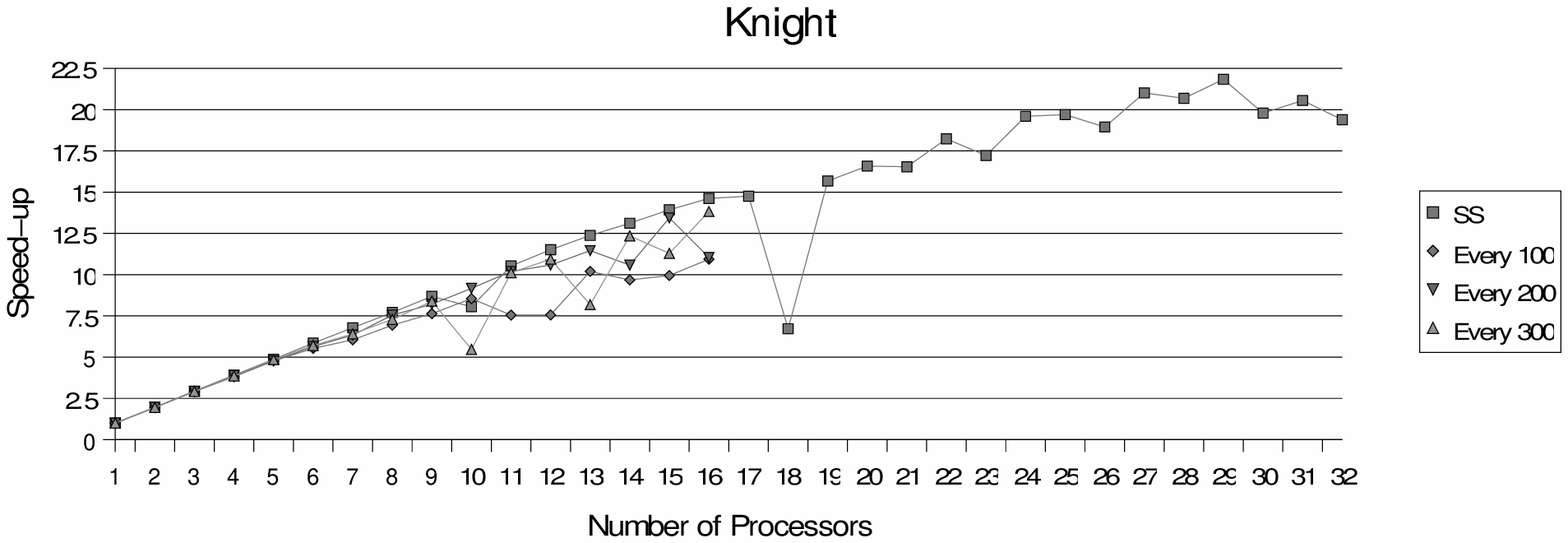,width=.53\textwidth}
\end{minipage}
\begin{minipage}[b]{\textwidth}
\psfig{figure=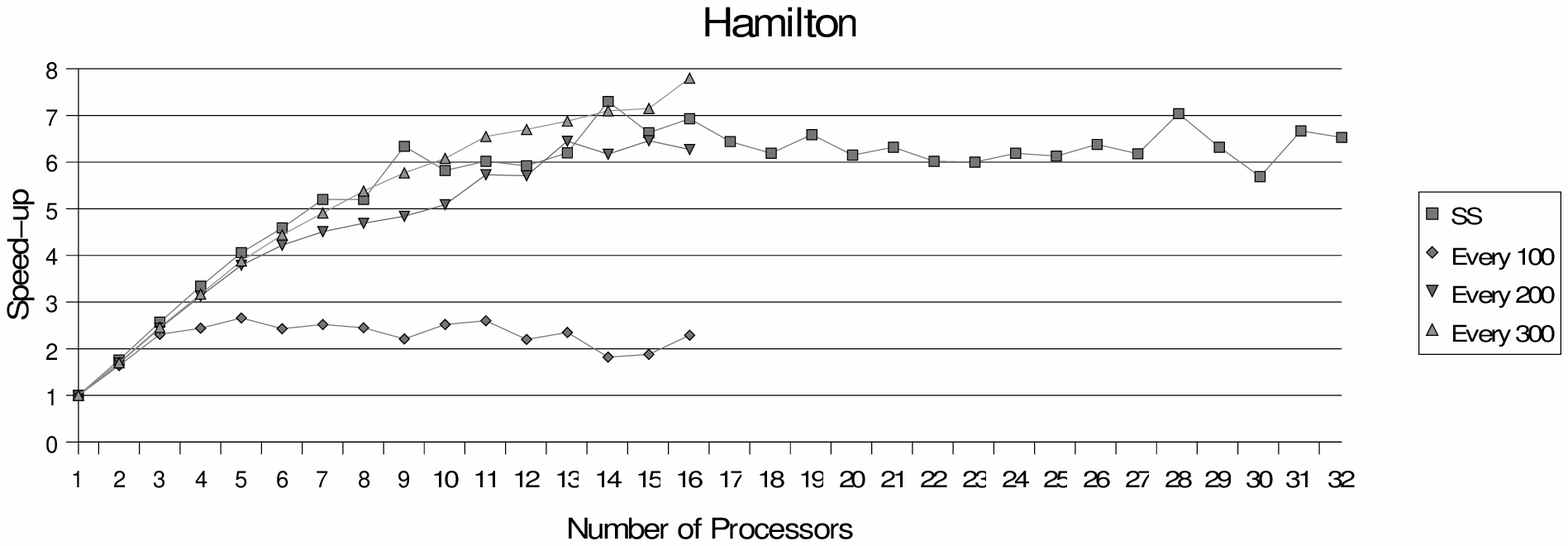,width=.53\textwidth}
\psfig{figure=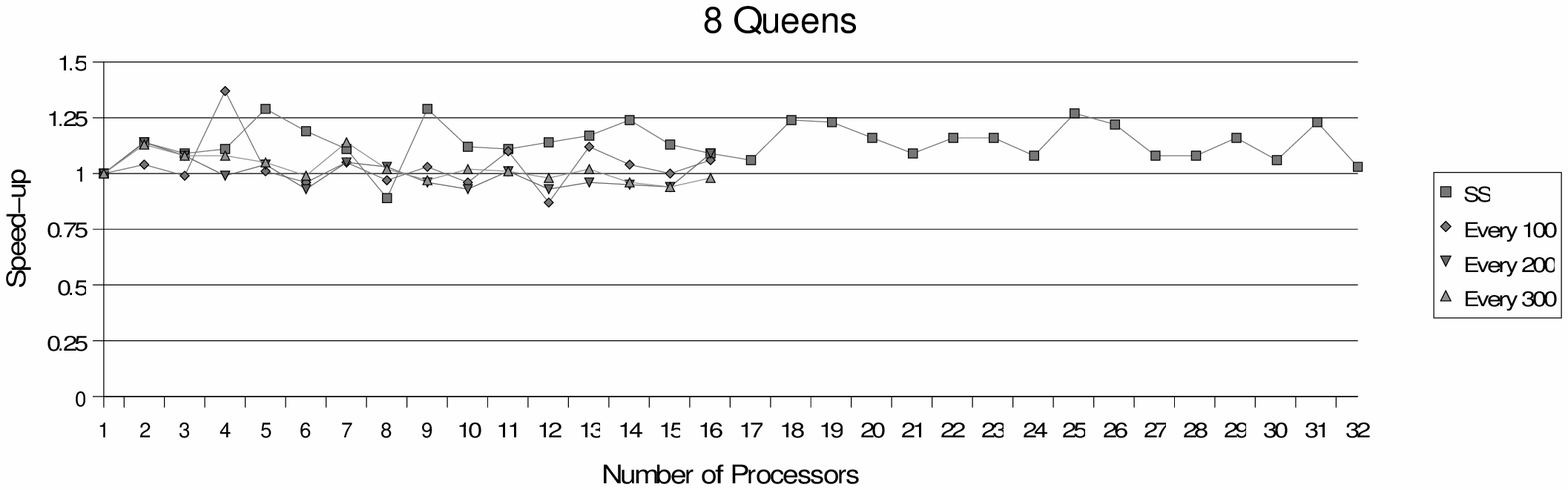,width=.53\textwidth}
\end{minipage}
\begin{minipage}[b]{\textwidth}
\psfig{figure=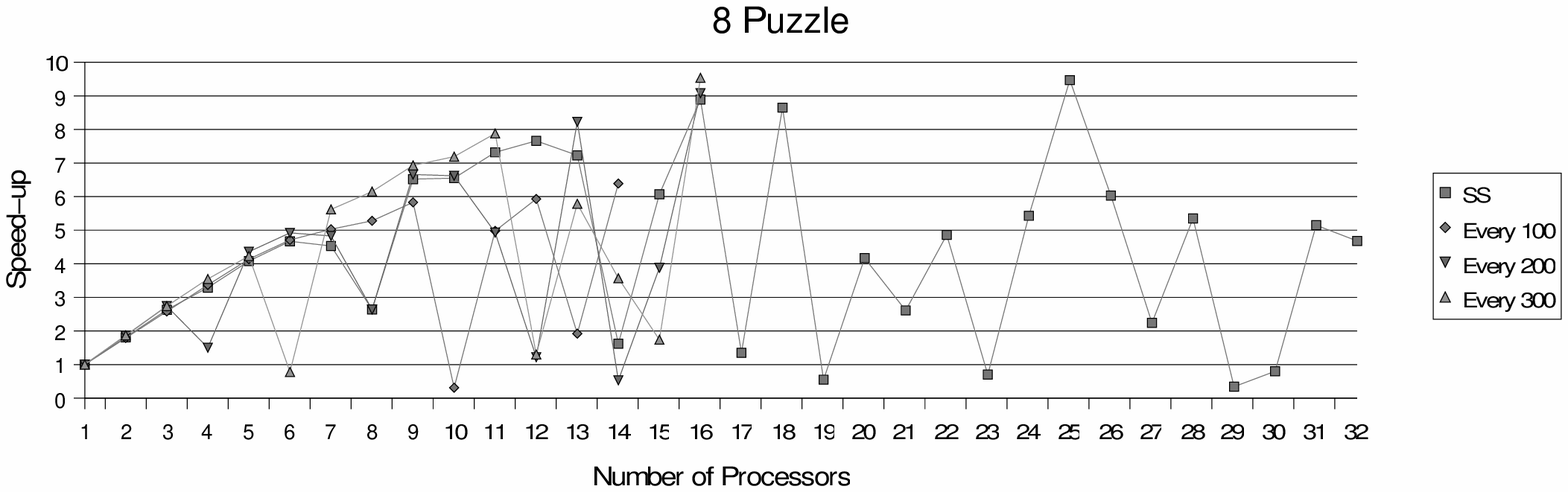,width=.53\textwidth}
\psfig{figure=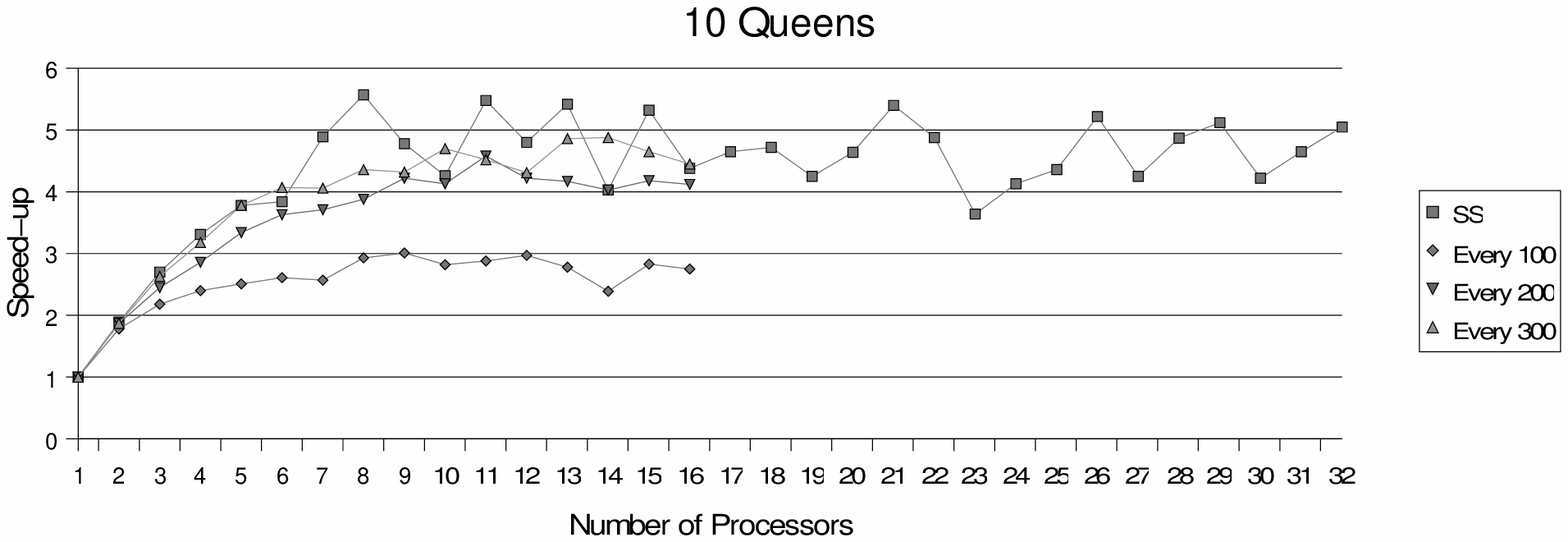,width=.53\textwidth}
\end{minipage}
\begin{minipage}[b]{\textwidth}
\psfig{figure=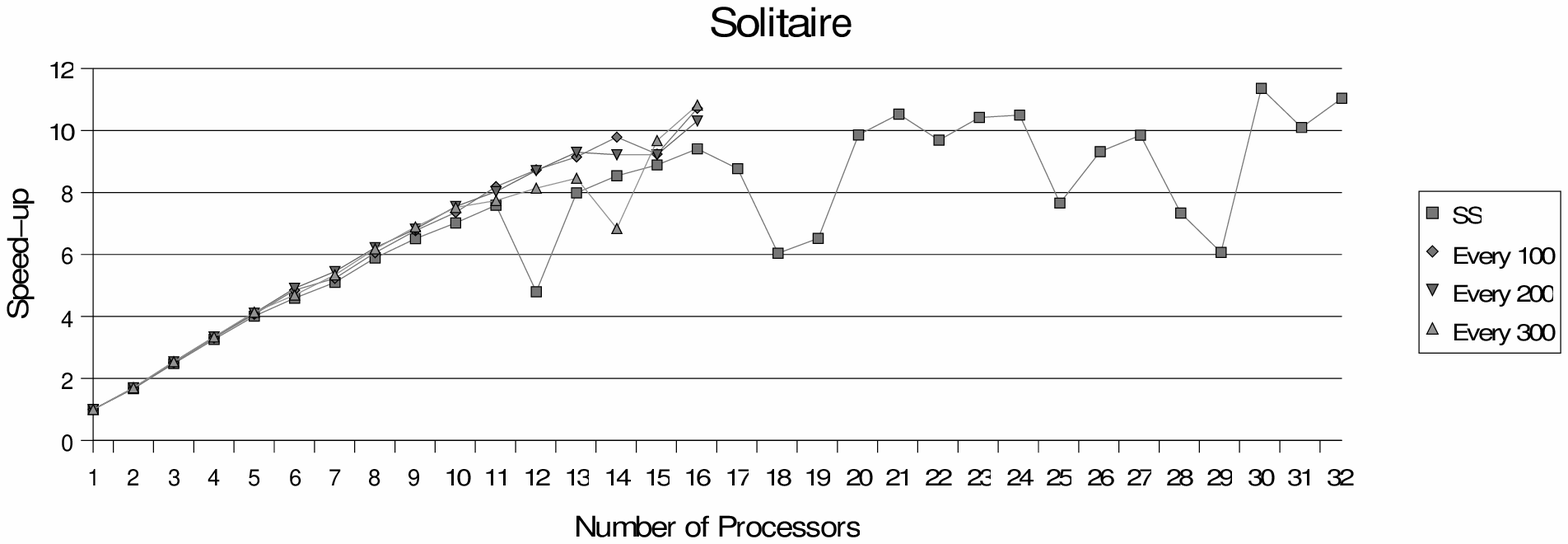,width=.53\textwidth}
\psfig{figure=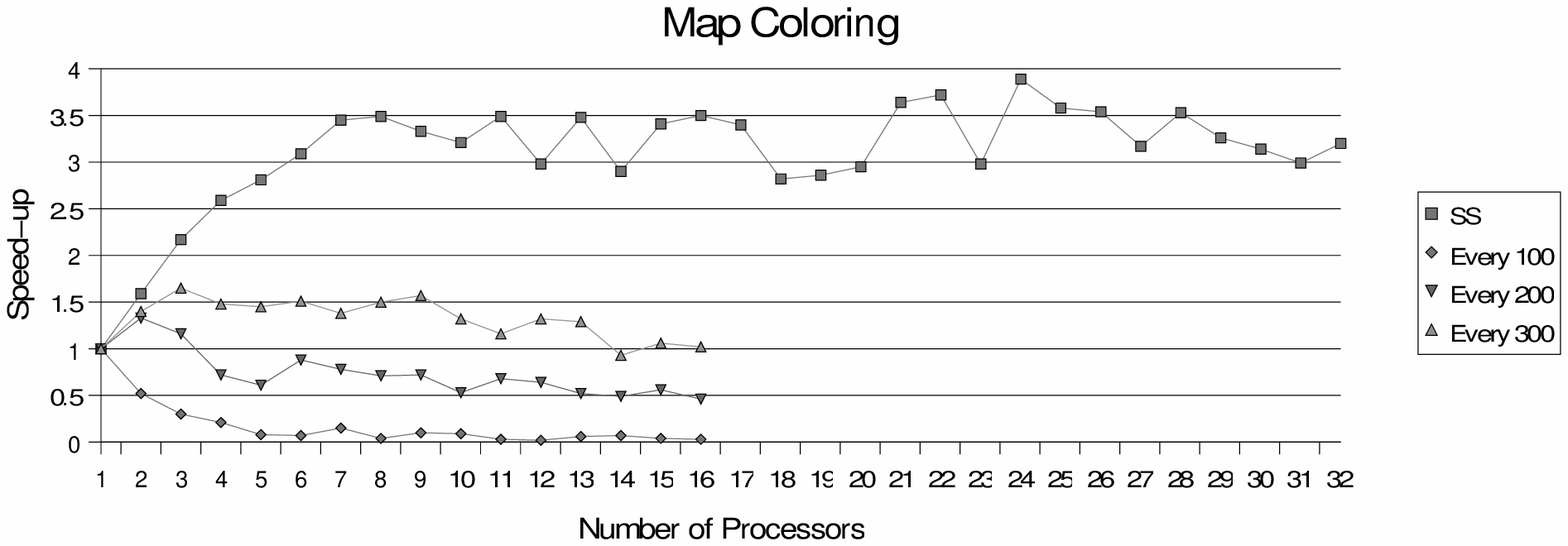,width=.53\textwidth}
\end{minipage}
\caption{Incremental Stack-Splitting vs. Propagation of Load Information}
\label{incre-prop}
\end{center}
\end{figure}

Regarding the second parameter, we are currently propagating load information only
in presence of a sharing operation. We tried to increase the frequency of propagation
of load information, hoping to provide agents with a more accurate view of the
load in the system. The results from this experiment are reported in Figure~\ref{incre-prop}.
As we can see, with the exception of \emph{Hamilton}, in all other cases increasing the
frequency leads only to a higher message traffic without any apparent advantage. In particular,
the higher the frequency, the lower is the resulting speedup.

\begin{figure}
\begin{center}
\begin{minipage}[b]{\textwidth}
\psfig{figure=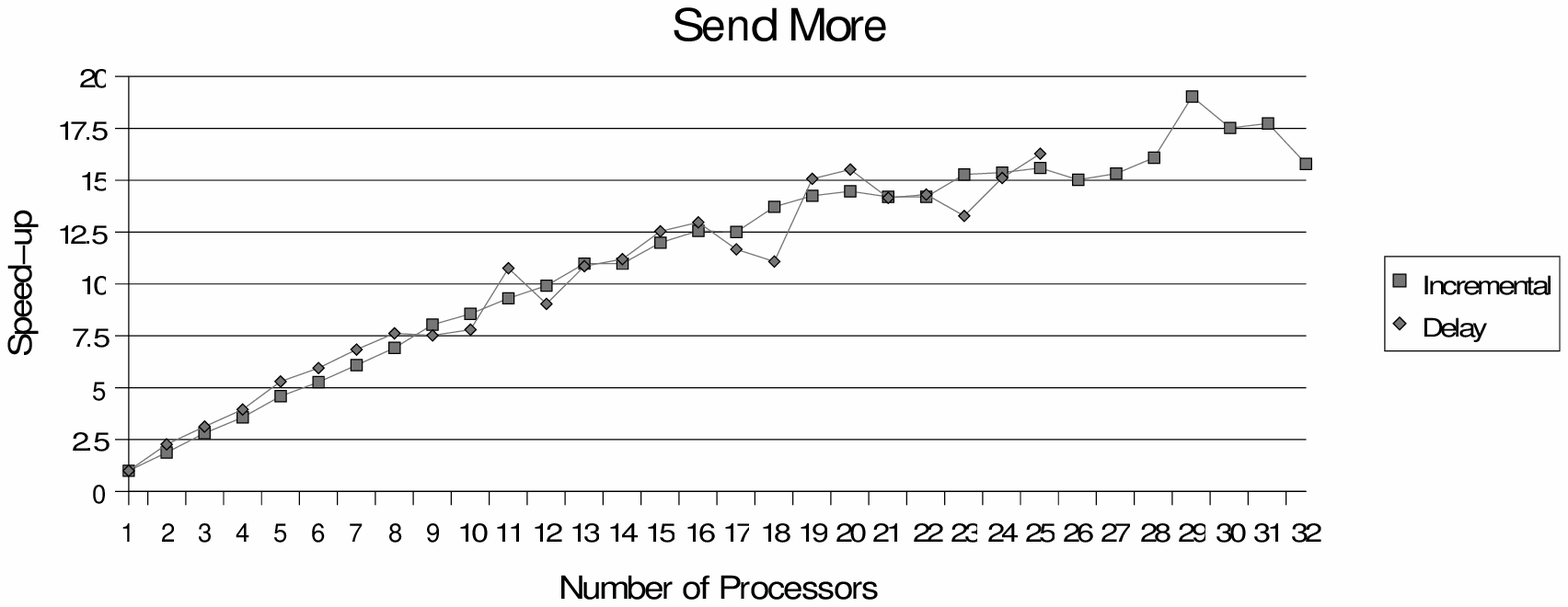,width=.53\textwidth}
\psfig{figure=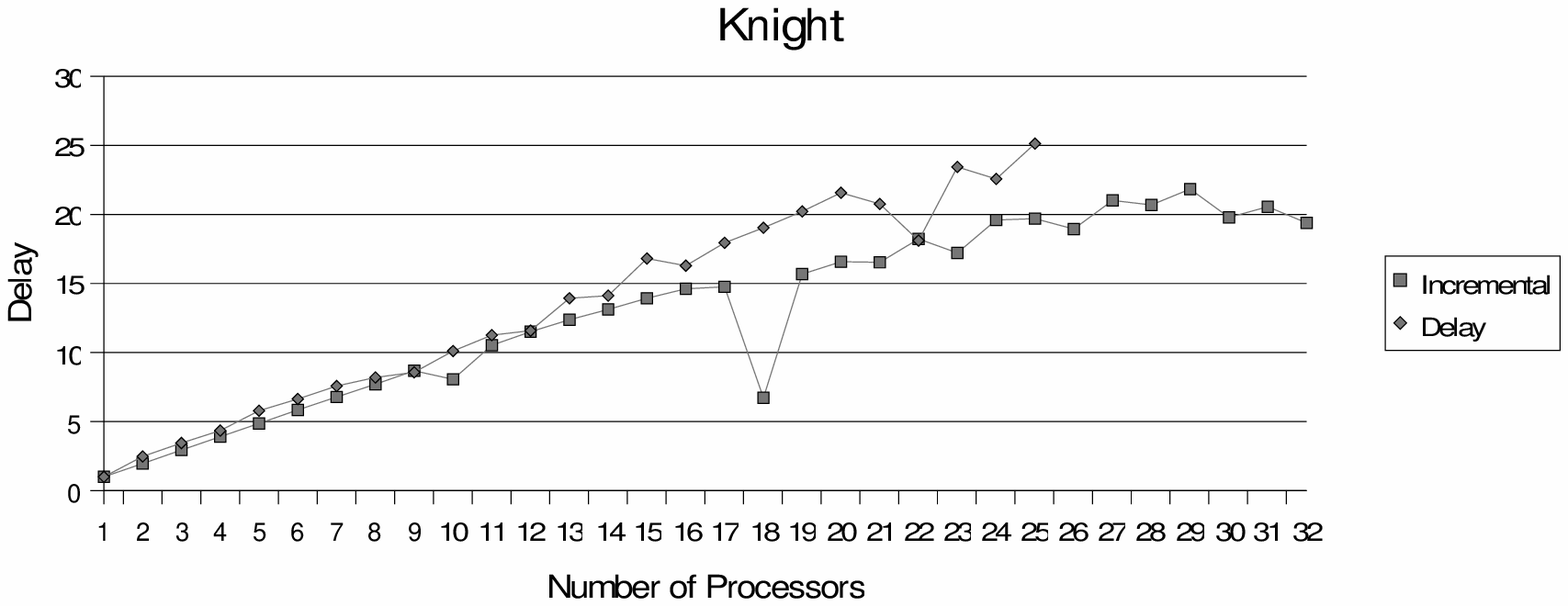,width=.53\textwidth}
\end{minipage}
\begin{minipage}[b]{\textwidth}
\psfig{figure=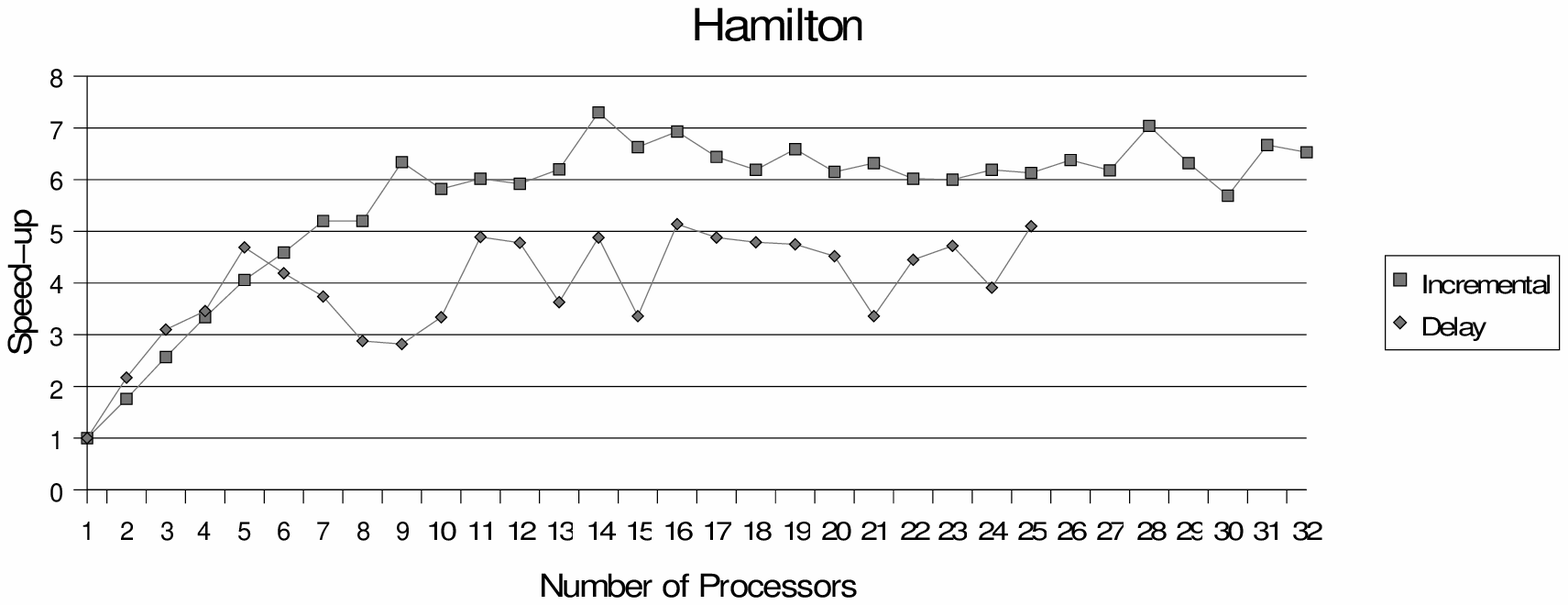,width=.53\textwidth}
\psfig{figure=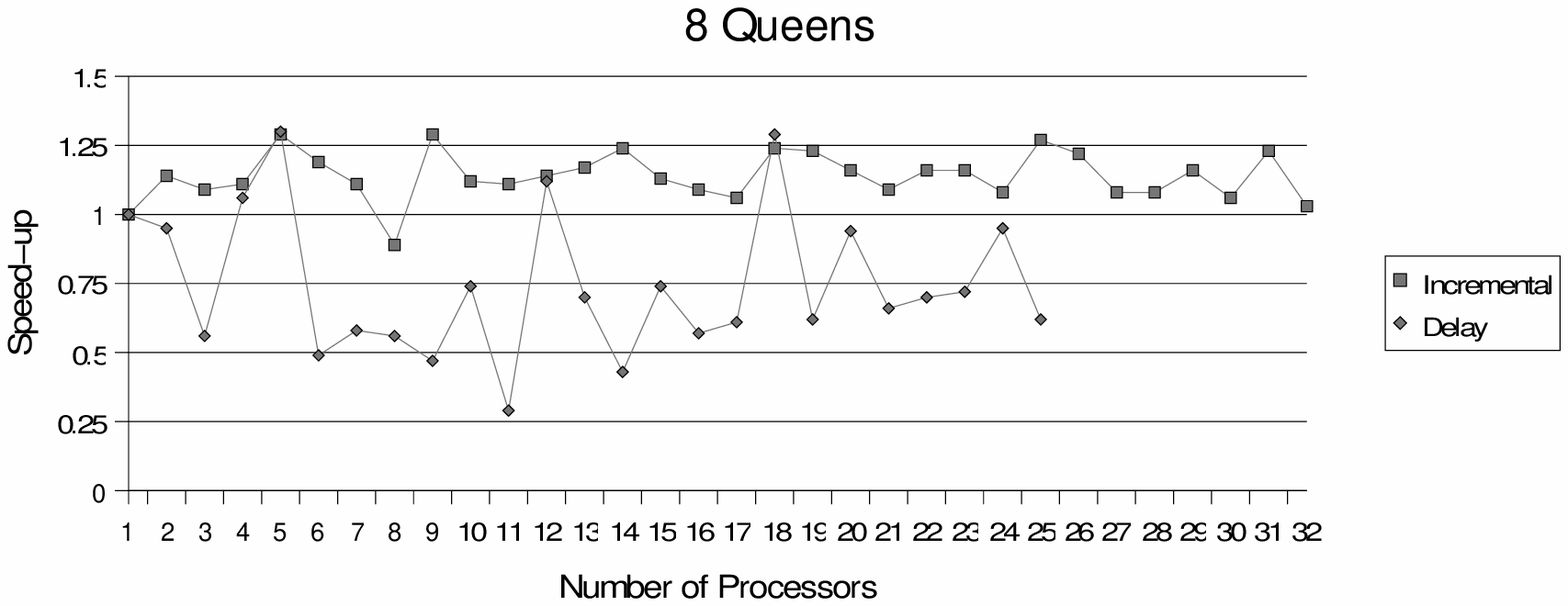,width=.53\textwidth}
\end{minipage}
\begin{minipage}[b]{\textwidth}
\psfig{figure=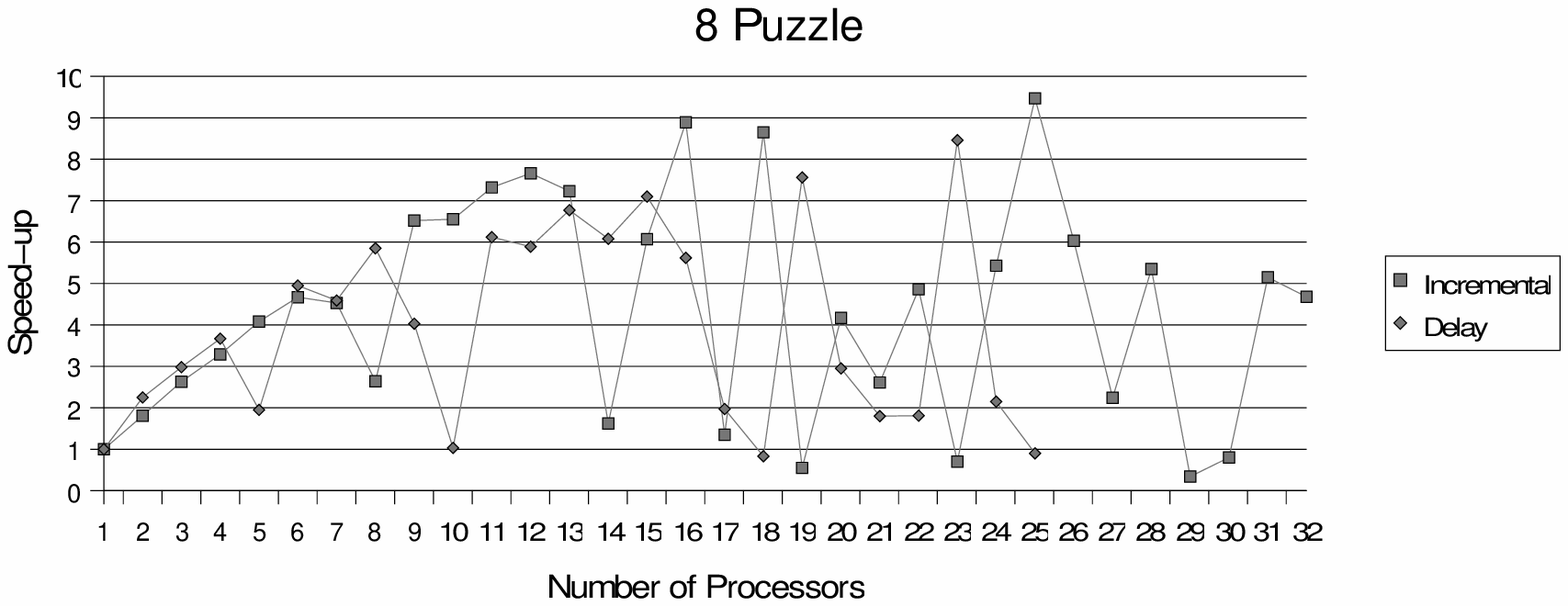,width=.53\textwidth}
\psfig{figure=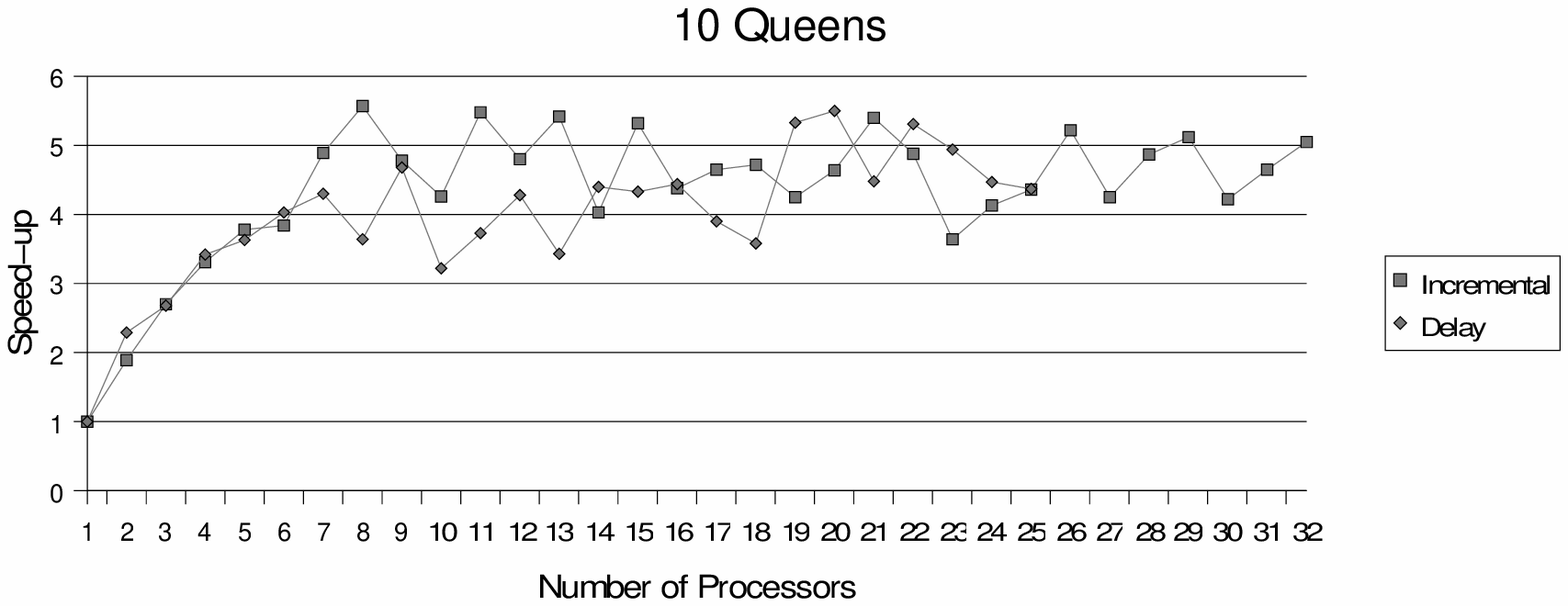,width=.53\textwidth}
\end{minipage}
\begin{minipage}[b]{\textwidth}
\psfig{figure=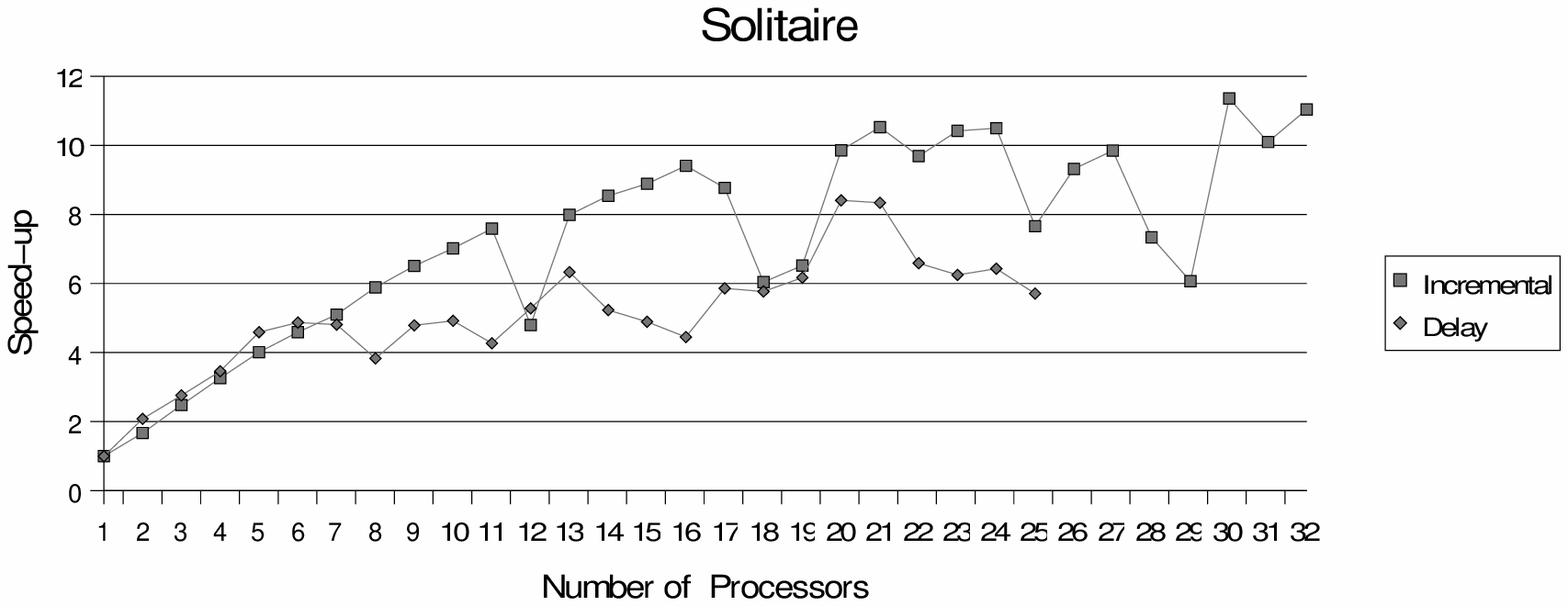,width=.53\textwidth}
\psfig{figure=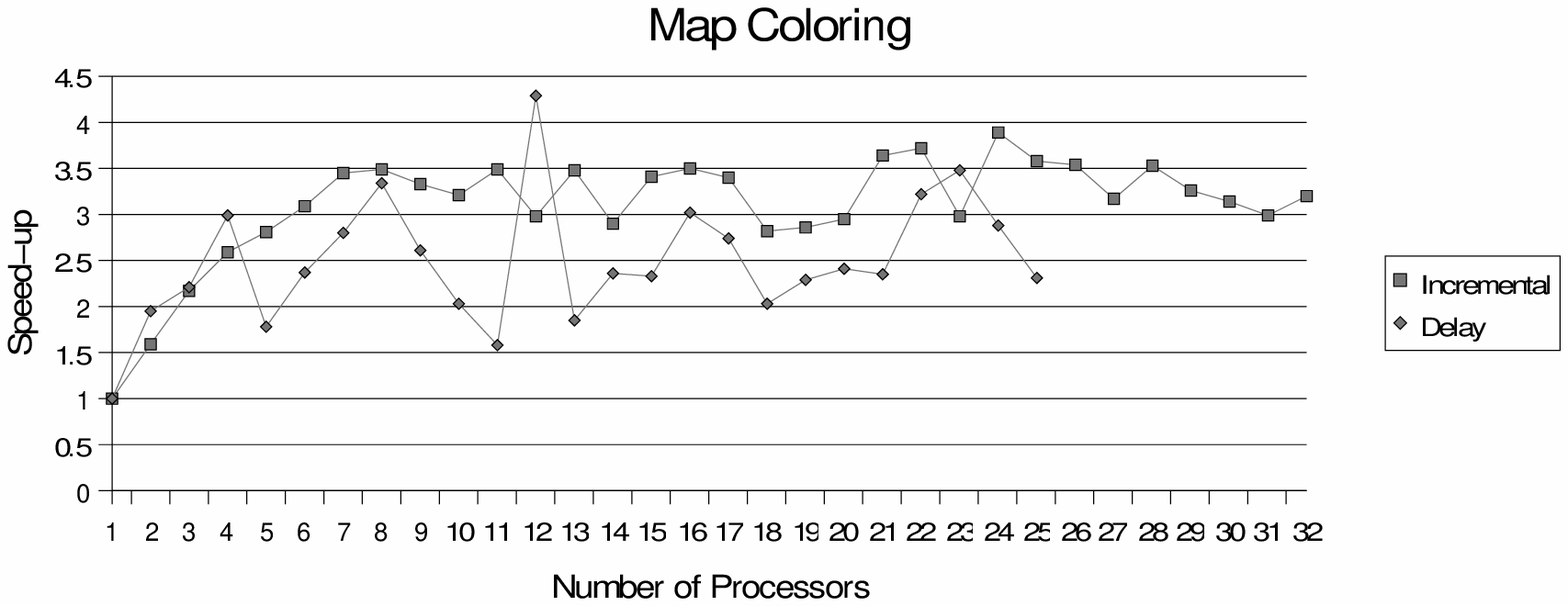,width=.53\textwidth}
\end{minipage}
\caption{Incremental Stack-Splitting vs. Delay Termination} 
\label{incre-delay}
\end{center}
\end{figure}

The last optimization that we tried concerns the check for termination of the
computation.
In our incremental stack-splitting system, once an agent finds that there
is no one to ask for work, it goes into a dead-end loop just waiting for the halt signal. 
Therefore, we modified our system to let an idle agent in this situation get out of 
this dead-end loop once it finds that its load vector has been updated so that it can go back
to life and ask for work.  We call this version \emph{delay termination}.
However, we still observed (see Figure~\ref{incre-delay}\footnote{Observe that some of the
experiments have been limited to smaller number of processors due to the
previously mentioned hardware problems.}) 
that our incremental stack-splitting 
system obtains higher speedups than using the delay termination version. 
This is probably due to the reason that, in most of these benchmarks, bringing 
the agents back leads to additional traffic of sharing requests, 
while actual work does not become available for sharing.
However, in general, we believe that the delay termination version ought to 
work better because it results in more agents participating in the computation.

\subsection{Order-sensitive Computations}

We implemented the techniques to handle \osc described 
in Section \ref{sideff}
in our PALS Prolog system and
tested it with the
 \emph{Stable}, \emph{Knight},
 \emph{9 Costas}, \emph{Hamilton},
 \emph{10 Queens}, and 
 \emph{Map Coloring} benchmarks.
These benchmarks compute all solutions and execute side-effect 
predicates, e.g., \emph{write} to describe the
 computations. The number of solutions for each benchmark
are reported in Table~\ref{bench}.

\begin{figure}[htb]
\begin{center}
\begin{minipage}[b]{.495\textwidth}
\begin{minipage}[b]{\textwidth}
\centerline{\psfig{figure=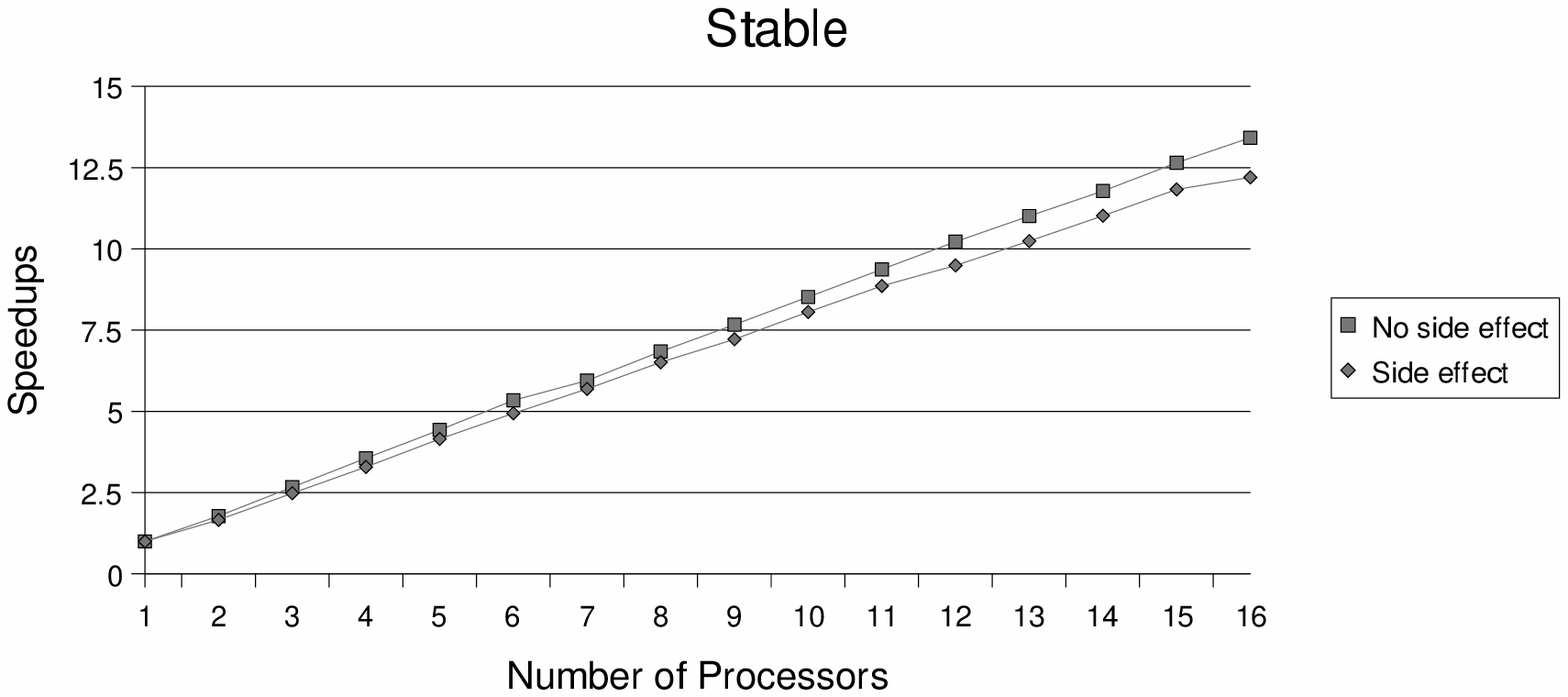,width=\textwidth}}
\end{minipage}
\begin{minipage}[b]{\textwidth}
\centerline{\psfig{figure=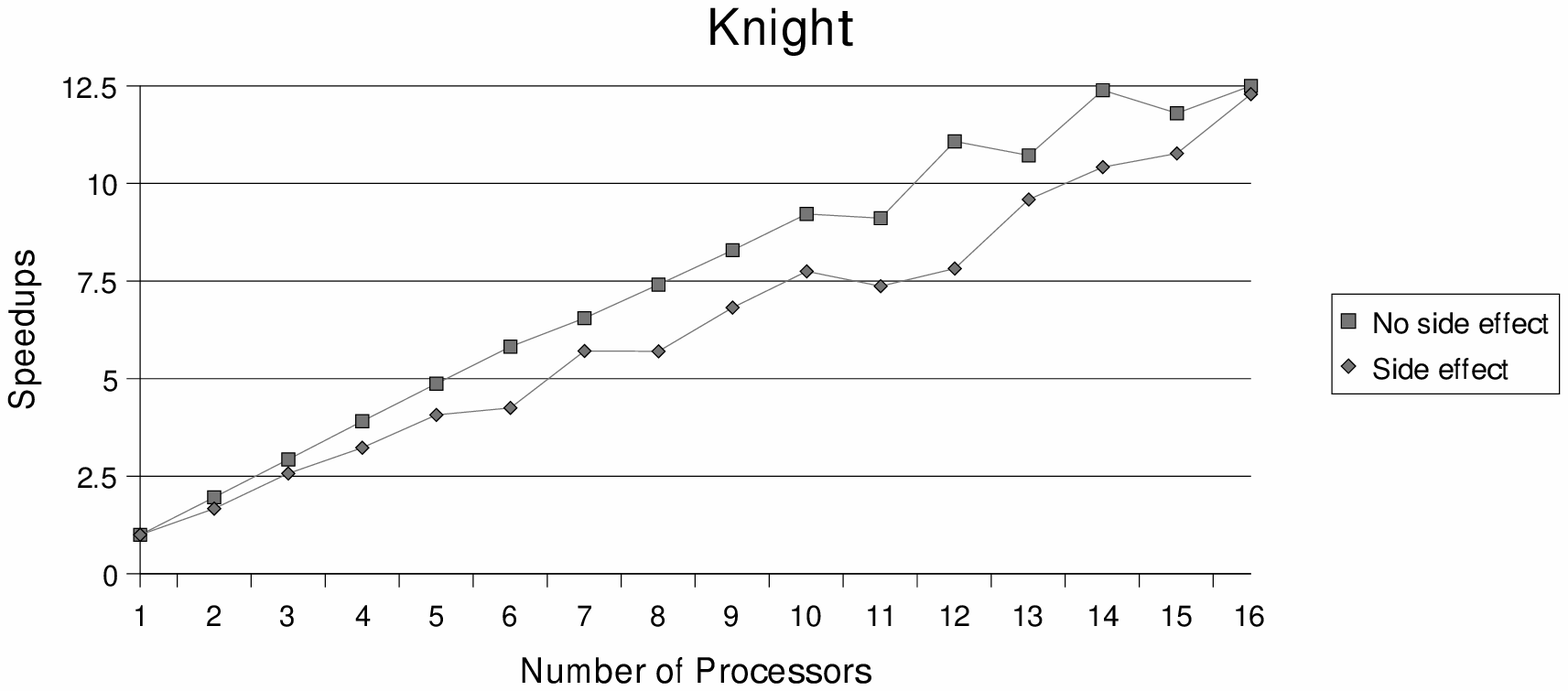,width=\textwidth}}
\end{minipage}
\begin{minipage}[b]{\textwidth}
\centerline{\psfig{figure=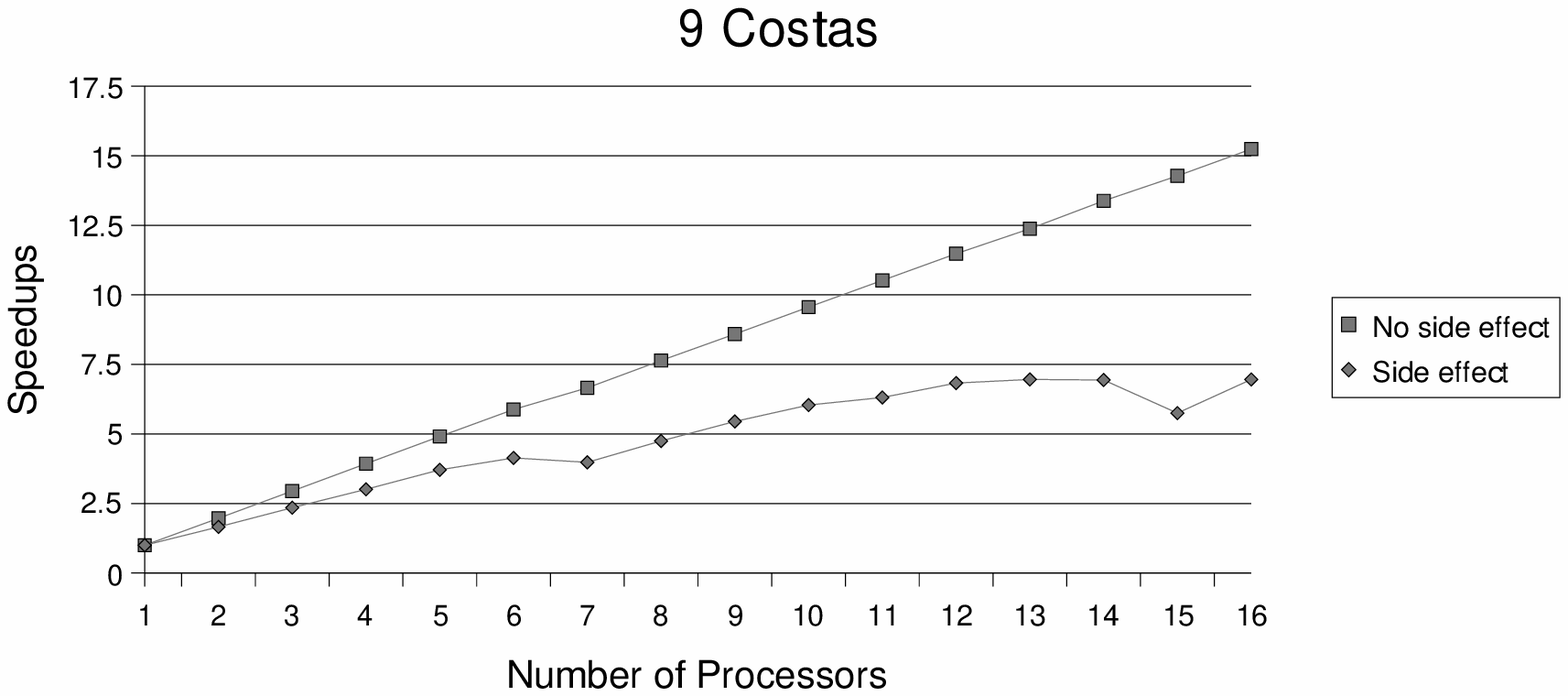,width=\textwidth}}
\end{minipage}
\end{minipage}
\begin{minipage}[b]{.495\textwidth}
\begin{minipage}[b]{\textwidth}
\centerline{\psfig{figure=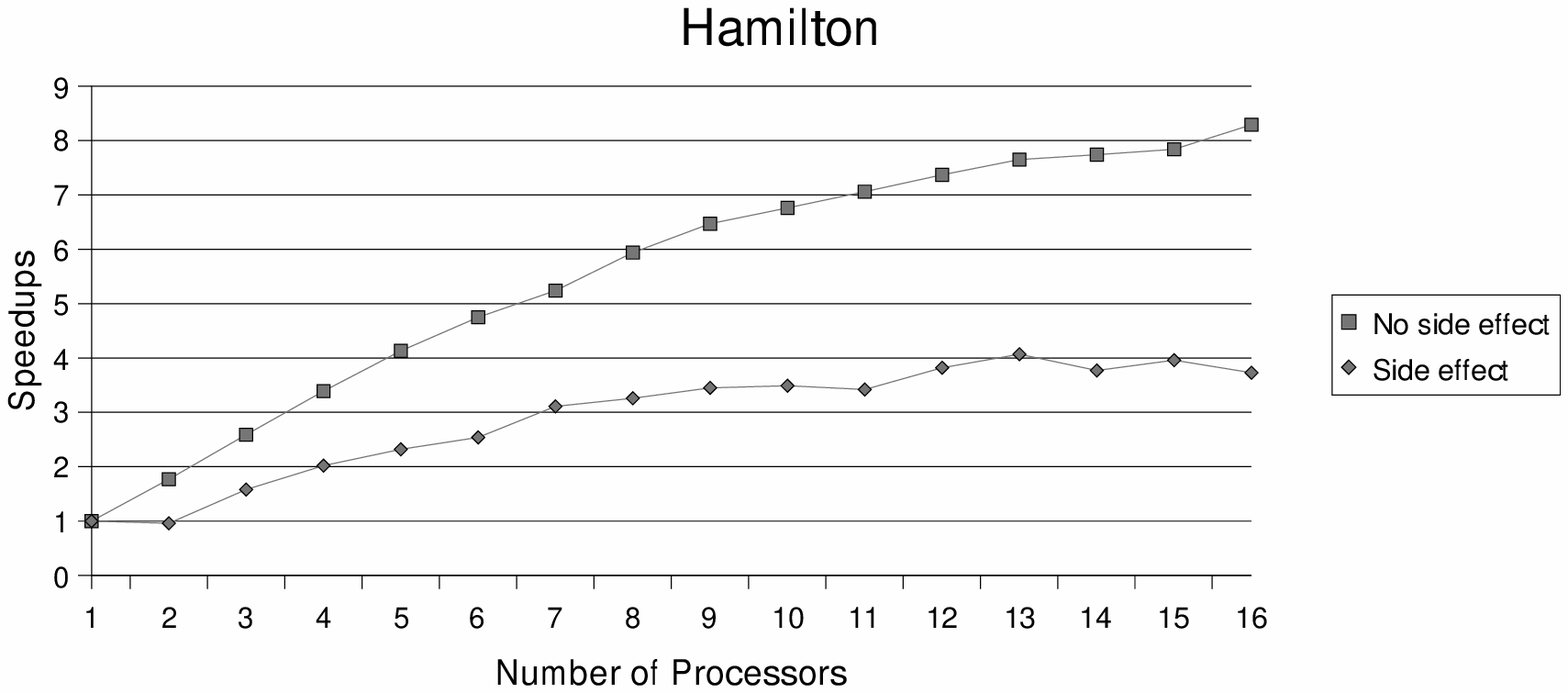,width=\textwidth}}
\end{minipage}
\begin{minipage}[b]{\textwidth}
\centerline{\psfig{figure=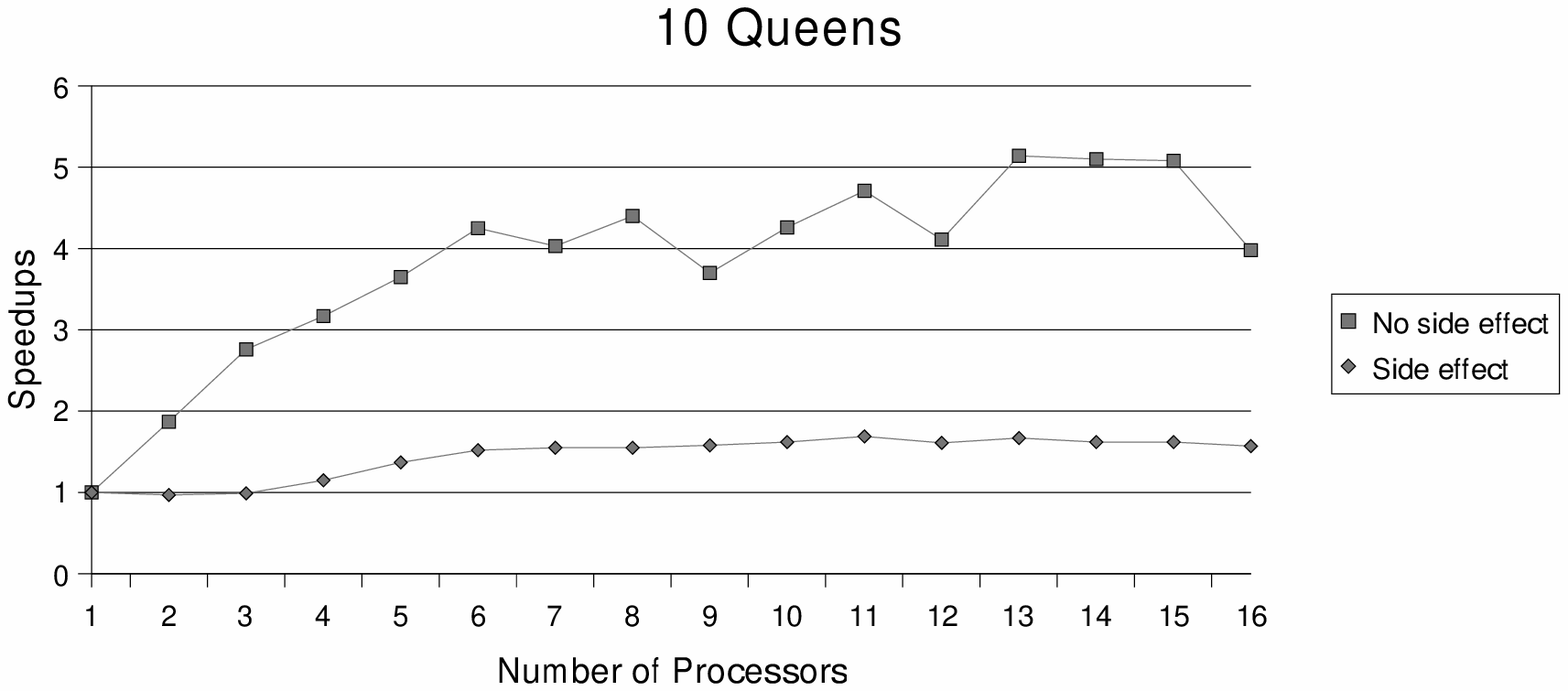,width=\textwidth}}
\end{minipage}
\begin{minipage}[b]{\textwidth}
\centerline{\psfig{figure=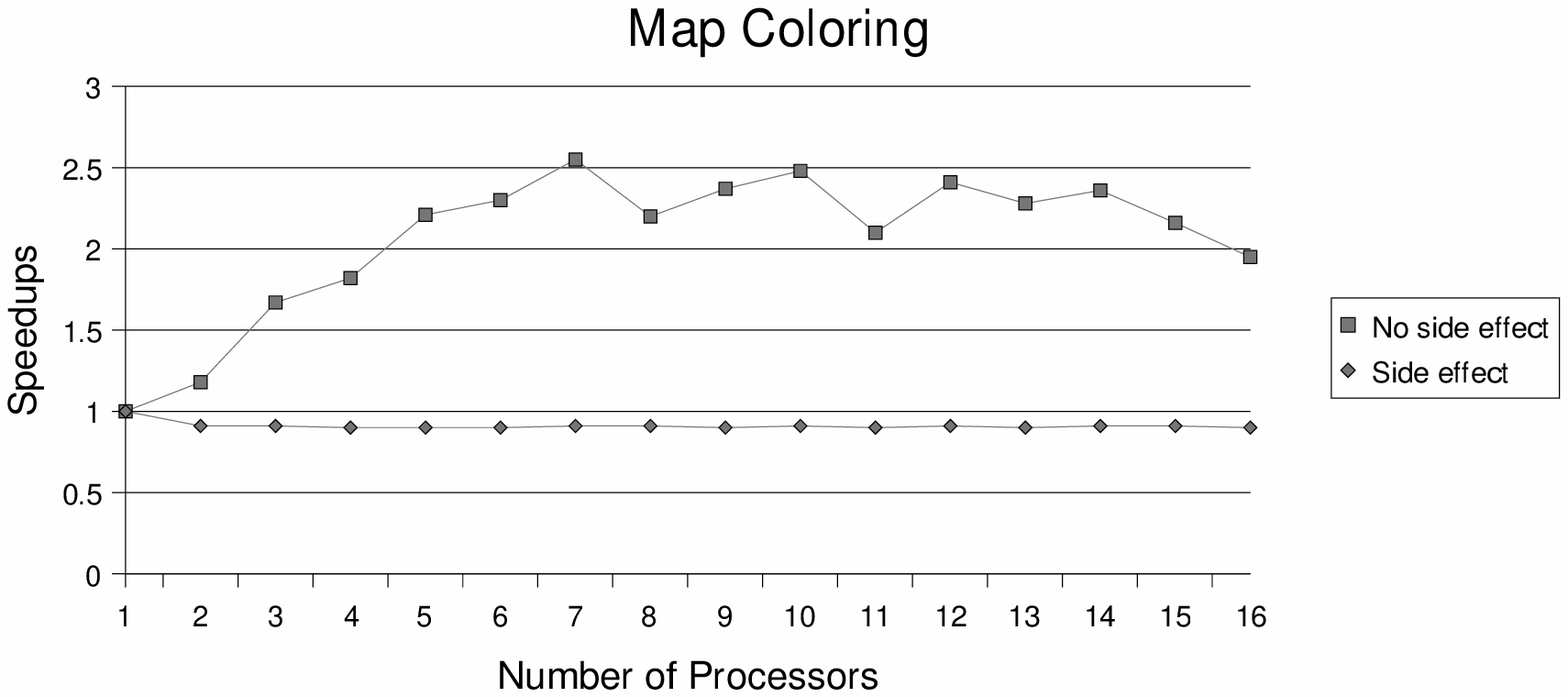,width=\textwidth}}
\end{minipage}
\end{minipage}
\caption{No Side-Effects vs. Side-Effects}
\label{side_effect2}
\end{center}
\end{figure}

Figure~\ref{side_effect2} shows the speedups obtained by 
this technique under the label \emph{side-effect}. 
The figures also show the speedups obtained when running these benchmarks without
treating the \emph{write} predicate as a side-effect but using the
stack splitting approach
described in Section~\ref{sidefex}.  
Two main observations arise from these experiments. First of all,
the speedups obtained using the modified scheduling scheme are not that different 
from those observed in our previous experiments \cite{karen-phd}; this means that 
the novel splitting strategy does not deteriorate the parallel performance of the system. 
For benchmarks with substantial running time and with the fewest number 
of printed solutions (\emph{Knight}, \emph{Stable}) the speedups are very good and 
close to the speedups
obtained without  handling  $\cal OSC$.
We also observe that for benchmarks with smaller
running time but larger number of side-effects (\emph{Hamilton})
the speedups
are still good but less close to the speedups obtained without side-effects.
Note that for benchmarks with small running time and the greatest number of 
printed solutions 
(\emph{Map Coloring}, \emph{10 Queens}), the speedups deteriorate significantly and 
may be less than $1$.  
This is not surprising; the presence of large numbers of
side-effects (proportional to the number of solutions) implies
the introduction of a large sequential component in the computation, leading to
reduced speedups.
\emph{9 Costas}  has the largest
number of solutions, but its speedups are  good.

\begin{table}[htb]
\begin{center}
\begin{tabular}{c|c|c}
\hline\hline
{\bf Benchmark} & {\bf Timings} & {\bf Number of solutions}\\
\hline\hline
\emph{9 Costas}        & 412.579 & 760 \\
\hline
\emph{Knight}          & 159.950 & 60 \\
\hline
\emph{Stable}          & 62.638  & 2 \\
\hline
\emph{10 Queens}       & 4.572 & 724 \\
\hline
\emph{Hamilton}        & 3.175 & 80 \\
\hline
\emph{Map Coloring}    & 1.113 & 2594 \\
\hline\hline
\end{tabular}
\end{center}
\caption{Benchmarks (Time in sec.)}
\label{bench}
\end{table}

\begin{figure}[htb]
\begin{center}
\begin{minipage}[c]{.42\textwidth}
\centerline{\psfig{figure=side/data/compa.eps,width=\textwidth}}
\end{minipage}
\hspace{.01\textwidth}
\begin{minipage}[c]{.42\textwidth}
\centerline{\psfig{figure=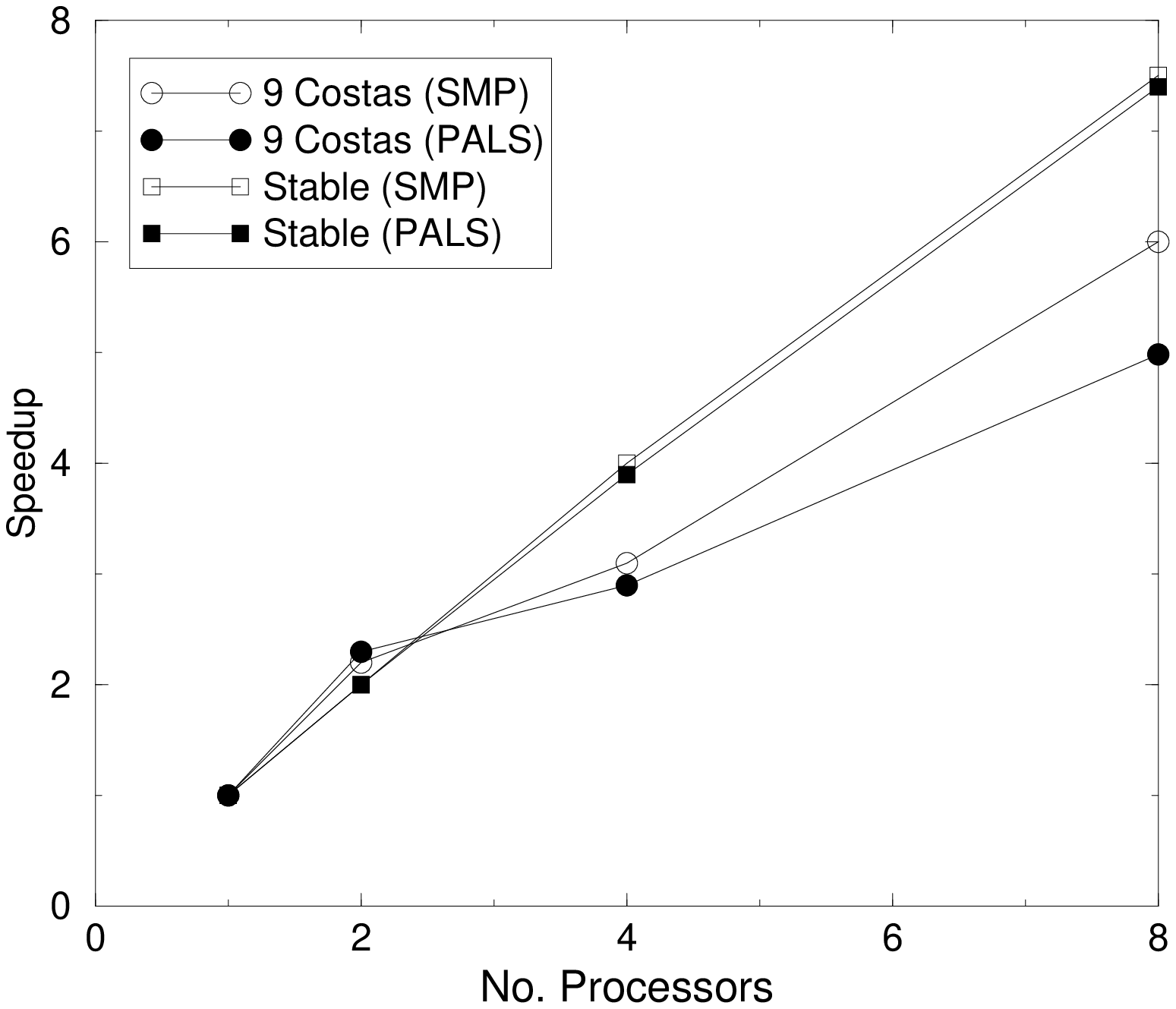,width=\textwidth}}
\end{minipage}
\end{center}
\caption{Comparison with MUSE}
\label{comps}
\end{figure}

The results obtained are consistent with our belief that DMP implementations
should be used for programs with coarse-grained parallelism and
a modest number of $\cal OSC$. Coarse-grained computations
are even more important if we want to handle large numbers of 
side-effects where it is 
necessary that the \osc be spaced far apart.
For programs with small-running times there is not enough work to offset the cost 
of exploiting parallelism and even less for handling  
$\cal OSC$.
Nevertheless, our system is reasonably efficient given that it produces good speedups
for large and medium size benchmarks with even a considerable number of 
$\cal OSC$, and
 produces no slow downs except for benchmarks with 
huge numbers of side-effects and
small running times. Even in presence of 
$\cal OSC$,
the parallel overhead observed is substantially low---on average $5.5\%$ and seldomly
over $10\%$ (it is slightly higher than what described in the previous sections, due
to some additional tests required for checking presence of messages related to 
$\cal OSC$).
Figure~\ref{comps} compares with the speedups for some benchmarks
obtained using a variant of the
MUSE system \cite{muse-journal} on  SMP (i.e., a highly optimized  stack-copying
system on shared-memory platform).\footnote{This is the original version of MUSE
with bottom-most scheduling and no suspensions, modified from version 14.07 of
MUSE.} The
results highlight the fact that, for benchmarks with significant running time, our
methodology is capable of approximating the best behavior on SMPs.

\section{Optimizations and Discussion}
\label{costs}

In this section we discuss some limitations of the 
current stack-splitting scheme and some possible
optimizations.

\subsection{Shared Frames and Distributed Computations}
The adoption of stack-splitting releases the system
from the need of keeping shared frames to support 
sharing of work.
The shared frame used in the stack-copying technique on
shared-memory platforms
is also where global information
related to scheduling is kept. The shared 
frames provide a globally accessible description of the
or-tree, and each shared frame keeps information regarding
which agent is working in which part of the tree. This 
last piece of information is needed
to support the kind of scheduling typically used in 
stack-copying systems---work is taken from the agent
that is ``closer'' in the computation tree, thus reducing
the amount of information to be copied---since the 
difference between
the stacks is minimized. The shared
nature of the frames ensures accessibility of
this information to all agents, providing
a consistent picture of the computation. 

However, under stack-splitting the shared
frames no longer exist;  scheduling and work-load information
has to be maintained in some other way. While we have
already described how to maintain work-load information in 
a distributed setting, through the use of work-load vectors,
we did not discuss how to provide agents with knowledge of their
relative positions in the computation tree.
This type of information could
be kept in a global shared area similar to the 
case of SMPs---e.g., by building a centralized representation 
of the or-tree---or distributed over multiple agents and accessed by 
message
passing in case of DMPs.
The maintenance of global scheduling information represents
a problem which is orthogonal to the environment representation.
This means that scheduling  management in a DMP
 will anyway require communication between agents.
%Nevertheless,  the
%use of stack-splitting allows scheduling-on-bottom-most and
%is expected to reduce the amount of scheduling communication needed.

Shared frames are also employed in MUSE \cite{muse-journal} 
to detect the Prolog order
of choice-points, needed to execute
order-sensitive  predicates (e.g., side-effects, extra-logical
predicates) in the correct order. As in the case of scheduling,
some information regarding
global ordering of choice-points needs to be maintained
to execute order-sensitive
predicates in the correct order---see Section \ref{sideff}.
Thus, stack-splitting does not completely remove the need of
a shared description of the or-tree. The
use of stack-splitting can mitigate the impact of accessing
shared resources---e.g., stack-splitting allows 
scheduling on bottom-most which, in general,  leads to a reduction
of the number of calls to the scheduler.

\subsection{The Cost of Stack-Splitting}

The stack-copying operation in  Stack-Splitting
is slightly more involved than in 
stack-copying on shared-memory platforms. In MUSE,
the original choice-point stack
is traversed 
and the choice-points transferred to the shared area. This
operation involves only those choice-points that have never
been shared before---shared choice-points already
reside in the global shared area. For this
reason the actual \emph{sharing} of the choice-points is performed
by the \emph{active-agent} (i.e., the agent that is 
providing work to the idle agent)---which is  forced to interrupt its
regular computation to assist the sharing process. 
The
actual copying of the stack 
takes place only after the choice-points have been copied
to the shared memory area.

In the stack-splitting technique, once the copying is 
completed, the actual sharing (i.e., transferring of 
choice-points to a shared area) is replaced 
by a
phase of splitting, performed by both  agents, where
they traverse 
the copied choice-points, completing the splitting of the untried 
alternatives.
In the case of SMP implementations, this operation
is expected to be considerably cheaper than transferring the
choice-points to the shared area---and indeed our experimental
studies have highlighted this by denoting improved performance
of stack-copying on SMPs.
The actual splitting can
be represented by a simple pair of indices that refer to the list
of alternatives---which, in a SMP system like MUSE, is static and 
shared by all the agents.
In the case of DMP implementations, the situation
is similar: since each agent maintains a local copy of the code,
the splitting can be performed by communicating to the copying agent
which alternatives it can execute for each choice-point (e.g., 
a pair of pointers to the list of alternatives). It is
simple to encode such information within the choice-point itself
during copying.

In both cases we expect the sharing operation to have comparable
complexity; a slight delay may occur in  stack-splitting, 
due to the traversal of the choice-point stack performed by each
agent. On the other hand, in stack-splitting
the two traversals---one in the \emph{idle-agent}
and one in the \emph{active-agent}---can be
overlapped.
%, while in the MUSE scheme the \emph{idle-agent}
%is suspended until the 
%\emph{active-agent} has completed the sharing
%operation.  
However, 
if the stack being copied, S$_o$, is itself
a copy of some other stack, then unlike regular
stack-copying (where once a choice-point is shared---i.e., moved
to a shared area---it will not have to be shared again), we may still
need to traverse both the source and target
stacks and split the choice-points (even those that have been acquired
through previous sharing operations). The presence of this
additional step depends on the policy adopted for the partitioning
of the alternatives between agents. It is, for example, required
if we  adopt a policy which assigns half of the alternatives
to each of the agents.	
In such cases,
the cost of sharing  will be slightly more than the cost of
regular stack-copying.

Once an agent selects new work,
it will look for work again only after it finishes
the exploration of all alternatives
acquired via stack-splitting.

\subsection{From Vertical to Horizontal Splitting}
As we mentioned earlier, different splitting modalities
can be envisioned, e.g., horizontal vs. vertical splitting.
Horizontal splitting, which is useful for programs having
choice-points with many alternatives, incurs a linear cost due to
the need of traversing a linear list of alternatives (provided
by the WAM representation of procedures) to perform the partition. 
The cost incurred in splitting the untried
alternatives between the copied stack and the
stack from which the copy is made, can be 
eliminated by amortizing it over the 
operation of picking untried alternatives.
 Let us assume that
the untried alternatives are evenly split using
horizontal splitting (as in Figure \ref{stacksplit}).

\begin{figure}[htb]
\centerline{\psfig{figure=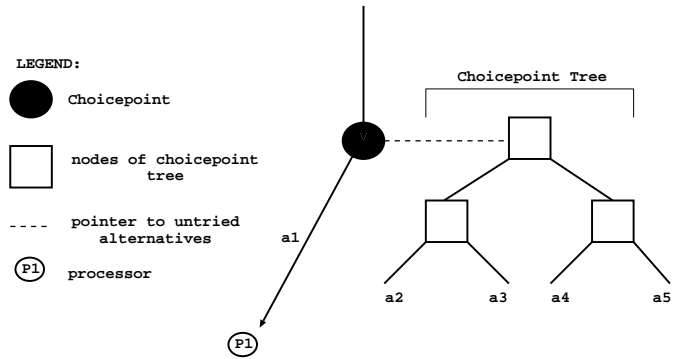,width=0.70\textwidth}}
\caption{Amortizing Splitting Overhead}
\label{bintree}
\end{figure}

In the modified approach,
no traversal and modification of the choice-points
is done during copying. The untried
alternatives are organized as a binary tree
 (see Figure \ref{bintree}). 
The binary
alternatives can be efficiently maintained in
an array, using standard techniques found
in any data-structure textbook.  In addition, each
choice-point maintains the ``copying distance'' from
the very first original choice-point as a bit
string. This number is initially 0 when the 
computation begins. When stack-splitting
takes place and a choice-point whose bit string is
$n$ is copied from, then the new choice-point's bit string
is $n1$ ($1$ appended to the bit string $n$), while the old
choice-point's bit string is changed to $n0$ ($0$ tagged
to bit string $n$). When an agent backtracks 
to a choice-point, it will use its bit string
to navigate in the tree of untried
alternatives, and find the alternatives that it
is responsible for. For example, if the bit-string
of an agent is 10, then all the
alternatives in the left subtree of the right subtree
of the or-tree are to be executed by that agent.
This scheme (originally proposed in~\cite{iclp99}) has
been introduced as part of the YapDss implementation 
\cite{fernando}.

However, it is not very clear which of the two 
strategies---incurring cost of splitting at copying time 
{\it vs} amortizing the cost over the selection of
untried alternatives---would be 
more efficient. In case of amortization,
the cost of picking an alternative from a choice-point is
now slightly higher, as the binary tree of choice-points
needs to be traversed
to find the right alternative.

\bigskip

Stack-splitting essentially performs semi-dynamic
work distribution, as the untried alternatives are
split at the time of picking work. If the choice-points
that are split are balanced, then we can expect good
performance. Thus, we should expect to see good performance
when the choice-points generated by the computation that are
parallelized contain a large number of alternatives. This is
the case for applications which fetch data from databases and for
most generate \& test type of applications.

For choice-points with a small number of alternatives,  
stack-splitting  is more susceptible to problems
created by the semi-dynamic
work distribution strategy that implicitly results from
it: for example,
in cases where OP is extracted from choice-points 
with only two
alternatives. Such choice-points arise quite frequently,
from the use of predicates like
{\sf member} and {\sf select}:

\bigskip

\begin{minipage}[b]{.47\textwidth}
\noindent{\sf member(X,[X$\mid$\_]).}

\noindent{\sf member(X,[\_$\mid$Y]) :- member(X,Y).}
\end{minipage}
\begin{minipage}[b]{.55\textwidth}
\noindent{\sf select(X,[X$\mid$Y],Y).}

\noindent{\sf select(X,[Y$\mid$Z],[Y$\mid$R]) :- select(X,Z,R).}
\end{minipage}

\bigskip

Both these predicates generate choice-points with only two
alternatives---thus, at the time of sharing, a single alternative
is available in each choice-point. The different alternatives
are spread across different choice-points. Stack-splitting would 
assign all the alternatives to the copying agent, thus 
leaving the original agent without local work. However, 
the problems raised by such situations can be solved using
a number of techniques:
\begin{itemize}
\item Use knowledge about the inputs and
	partial evaluation, or automatic optimizations
	(e.g., \emph{Last Alternative Optimization (LAO)} \cite{lao-spdp})
	 to collapse
	 the different choice-points into a single
	one.
\item  Use more complex splitting strategies, e.g.,
if a choice-point has odd number of untried alternatives
remaining ($2n+1$), then one agent will be
assigned $n$ alternatives and the other $n+1$. The agent
which gets $n$ and the agent which gets $n+1$ can
be alternated for the different choice-points encountered in the
stack, thus ensuring that no processor is left completely without
work.
\item  Perform a {\it vertical} splitting of the choice-points;
\end{itemize}
Additionally, observe that the splitting strategy adopted (e.g.,
horizontal splitting, vertical splitting) can be changed depending
on the specific structure of the computation. For example,
along these lines Rocha et al. \cite{fernando} have recently
proposed a splitting strategy---\emph{diagonal splitting}---that
combines vertical and horizontal splitting and performs 
well for certain classes of benchmarks.

\section{Conclusion}
\label{concl}

In this paper, we presented a technique called stack-splitting
for implementing OP and discussed its advantages
and disadvantages. We showed how stack-splitting can be extended
to incremental stack-splitting which incrementally copies the
difference of two stacks. Implementations on both a shared memory
multiprocessor and a distributed-memory multiprocessor were realized
and reported. Our DMP implementation is the first ever implementation
of a Prolog system on a Beowulf architecture.

Stack-splitting is an extension of
stack-copying. Its
main advantage, compared to
other  techniques for implementing OP, is that
it allows large grain-sized work to be picked up by idle agents
and executed efficiently without incurring excessive
communication overhead.
The technique bears some similarity to the Delphi
model \cite{delphi} used in parallel execution of Prolog
 (the Delphi model was not the inspiration for our 
stack-splitting technique),
where computation leading to a 
goal with multiple alternatives
is replicated in multiple agents, and each agent
chooses a different alternative when that goal is
reached. Instead
of recomputing we use stack-copying, which, we believe,
is more efficient---and the existing literature has
indicated this is the case for shared-memory implementations
of Prolog \cite{parallellp-survey}. In a separate work
\cite{parco}, we also showed how 
stack-splitting can be used for implementing 
non-monotonic reasoning
systems under stable models semantics---by exploiting or-parallelism
from a careful implementation of the Davis-Putnam procedure and
using stack-splitting to transfer atom-split operations between
processors. Also in this case,
copying with stack-splitting
provides a superior performance than recomputation.

The current implementation of stack-splitting in the PALS
system is stable, and work is in progress to evaluate its
performance on larger applications. A number of issues
are still open, and they will be addressed as future work. First of
all, it is clear from our experience that the giving the ability
to the programmer to supply information about the program can
greatly affect parallel performance; we are currently working
in developing tools to analyze parallel executions of PALS (e.g.,
through visualization of the parallel computation) and support
user-annotations to guide exploitation of parallelism. 
%guo
Work is also in progress in supporting order-sensitive 
control predicates (e.g., pruning predicates) in  PALS, and
developing adaptive scheduling heuristics, 
which take advantage of knowledge of the structure of the computation
to improve distribution of work,

\section*{Acknowledgments}

Thanks to C. Geyer, L. Castro, V. Santos Costa, F. Silva, M. Carro, and M.
Hermenegildo for discussions on implementation of distributed LP
systems. 

The authors wish to thank the anonymous referees for their thorough reviews
and their insightful comments.

Pontelli and Villaverde have been been supported by NSF grants 
CNS-0220590, CNS-0454066, and HRD-0420407, Guo by 
NSF Nebraska EPSCoR grant, and Gupta by NSF grant 
CNS-0130847 and grants from the US Environmental 
Protection Agency.

%The authors thank P. Magnusson and Virtutech for
%providing the SimICS simulator.

%ipps: Manuel's tool for detecting par. in fortran prog.
%inductive lp apps are perhaps a good source of or-parallelism

\small
\bibliographystyle{acmtrans}

\end{document}